\documentclass[usenatbib]{mnras}
\usepackage{graphicx}
\usepackage{txfonts}
\usepackage{rotating}
\usepackage{accents}
\usepackage{float}
\usepackage{textcomp}
\usepackage{upgreek}
\usepackage{epstopdf}
\usepackage{color}
\usepackage{pdflscape}
\usepackage{appendix}

\begin{document}

\title[MEGARA-GTC Stellar Spectral Library (I)]
{MEGARA-GTC Stellar Spectral Library (I)}

\author[Garc{\'\i}a-Vargas et al. ]
{M.L. Garc{\'\i}a-Vargas$^{1}$\thanks{E-mail:}, E. Carrasco$^{2}$, M. Moll\'{a}$^{3}$, A. Gil de Paz$^{4, 5}$, S. R. Berlanas$^{6, 7}$, 
\newauthor
N. Cardiel $^{4, 5}$, P. G\'{o}mez-Alvarez$^{1}$, J. Gallego $^{4, 5}$, J. Iglesias-P\'{a}ramo$^{8, 9}$, R. Cedazo$^{10}$,
\newauthor
 S. Pascual $^{4, 5}$, A. Castillo-Morales $^{4, 5}$,  A. P\'{e}rez-Calpena $^{1}$, I. Mart{\'\i}nez-Delgado$^{1}$
\\
$^{1}$ FRACTAL S.L.N.E., Calle Tulip{\'a}n 2, portal 13, 1A, E-28231 Las Rozas de Madrid, Spain \\
$^{2}$ Instituto Nacional de Astrof{\'\i}sica, {\'O}ptica y Electr{\'o}nica, INAOE, Calle Luis Enrique Erro 1, C.P. 72840 Santa Mar{\'\i}a Tonantzintla, Puebla, Mexico\\
$^{3}$ Dpto. de Investigaci\'{o}n B\'{a}sica, CIEMAT, Avda. Complutense 40, E-28040 Madrid, Spain\\
$^{4}$ Dpto. de F{\'\i}sica de la Tierra y Astrof{\'\i}sica, Fac. CC. F{\'\i}sicas, Universidad Complutense de Madrid, Plaza de las Ciencias, 1, E-28040 Madrid, Spain \\
$^{5}$ Instituto de F{\'\i}sica de Part{\'\i}culas y del Cosmos, IPARCOS, Fac. CC. F{\'\i}sicas, Universidad Complutense de Madrid, Plaza de las Ciencias 1, E-28040 Madrid, Spain\\
$^{6}$ Instituto de Astrof{\'\i}sica de Canarias, Calle V{\'\i}a L\'{a}ctea s/n, E-38205 San Crist{\'o}bal de la Laguna, Santa Cruz de Tenerife, Spain\\
$^{7}$ Dpto. de Astrof{\'\i}sica, Universidad de La Laguna, E-38205 San Crist{\'o}bal de la Laguna, Santa Cruz de Tenerife, Spain\\
$^{8}$ Instituto de Astrof{\'\i}sica de Andaluc{\'\i}a, IAA-CSIC, Glorieta de la Astronom{\'\i}a s/n, E-18008  Granada, Spain\\
$^{9}$ Estaci{\'o}n Experimental de Zonas {\'A}ridas, CSIC, Carretera de Sacramento s/n,  E-04120  Almer{\'\i}a, Spain\\
$^{10}$ Dpto. de Ingenier{\'\i}a El{\'e}ctrica, Electr{\'o}nica, Autom{\'a}tica y F{\'\i}sica Aplicada, ETS de Ingenier{\'\i}a y Dise{\~n}o Industrial, UPM, Ronda de Valencia 3, E-28012 Madrid, Spain}

\date{Accepted Received ; in original form }
\pagerange{\pageref{firstpage}--\pageref{lastpage}} \pubyear{2019}

\maketitle
\label{firstpage}

\begin{abstract}
MEGARA (Multi Espectr{\'o}grafo en GTC de Alta Resoluci{\'o}n para Astronom{\'\i}a) is an optical (3650~--~9750\AA), fibre-fed, medium-high spectral resolution (R = 6000, 12000, 20000) instrument for the GTC 10.4m telescope, commissioned in the summer of 2017, and currently in operation. The scientific exploitation of MEGARA demands a stellar-spectra library to interpret galaxy data and to estimate the contribution of the stellar populations. This paper introduces the MEGARA-GTC spectral library, detailing the rationale behind the catalogue building. We present the spectra of 97 stars (21 individual stars and 56 members of the globular cluster M15, being both sub-samples taken during the commissioning runs; and 20 stars from our on-going GTC Open-Time program). The spectra have R~=~20000 in the HR-R and HR-I setups, centred at 6563 and 8633~\AA\ respectively. We describe the procedures to reduce and analyse the data. Then, we determine the best-fitting theoretical models to each spectrum through a $\chi^{2}$ minimisation technique to derive the stellar physical parameters and discuss the results. We have also measured some absorption lines and indices. Finally, this article introduces our project to complete the library and the database to make the spectra available to the community.
\end{abstract}


\section{Introduction}
\subsection{MEGARA: the new mid-resolution fibre-fed multi-object spectrograph for the GTC}\label{megintro}

MEGARA (Multi Espectr{\'o}grafo en GTC de Alta Resoluci{\'o}n para Astronom{\'\i}a) is an optical fibre-fed spectrograph for the Gran Telescopio CANARIAS, hereafter GTC (\url{http://www.gtc.iac.es}), a 10.4m telescope in La Palma (Canary Islands, Spain). The main instrument characteristics are summarised in Table~\ref{mainchar}. 

The instrument offers two spectroscopy modes: bi-dimensional, with an Integral Field Unit (IFU); and Multi-Object (MOS). The IFU, named as Large Compact Bundle (LCB), provides a Field Of View (FOV) of 12.5$\arcsec~\times$~11.3$\arcsec$, plus eight additional minibundles, located at the edge of the FOV, to provide simultaneous sky subtraction. 
The MOS assembly can place up to 92 mini-bundles covering a target area per minibundle of 1.6$\arcsec$ and a total coverage area on sky of 3.5$\arcmin$~$\times$~3.5$\arcmin$. The spatial sampling in both modes is 0.62$\arcsec$ per fibre\footnote{This size corresponds to the diameter of the circle on which the hexagonal spaxel is inscribed}, being each spaxel projected thanks to the combination of a 100~$\upmu$m core fibre coupled to a microlens that converts the f/17 entrance telescope beam to f/3, for maximum efficiency. 
The spaxel projection is over sized relative to the fibre core to allow for a precise fibre-to-fibre flux uniformity. A fibre link 44.5m long drives the science light, coming  from the Folded Cassegrain F focal plane, into the spectrograph, hosted at the Nasmyth A platform.

The spectrograph is an all-refractive system (with f/3 and f/1.5 focal ratios for collimator and camera,  respectively) and includes a set of 18 Volume Phase Holographic (VPH) gratings placed at the pupil in the collimated beam. These gratings offer three spectral modes with different Resolving power, R, labelled as Low Resolution (LR), R(FWHM)~$=$~6000; Medium Resolution (MR), R(FWHM)~$=$~12000; and High Resolution (HR), R(FWHM)~$=$~20000. 

The different MEGARA setups in terms of wavelength coverage and linear reciprocal dispersion are shown in Table~\ref{setups}. The columns indicate: (1) setup or name of the VPH; (2) the shortest wavelength (\AA) of the central fibre spectrum; (3) the shortest  wavelength (\AA) common to all spectra; (4) the central wavelength (\AA); (5) the longest wavelength (\AA) for all spectra; (6) the longest wavelength (\AA) of the central fibre spectrum and (7) constant linear reciprocal dispersion (\AA\,pix$^{-1}$). The range between $\lambda_{min}$ and $\lambda_{max}$ is common to all fibres while the wavelengths $\lambda_{min,1}$ and $\lambda_{max,1}$ are not reachable in the fibres located at the centre and edges of the pseudo-slit,  respectively. 
\begin{table}
\caption{Main characteristics of MEGARA LCB IFU and MOS modes.}
\label{mainchar}
\begin{tabular}{lccc}
\hline
Telescope & \multicolumn{3}{c}{GTC} \\
Plate Scale & \multicolumn{3}{c}{0.824 \,mm\,arcsec$^{-1}$}\\
 & Observing mode & LCB  & MOS \\
& No. of fibres & 623 & 644\\
& Spaxel & 0.62\,$\arcsec$ &  0.62\,$\arcsec$ \\
& FOV & 12.5$\,\arcsec \times 11.3\,\arcsec$ &
3.5\,$\arcmin \times 3.5\, \arcmin$ \\
 &$\Delta \lambda\ (EED_{80})$ & 4.0\,pix& 4.0\,pix \\
 & $\Delta \lambda\ (FWHM)$ & 3.6\,pix & 3.6\,pix \\ 
  & LR & 6000 & 6000\\
Resolving power & MR & 12000 & 12000\\
  & HR & 20000 & 20000\\
Gratings (VPH) & \multicolumn{3}{c}{6 LR, 10 MR and 2 HR} \\
Wavelength coverage  & \multicolumn{3}{c}{3650\ -- 9750\,\AA} \\
Detector & \multicolumn{3}{c}{e2V 1 $\times$ 4k $\times$ 4k, 15$\upmu$m pixel size, AR coated }\\
\hline
\end{tabular}
\end{table}

\begin{table}
\scriptsize
\caption{MEGARA spectral configurations as measured at the GTC.  See column description in subsection~\ref{megintro}.}
\label{setups}
\begin{tabular}{lcccccc}
\hline
VPH & $\lambda_{min,1}$(\AA) & $\lambda_{min}$(\AA) & $\lambda_{c}$(\AA) & $\lambda_{max}$(\AA) & 
$\lambda_{max,1}$(\AA) & \AA/pix \\
LR-U  & 3640.0 & 3654.3 & 4035.8 & 4391.9 & 4417.3 & 0.195 \\
LR-B  & 4278.4 & 4332.1 & 4802.1 & 5200.0 & 5232.0 & 0.230 \\
LR-V  & 5101.1 & 5143.7 & 5698.5 & 6168.2 & 6206.0 & 0.271 \\
LR-R  & 6047.6 & 6096.5 & 6753.5 & 7303.2 & 7379.9 & 0.321 \\
LR-I  & 7166.5 & 7224.1 & 8007.3 & 8640.4 & 8822.3 & 0.380 \\
LR-Z  & 7978.4 & 8042.7 & 8903.4 & 9634.9 & 9692.6 & 0.421 \\
MR-U  & 3912.0 & 3919.8 & 4107.6 & 4282.2 & 4289.1 & 0.092 \\
MR-UB & 4217.4 & 4226.4 & 4433.7 & 4625.8 & 4633.7 & 0.103 \\
MR-B  & 4575.8 & 4585.7 & 4814.1 & 5025.1 & 5033.7 & 0.112 \\
MR-G  & 4952.2 & 4963.2 & 5214.0 & 5445.0 & 5454.6 & 0.126 \\
MR-V  & 5369.0 & 5413.1 & 5670.4 & 5923.9 & 5659.6 & 0.135 \\
MR-VR & 5850.2 & 5894.2 & 6171.7 & 6448.5 & 6468.5 & 0.148 \\
MR-R  & 6228.2 & 6243.1 & 6567.3 & 6865.3 & 6878.3 & 0.163 \\
MR-RI & 6748.9 & 6764.6 & 7117.1 & 7440.9 & 7454.5 & 0.172 \\
MR-I  & 7369.4 & 7386.5 & 7773.1 & 8128.0 & 8142.8 & 0.189 \\
MR-Z  & 8787.9 & 8810.5 & 9274.8 & 9699.0 & 9740.2 & 0.220 \\
HR-R  & 6397.6 & 6405.6 & 6606.5 & 6797.1 & 6804.9 & 0.098 \\
HR-I  & 8358.6 & 8380.2 & 8633.0 & 8882.4 & 8984.9 & 0.130 \\
\hline
\end{tabular}
\end{table}
\normalsize

A subset of 11 VPHs are simultaneously mounted on the instrument wheel so that are available in the same observing night. We emphasise that the whole optical wavelength range is covered at R~$=$~6000 and 12000 (FWHM) while R~$=$~20000 is only available around two specific ranges defined by HR-R and HR-I.  The scientific data are recorded by means of a deep-depleted e2V 4096~pix~$\times$~4096~pix detector with 15$\upmu$m-pixel pitch. MEGARA is completed with the instrument control, fully integrated in the GTC Control System (GCS), and a set of stand-alone software tools to allow the user to prepare, visualise and fully-reduce the observations. A complete information and final instrument performance on the GTC based on commissioning results can be found in \citet{carspie18}, \citet{gilspie18}, \citet{dull19} and Gil de Paz et al. (in prep.). 

\subsection{Population Synthesis models and importance of stellar libraries}
\label{section1.2}

Population Synthesis models have proven to be key to derive the star formation histories of galaxies when used to interpret the observations. The integrated properties of a galaxy can be modelled, through a technique as the ones used in {\sc STARLIGHT} \citep{cid05}, or {\sc FADO} \citep{gomes18} codes,  as a combination of Single Stellar Populations (SSP), the building blocks of the population synthesis technique. There have been many studies devoted to the computation of SSP integrated properties, e.g.~\citet{mhk91,gvd94,gvbd95,lei99,bc03,gon05,fritze06,elst09,mar11, vaz16, vg17, mar19}, and our own {\sc PopStar} models \citep{mol09,mman10,gvargas13}.

There are important differences among the available SSP models due to the use of different stellar tracks (so that the isochrones), different input physics and computational algorithms, inclusion of nebular emission and/or dust, and, overall, different stellar atmosphere libraries with their particular wavelength coverage and spectral resolution. 
Stellar libraries are a key piece in the SSP code building. A spectral library is a collection of star spectra sharing the same wavelength range and resolution. The stars comprising these libraries are classified according to the main stellar atmospheric parameters governing their spectral energy distribution, namely effective temperature ($\rm T_{eff}$), surface gravity, ($\rm \log{g}$) and metallicity, usually given in terms of iron, [Fe/H], or all-metal, $\rm \log{(Z/Z_{\sun})}$ abundance; hereinafter, we will use $\rm [M/H]$ for metallicity. In order to reproduce the synthetic spectra as accurate as possible, the spectral star library should cover a wide range in all the three parameters. 

The stellar libraries can be classified in empirical and theoretical, being the first ones based on observed data while the second ones are built with stellar atmosphere and radiative-transfer processes models computed as a function of physical parameters.  
\begin{table*}
\caption{Existing stellar libraries used for MEGARA-GTC Library input catalogue.}
\label{summary}
\begin{tabular}{lrrrcl}
\hline
Library & Resolving Power & Spectral Range & Number & MEGARA setup & Reference \\
INDO-US & 5000 & 3460 - \hspace{2pt} 9460 & 1237 & LR & \citet{val04} \\
MILES   & 2100 & 3520 - \hspace{2pt}  7500 & 987 & LR & \citet{pat06}\\
NGSL    & 1000 & 1670 - 10250 & 374 & LR & \citet{gregg06}\\
STELIB  & 1600 & 3200 - \hspace{2pt}	9300 & 249 & LR & \citet{lebor03}\\
ELODIE low  & 10000 & 3900 - \hspace{2pt}	6800 & 1388 & MR & \citet{pru01,pru04}\\
FOE     & 12000 & 3800 - 10000 & 125 & MR & \citet{mon99}\\
X-SHOOTER & 10000 & 3000 - 25000 & 379 & MR & \citet{chen12, chen14}\\
ELODIE high & 42000 & 3900 - \hspace{2pt}	6800 & 1388 & HR & \citet{pru01,pru04}\\
UES    & 55000 & 4800 - 10600 & 83 & HR & \citet{mon98}\\
UVES-POP & 80000& 3070 - 10300 & 300 & HR & \citet{bag03}\\
\hline
\end{tabular}
\end{table*}
The theoretical libraries have the advantages of covering a very wide range of selectable stellar parameters, in particular abundance, and providing noise-free spectra. However, these libraries require a wide and reliable database of both atomic and molecular lines opacity, not always complete or available; and suffer from systematic potential uncertainties coming from the atmosphere models limitations (convection properties, line-blanketing, expansion, non-radiative heating, non-Local-Thermodynamic-Equilibrium -- non-LTE -- effects etc.). Examples of theoretical libraries used for stellar synthesis population models are those from \citet{kur93, coe05, mar05, mun05, rm05, f06, coe07, gus08, l10, pa10, s10, k11, dl12, coe14} and  \citet{b17} among others. 

The empirical stellar libraries have the advantage of being composed by real observed stars. A very good review of empirical libraries has been recently compiled by \citet{yan191} and \citet{yan192}, which present the MaStar library (MaNGA project). However, there are some  shortcomings and limitations. The existing libraries have relatively low resolution and a limited coverage in terms of the parameter space and are constrained to the ranges of $\rm T_{eff}$, $\rm \log{g}$ and $\rm [M/H]$ spanned by the stars in the Milky Way galaxy and its satellites. Moreover, these existent libraries are often biased to the brightest stars (to save observing time) and/or the most frequent types (associated to the length of the evolutionary stage of each star type). For these reasons, the empirical libraries are short in low metallicity stars, in general, and also deficient in cool dwarfs, and other types of stars not so numerous, which however can make a big difference in the composed synthetic spectrum. This is the case of Wolf-Rayet (WR), Luminous Blue Variable (LBV) and thermally-pulsing Asynthotic Giant Branch (AGB) stars, whose high luminosity in certain wavelength intervals could even dominate completely the spectrum. For these reasons, few SSP codes include an empirical stellar library to model the atmospheres. 

There is a final consideration in the use of libraries when the resulting synthetic spectra are intended to be compared to observations and is that in both theoretical and empirical-based spectra, the instrument characteristics (noise, instrumental specific effects, spectral resolution, and even data reduction), introduce another level of uncertainty in the observed spectra. The exception is an empirical library obtained from star observations with the same instrument. This is the case of the MEGARA-GTC spectral library, a total instrument-oriented catalogue, crucial for creating the necessary synthetic templates to interpret the observations taken with MEGARA at the same instrumental setup.

We describe the construction of our catalogue in Section \ref{MEGARA Stellar Library}. Section \ref{data} summarises the observations of the sample presented in this work in subsections \ref{Commissioning single-star Observations}, \ref{Commissioning M15 MOS Observations} and \ref{filler} for commissioning single stars -- COM hereafter -- , M15 members, and Open-Time (OT hereafter) stars sub-samples respectively, while subsection \ref{datareduction} is devoted to the data reduction process. The spectra analysis is described in Section \ref{spectra-analyses}, dedicating subsections \ref{star-par} to \ref{ace} to the derivation of the stellar parameters and their analysis; and subsections \ref{HRRspec} and \ref{HRIspec} to the description of the absorption line spectra in HR-R and HR-I setups, respectively. Section \ref{database} contains an overview of the MEGARA-GTC database tool. Our conclusions are summarised in Section \ref{conclusions}. 

\begin{figure*}
\includegraphics[width=0.24\textwidth,angle=-90]{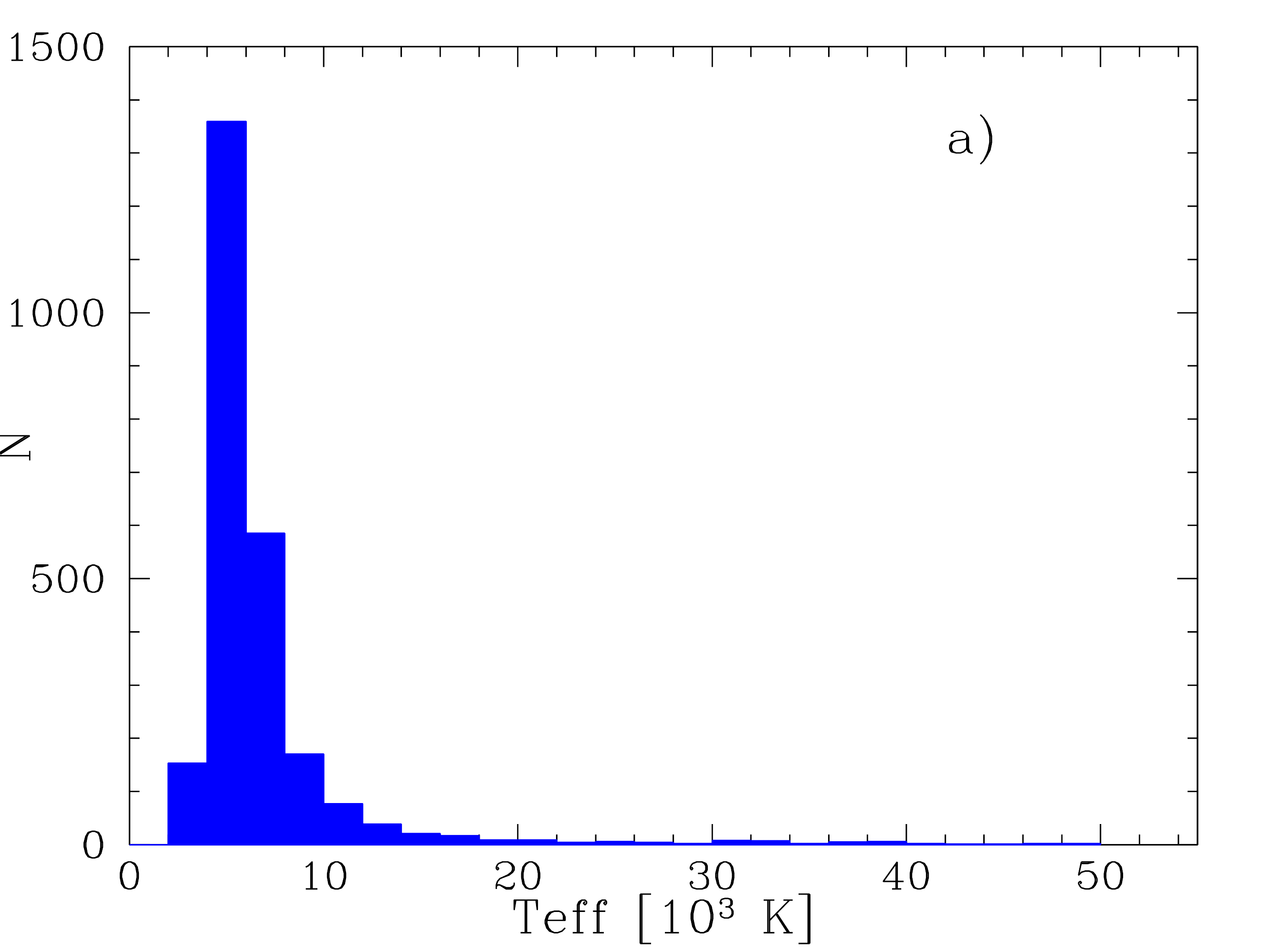}
\includegraphics[width=0.24\textwidth,angle=-90]{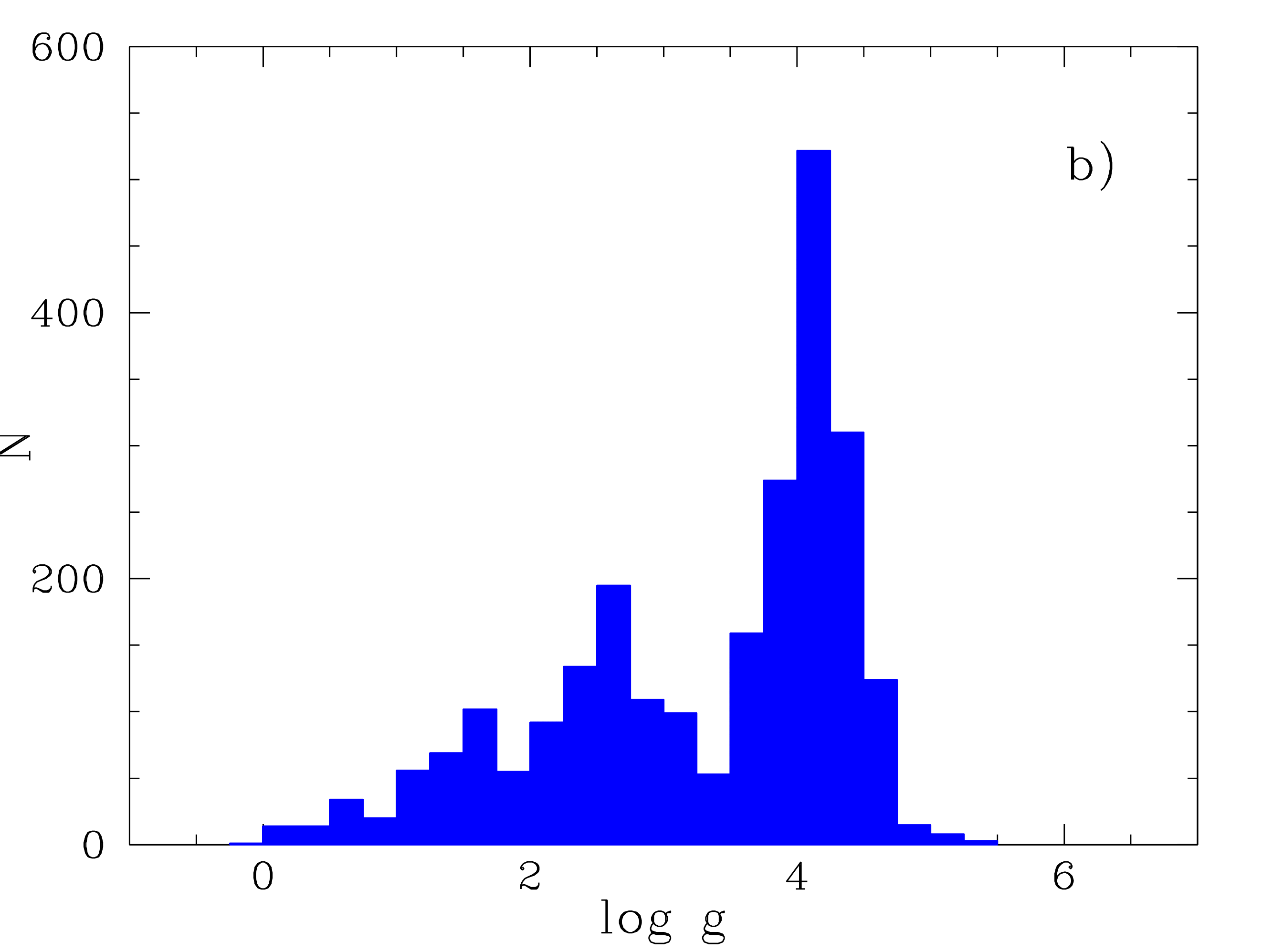}
\includegraphics[width=0.24\textwidth,angle=-90]{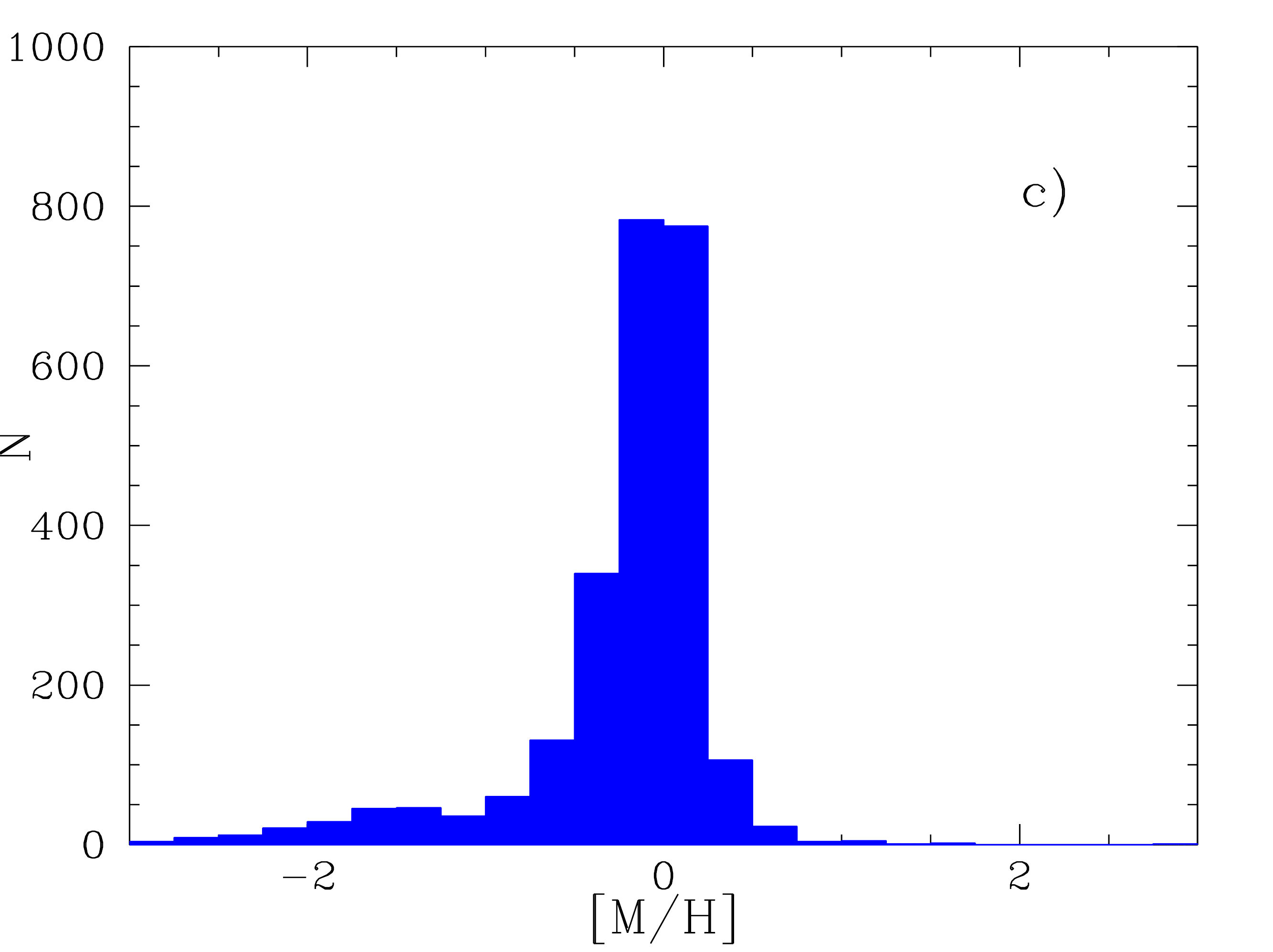}
\caption{Histogram representing the number of stars in the current baseline library - without including the extension of hot stars - as a function of $\rm T_{eff}$ (a); $\rm \log{g}$ (b); and $\rm [M/H]$ (c).}
\label{histograms}
\end{figure*}

\section{MEGARA Stellar Library}\label{MEGARA Stellar Library}

The MEGARA-GTC basic stellar library currently contains 2983 stars covering wide ranges in $\rm T_{eff}$, $\rm \log{g}$ and abundance $\rm [M/H]$, whose histograms are shown in Figure~\ref{histograms}. To define the catalogue, we made a compilation of observed stars from different libraries whose main characteristics and references are summarised in Table~\ref{summary}. We selected libraries whose spectral resolution was comparable to that of MEGARA at LR, MR and HR modes, with resolving power of 6000, 12000 and 20000 respectively. For this search, we used databases from SAO/NASA ADS and ArXiv. The INDO-US library \citep{val04} and ELODIE low-resolution \citep{pru01, pru04} fit well the MEGARA-LR spectral setups; X-SHOOTER \citep{chen12,chen14} and FOE \citep{mon99} have similar spectral resolution than MEGARA-MR and, finally, UVES-POP \citep{bag03}, ELODIE high-resolution \citep{pru04} and UES \citep{mon98} surpass MEGARA-HR spectral resolution. The final library has been produced avoiding any duplicated stars when coming from more than one catalogue.
\begin{table*}
\caption{MEGARA-GTC Stellar Library Catalogue sample. Table description is in section~\ref{MEGARA Stellar Library}.}
\label{catalogue}
\begin{tabular}{cccrrlccccrrrc} 
\hline
Name & RA  & Dec   & pmRA  & pmDec & Sp.Type &   V & R & I & J & T$_{\rm eff}$ & $\log{g}$ & $\rm [M/H]$ & Catalogue \\
 & hh:mm:ss.s & dd:dd:ss.s & masyr$^{-1}$ & masyr$^{-1}$ &  &  &  & &  &  &  &  &   \\
\hline
HD~006229   &  01:03:36.5   &  $\phantom{-}$23:46:06.4   &   $\phantom{-}$14.592	&  $-$20.505	& G5~IIIw		& 8.6	&	&	& 7.1  &  5218 & 3.00 & -1.09     &  X-SHOOTER	\\
HD~006397   &  01:05:05.4   &  $\phantom{-}$14:56:46.1   &    $\phantom{-}$8.265  &   $\phantom{-}$53.750	& F5~III			& 5.6	&	&	&       &       &          &          &             \\
HD~006461   &  01:05:25.4   &  $-$12:54:12.1   &   $\phantom{-}$62.973  &   $\phantom{-}$50.091	& G2~V				& 7.7	&	&	& 6.1  &       &      &               &             \\
HD~006474   &  01:07:00.0   &  $\phantom{-}$63:46:23.4   &   $-$2.077  &   $-$0.304	& G4~Ia			& 7.6	&	&	& 4.8  &       &      &             &             \\
HD~006482   &  01:05:36.9   &  $-$09:58:45.6   &  $-$31.450  &  $-$34.294	& K0~III				& 6.1	&	&	& 4.4  &       &      &              &             \\
HD~006497   &  01:07:00.2   &  $\phantom{-}$56:56:05.9   &   $\phantom{-}$94.445  & $-$108.658	& K2~III				& 6.4	&	&	& 4.7  &       &      &            &             \\
HD~006582   &  01:08:16.4   &  $\phantom{-}$54:55:13.2   & $\phantom{-}$3422.230  & $-$1598.930	& G5~Vb			&	 5.1	& 4.7	& 4.4	& 4.0  &  5320 & 4.49 &  $-$0.76  &     ELODIE low \\
HD~006695   &  01:07:57.2   &  $\phantom{-}$20:44:20.7   &   $\phantom{-}$80.020  &  $-$94.096	& A3~V			&	5.6	&	&	&      &  8266     &  3.91    &      $-$0.46     &  ELODIE low \\
HD~006715   &  01:08:12.5   &  $\phantom{-}$21:58:37.2   &  $\phantom{-}$400.593  &  $-$46.588	& G5			&	 7.7	& 7.2	& 6.9	& 6.3  &  5652     &  4.40    &      $-$0.20     &  ELODIE low \\
HD~006734   &  01:08:00.0   &  $\phantom{-}$01:59:35.0   &  $\phantom{-}$145.370  & $-$437.902	& K0~IV			&	 6.5	& 5.9	& 5.5	& 4.9  &  4934 & 3.18 & -0.58  &     MONTES     \\
\hline
\end{tabular}
\end{table*}

The GTC at the Observatorio del Roque de los Muchachos (ORM) has geographical coordinates of 28$^{\circ}$ 45$\arcmin$ 25$\arcsec$ N Latitude, 17$^{\circ}$ 53$\arcmin$ 33$\arcsec$ W Longitude and 2396m altitude. The catalogue coordinates limits have been selected assuming some margin over the GTC operational restrictions (the minimum declination reachable is -35$^{\circ}$ and the lowest elevation is +25$^{\circ}$), to what we added a safety margin resulting a declination range for the star library of -20$^{\circ}~\leq$~$\delta~\leq$~+89$^{\circ}$.  All coordinates and star data have been checked by parsing the catalogue against SIMBAD4 release 1.7 database.

Table~\ref{catalogue} shows some rows and columns, as an example of the MEGARA-GTC library information. The catalogue will be published with the first release and the observations will be available as soon as the stars are observed and pass our data reduction and quality control processes. The selected columns include the star name, Right Ascension (RA) and Declination (Dec) equatorial coordinates J2000.0, RA and Dec proper motions in masyr$^{-1}$, spectral type and luminosity class, and referenced Johnson magnitudes V, R, I and J whenever available. Most stars come from other spectral libraries. In these cases, values for $\rm T_{eff}$, $\rm \log{g}$ and $\rm [M/H]$ (usually meaning [Fe/H]) exist and are displayed in columns 11, 12 and 13 respectively, together with the original catalogue's name, shown in the last column. The full catalogue contains also additional information of the stars that have been compiled from the literature. The results for the first library  release is in preparation \citep{cmgbg20}; hereafter Paper~II. However, the  assignment of the stellar parameters - initially obtained from the literature - depends on the wavelength range and resolution of each spectrum. For this reason, one of the goals of this work is to propose a method to uniformly determine the stellar parameters of our sample from MEGARA HR spectra (see subsection~\ref{star-par}) and to apply it to all the stars in the library as soon as they are observed. 

The dominant spectral types in the basic catalogue are G stars (751), K (705) and F (630), followed by A (335), B (326), M (153), and O (49). We have a minimum number of L, S, R stars and only 1 WR. We have added an extension to the library with a hot star catalogue composed by 199 Galactic O and B stars from the IACOB database \citep{simondiaz11}, plus WR stars and LBV stars (or candidates), with the restriction limits described above to be observed with the GTC. The WR sub-sample includes 166 galactic \citep[compiled by][]{crowther18}, 14 in M81 \citep{gomezgon16}, 205 in M33 \citep{neu11} and 53 in M31 \citep{neu12}. We have some of these WR stars scheduled to be observed. LBV stars are rare and only few are properly confirmed. Our LBV sub-sample includes 8 stars in our Galaxy (plus 4 candidates), 6 in M31, 4 in M33, 1 in M~81 (plus 1 candidate) and 2 in NGC~2403. 
 The observational program is on-going and has been awarded GTC Open time in three consecutive semesters and we have submitted a proposal for the fourth one (see subsection~\ref{filler}). 
The MEGARA-GTC library's composition might evolve, so the catalogue will be updated as far as the project progresses to have the most complete database possible with the available GTC time.

\begin{table*}
\caption{Sub-sample of 21 stars observed with the HR-I spectral configuration during MEGARA commissioning between June and August 2017. Johnson magnitudes are shown for the stars if available (Note: BD+12~237 is also known as Feige 15 and BD+40~4032 is also known as HD~227900).}
\label{commstars}
\begin{tabular}{lcrrrrrcl}
\hline
Star Name & Sp-Type / Lum-class & U & B & V & R & I & Date & Texp (s) \\
Schulte~9	& 	O4.5If C		&	13.4&	12.8	&	11.0	&	11.0	& 	-	&	30/07/2017	&	$3 \times\ 200$\\
HD~192281	& 	O4.5VC			& 	7.3	&	7.9		&	7.6		&	-		&	-	&	28/08/2017	&	$3 \times\ 25$\\
BD+25~4655	& 	sdO6 C			& 	8.3	&	9.4		&	9.7		&-			&	-	&	01/07/2017	&	$3 \times\ 30$\\
HD~218915	& 	O9.2Iab			& 	6.3	&	7.2		&	7.2		&	-		&	-	&	31/08/2017	&	$1 \times\ 30$\\
BD+40~4032 	& 	B2III D			& 	5.0	&	10.8	&	10.6	&	-		&	-	&	28/08/2017	&	$3 \times\ 100$\\
BD+33~2642	&   O7pD (B2IV)		& 	9.8	&	10.8	&	10.7	&	10.9	&	11.0&	29/06/2017	&	$3 \times\ 300$\\
HD~220575	& 	B8IIIc			&	6.4	&	6.7		&	6.7		&	-		& 	-	&	31/08/2017	&	$2 \times\ 15$\\
BD+42~3227	& 	A0 D			& 	-	&	10.1	&	10.1	&	-		&	-	&	23/08/2017	&	$3 \times\ 60$\\
BD+12~237 	& 	sdA0IVHe1 B		& 	-	& 	10.4	&	10.2	& -			& 	-	& 	29/08/2017	&	$3 \times\ 90$\\
BD+17~4708	& 	sdF8 D			& 	9.7	&	9.9		&	9.5		&	9.0		& 	8.7	&	30/07/2017	&	$3 \times\ 45$\\
HD~026630	& 	G0Ib B			& 	5.8	&	5.1		&	4.2		&	3.4		&	2.8	&	29/08/2017	&	$1 \times\ 20 + 1 \times\ 10$\\
HD~216219	& 	G1II-III-Fe-1	& 	-	&	8.1		&	7.4		&	-		& 	-	&	30/06/2017	&	$3 \times\ 150$\\
HD~011544	& 	G2Ib C			& 	-	&	8.0		&	6.8		&	-		&	-	&	23/08/2017	&	$3 \times\ 20$\\
HD~019445	& 	G2VFe-3 C		& 	8.3	&	8.5		&	8.1		&	7.6		&	7.3	&	23/08/2017	&	$3 \times\ 30$\\
HD~020123	& 	G5Ib-IIa C		& 	7.0	&	6.2		&	5.0		&	-		&	-	&	23/08/2017	&	$3 \times\ 15$\\
HD~224458	& 	G8III C			&	-	&	9.3		&	8.3		&	-		& 	-	& 	30/06/2017	&	$3 \times\ 150$\\
HD~220954	& 	K0.5III			&	6.4	&	5.4		&	4.3		&	3.5		& 	-	&	31/08/2017	&	$1 \times\ 5$ \\
HD~025975	& 	K1III C			& 	7.8	&	7.0		&	6.1		&	5.5		&	5.0	&	29/08/2017	&	$1 \times\ 60$\\
HD~027971	& 	K1III C			& 	-	&	6.3		&	5.3		&	-		&	-	&	29/08/2017	&	$1 \times\ 60$\\
HD~174350	& 	K1III C			& 	-	&	9.1		&	7.9		&	-		&	-	&	30/06/2017	&	$3 \times\ 100$\\
HD~185622	& 	K4Ib C			& 	-	&	8.3		&	6.3		&	-		&	-	&	24/06/2017	&	$2 \times\ 20$\\
\hline
\end{tabular}
\end{table*}

\begin{figure*}
\includegraphics[width=0.48\textwidth,angle=0]{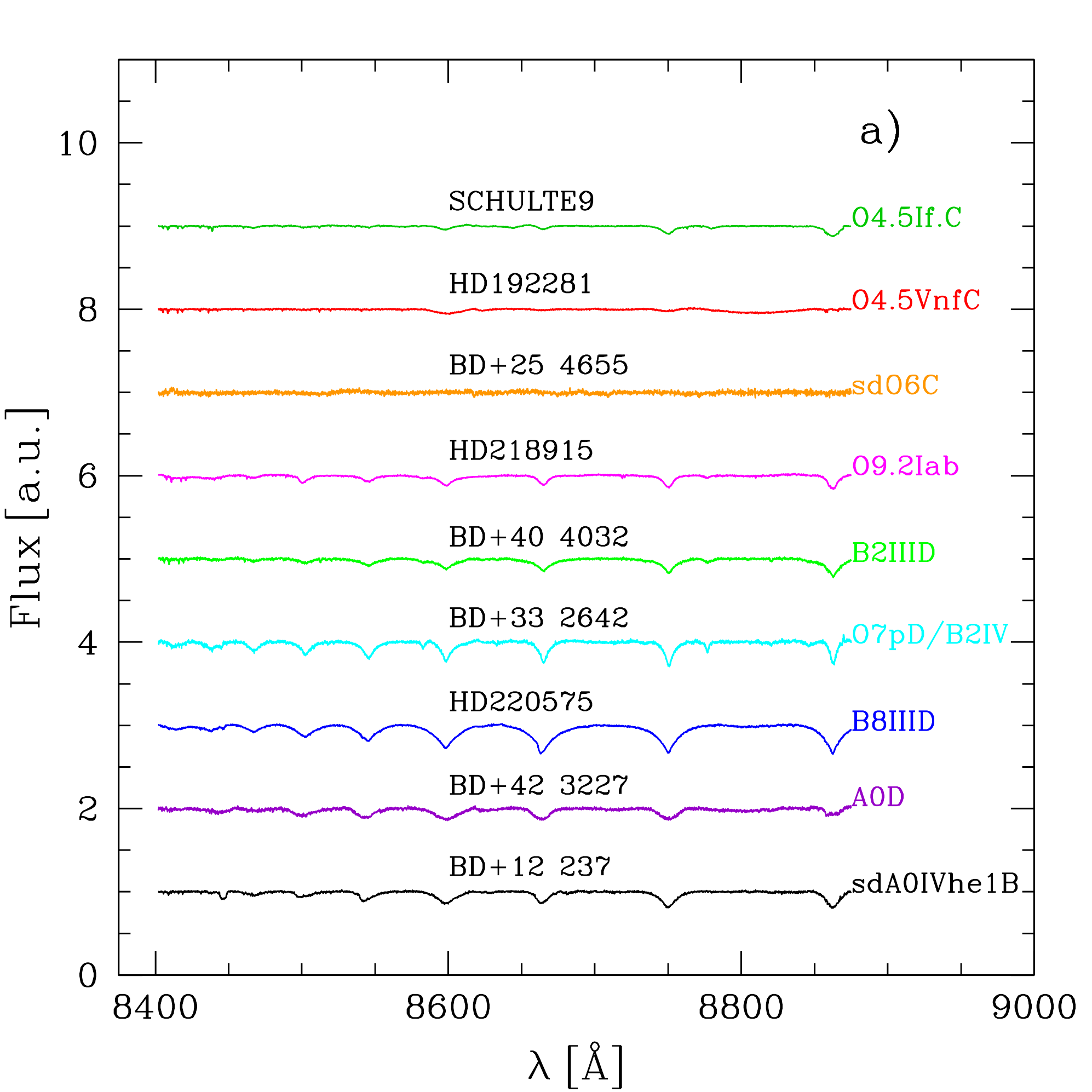}
\includegraphics[width=0.48\textwidth,angle=0]{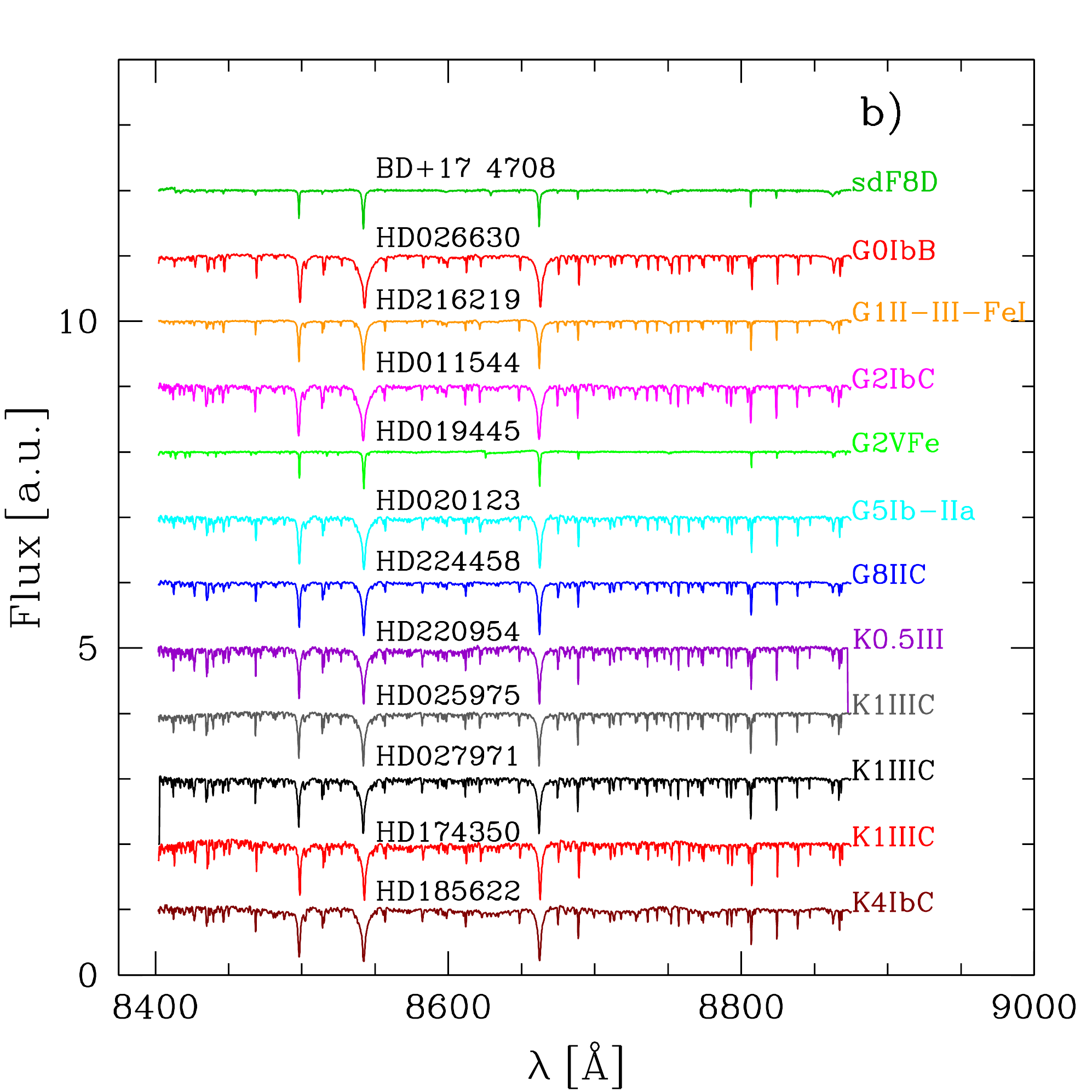}
\caption{Plot showing HR-I spectra in arbitrary units for: a) the hottest stars of the COM sub-sample, being nine stars with spectral types from O (top) to A (bottom); b) the twelve coolest stars of the COM sub-sample, with spectral types from F (top) to K (bottom).}
\label{extractedspectra}
\end{figure*}

In terms of the spectral region, we have prioritised the library in HR-R and HR-I (R~$\simeq$~20000) since there are not any published empirical catalogues covering these spectral ranges and resolution while however are highly demanded. As of today, R~$\simeq$~20000 is not being offered in any other Integral Field Spectroscopy (IFS) instrument, with a combination of efficiency and telescope collecting area high enough to study gas and stellar populations physical properties and kinematics in external galaxies, something particularly powerful in the study of dwarf and face-on disk galaxies. The HR-R setup is centred at H${\alpha}$ at $z=0$, so that observations with this grating will provide stellar templates to support subtraction of the underlying stellar population in nearby star-forming galaxies with the same spectral resolution than the gas spectrum. Also, the lines He{\textsc ii} 6678 and He{\textsc ii} 6683\AA\ will be crucial for classification of massive and hot stars. The HR-I setup, centred in the brightest line of {Ca}{\textsc ii} triplet at 8542.09\AA, will trace the presence of both very young (through the Pa series) and intermediate to old (through the Ca{\textsc ii}, Mg{\textsc i}, {Na}{\textsc i}, and {Fe}{\textsc i} features) populations in nearby galaxies. However, the absence of {He}{\textsc i} and {He}{\textsc ii} lines will add uncertainty when determining the $\rm T_{eff}$ of hot stars on the basis on their HR-I spectrum solely. The spectra of these setups also contain important features used for abundance determination. A more detailed description of the observations in these spectral ranges is given in sections~\ref{HRRspec} and \ref{HRIspec} for HR-R and HR-I, respectively.

Our team is currently developing a new grid of {\sc PopStar} evolutionary synthesis models based on high resolution theoretical spectra (Moll{\'a} et al. in prep.). {\sc PopStar} code is being, therefore, prepared to include HR spectra as those ones resulting from MEGARA stellar library observations and, in particular, HR-R and HR-I setups. This will allow us to obtain both SSPs and composed population models, to be used as stellar-population templates to interpret MEGARA data. 

\section{MEGARA-GTC Library Observations}
\label{data}

We present in this paper a sub-sample of 97 stars that will form part of the first release of the MEGARA-GTC Library. During MEGARA commissioning at the GTC (June - August 2017), we started a pilot program whose goal was to obtain the first observations of the MEGARA-GTC Library using HR-I. The subsection~\ref{Commissioning single-star Observations} is devoted to describe these observations. We also observed the centre of the M15 cluster in both HR-R and HR-I configurations, with the MEGARA MOS mode using the robotic positioners. These observations are described in subsection~\ref{Commissioning M15 MOS Observations}. Finally, we include in this paper 20 stars, also observed in HR-R and HR-I, belonging to the basic library and observed with the filler GTC-program described in~\ref{filler}. 

\subsection{Commissioning single-star LCB Observations}\label{Commissioning single-star Observations}

During the MEGARA commissioning phase, we observed 21 single stars from the MEGARA-GTC catalogue described in section~\ref{MEGARA Stellar Library} with the LCB mode in HR-I setup. Observations were carried out during twilight time or under-optimum observing conditions, in which the other commissioning tests could not be done. The exposure times ranged between 5 and 900s. The name and properties of these COM stars are listed in Table~\ref{commstars}. The 21 HR-I 2D spectra have been fully reduced as described in subsection~\ref{datareduction}. Once the {\it final} (sky subtracted) flux-calibrated spectra were obtained, we used the MEGARA Quick Look Analysis (QLA) tool \citep{gomezalv18} to extract the 1D spectra by integrating 3 rings on sky-projection (37 spaxels). Figure~\ref{extractedspectra} shows the spectra of this COM sub-sample for the hottest (panel a) and the coolest (panel b) stars, respectively.

\begin{figure}
\includegraphics[width=0.45\textwidth,angle=0]{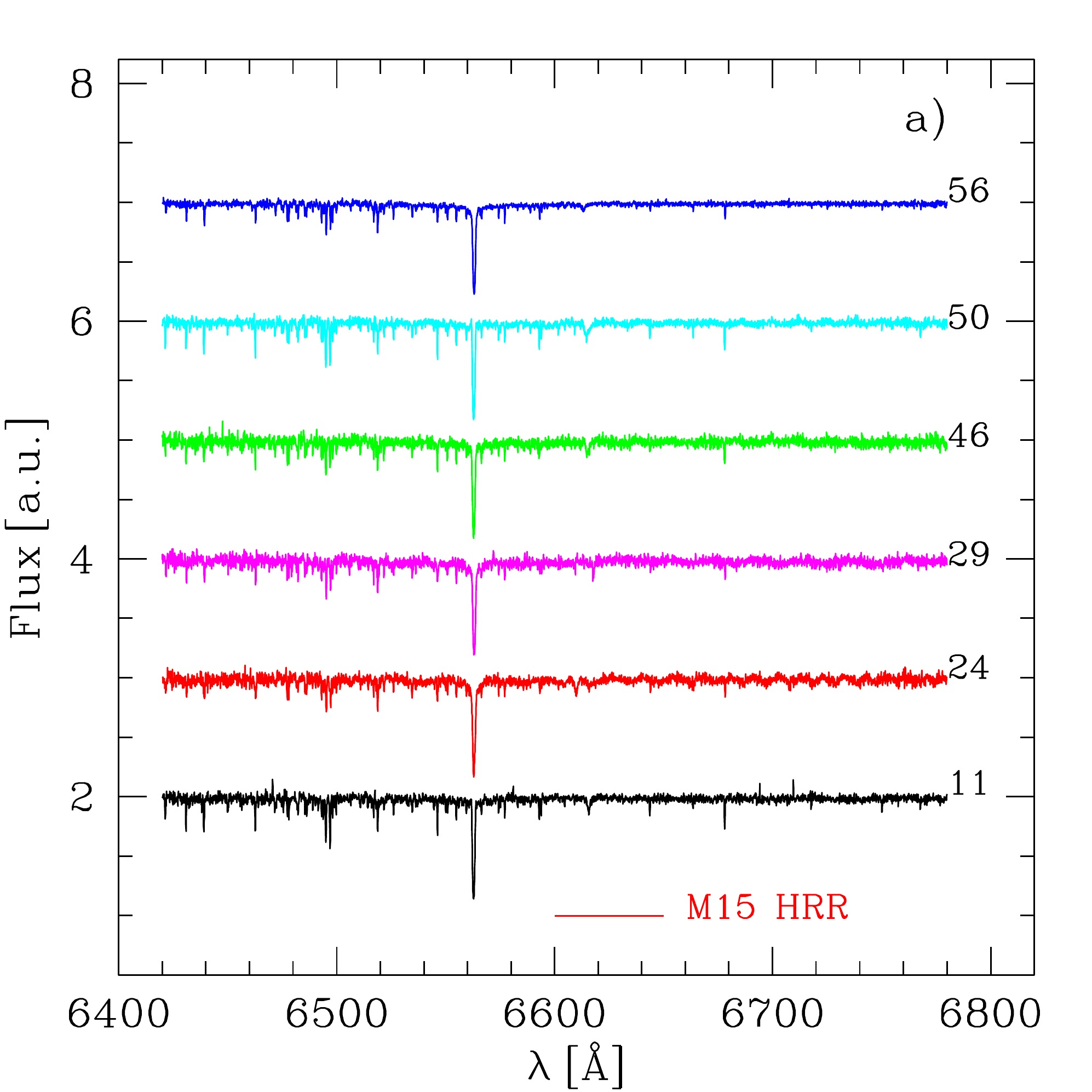}
\includegraphics[width=0.45\textwidth,angle=0]{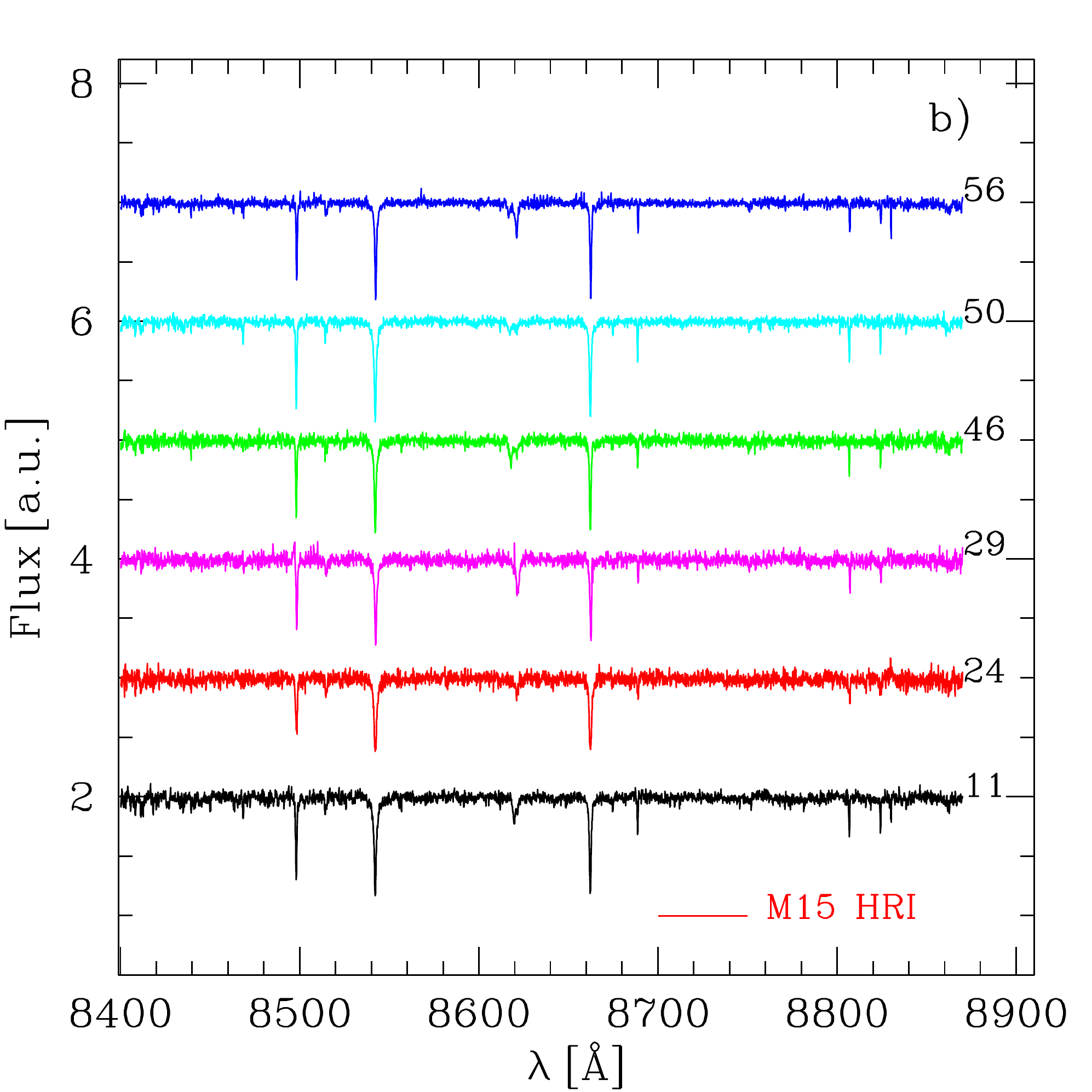}
\caption{Normalized spectra in arbitrary units for 6 selected stars from M15 in two different setups: a) in HR-R; b) in HR-I.}
\label{M15-sel}
\end{figure}

\subsection{Commissioning M15 MOS Observations}\label{Commissioning M15 MOS Observations}

We also observed the centre of the M15 globular cluster in both HR-R and HR-I spectral configurations. The observations of  1800s (3 x 600s) in each configuration were taken on 24 August 2017, during MEGARA commissioning, with the MOS mode and excellent seeing (0.4 arcsec). The pointing coordinates were chosen to accommodate as many stars as possible in the 88 positioners available. These coordinates were $\alpha~$(J2000.0 FK5) = 21h:29m:58.147s and $\delta~$(J2000.0 FK5) = 12$^{\circ}$10$\arcmin$09.62$\arcsec$ with an Instrument Position Angle, IPA, of 196.146$^{\circ}$. Calibrations for both setups were taken with the same MOS configuration to correct from Bias, Flat Field and Wavelength Calibration following the data reduction steps described in section~\ref{datareduction}. Correction for Atmospheric Extinction and Flux Calibration were done by observing two standard stars with the IFU (in both setups). From these observations, we obtained the master sensitivity curve to correct all the 1D star spectra from spectral response and flux calibration. After performing sky-subtraction with sky spaxels, we used the QLA tool to extract the 1D star spectra. We finally obtained 56 stellar spectra in each setup to be used in this work. These stars (numbered from 1 to 56) had not previous stellar parameter determination so that we have applied the method described in this paper (see section~\ref{M15-res}) to obtain them. Figure~\ref{M15-sel} shows the spectra of 6 of these 56 M15 stars in HR-R (a) and HR-I (b). Each star is plotted with a different color and with a shift for the sake of clarity. From the observations of HR-R and HR-I we have derived a radial velocity of -106.9 kms$^{-1}$ (after the de-convolution with the instrumental profile), compatible with the value commonly adopted in the literature (v~=~-106.6 $\pm$ 0.6 kms$^{-1}$ from SIMBAD).

\subsection{Filler program to observe MEGARA-GTC Library}
\label{filler}

We have proposed a filler-type program during the last 3 semesters in the {\it Call for Proposals for GTC Open Time} to observe stars for the MEGARA-GTC Library. We have already been awarded 175 hours of observing time (programs 35-GTC22/18B, 61-GTC37/19A and GTC33-19B, PI: Moll\'a), and it is our intention to complete this project until having a high number of stars to do a precise enough evolutionary synthesis code. The motivation of this program is to finalise, in the shortest timescale possible, the MEGARA-GTC spectral library. To delimit the goal and, consequently, the telescope time, this filler proposal is focused on the highest spectral resolution configurations: HR-R and HR-I at R~$\sim$~20000, for the reasons formerly described in section~\ref{MEGARA Stellar Library}. In the moment of submitting this paper, we have more than 260 stars observed and reduced in both setups.

\begin{figure*}
\includegraphics[width=0.95\textwidth,angle=0]{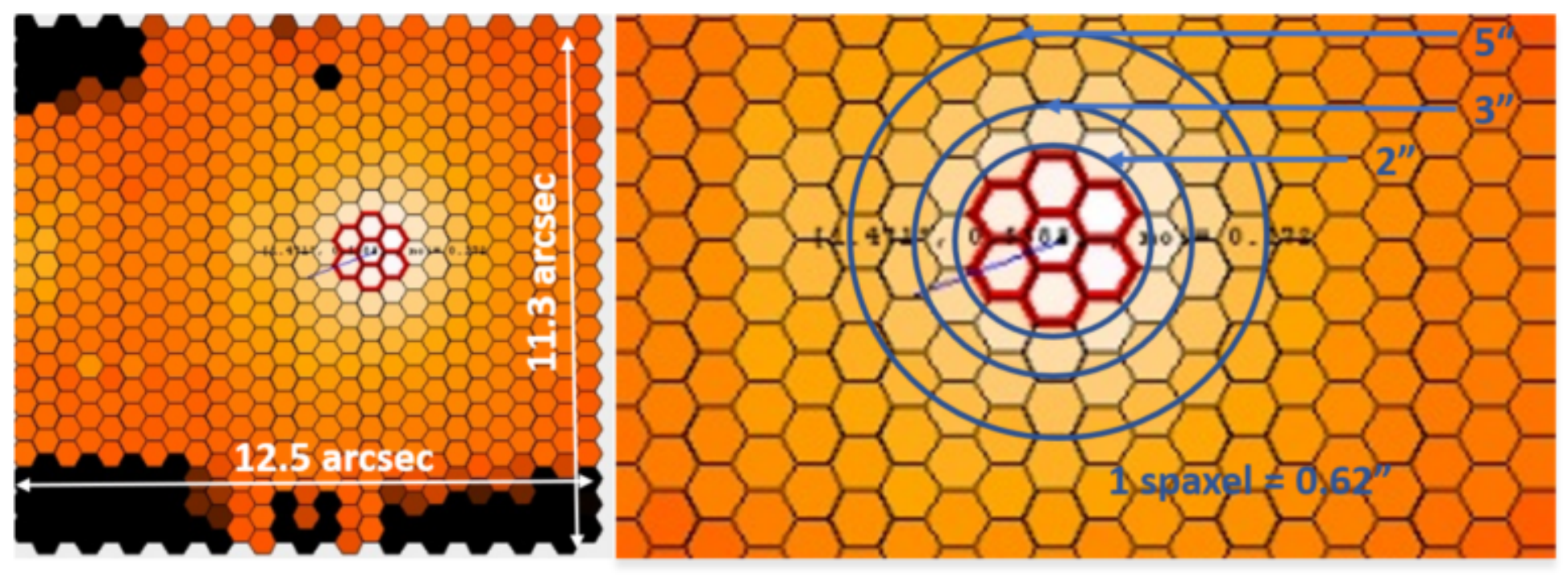}
\caption{Reconstructed image of the star HD~192281 observed with MEGARA LCB mode in HR-I at the GTC during the commissioning, 2 August 2017.}
\label{staronlcb}
\end{figure*}

GTC filler programs require relaxed observing conditions to be executed even when no other approved program in the other regular (higher-priority) bands fits. The MEGARA-GTC Library program fits perfectly as a filler since the star observations can be carried out under almost any conditions, in particular with any seeing, which gives the program high flexibility. On the one hand, the spectral resolution is preserved since the stop is at the fibre so the slit width remains constant and the resolving power on the detector will not change with seeing. On the other hand, flux can be recovered by adding IFU spaxels on sky (fibres at the detector) so that flux is always guaranteed regardless the seeing value. This is explained in Figure~\ref{staronlcb}, which shows the image of a real star observed with MEGARA LCB mode at the GTC during the commissioning. This image is taken directly from the QLA tool available at GTC. The left image shows the full LCB while the right image shows a zoom. The resulting observation is completely valid regardless the seeing conditions. Different circles indicate different seeing values aperture all in the filler range: 2$\arcsec$, 3$\arcsec$ and 5$\arcsec$ are shown as examples. The spaxel size is 0.62$\arcsec$ on sky.

\begin{table*}
\caption{ Sub-sample of 20 OT stars observed with HR-R and HR-I setups from the program 35-GTC22/18B.}
\label{opentimestars}
\begin{tabular}{lcrrrrrcccrcrc}
\hline
Star Name & SpType & U & B & V & R & I &  Date & Texp (s) & Texp (s) & Teff  & logg & [Fe/H] & Ref\\
 &  & &  &  &  &  &  dd/mm/yy & HR-R & HR-I & Lit & Lit & Lit & \\
HD~147677	& K0III		& 	--	&	5.8	&	4.9	&	--	&	--&	23/08/18	&	$3 \times\ 20$ &	$1 \times\ 25$ &  4910 &	2.98 &	$-$0.08 &	INDO-US \\
HD~174912	& F8V	& 	--	&	7.7		&	7.1		&	6.8		&	6.5	&	21/08/18	&	$3 \times\ 20$ &	$3 \times\ 20$ &  5746 &	4.32 &	$-$0.48 &	MILES \\
HD~200580	& F9V	& 	--	&	7.9		&	7.5		&	7.0	& 6.8	&	20/08/18	&	$3 \times\ 20$ &	$3 \times\ 20$ &  5774 &	4.28 &	$-$0.65 &	MILES \\
HD~206374	& G8V		& 	--	&	8.2		&	7.5		&	7.0		&	6.7	&	22/08/18	&	$3 \times\ 20$ &	$3 \times\ 20$ &  5622 &	4.47 &	0.00    &	ELODIE \\
HD~211472	& K1V			& 	--	&	8.3		&	7.5		&	7.6		& 6.6	&	22/08/18	&	$3 \times\ 20$ &	$3 \times\ 20$ &  5319 &	4.40 &	$-$0.04 &	ELODIE \\
HD~218059	& F8V	& 	--	&	7.5		&	7.1		&	6.8		&	6.6	&	22/08/18	
&	$3 \times\ 20$ &	$3 \times\ 20$ &  6253 &	4.27 &	$-$0.27 &	ELODIE \\
HD~220182	& K1V	& 	-	&	8.2		&	7.4		&	6.8		& 	6.5	&	22/08/18	&	$3 \times\ 20$ &	$3 \times\ 20$ &  5372 &	4.31 &	0.00    &	ELODIE \\
HD~221585	& G8IV	& 	-	&	8.2		&	7.4		&	7.0		& 	6.6	&	22/08/18	&	$3 \times\ 20$ &	$3 \times\ 20$ &  5352 &	4.24 &	0.27    &	ELODIE \\
HD~221830	& F9V	&	--	&	7.5		&	6.9		&	6.5		& 	6.2	&	22/08/18	&	$3 \times\ 20$ &	$3 \times\ 20$ &  5688 &	4.16 &	$-$0.44 &	MILES \\
BD+083095 	& G0V	& 	--	& 	10.6	&	10.0	& 9.8		& 	9.6	& 	22/02/19	&	$3 \times\ 90$ &	$3 \times\ 90$ & 5728	& 4.12	& $-$0.36	& INDO-US\\
HD~100696	& K0III		& 	--	&	6.2		&	5.2	&	4.6		& 	4.1 &	10/02/19	&	$1 \times\ 30$ &	$1 \times\ 20$ & 4890	& 2.27	& $-$0.25	& INDO-US\\
HD~101107	& F2II-III	& 	--	&	5.9	& 5.6 & 5.4 & 5.2	&	10/02/19	&	$1 \times\ 40$ &	$1 \times\ 40$ & 7036 & 4.09 & $-$0,02 & NGSL\\
HD~104985	& G8III		& 	--	&	6.8	&	5.8	&	5.2	&	4.7 &	12/04/19	&	$1 \times\ 40$ &	$1 \times\ 30$ &  4658 &   2.20  & $-$0.31   & INDO-US \\
HD~113002 	& G2II-III	& 	9.7	&	9.5	&	8.7	& 8.3	&	8.0	&	03/02/19	&	$3 \times\ 40$ &	$3 \times\ 40$ &  5152 & 2.53  &	$-$1.08   & NGSL    \\
HD~115136	& K2III	& -- &	7.7	&	6.5	& 5.8 &	5.3	&	12/04/19	&	$3 \times\ 15$ &	$1 \times\ 40$ &  4541 & 2.40  &	0.05    & INDO-US \\
HD~117243	& G5III	& -- &	9.0	&	8.3		& 7.9	& 7.6 &	29/01/19	&	$3 \times\ 30$ &	$3 \times\ 30$ &  5902 & 4.36  &	0.24    & INDO-US \\
HD~131111	& K0III		& -- &	6.5		&	5.5		&	4.8	&	4.4 &	03/03/19	&	$1 \times\ 30$ &	$1 \times\ 20$ &  4710 & 	3.11  &	$-$0.29	& INDO-US \\
HD~131507	& K4III		& 	--	&	6.9		&	5.5		&	4.6		& 3.8 &	03/03/19	&	$1 \times\ 30$ &	$1 \times\ 15$ &  4140 & 	1.99  &	$-$0.20	& INDO-US \\
HD~144206	& B9III		& 	4.3	&	4.6	&	4.7		&	--		&	--	&	27/09/18	&	$3 \times\ 20$ &	$3 \times\ 20$  & 11957 & 	3.70  &	$-$0.17	& Stelib  \\
HD~175535	& G7IIIa	& 	--	&	5.8	&	4.9		&	--		&	-- &	07/09/18	&	$3 \times\ 15$ &	$3 \times\ 10$ &  5066 & 	2.55  &	$-$0.09	& MILES   \\
\hline
\end{tabular}
\end{table*}

\begin{figure}
\centering\includegraphics[width=0.48\textwidth,angle=0]{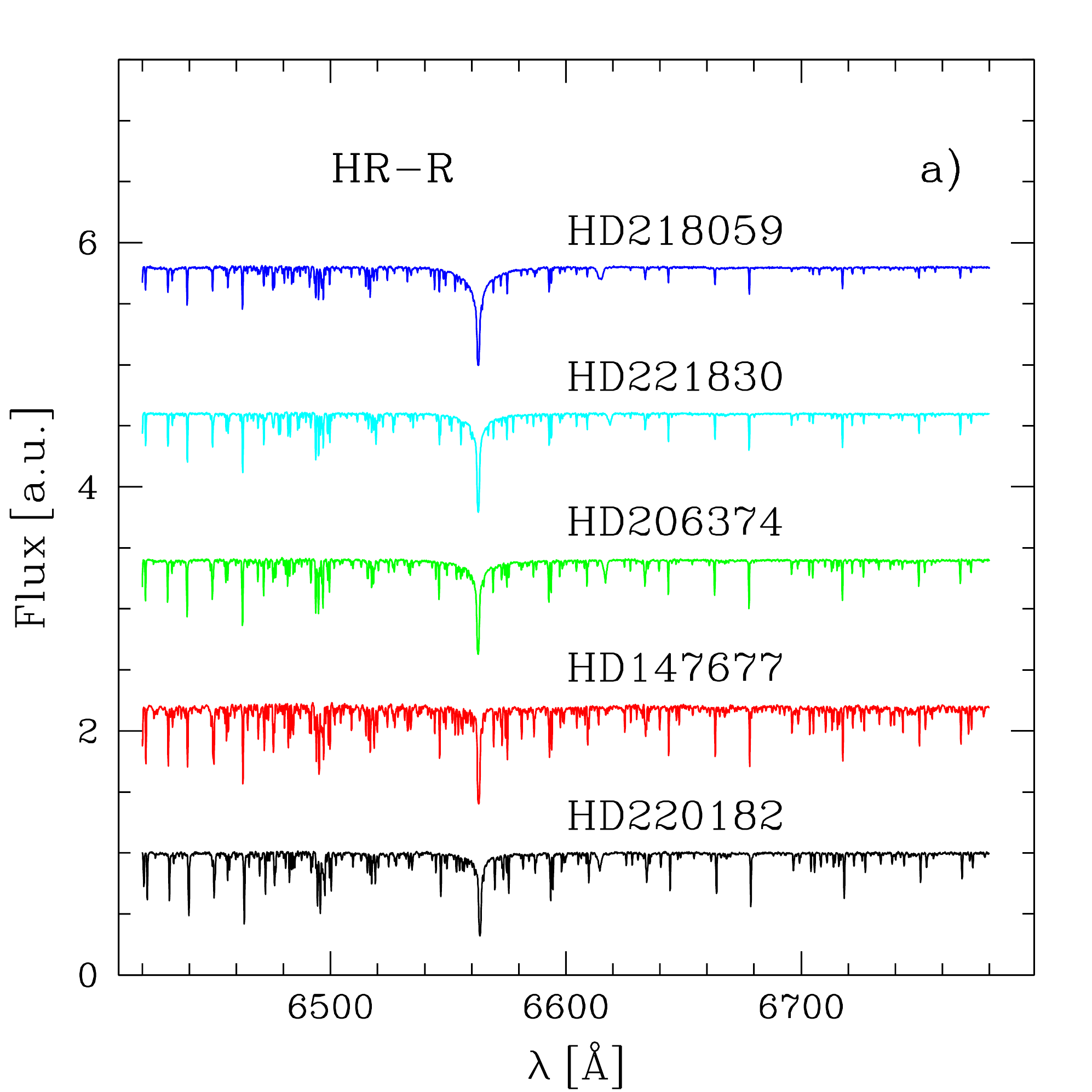}
\centering\includegraphics[width=0.48\textwidth,angle=0]{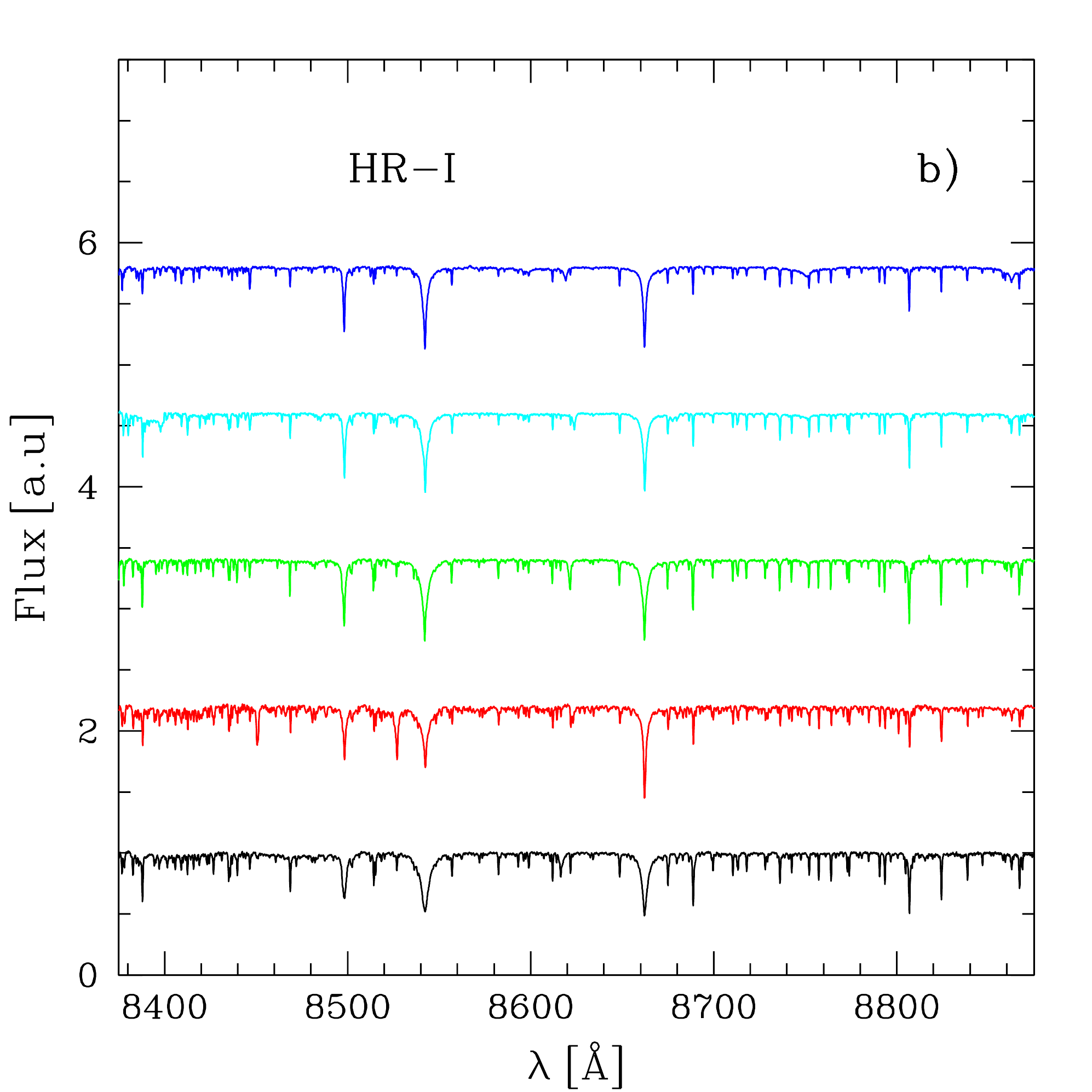}
\caption{Flux of normalized spectra in arbitrary units for 5 stars, HD~220182, HD~147677, HD~206374, HD~221830 and HD~218059, from bottom to top, in two different setups: a) HR-R and b) HR-I.}
\label{2018B}
\end{figure}

Stars can be observed in bright conditions (most of the pilot commissioning observations presented in this paper were carried out during the twilight). Therefore, this program can run up to 1~hour after astronomical twilight every night, especially when observing HR-I (very red setup). Finally, photometric conditions are not required. At least a standard star for flux calibration is taken  every night even for a filler-type program according to GTC observing policies, but even in absence of flux calibration the same day, we can use the sensitivity curve taken in another night to correct for the instrument response curve. Therefore, even in absence of a proper (same-night) flux calibration, the goals of determining the stellar parameters and measuring equivalent widths and indices can be completed for the whole stellar library with a high degree of reliability.  

The program is not very demanding in terms of telescope operation. Most of the targets are bright and an accurate pointing in the centre of the LCB is not needed. We have included a large number of short-exposure time observations that can be executed between observing gaps of other programs, without {\it idle}, expensive, telescope time. This filler program can even be executed in absence of A\&G operation or for high levels of sky brightness. Several strategies have been discussed with GTC staff and are applied to minimise overheads and to increase program efficiency. On the one hand, the large number of stars in the database makes the GTC staff's choice very easy and several targets can be selected in the same sky field, reachable with minimum telescope re-pointing. On the other hand,  a common {\it Observing Block}, OB, with both HR-R and HR-I observations of each star is entering in the GTC Phase-2 tool, guaranteeing a stable database population of observations in both setups. The on-target time per spectral configuration has been estimated with the MEGARA Exposure Time Calculator (ETC) tool to have Signal to Noise Ratio (SNR) between 20 and 300. 

The observations are carried out in queue mode with the following strategy: (a) search in the GTC Phase-2 the most appropriate target according to visibility and priority, (b) configure MEGARA while slewing; (c) acquire the target with LCB-Image mode, and (d) carry out HR-R and then HR-I on the same target. For calibration purpose, we have requested halogen and wavelength calibration lamp images in DayTime and at least one standard star in NightTime, whenever possible. The data are reduced with the MEGARA Data Reduction Pipeline (MEGARA DRP) by applying the different recipes as described in subsection~\ref{datareduction}. Fully reduced and calibrated products are obtained soon after being received and frequent releases shall be delivered to the community, being the first one planned by the first semester of 2020. 
As the project progresses we update the targets in the GTC Phase 2 and manage the priority levels to populate all physical parameters regions of the stellar library. This strategy is possible thanks to the high flexibility of the GTC-Phase 2 queued-observations to accommodate changes within the allocated time. We have included in this paper a sub-sample of 20 OT stars observed in both HR-R and HR-I, as part of our program 35-GTC22/18B. The summary of the star data and observations are given in Table~\ref{opentimestars}. Figure~\ref{2018B} shows the spectra of five of these stars in HR-R, panel a), and HR-I, panel b). The stars are, from bottom to top, HD~220182, HD~147677, HD~206374, HD~221830 and HD~218059, corresponding to spectral types and luminosity classes of K1~V, K0~III, G8~V, F9~V and F8~V respectively. 

\subsection{Data Reduction}\label{datareduction}

All observations were taken with their corresponding calibrations. Whenever the exposure time was longer than 30s, it was divided into 3 identical exposures to facilitate cosmic rays removal. Halogen and ThNe lamp calibrations were taken in DayTime for tracing, flat field, and wavelength calibration. Standard stars observations for flux calibration were taken during NightTime. Twilight images were taken for the commissioning observations only.

MEGARA DRP is a python-based software tool operating in command-line \citep{carandpas18, pasetal18}. The DRP uses a file called \textit{control.yaml} that includes all relevant information needed for MEGARA data reduction such as the data directories, the polynomial degree and number of spectral lines used for wavelength calibration and the site's extinction curve. Once MEGARA DRP is installed, the user has to do a local copy of the complete calibrations file tree. The MEGARA team has made available a complete set of calibration files ready-to-use for the 36 MEGARA configurations (defined as the combination of the observing mode - LCB or MOS - and the selected VPH setup - from a total of 18 gratings). These calibration files can be substituted by updated ones whenever available. The pipeline operates in a cascade mode so that each step requires processed data with the previous recipes. Each routine has its own input file in which the images to be processed are identified together with the specific parameters needed for that recipe. Once executed, each DRP routine produces a set of output images and quality control files allowing a full tracking of the reduction process.

The data reduction starts generating a Master Bias with the {\it MegaraBiasImage} routine and bias images. In further steps (except the Bias itself) all the calibration files have to be taken with the same instrument setup. The cosmic rays removal routine is automatically applied by default to eliminate the undesired cosmic rays registered on detector by the combination of several images of the same target whenever available. 

The {\it MegaraTraceMap} recipe uses the halogen lamp images to find the position of the illuminated fibres on the detector, storing the information in a {\it .json} formatted file and producing a region {\it .reg} ds9 file \citep{ds9} with the trace identification. This ds9 region can be overlapped on any images (wavelength calibration lamp, twilight, standard star and target star) to determine in each case the offset (if needed) between the reference traced fibres and the actual position. This information allows the matching between the fibres and the traces when extracting the fibre spectra. Although these offsets are predictable as a function of the temperature change, the recommendation is to check the offset position and apply that as a parameter when calling the subsequent pipeline recipes. The {\it MegaraModelMap} recipe, starting from the results of the {\it MegaraTraceMap} and taken the halogen images as input, produces an optimised extraction of the fibre spectra. 

The {\it MegaraArcCalibration} recipe uses the lamp wavelength calibration images, their offset value and the output of the {\it MegaraModelMap} to produce a wavelength calibration whose parameters are stored in a new {\it .json} file. After wavelength calibration, the {\it MegaraFiberFlatImage} recipe is used to correct for the global variations in transmission between fibres while the {\it MegaraTwilightFlatImage} recipe corrects for the response introduced by the fibre flat (illumination correction). 

\begin{figure*}
\includegraphics[width=0.498\textwidth,angle=0]{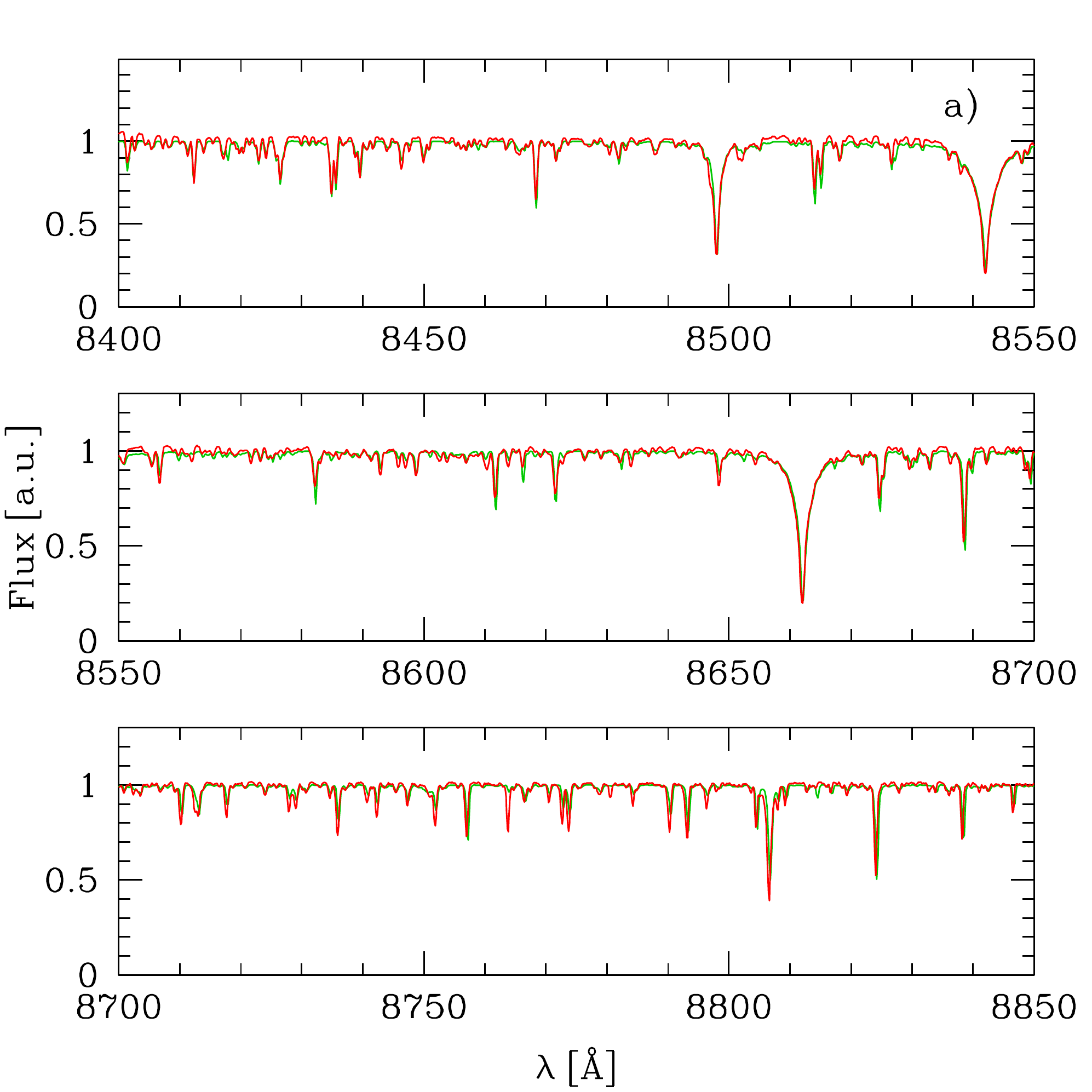}
\includegraphics[width=0.498\textwidth,angle=0]{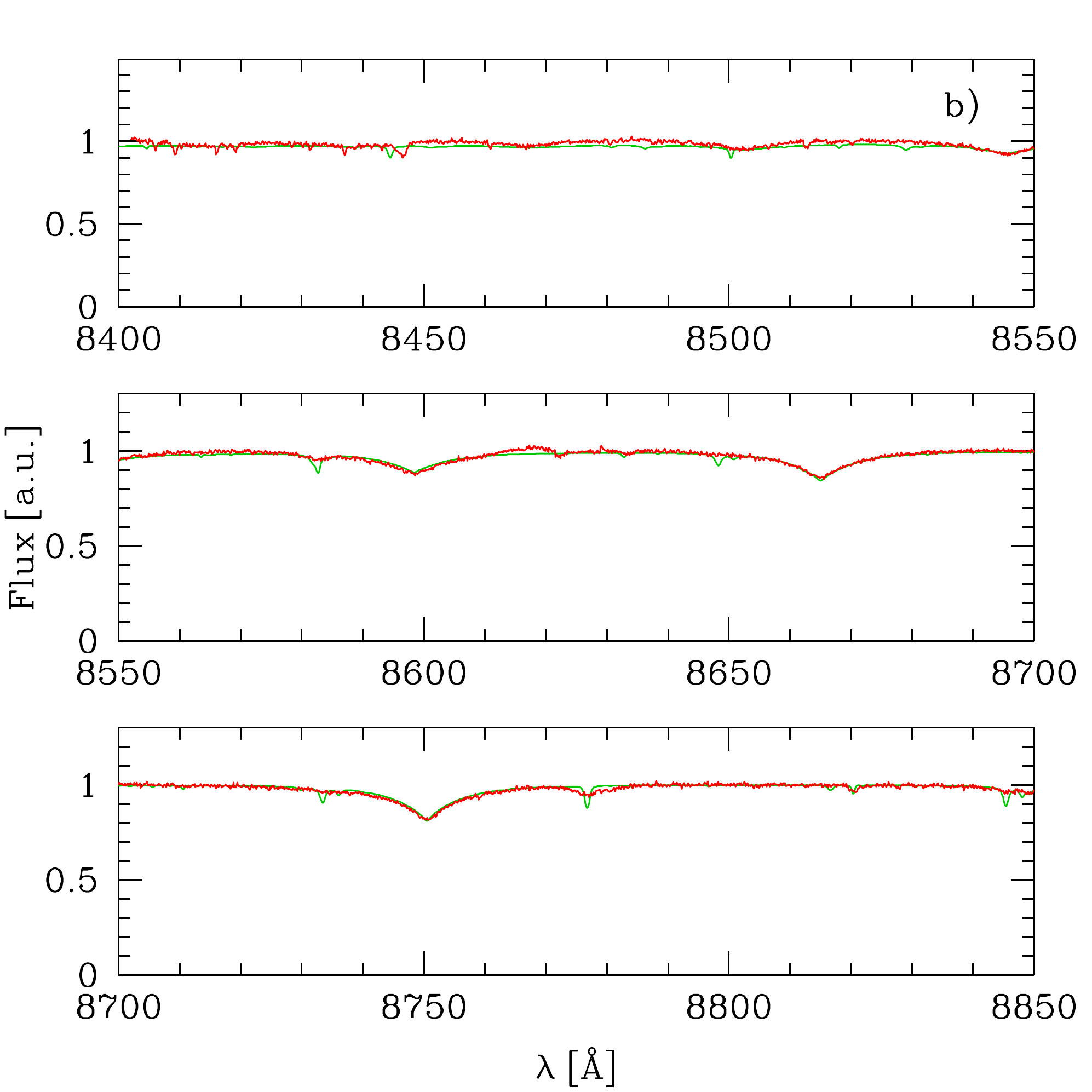}
\caption{Panel a) shows the MEGARA HR-I normalized spectrum of the cool giant K1~III star HD~025975 while panel b) displays the one for the hot B2~IIID star BD+40 4032. Observed spectra are plotted in red while the MUN05 model which gives the best fit is overplotted in green. The $\chi^{2}_{min}$ model for HD~025975 has $\rm T_{eff}$~=~4750~K, $\rm \log{g}$~=~3.0, and $\rm [M/H]$~=~$-$0.5, while the one for BD~+40 4032 has $\rm T_{eff}$~=~28000~K, $\rm \log{g}$~=~3.5 and $\rm [M/H]$~=~0.0. Each star spectrum is divided in three panels to show the fitting details.}
\label{fits-detail}
\end{figure*}

Flux calibration is applied with the DRP by comparing the reduced standard star spectrum and the corresponding reference template. The {\it MegaraLcbAcquisition} recipe delivers the position of the standard star on the LCB. This routine uses all the calibrations from the former steps to reduce the standard star images. Once the position of the standard star on the LCB is known, the {\it MegaraLcbStdStar} routine produces the {\it Master Sensitivity} curve by comparing the 1D flux spectrum of the standard star (corrected from atmospheric extinction) with its tabulated flux-calibrated template. This sensitivity curve also corrects from the spectral instrument response (mostly dominated by VPH transmission and detector quantum efficiency), so that this step is needed even when non-photometric conditions prevents a reliable flux calibration.

Once all the calibrations files are obtained and properly placed on the calibration tree, the scientific observations (MEGARA-GTC library stars) are processed with the recipe {\it MegaraLcbImage} or {\it MegaraMosImage}, producing the Row-Stacked-Spectra (RSS) file  with the individual flux-calibrated spectra for all fibres (corrected from extinction and flux/spectral response). For the individual stars and the LCB mode, we have used the automatic sky subtraction done by MEGARA DRP resulted from the median of the signal of all the 8 sky-minibundles. In the MOS M15 images, we have used the optimised sky subtraction procedure offered by the QLA, which allows the selection of a customised combination of individual spaxels for sky subtraction. In this case, we took a total of 9 sky-spaxels (4 in positioner 26 and 5 in positioner 36).

One of the critical steps before starting the model fitting is the normalisation of the observed star spectrum, which needs a reliable continuum fitting that takes into account the proper spectral windows and avoids confusion with the high number of spectral lines due to the high resolution. For that purpose, we have made use of the fitting technique by \citet{car09}, who described a generalised least-squares method that provides boundary functions for arbitrary data sets. In particular, this technique can be employed to determine the upper boundary of a particular spectrum. The method is based on the asymmetric treatment of the data on both sides of the boundary. When applied to a particular spectrum, the upper boundary becomes an excellent fit to the expected continuum. This is especially so when using adaptive splines as the mathematical function for the boundary. The flexibility has been improved a step further by splitting the fitted wavelength range of each spectrum into smaller intervals and by smoothly merging the independent adaptive splines to obtain a single continuum fit for the whole spectral range. Before the fitting process, a median filter of a few tens of pixels has been applied to each spectrum to minimize the bias that the data random noise may introduce in this kind of asymmetric fit. Although the fitting process, carried out using the public software
\texttt{boundfit}\footnote{\url{https://boundfit.readthedocs.io/en/latest/}}, has been completely automatized, all the resulting fits were visually inspected.

\section{Spectra Analysis}\label{spectra-analyses}
\subsection{Estimates of the stellar physical parameters}
\label{star-par}

\begin{figure}
\includegraphics[width=0.48\textwidth,angle=0]{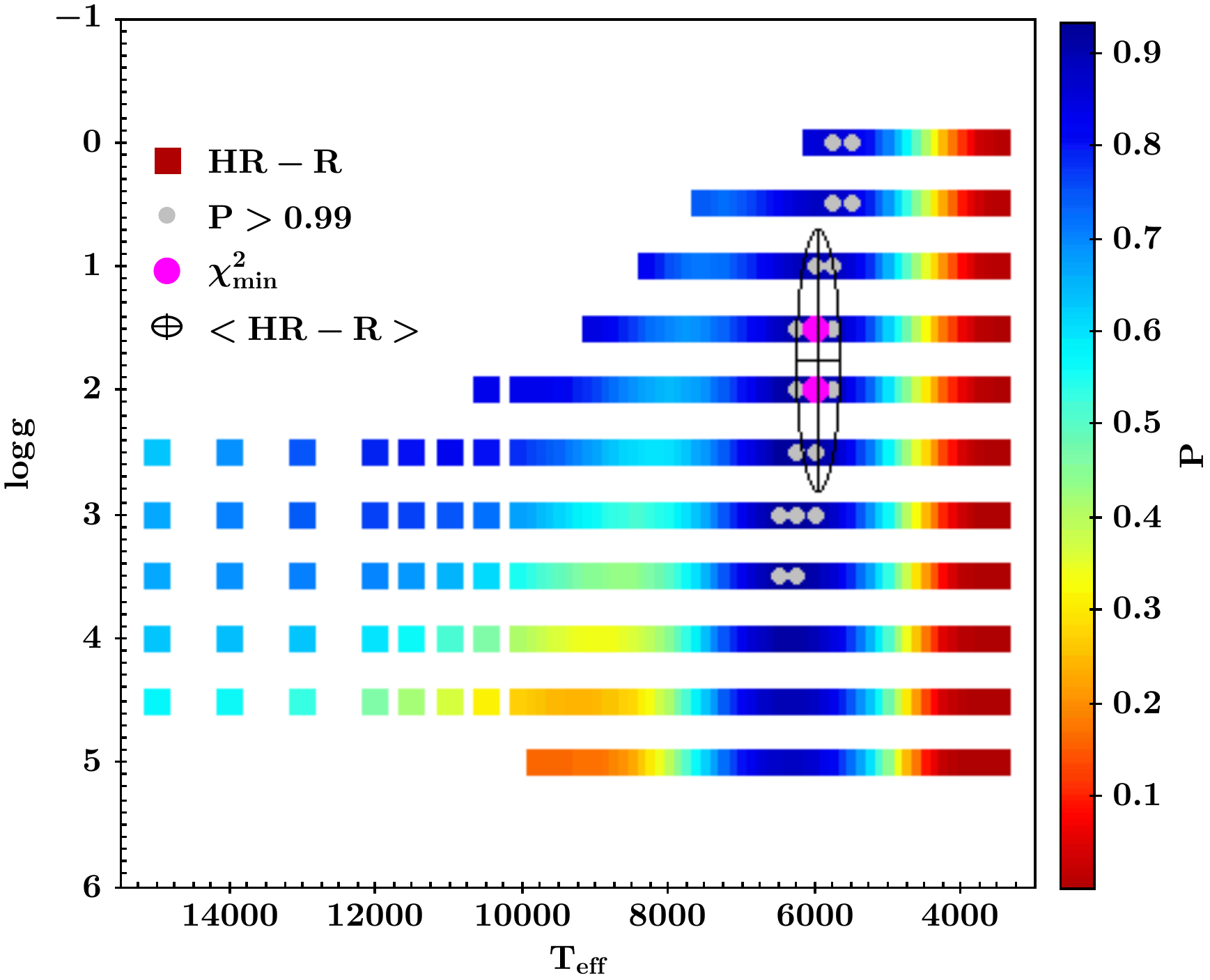}
\includegraphics[width=0.48\textwidth,angle=0]{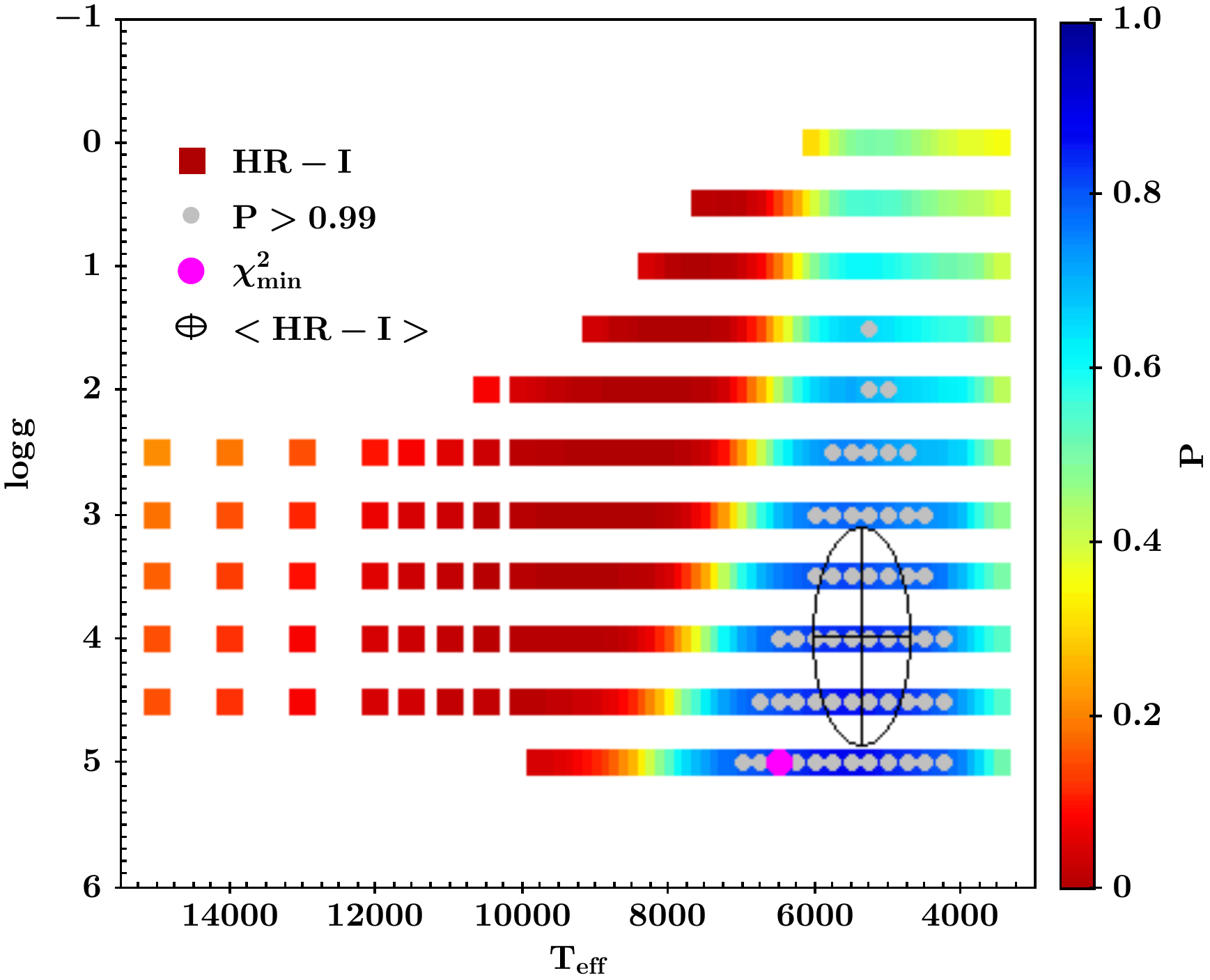}
\includegraphics[width=0.48\textwidth,angle=0]{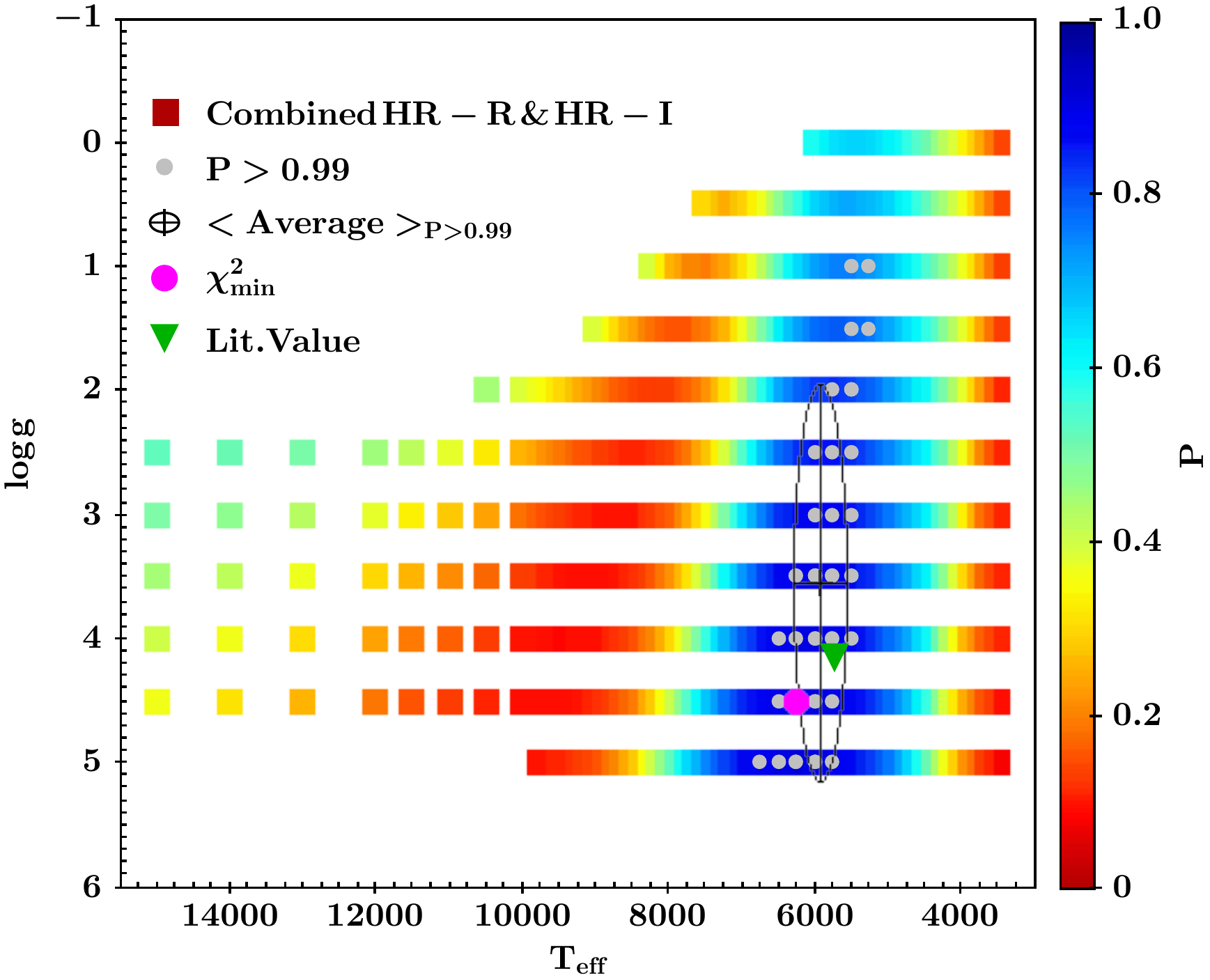}
\caption{Probability maps of models (representing $\rm \log{g}$ in y-axis and $\rm T_{eff}$ in x-axis as in a H-R diagram) fitted to the observed spectrum in HR-R (top panel), HR-I (medium panel), and the combined HR-R and HR-I spectrum of the OT star BD+08 3095 (bottom panel). The colour scale indicates the probability $P$ obtained from the reduced $\chi^{2}_{red}=\chi^{2}-\chi^{2}_{min}$, whose value gives the closeness to the $\chi^{2}_{min}$ model. The selected models with P~$\ge$~0.99 are represented with grey dots over-plotted to the blue region, and the magenta large dot is the model of $\chi^{2}_{min}$ ($P=1$). Over each diagram we have also plot the averaged value, with its dispersion, with a large black cross and an ellipse. The green triangle in the bottom panel indicates the values given by the literature.}
\label{prob}
\end{figure}

This section describes the technique we use to estimate the physical stellar parameters, effective temperature, $\rm T_{eff}$, surface gravity, $\rm \log{g}$, and metallicity, $\rm [M/H]$, of the stars in our sample. We apply a $\chi^{2}$ technique to compare the observed spectra with each of the modelled spectra from a theoretical grid. This method has been proven to be successful in the physical stellar parameters determination, e.g. ULyss code by \citet{kolevaetal09}, MA$\chi$ by \citet{jofreetal10} and {\sc SP\_ACE} by \citet{boecheetgrebel6}. 

We use the models by \citet{mun05}, hereinafter MUN05, for the spectra comparison. MUN05 presents a complete library of synthetic spectra based on Kurucz's covering the 250~--~10500~\AA\ range at  different values of the spectral resolution, from which we have selected the ones at R$=$20000. This theoretical stellar library has 71754 wavelength values, having selected for the fitting the ranges corresponding to the MEGARA gratings HR-R, 6420~--~6800~\AA, and HR-I, 8400~--~8850~\AA. The effective temperature, $\rm T_{eff}$, ranges between 3500 and 47500~K, with a 250~K step for models between 3500 and 10000~K, having less resolution for temperature hotter than 10000~K, leading to a total of 58 models with different values of $\rm T_{eff}$. The gravity ranges from 0.0 to 5.0~dex, with 0.5~dex step, giving 11 models with different values of $\rm \log{g}$ for any value of $\rm T_{eff}$, except for the hottest models that have a smaller number of models with different $\rm \log{g}$. Finally, the abundance varies from $\rm [M/H]$~=~-2.5 to $+0.5$~dex, with $+0.5$~dex, step, giving a total of 7 abundance values. The total number of models is 2665, with an average number of $\sim$~380 different models for each metallicity. 

The $\chi^{2}$ technique uses the flux-normalised spectra of both, observations and models, and computes the differences at any given wavelength on a certain range with the well--known equation:
\begin{equation}
\label{chi}
\chi^{2}=\sum_{i=1}^{nl}\frac{(F_{mod}(\lambda)-F_{obs}(\lambda))^{2}}{\sigma^{2}},
\end{equation}
where $F_{mod}$ and $F_{obs}$ are the model and the observed normalised fluxes respectively, $nl$ is the number of available wavelengths and $\sigma$ is the flux error, which we have calculated with the continuum SNR averaged over the whole spectral range. We have obtained the modelled spectra with the same spectral sampling than the observed ones by polynomial interpolation, and then we have computed the $\chi^{2}$ (Eq.~\ref{chi}) by comparing the observed spectrum with each single modelled spectra in the theoretical grid. Then, we have assigned to each observed star the physical stellar parameters corresponding to the model that gives the minimum $\chi^{2}$, and we have labelled these values $\rm T_{eff}$, $\rm \log{g}$ and $\rm [M/H]$. As an example of these fitting results, Figure~\ref{fits-detail} shows the detailed spectra of the $\chi^{2}_{min}$ model in three consecutive spectral ranges within the HR-I setup for a cool star (left panel) and a hot star (right panel) of the COM sub-sample.

When the $\chi^{2}$ technique is used, a likelihood or confidence level, $P$ for a $\chi^{2}$ distribution is obtained, given by:
\begin{equation}
\label{p}
P=1-\alpha (\chi^{2}< x)=1-\int_{0}^{x}{\chi^{2}_{k}du} = \\
1-\int_{0}^{x}{\frac{u^{k/2}e^{-u/2}}{2^{k/2}\Gamma(k/2)}du}.
\end{equation}

The minimum $\chi^{2}$ technique described above gives the most likely model, obtaining the most probable stellar parameters. However, when analysing the table with the $\chi^{2}$ values obtained from the fitting of each single model of the theoretical grid to a given observed star, we usually find several models with similar likelihood, which, therefore, still provide a good fit to the observational data. We have performed an analysis of the $\chi^{2}$ results of all models of each observed spectrum, to find those with likelihood similar to the corresponding one for $\chi^{2}_{\rm min}$. We, thus,  derive the likelihood contours around the best-fit model that define the parameter region offering results within a given confidence level, $L_{sel}$. To select models within a region $R_{\nu,\alpha}$ \citep{avni76}:
\begin{equation}
    \chi^{2}-\chi^{2}_{min}\le \Delta(\nu,\alpha),
\end{equation}
it is necessary to define, $\nu$, which is the number of free parameters (3 in our case), and $\alpha$, the significance level. We have considered all the models with $\alpha~=~0.01$, implying differences as: 
\begin{equation}
\chi^{2}-\chi^{2}_{\rm min} \le \Delta(3,0.01)=0.115
\label{condition}
\end{equation}

\begin{figure*}
\includegraphics[width=0.30\textwidth,angle=0]{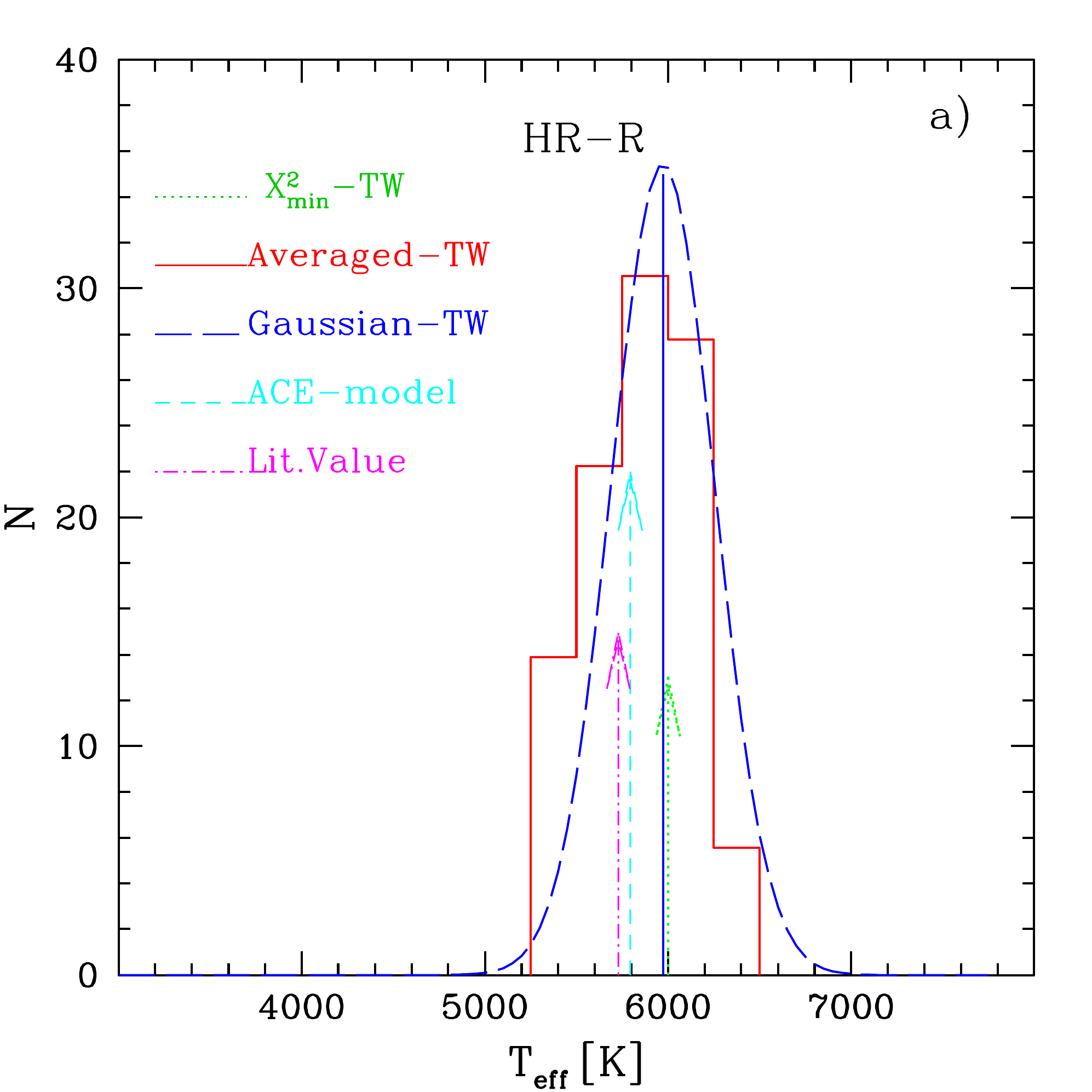}
\includegraphics[width=0.30\textwidth,angle=0]{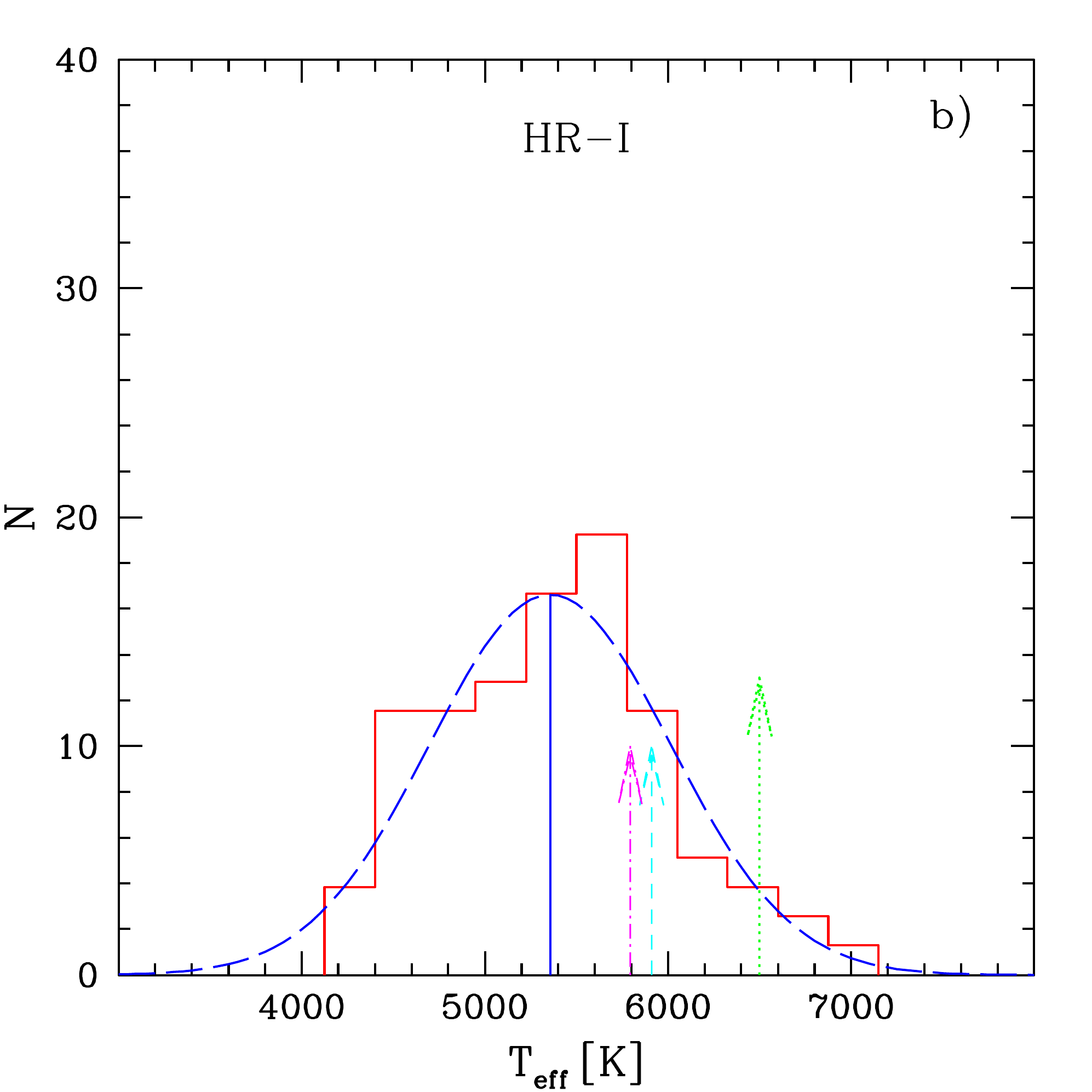}
\includegraphics[width=0.30\textwidth,angle=0]{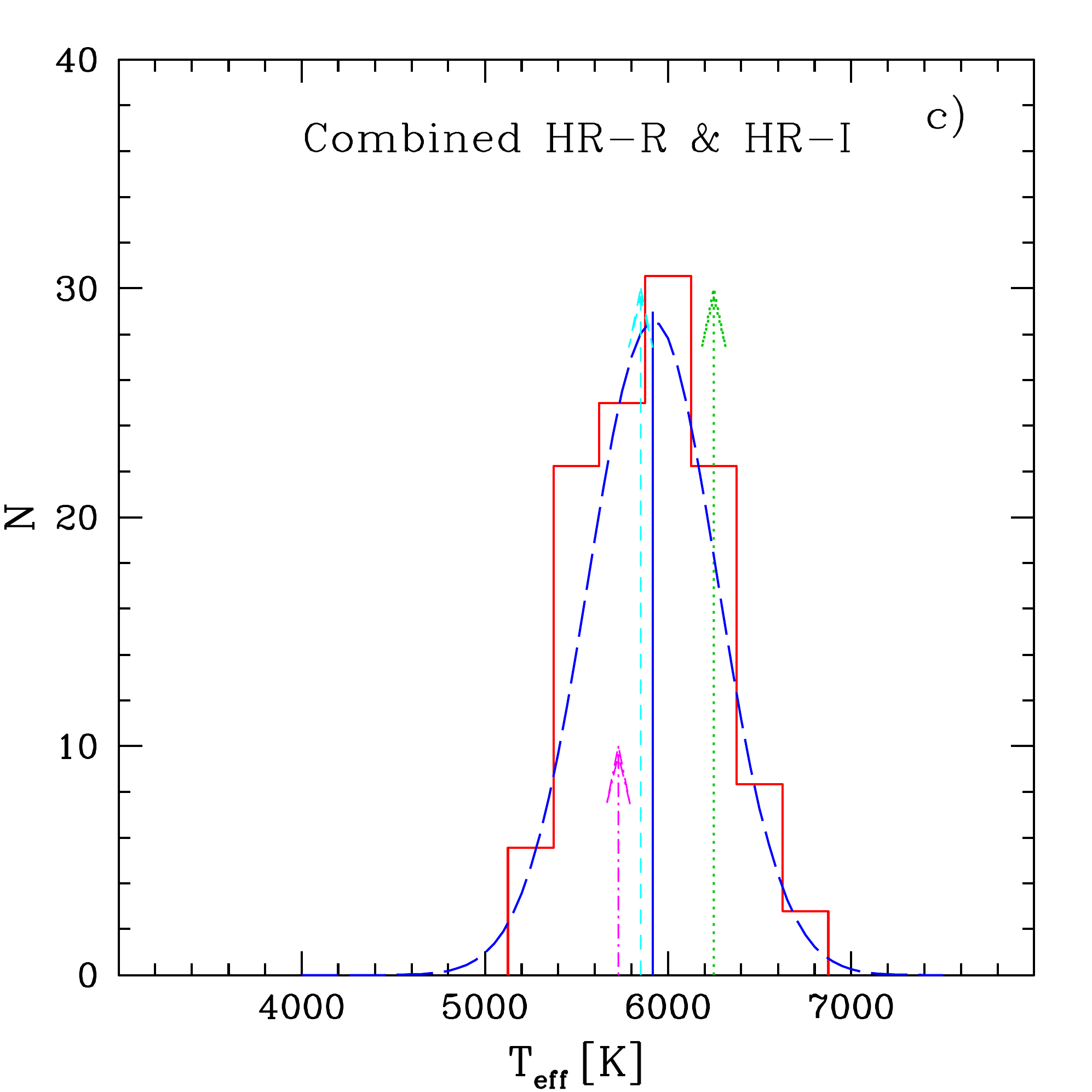}
\includegraphics[width=0.30\textwidth,angle=0]{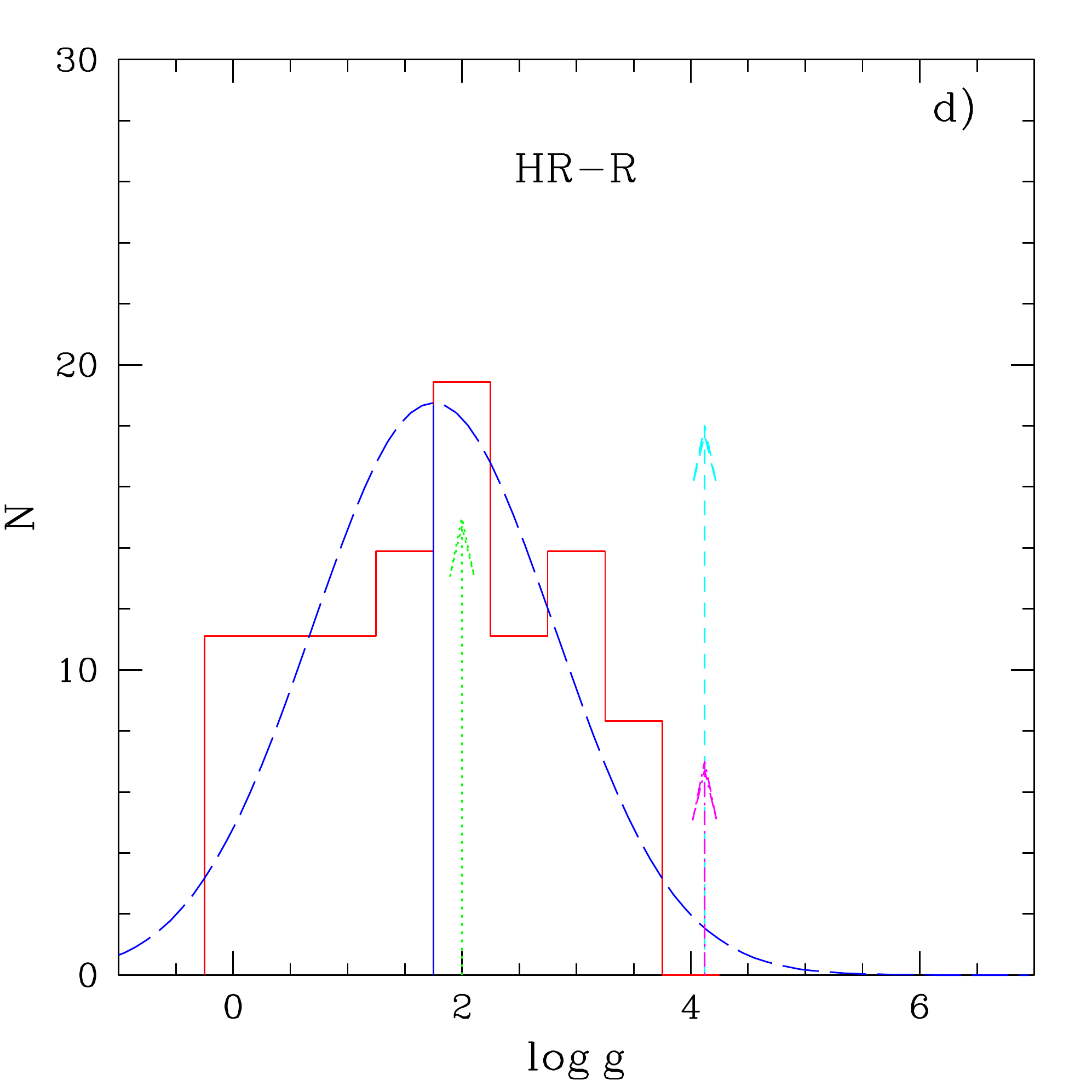}
\includegraphics[width=0.30\textwidth,angle=0]{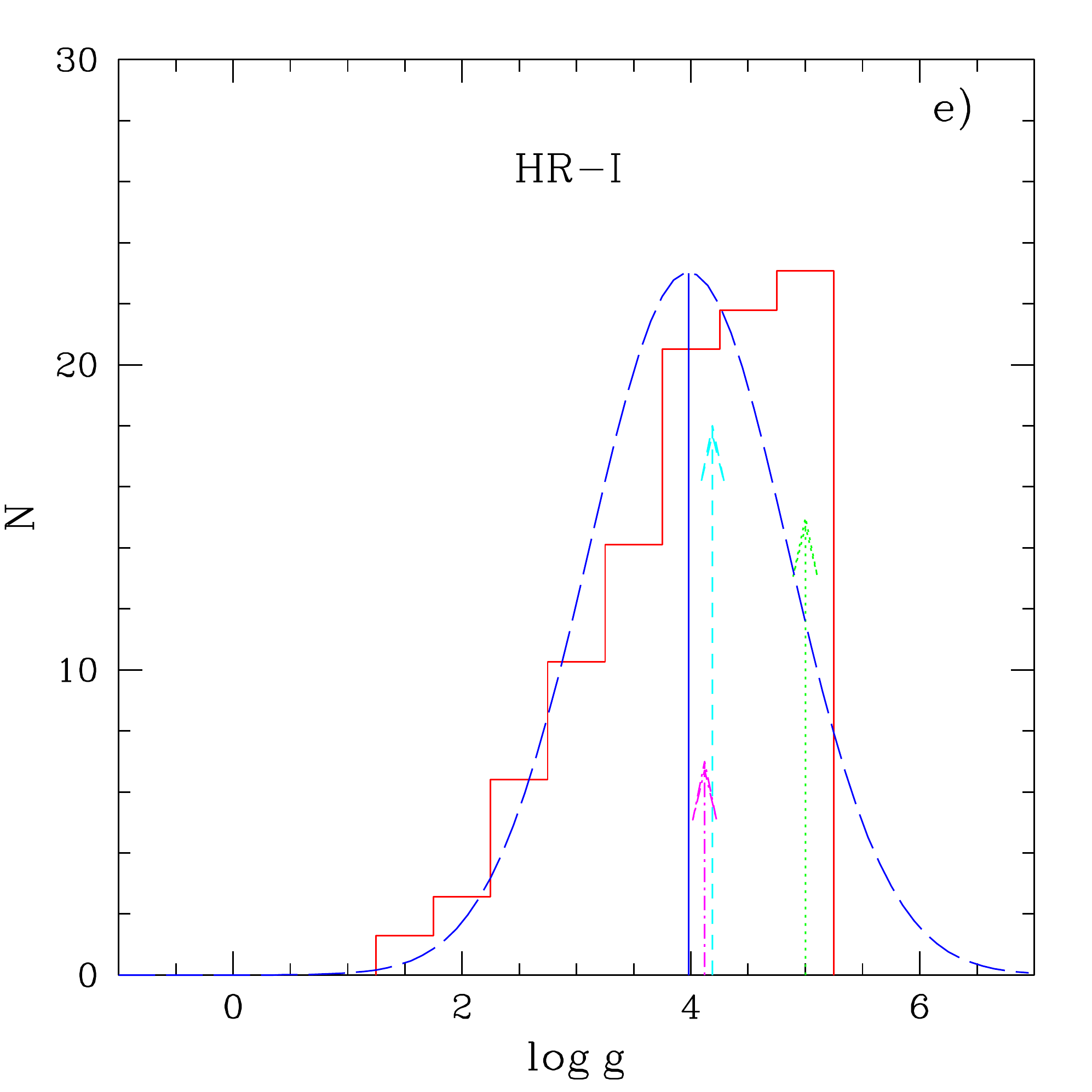}
\includegraphics[width=0.30\textwidth,angle=0]{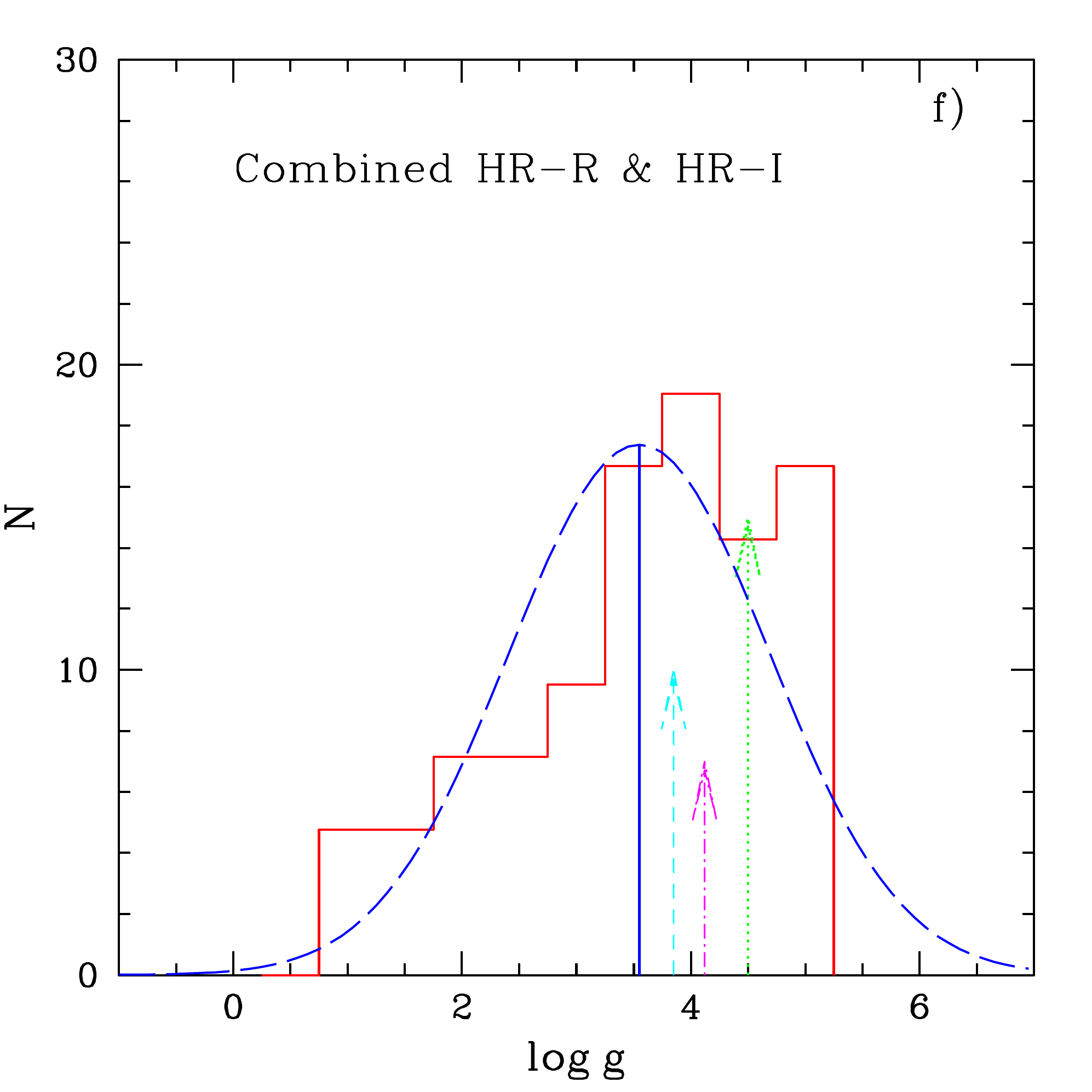}
\includegraphics[width=0.30\textwidth,angle=0]{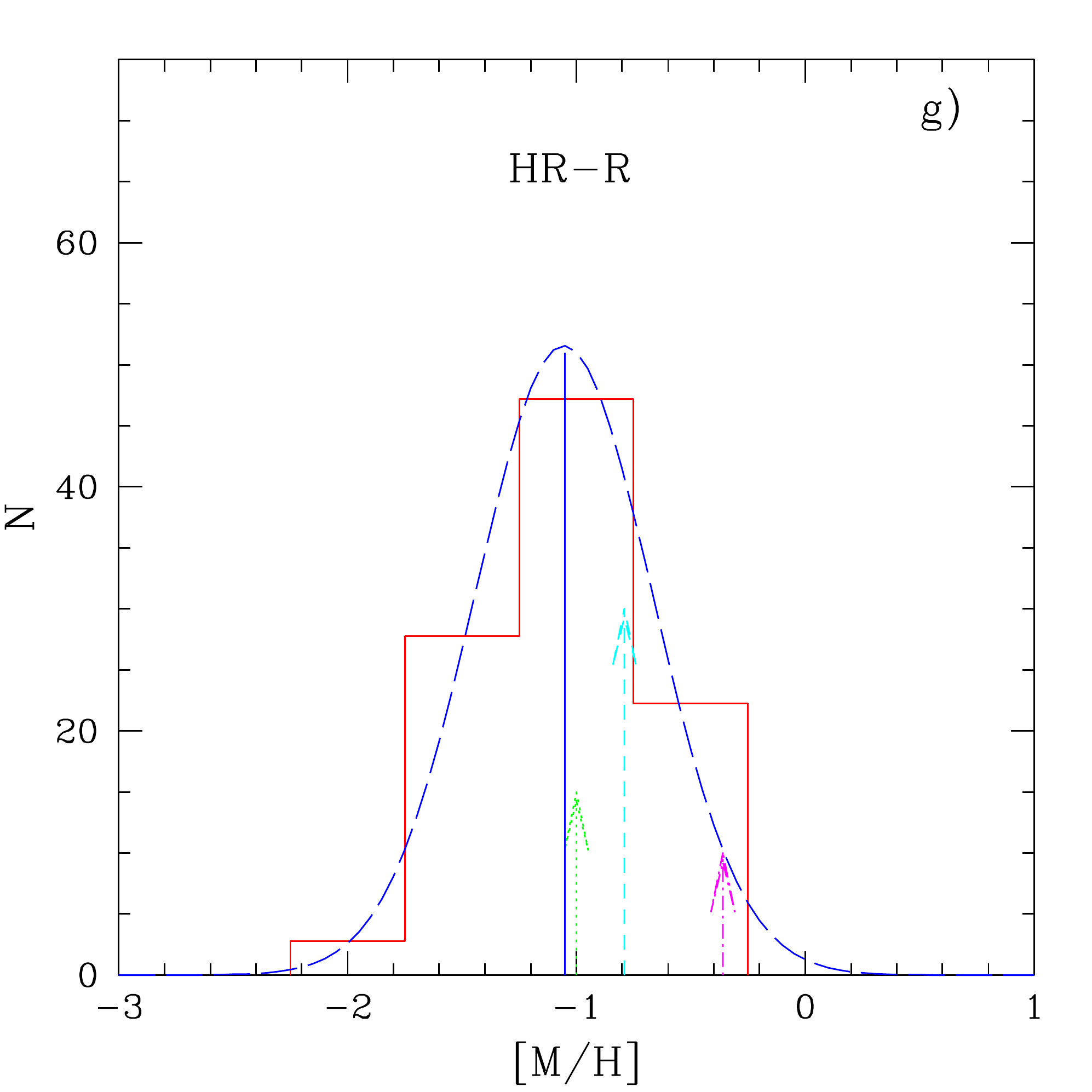}
\includegraphics[width=0.30\textwidth,angle=0]{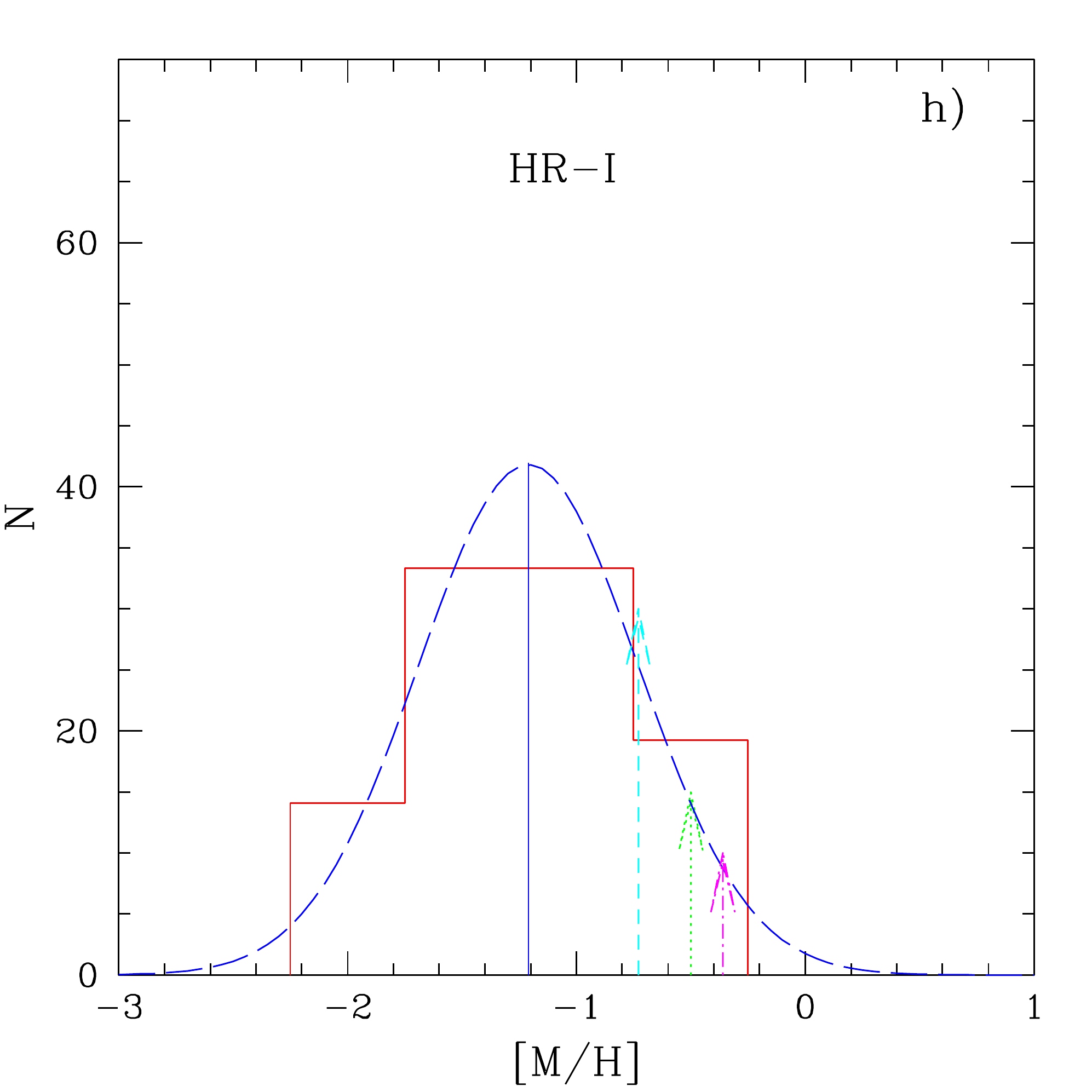}
\includegraphics[width=0.30\textwidth,angle=0]{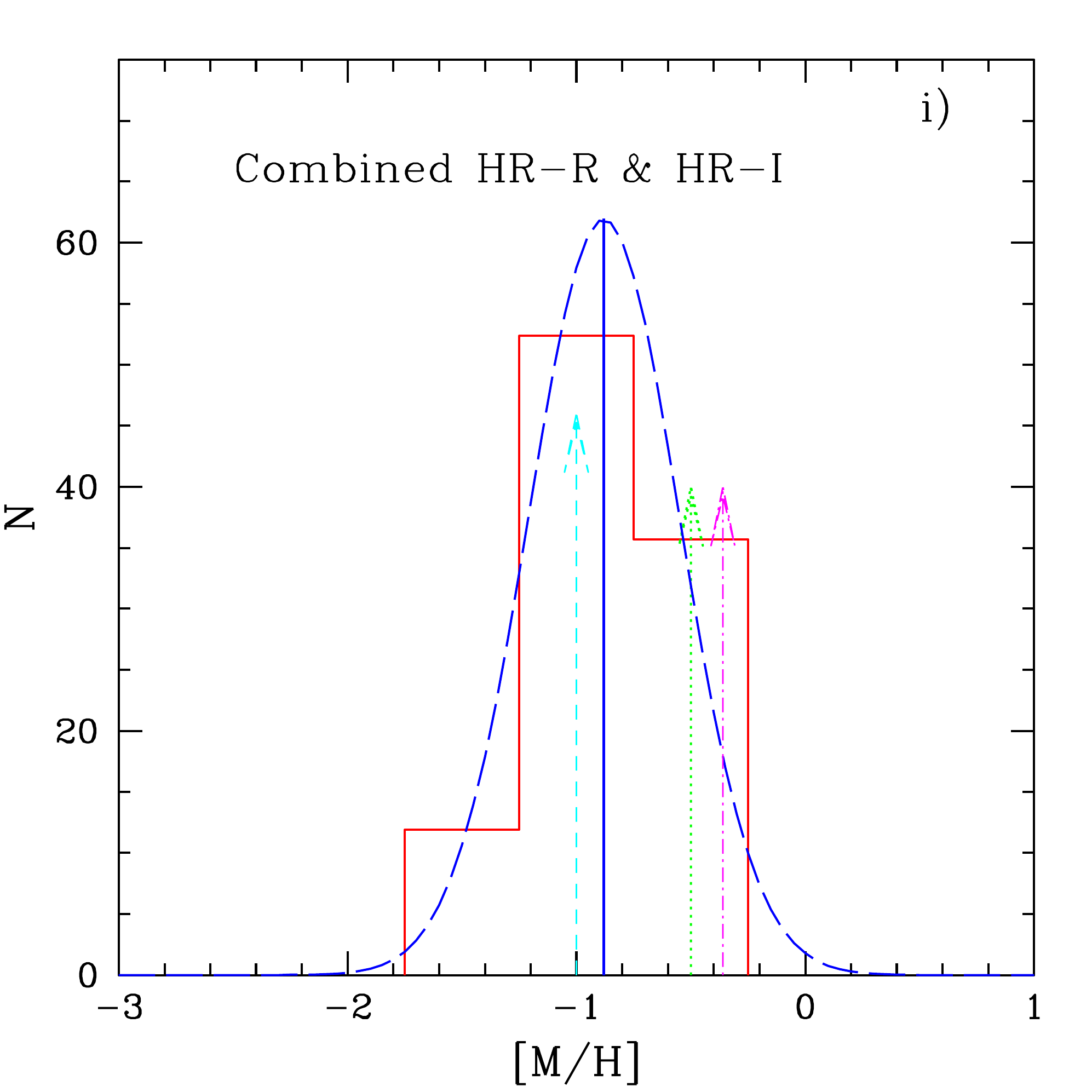}
\caption{Histograms of $\rm T_{eff}$, $\rm \log{g}$ and $\rm [M/H]$, at top, middle and bottom panels, for the selected models (P~$>$~0.99) fitted to HR-R (left panels), HR-I (center panels) and both combined into a single spectrum (right panels) for star BD$+$08~3095. Over each histogram in red, we add a Gaussian in blue dashed line which gives the averaged value, marked as a blue solid line. The value given by the $\chi^{2}_{min}$ model is plotted with a green dotted line arrow, the value from the literature is the magenta arrow and the cyan dashed line arrow indicates the result obtained when applying the {\sc SP\_ACE} model to the observed spectrum.}
\label{hist}
\end{figure*}

\begin{table*}
\scriptsize
\caption{Stellar parameters for the 21 commissioning star derived from the fitting with the theoretical models for the HR-I setup, and from the literature. Table description in subsection~\ref{com-res}.}
\label{parameters-com}
\begin{tabular}{lcrrccrccccccc}
\hline
Star & $\chi^{2}_{min}$ & P$_{\rm max}$ &  \multicolumn{3}{c}{Results with minimum $\chi^{2}$} & $N$ & \multicolumn{3}{c}{Average results and dispersion with $N$ models} & \multicolumn{3}{c}{Data from the literature} & Ref.
\\
Name & & & T$_{\rm eff}$ & $\log{g}$ & [M/H] & & $\langle \rm T_{ eff} \rangle$ & $\langle\log{g}\rangle$ & $\langle\rm [M/H]\rangle$ & T$_{\rm eff,lit}$ & $\log{g}_{\rm lit}$ & $[M/H]_{\rm lit}$ &  \\ 
\hline 
Schulte~9	&	0.06	&	0.996	&	37500	&	4.5	&	-2.5	&	200	&	32808	$\pm$	5320	&	4.41	$\pm$	0.59	&	-0.92	$\pm$	0.98	&	38520	&	3.57 	&		&	 (8. 9)	\\
HD~192281	&	0.13	&	0.988	&	40000	&	4.5	&	-2.5	&	118	&	35725	$\pm$	4839	&	4.60	$\pm$	0.45	&	-1.01	$\pm$	0.99	&	40800	&	3.73 	&		&	(7)	\\
BD~254655	&	0.08	&	0.994	&	40000	&	5.0	&	-2.5	&	134	&	35295	$\pm$	4583	&	4.67	$\pm$	0.37	&	-1.01	$\pm$	0.98	&		&		&		&		\\
HD~218915	&	0.06	&	0.996	&	24000	&	3.0	&	0.5	&	249	&	31347	$\pm$	5717	&	4.28	$\pm$	0.65	&	-0.96	$\pm$	0.99	&	31100	&	3.21  	&		&	(7)	\\
BD+40~4032	&	0.05	&	0.997	&	28000	&	3.5	&	0.0	&	305	&	29543	$\pm$	5199	&	4.21	$\pm$	0.65	&	-0.97	$\pm$	0.99	&	33813	&	3.13 	&	 $\phantom{-}$0.0 	&	(5)	\\
BD+33~2642	&	0.17	&	0.983	&	17000	&	2.5	&	0.0	&	145	&	21131	$\pm$	3797	&	3.13	$\pm$	0.37	&	-0.96	$\pm$	0.99	&		&		&		&		\\
HD~220575	&	0.23	&	0.974	&	18000	&	3.5	&	-2.5	&	219	&	18183	$\pm$	2936	&	3.36	$\pm$	0.51	&	-1.03	$\pm$	1.01	&	12293	&	3.70  	&	 $\phantom{-}$0.50 	&	(1)	\\
	&		&				&		&		&		&				&				&		&		&	12241 $\pm$ 402 	&	4.09 $\pm$ 0.13 	& 0.27 $\pm$ 0.15 	&	(4)	\\
BD+42~3227	&	0.07	&	0.995	&	28000	&	3.5	&	-0.5	&	349	&	28117	$\pm$	5177	&	4.12	$\pm$	0.69	&	-0.98	$\pm$	0.99	&		&		&		&		\\
BD+12~2237	&	0.29	&	0.962	&	28000	&	3.5	&	-2.5	&	286	&	25615	$\pm$	4735	&	3.77	$\pm$	0.59	&	-0.96	$\pm$	0.99	&		&		&		&		\\
BD+17~4708	&	0.02	&	0.999	&	5750	&	4.0	&	-1.5	&	285	&	5244	$\pm$	776	&	3.17	$\pm$	1.41	&	-1.91	$\pm$	0.55	&	5993	&	3.94 	&	 $-$1.65 	&	(3)	\\
HD~026630	&	1.03	&	0.794	&	5500	&	0.5	&	-0.5	&	14	&	5357	$\pm$	306	&	0.50	$\pm$	0.55	&	-0.54	$\pm$	0.31	&	5643	&	1.54  	&	$\phantom{-}$0.10	&	(6)	\\
HD~216219	&	0.05	&	0.997	&	5500	&	3.0	&	-0.5	&	179	&	5159	$\pm$	654	&	3.00	$\pm$	1.48	&	-0.93	$\pm$	0.56	&		&		&		&		\\
HD~011544	&	0.42	&	0.935	&	5000	&	1.0	&	-0.5	&	22	&	4943	$\pm$	336	&	1.02	$\pm$	0.75	&	-0.48	$\pm$	0.39	&		&		&		&		\\
HD~019445	&	0.07	&	0.996	&	5000	&	5.0	&	-2.5	&	238	&	5283	$\pm$	730	&	3.20	$\pm$	1.40	&	-2.00	$\pm$	0.50	&	5929	&	4.36  	&	$-$2.02 	&	(2)	\\
HD~020123	&	0.48	&	0.924	&	5000	&	1.5	&	-0.5	&	42	&	5101	$\pm$	445	&	1.43	$\pm$	0.94	&	-0.54	$\pm$	0.45	&	4901	&		&		&	(6)	\\
HD~224458	&	0.19	&	0.979	&	5000	&	2.0	&	-0.5	&	104	&	4892	$\pm$	558	&	2.40	$\pm$	1.43	&	-0.63	$\pm$	0.54	&	4722	&	2.20  	&	$-$0.50 	&	(1)	\\
	&		&				&		&		&		&				&				&				&	4819 $\pm$ 67 	&	2.29 $\pm$ 0.17 	&	 $-$0.44 $\pm$ 0.08	&	(4)	\\
HD~220954	&	0.22	&	0.974	&	4750	&	1.5	&	-0.5	&	49	&	4500	$\pm$	430	&	1.83	$\pm$	1.15	&	-0.51	$\pm$	0.47	&	4664	&	2.37 	&	 $-$0.10 	&	(4)	\\
	&		&		&		&		&		&		&				&				&				&	4731 $\pm$ 46 	&	2.61 $\pm$ 0.11 	&	0.07 $\pm$ 0.05 	&	(4)	\\
HD~025975	&	0.18	&	0.981	&	4750	&	3.0	&	-0.5	&	114	&	4752	$\pm$	563	&	3.11	$\pm$	1.35	&	-0.52	$\pm$	0.49	&	4941	&	3.40  	&	$-$0.20 	&	(6)	\\
HD~027971	&	0.29	&	0.962	&	4750	&	2.0	&	-0.5	&	62	&	4746	$\pm$	491	&	2.40	$\pm$	1.33	&	-0.45	$\pm$	0.52	&	4860	&	2.82  	&	$-$0.08 	&	(6)	\\
HD~174350	&	1.60	&	0.659	&	4750	&	0.0	&	-1.0	&	21	&	5190	$\pm$	353	&	0.81	$\pm$	0.72	&	-0.48	$\pm$	0.37	&	4537	&	2.56  	&	$-$0.02 	&	(2)	\\
HD~185622	&	0.24	&	0.972	&	5000	&	1.5	&	-0.5	&	42	&	4958	$\pm$	431	&	1.60	$\pm$	1.08	&	-0.55	$\pm$	0.47	&		&		&		&		\\
\hline
\end{tabular}
\footnotesize

\vspace*{0.2cm}
(1) \citet{cen01b}; (2) \citet{chen14}; (3) \citet{pru04}; (4) \citet{pru11}; (5) \citet{cam03}; (6) \citet{val04}; (7) Holgado et al 2019 (in prep); (8) \citet{blomme13}; (9) \citet{martins05}.
\normalsize
\end{table*}

\begin{figure*}
\includegraphics[width=0.24\textwidth,angle=-90]{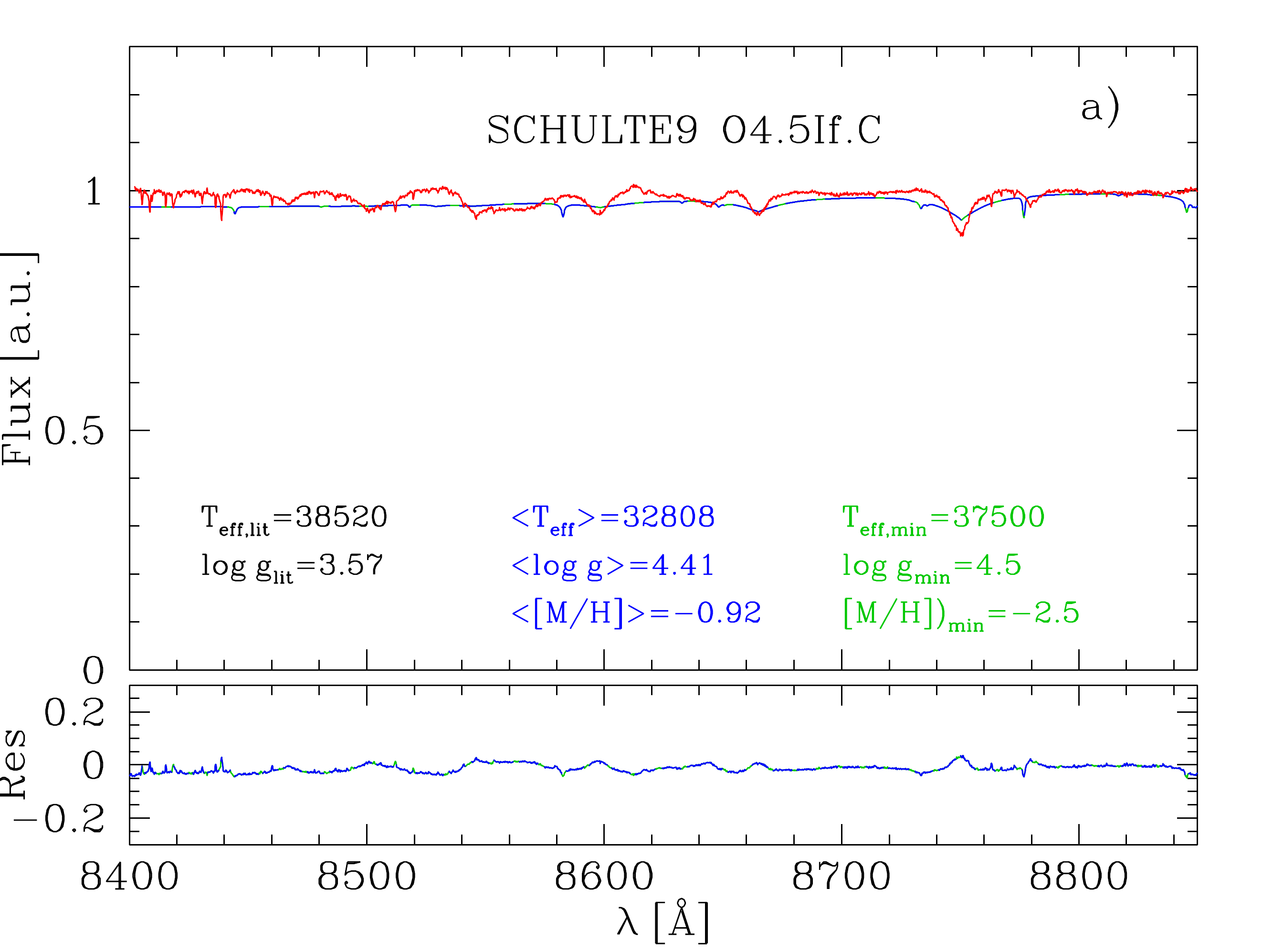}
\includegraphics[width=0.24\textwidth,angle=-90]{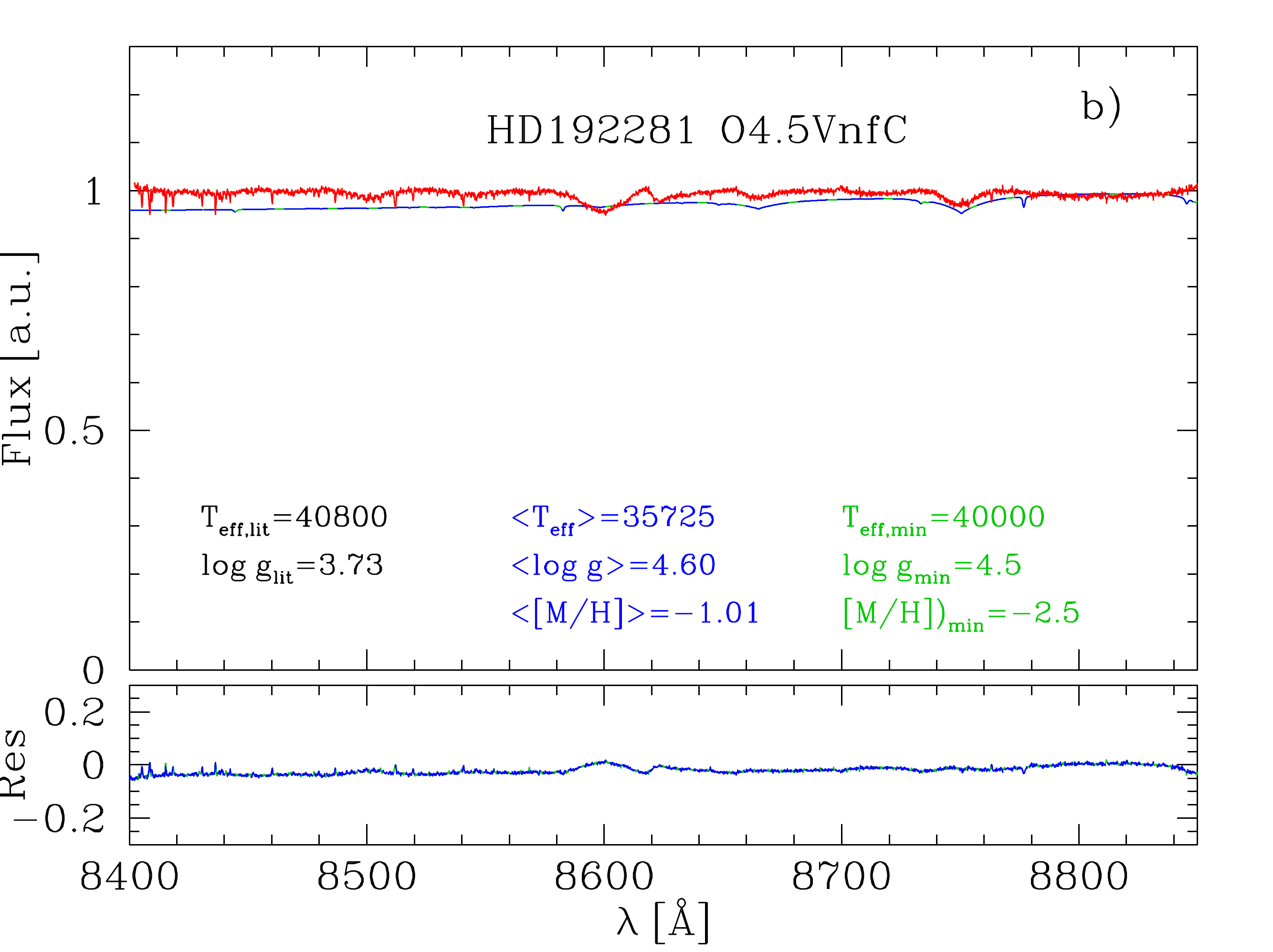}
\includegraphics[width=0.24\textwidth,angle=-90]{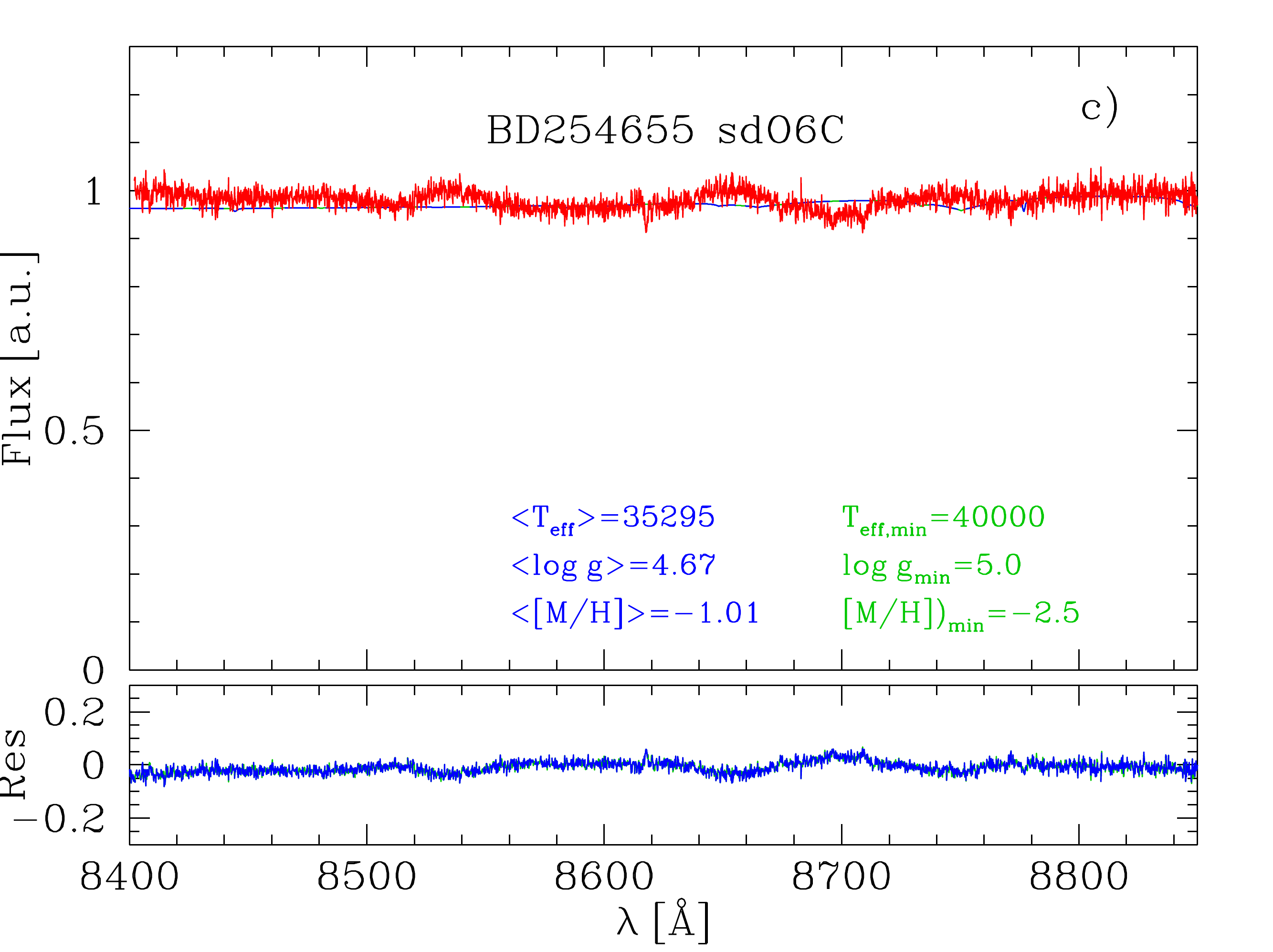}
\includegraphics[width=0.24\textwidth,angle=-90]{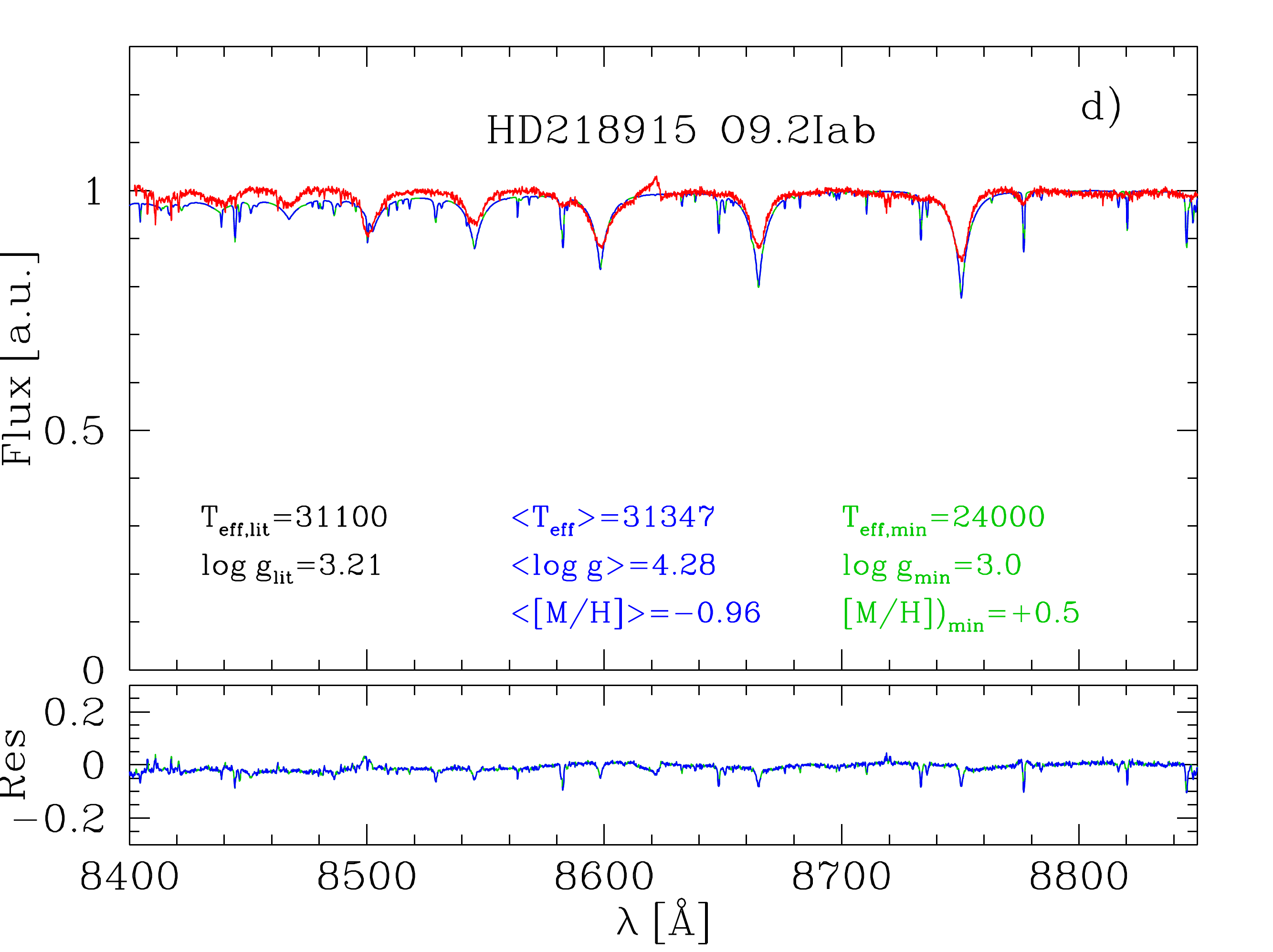}
\includegraphics[width=0.24\textwidth,angle=-90]{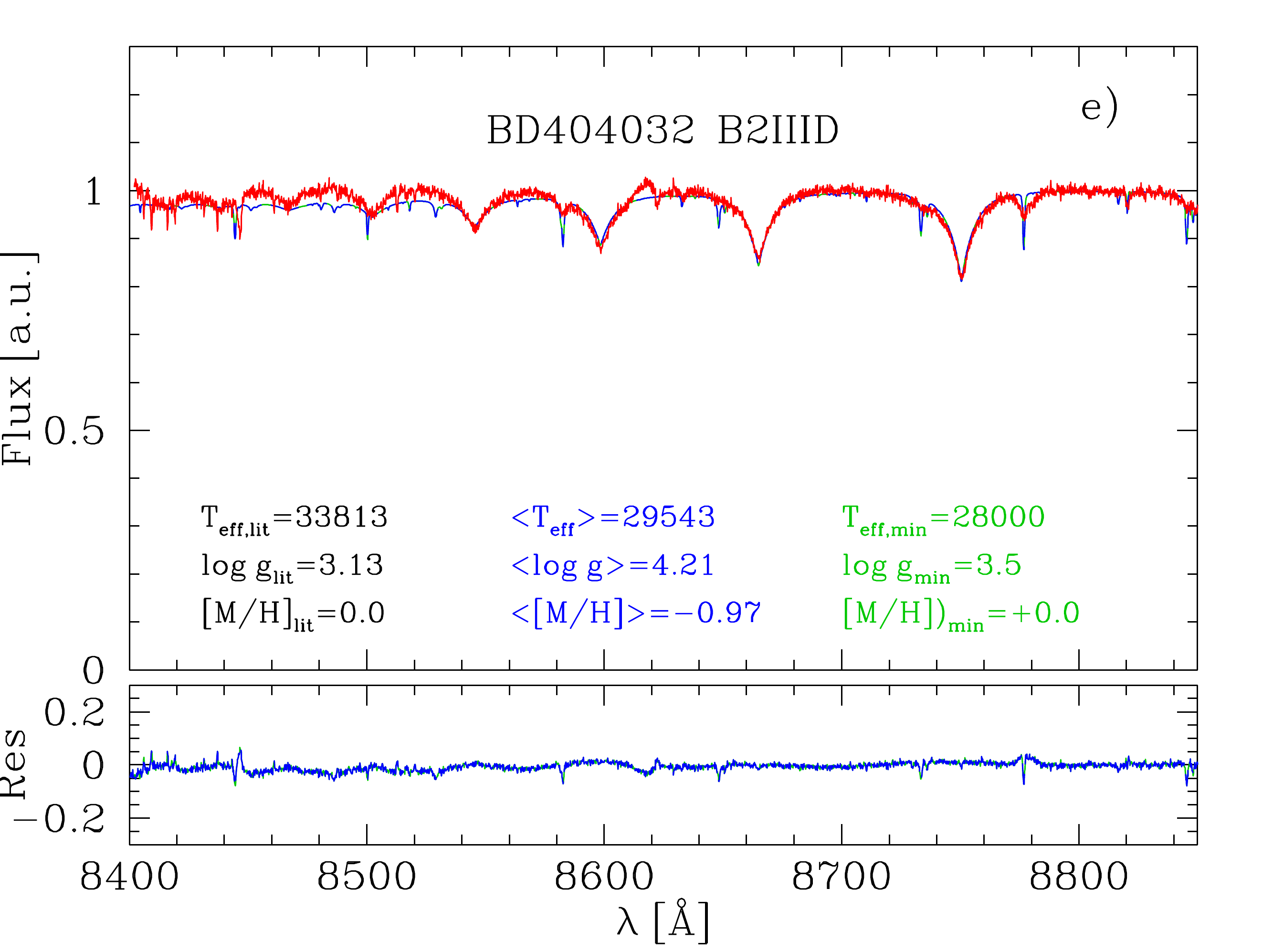}
\includegraphics[width=0.24\textwidth,angle=-90]{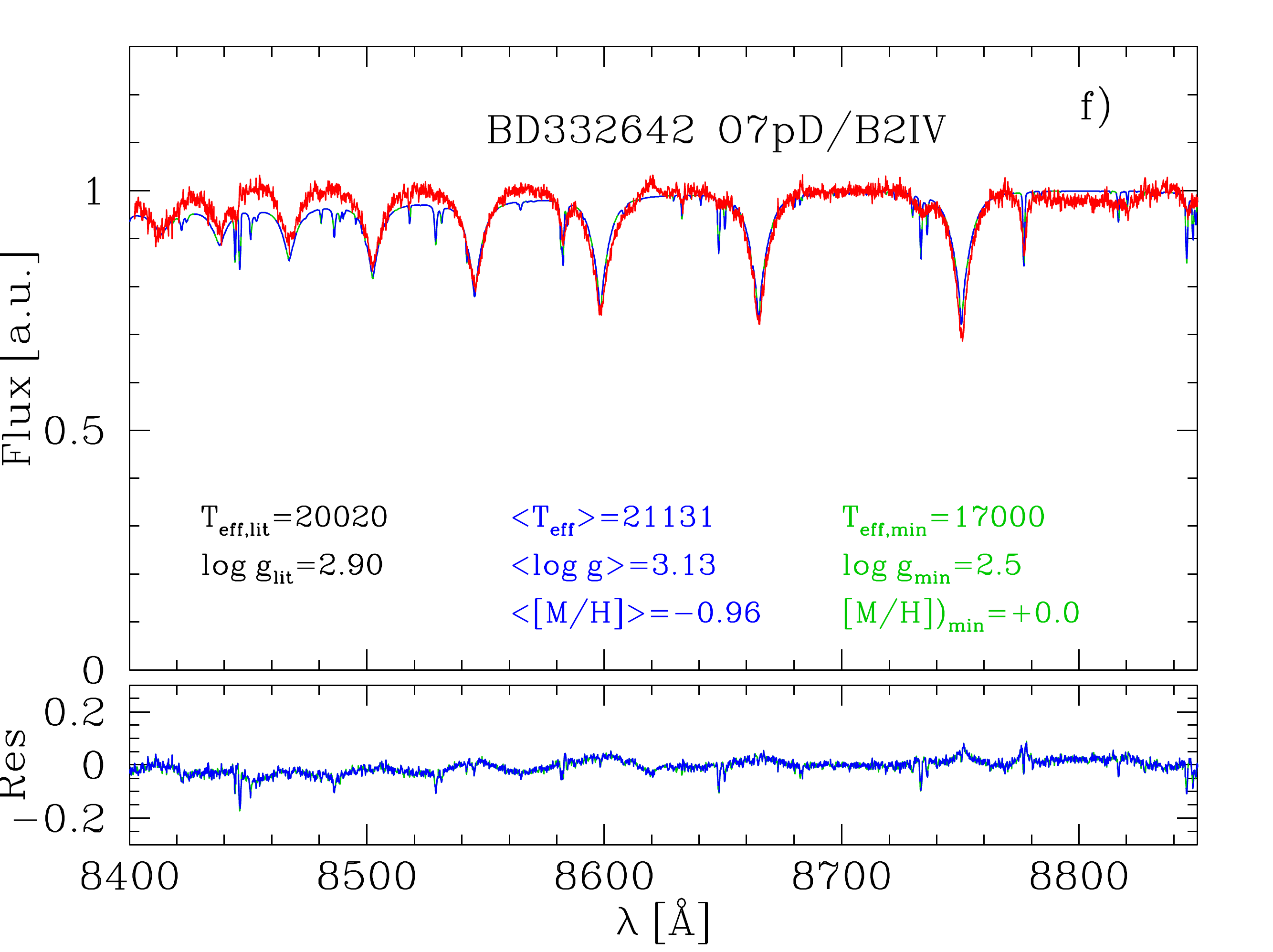}
\includegraphics[width=0.24\textwidth,angle=-90]{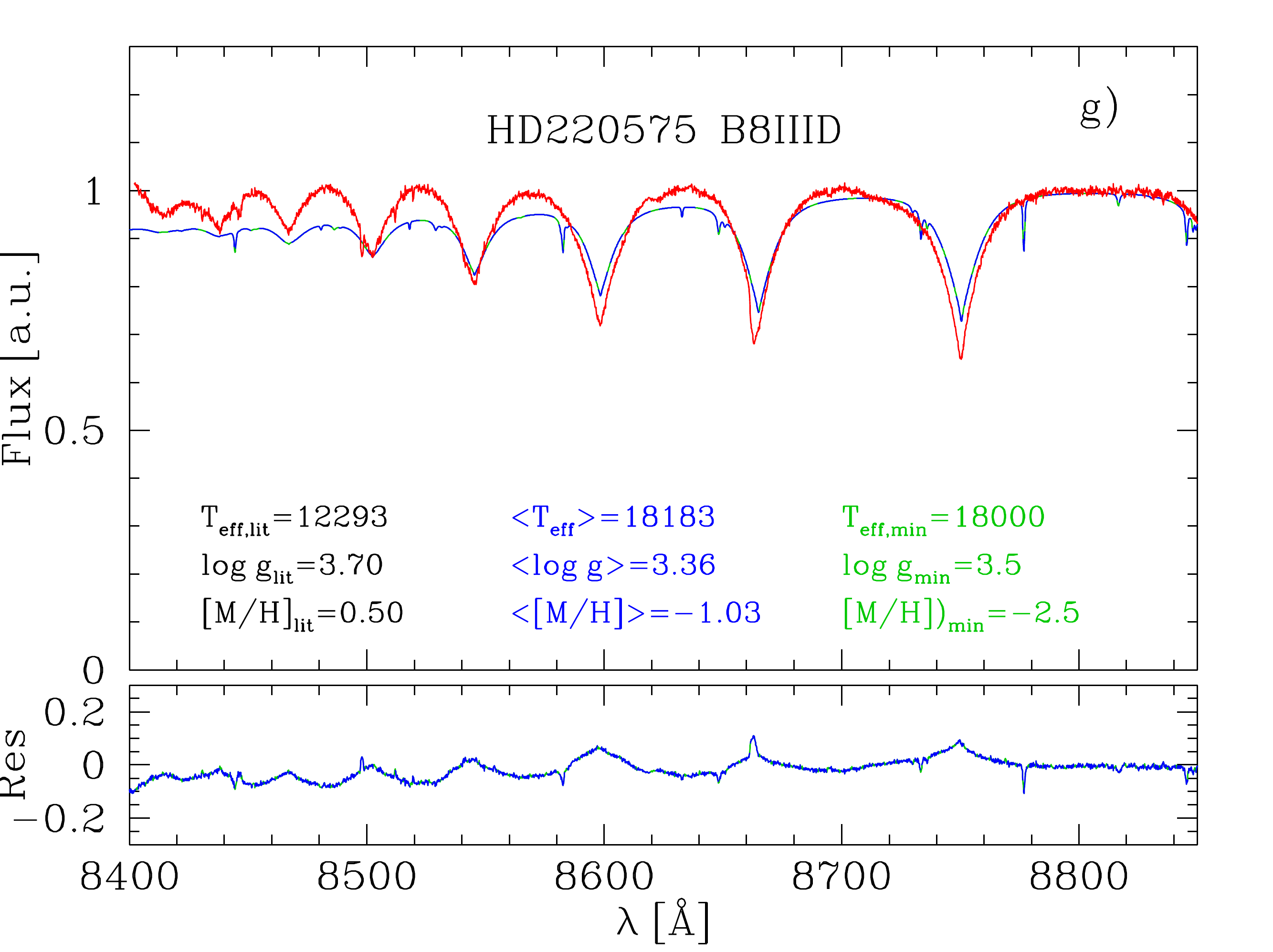}
\includegraphics[width=0.24\textwidth,angle=-90]{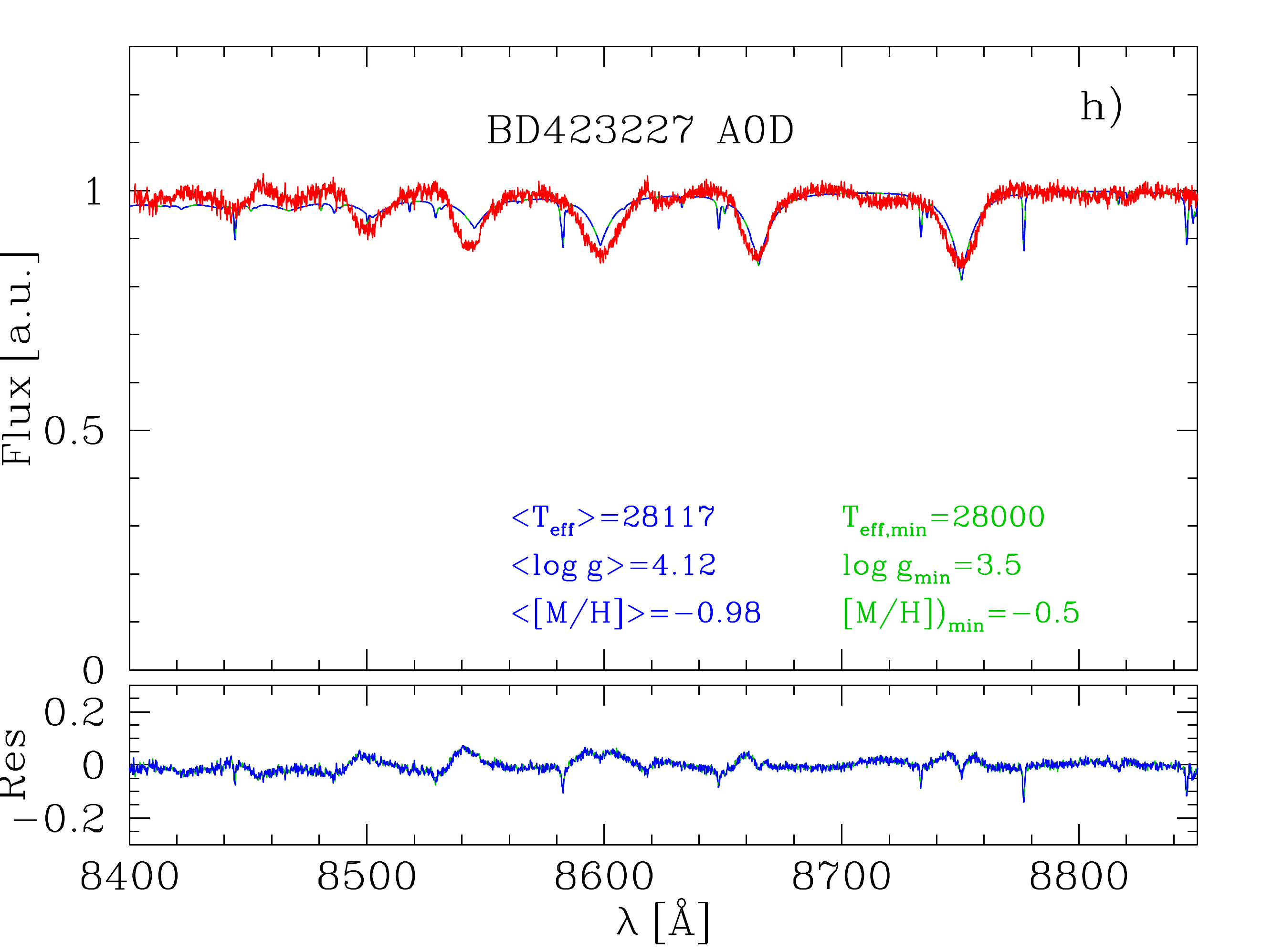}
\includegraphics[width=0.24\textwidth,angle=-90]{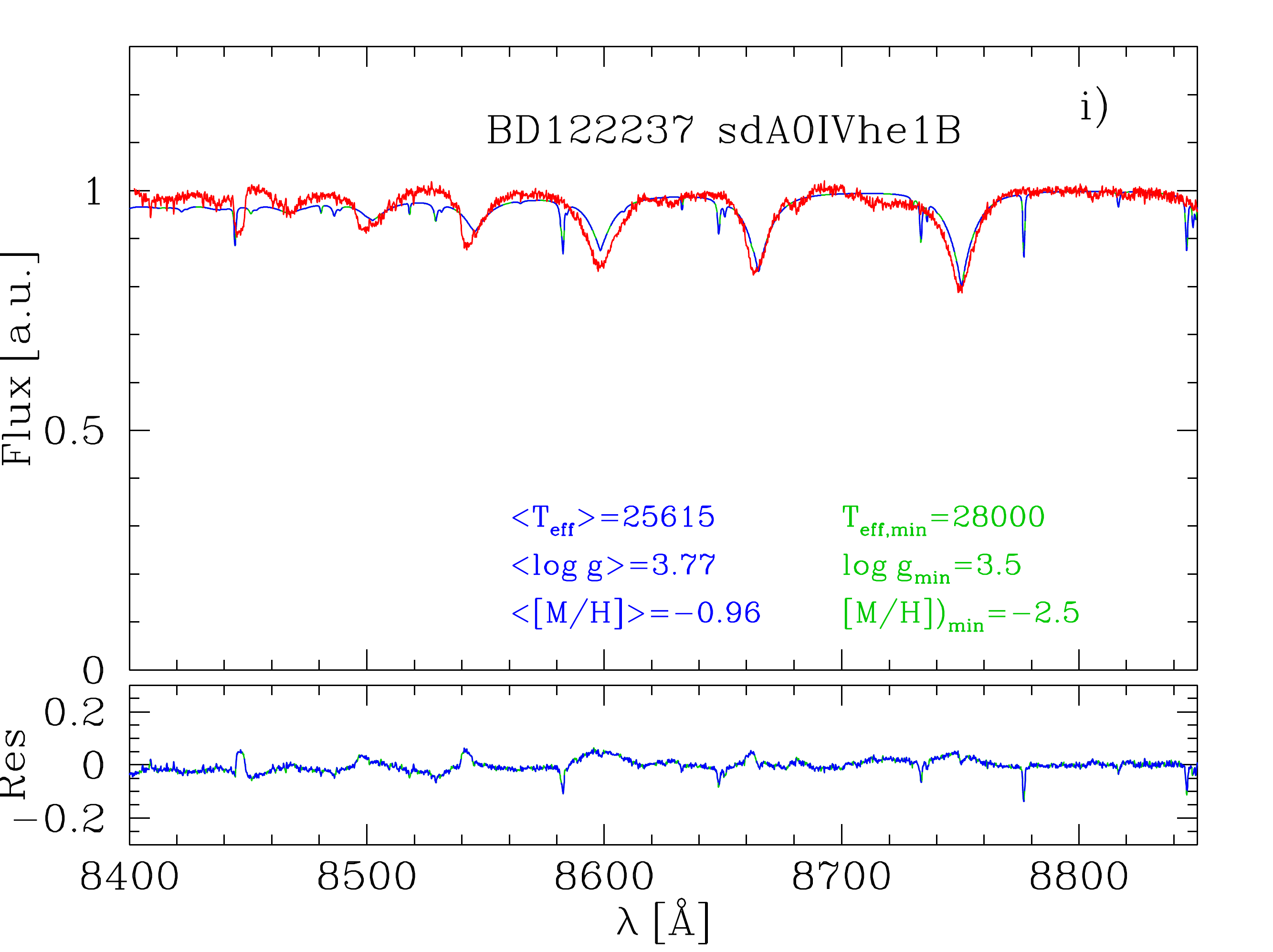}
\caption{The observed spectrum of the commissioning hot stars with the best-fitting MUN05 models over-plotted. The top panel shows the observed spectrum (red solid line), the averaged fitted model (blue long-dashed line) and the $\chi^{2}_{min}$ model (green short-dashed line). The bottom panel shows the residuals as Flux$_{obs}$ - Flux$_{mod}$ (blue and green lines as the corresponding fitted models).The stellar parameters derived from the literature (black, whenever available), from the averaged  model (blue) and from  $\chi^{2}_{min}$ model (green) are shown in each panel.}
\label{fits-hot}
\end{figure*}

\begin{figure*}
\includegraphics[width=0.24\textwidth,angle=-90]{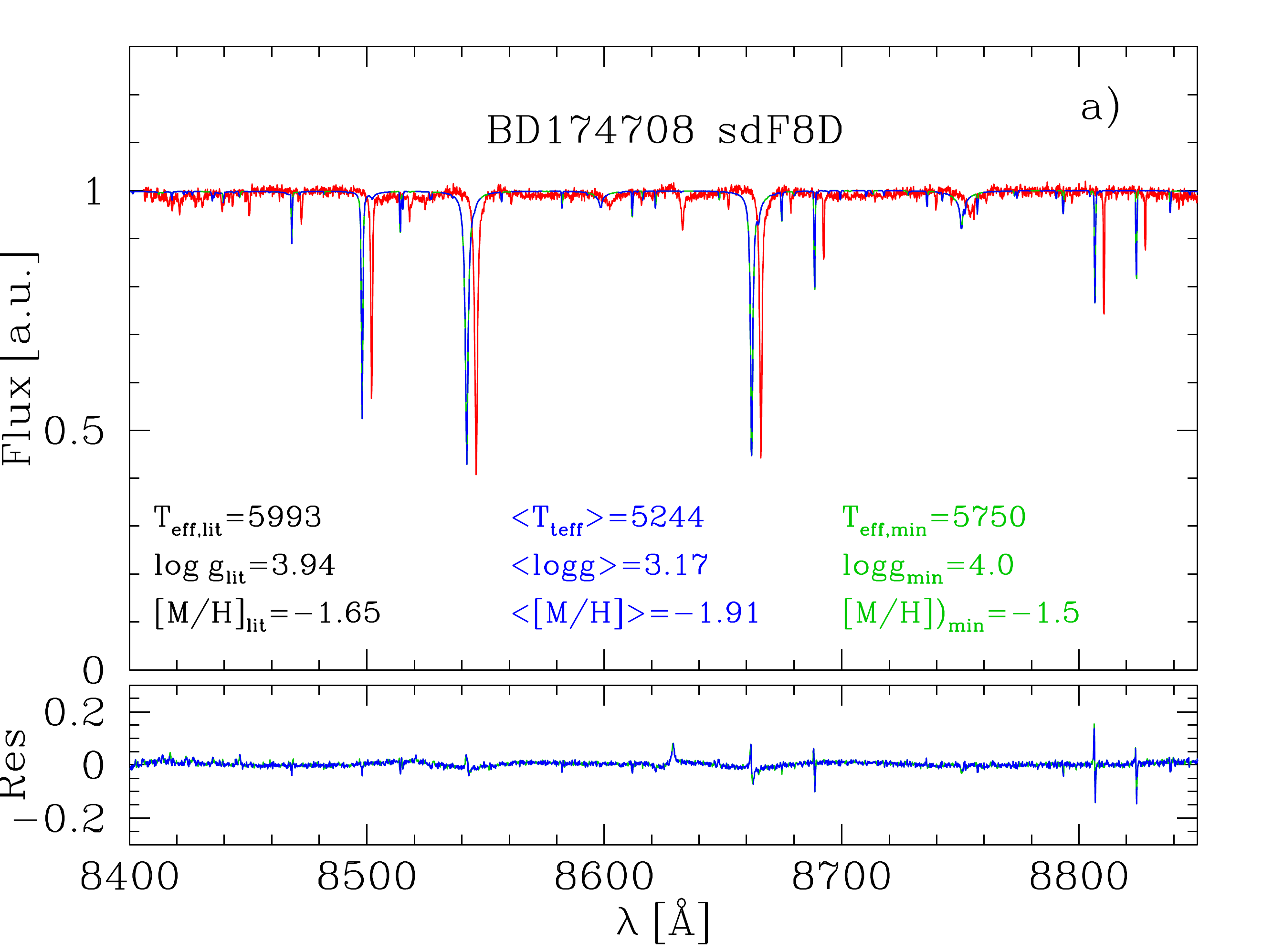}
\includegraphics[width=0.24\textwidth,angle=-90]{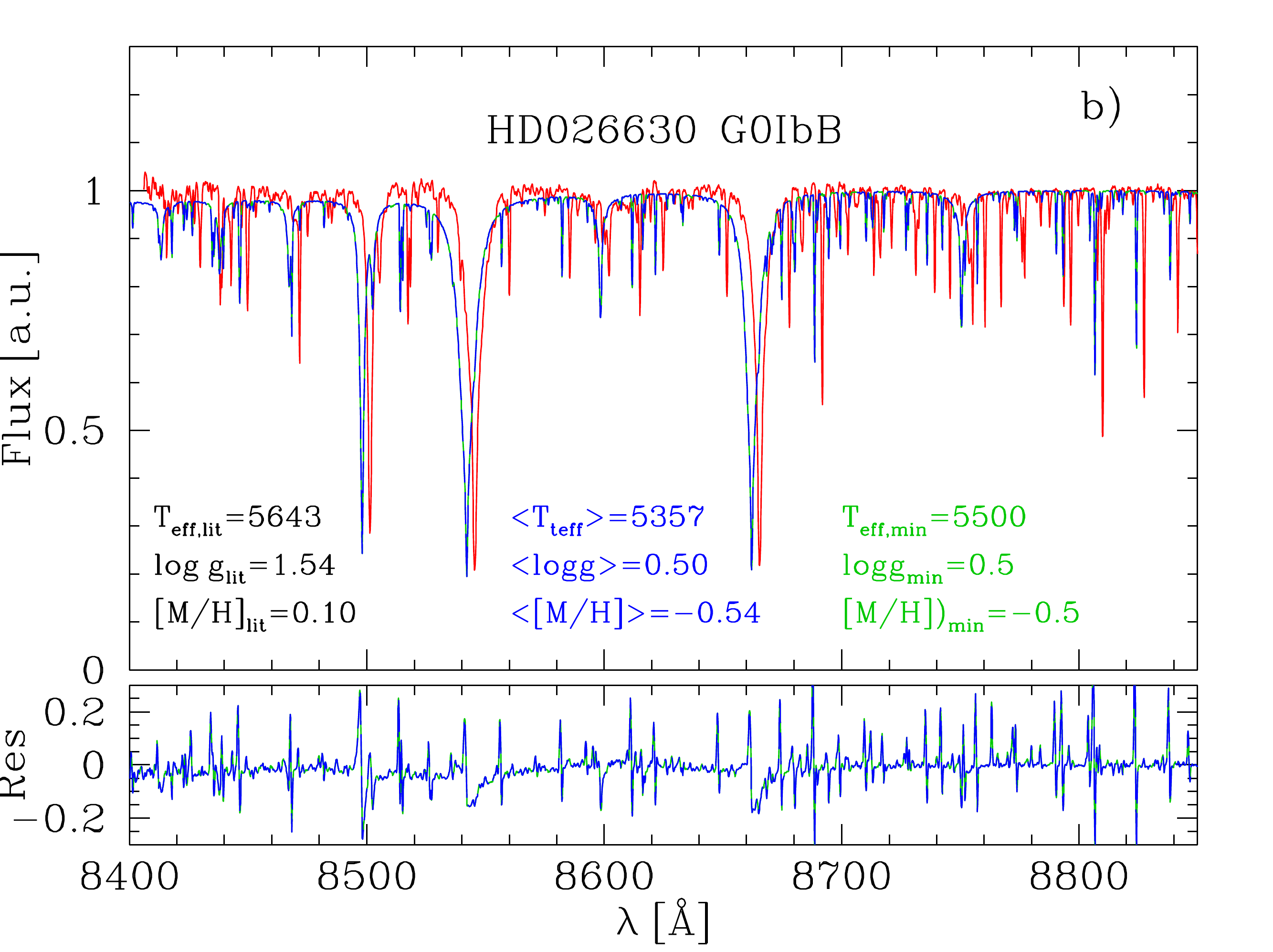}
\includegraphics[width=0.24\textwidth,angle=-90]{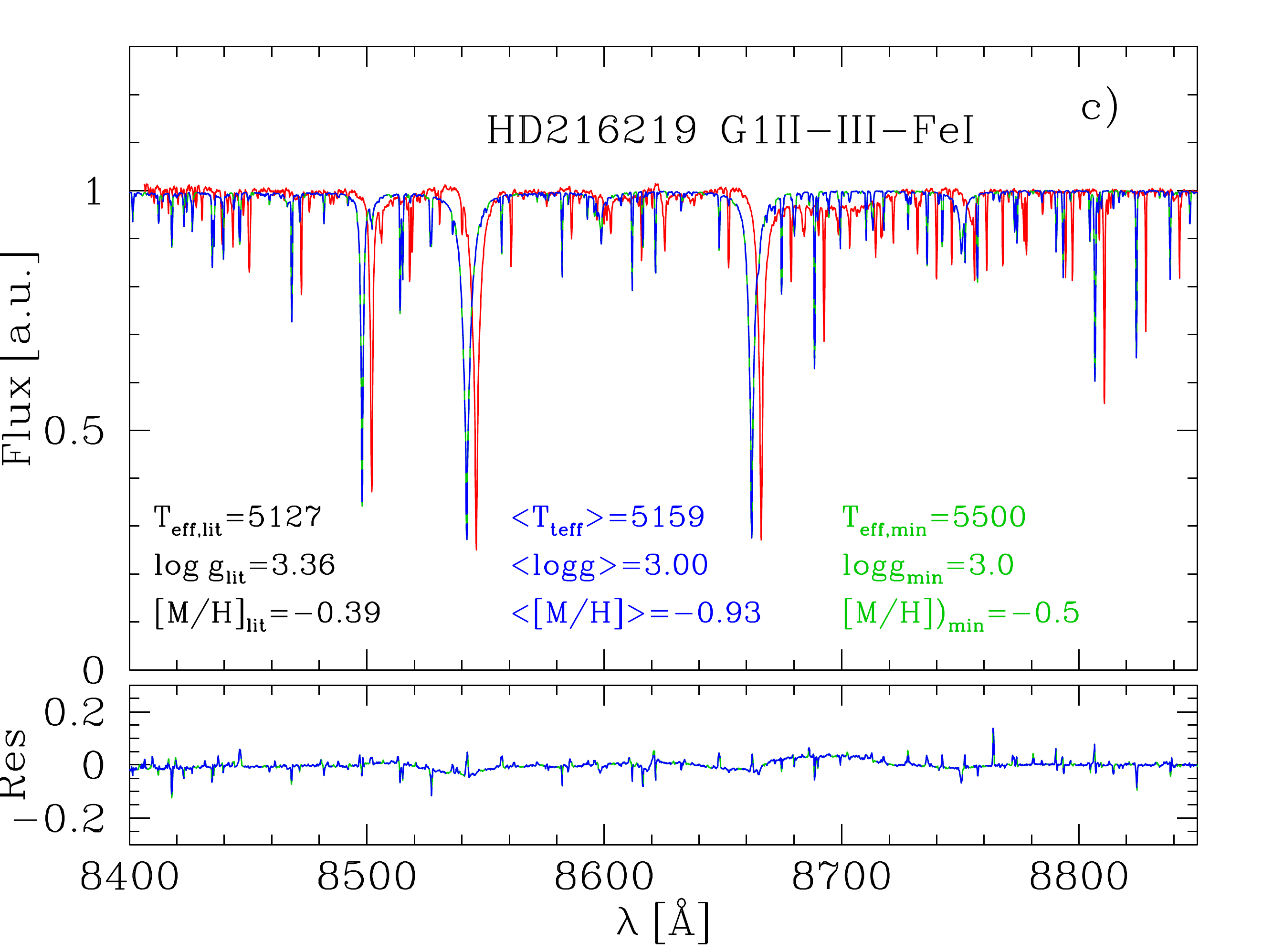}
\includegraphics[width=0.24\textwidth,angle=-90]{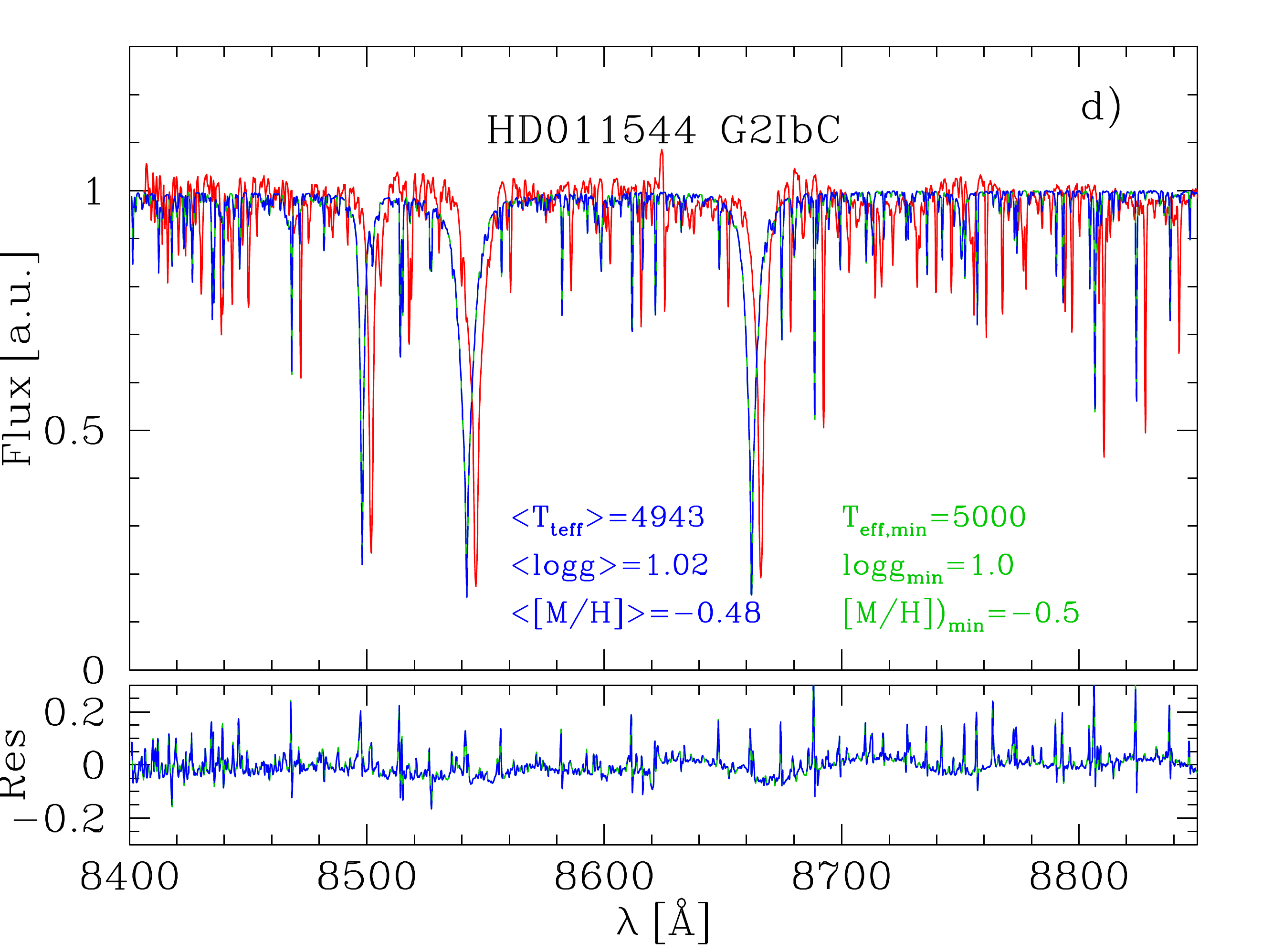}
\includegraphics[width=0.24\textwidth,angle=-90]{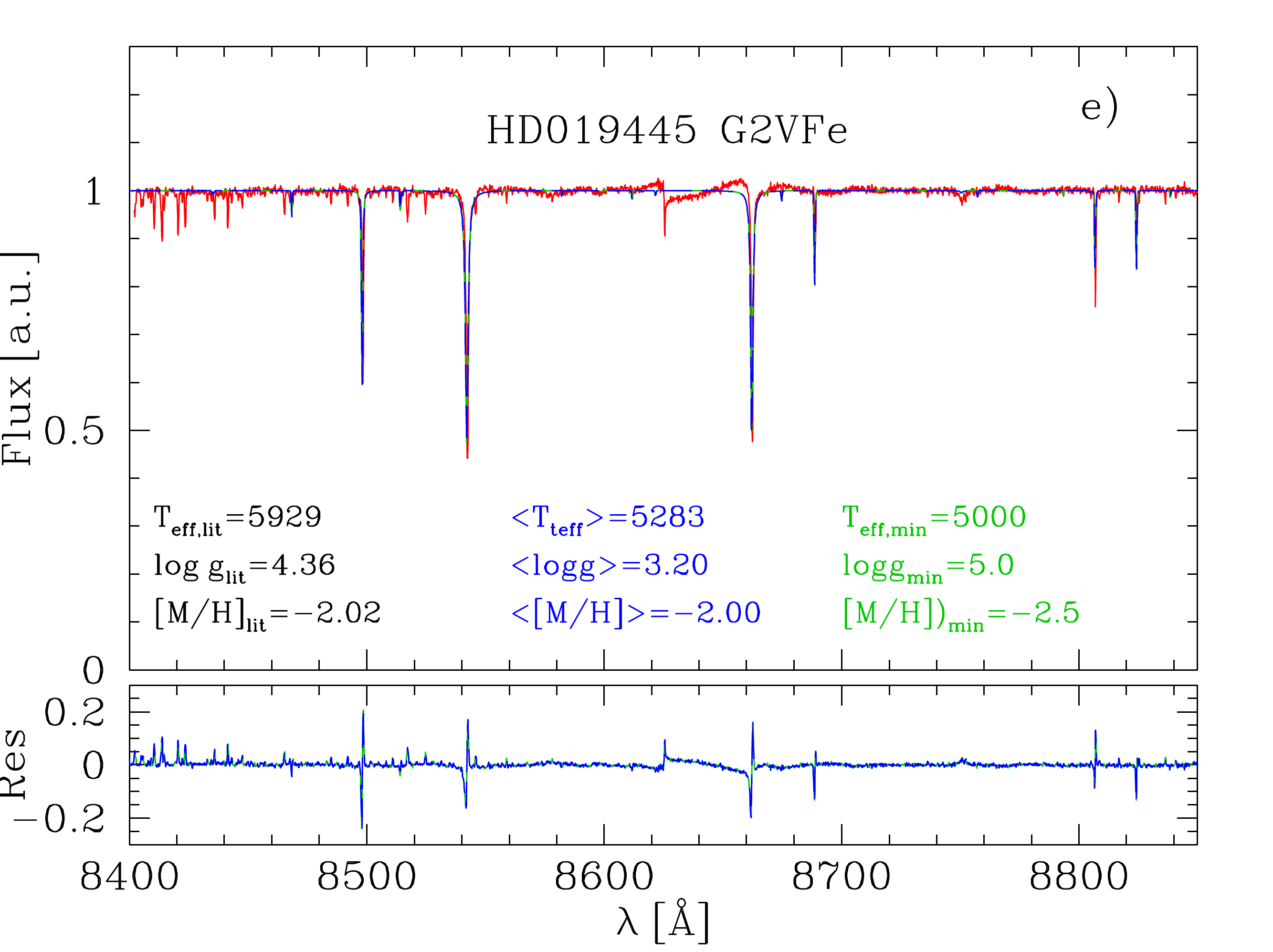}
\includegraphics[width=0.24\textwidth,angle=-90]{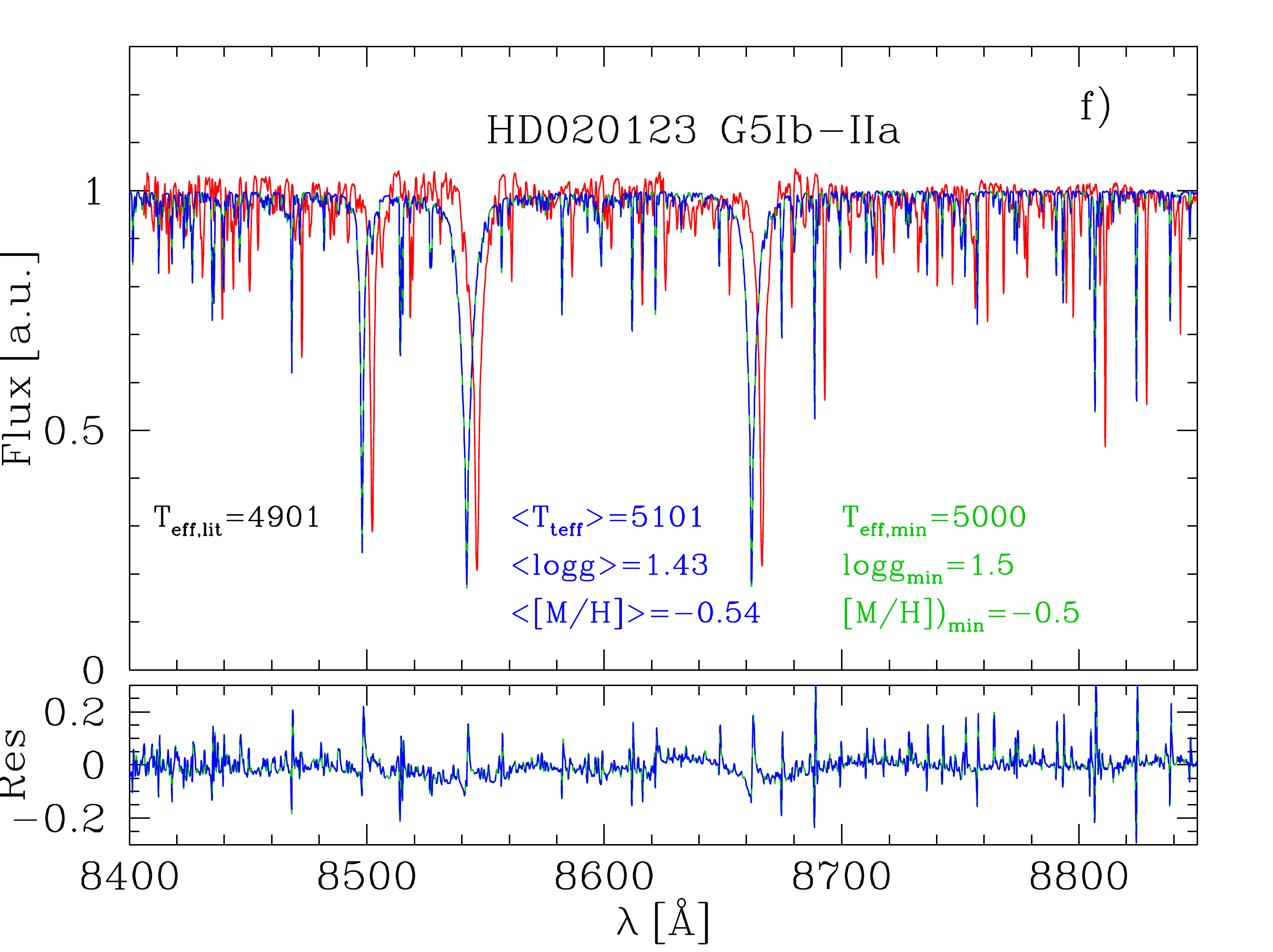}
\includegraphics[width=0.24\textwidth,angle=-90]{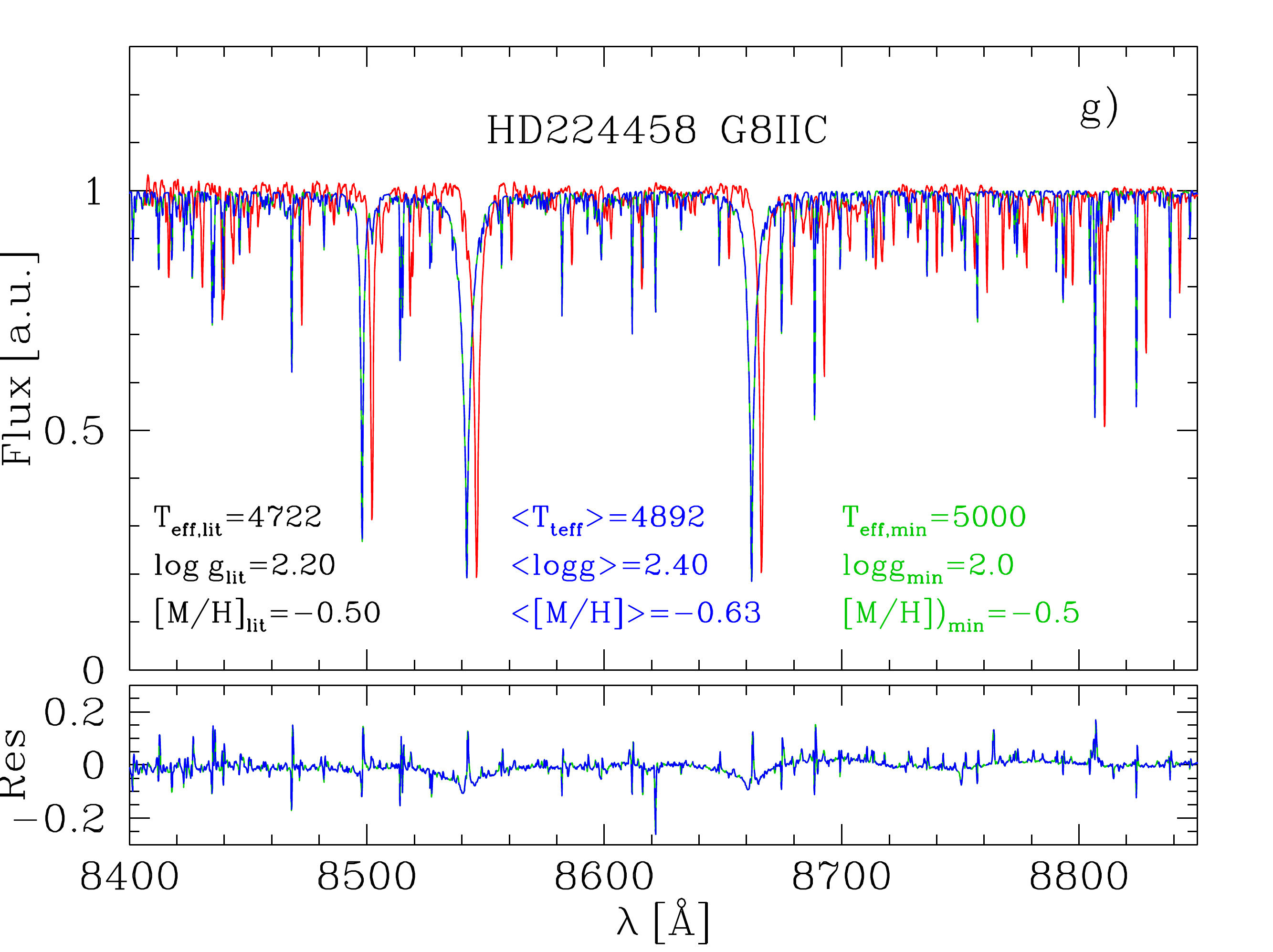}
\includegraphics[width=0.24\textwidth,angle=-90]{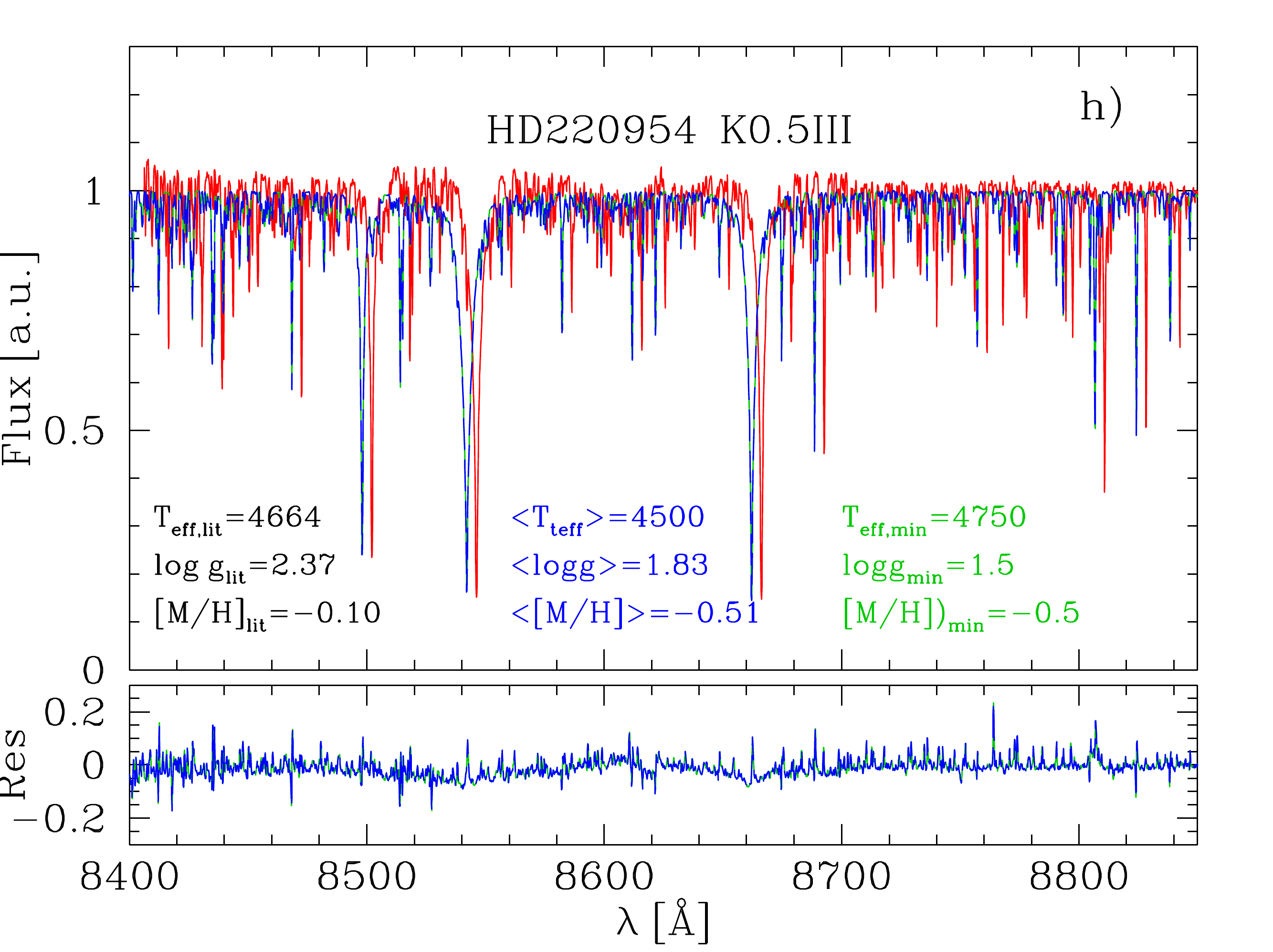}
\includegraphics[width=0.24\textwidth,angle=-90]{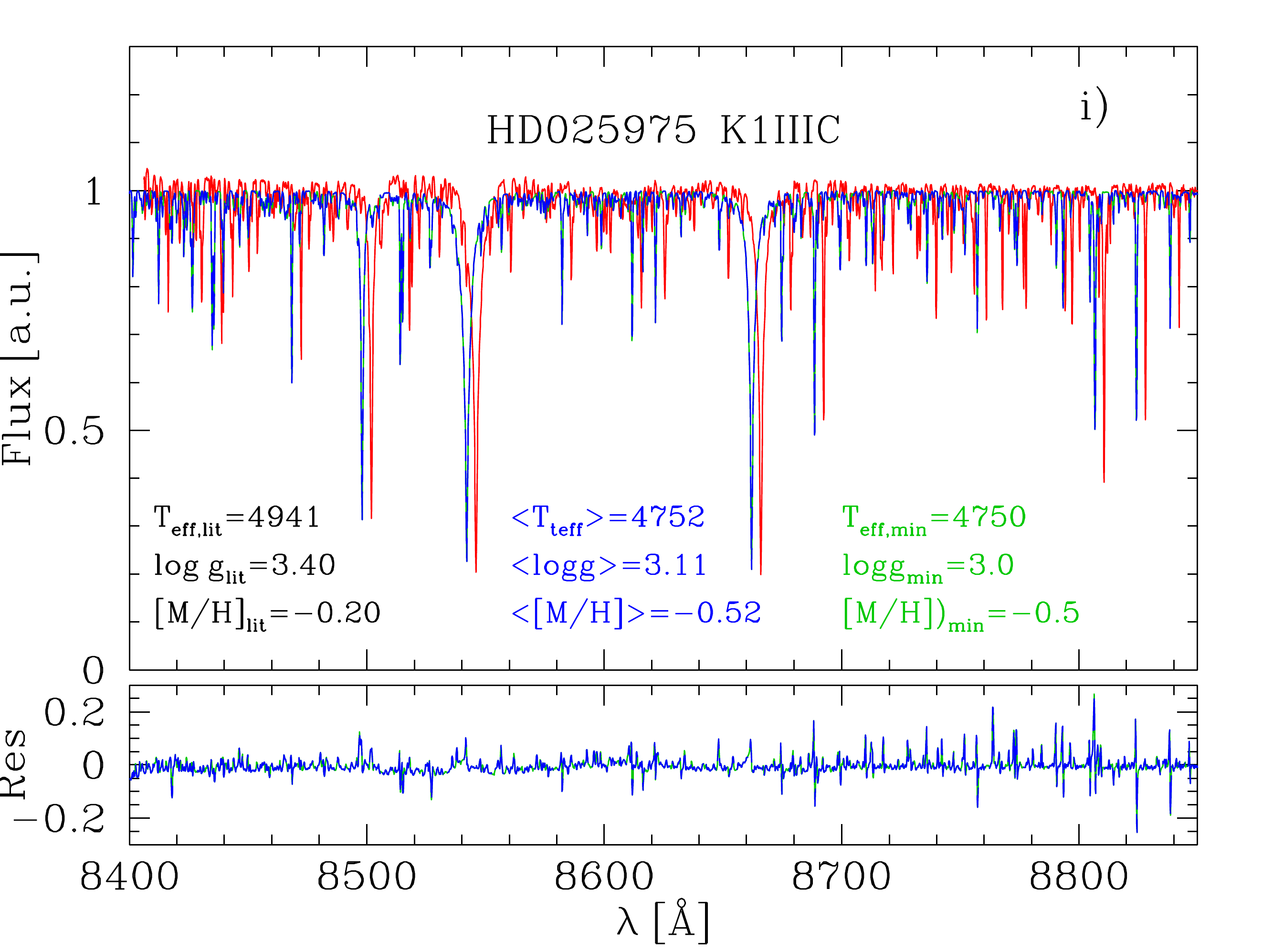}
\includegraphics[width=0.24\textwidth,angle=-90]{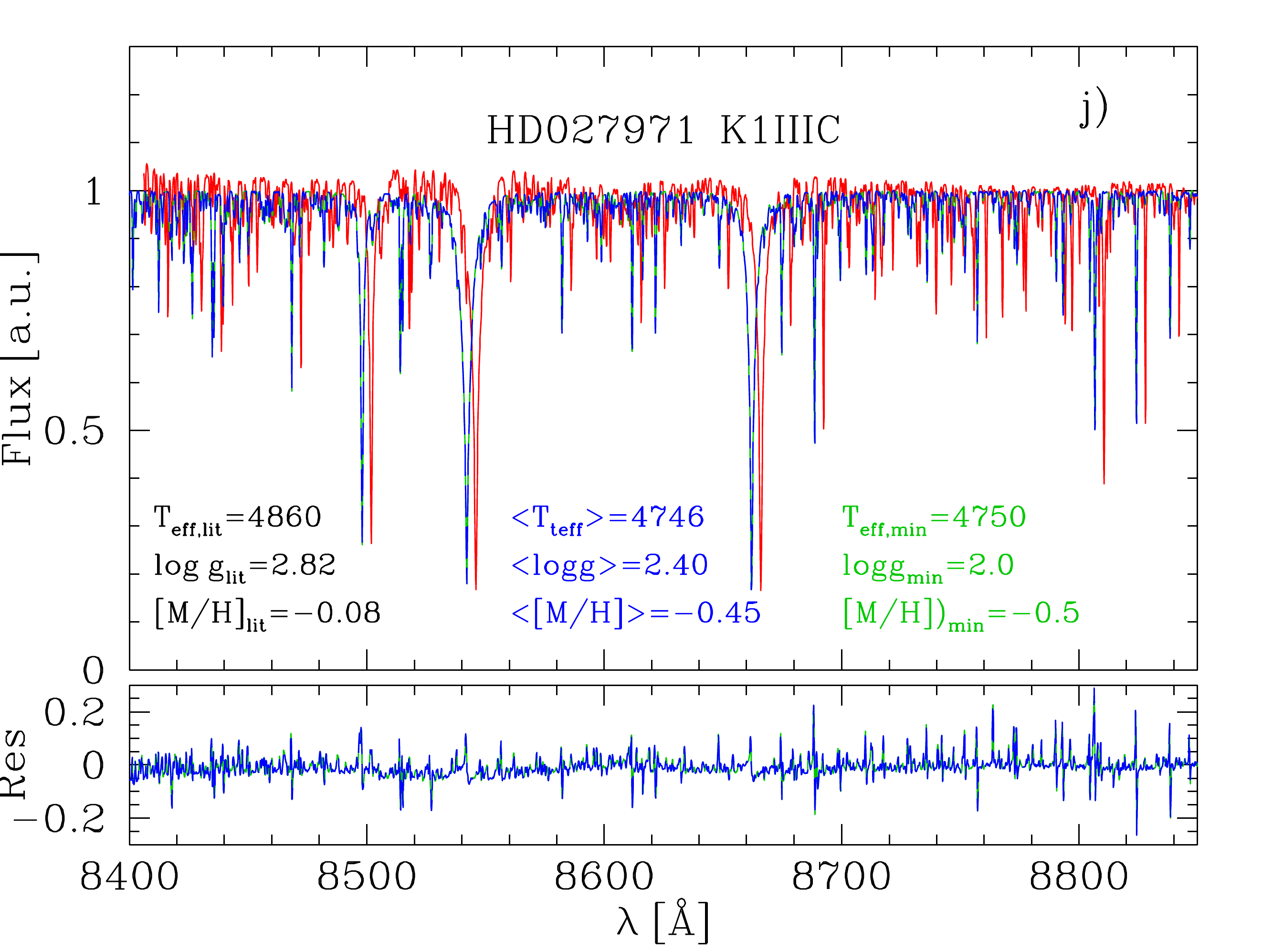}
\includegraphics[width=0.24\textwidth,angle=-90]{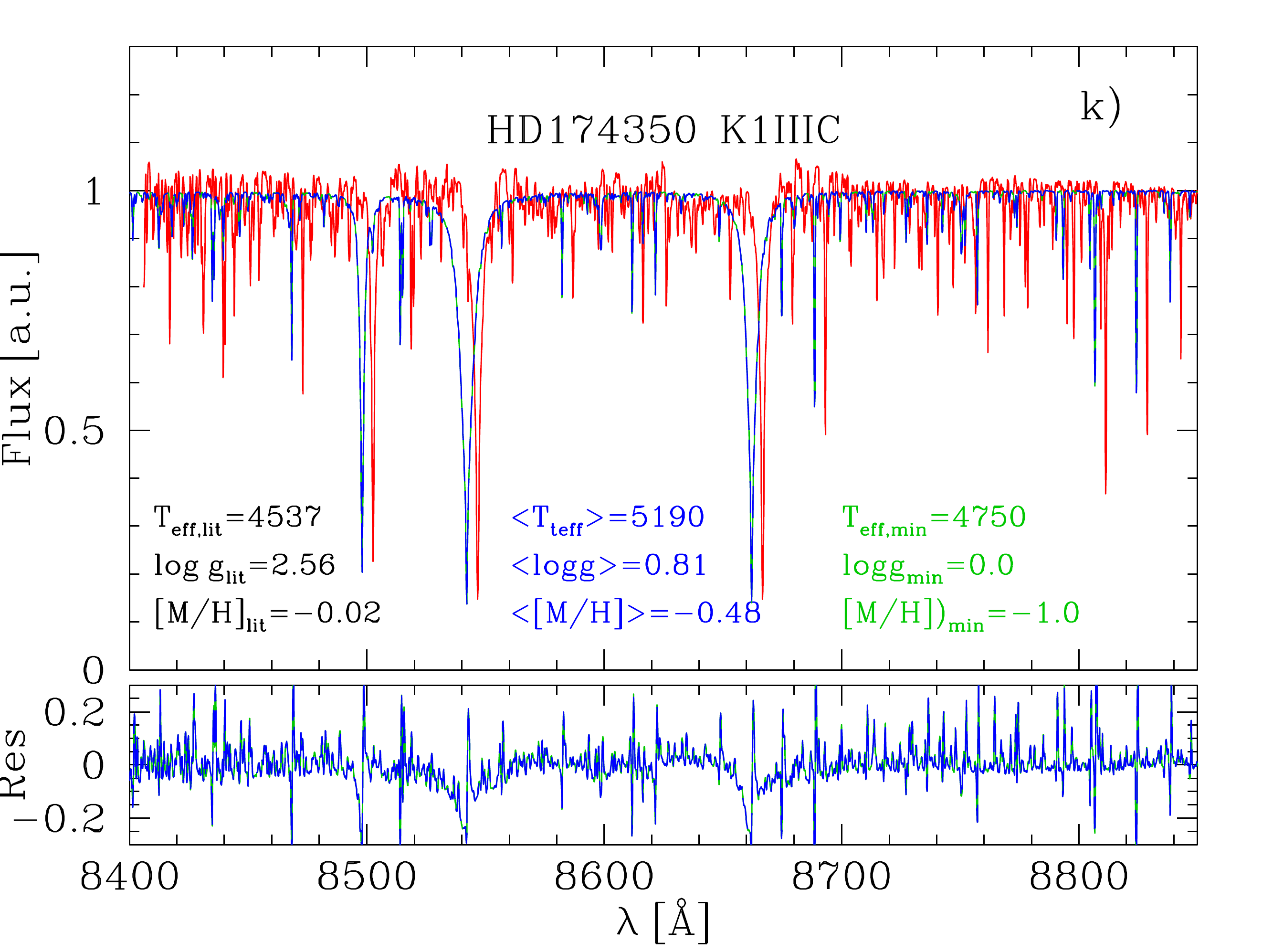}
\includegraphics[width=0.24\textwidth,angle=-90]{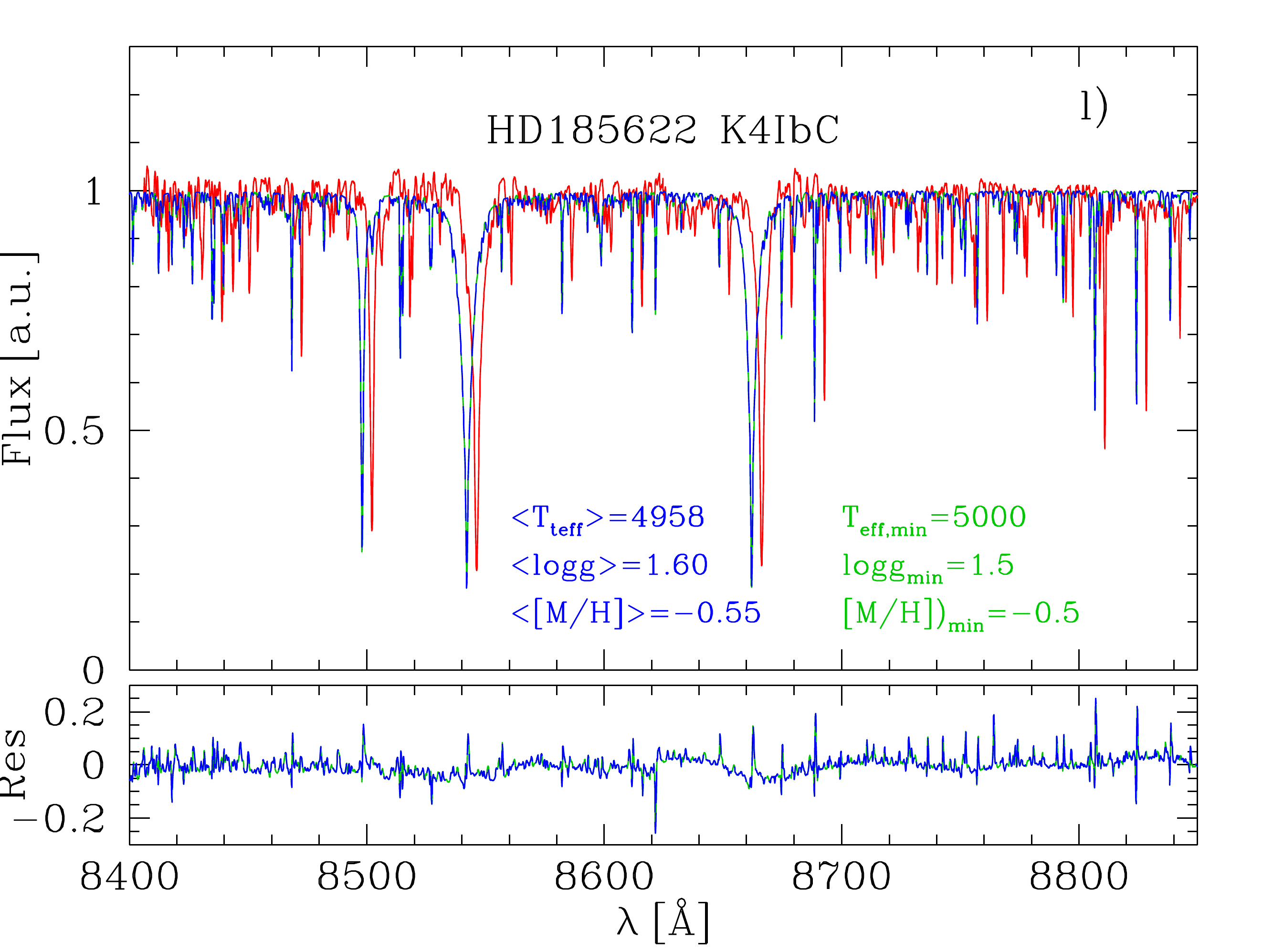}
\caption{The observed spectra of the commissioning cool stars with the best-fitting MUN05 models over-plotted. The top panel shows the observed spectrum (red solid line), the averaged fitted model (blue long-dashed line) and the $\chi^{2}_{min}$ model (green short-dashed line). The bottom panel shows the residuals as Flux$_{obs}$ - Flux$_{mod}$ (blue and green lines as the corresponding fitted models). The stellar parameters derived from the literature (black, whenever available), from the averaged  model (blue) and from  $\chi^{2}_{min}$ model (green) are shown in each panel.}
\label{fits-cool}
\end{figure*}

Every model within this contoured region has a likelihood $L_{sel}$ similar to the maximum $L_{max}=1$ (corresponding to the best model with $\chi^{2}=\chi^{2}_{\rm min}$), within 1 per cent error,  $L_{sel}\ge L_{max}-0.01$; meaning that the probability of choosing the best model will be better than 99 per cent when selecting the results within this region $R_{\nu,\alpha}$. The number of models, $N$, fulfilling Eq.~\ref{condition} is different, obviously, from star to star, as can be seen in the tables presented with the fitting results (Tables~\ref{parameters-com},~\ref{parameters-M15} and \ref{parameters-OT} for the COM, M15 and OT stars, respectively), since this depends on the characteristics of each spectrum.

We may apply this method to each observed spectrum either in the HR-R or in the HR-I setups. In fact, we have performed the calculations in each one separately and, then combining both spectra for each star in only one fit.  Figure~\ref{prob} shows, as an example, the results for the OT star BD$+$08~3095 observed in both HR-R and HR-I setups. This figure plots the resulting models in the $\rm T_{eff}$~--~$\rm \log{g}$ plane for all metallicities (although we have also tabulated our results separately for each value of $\rm [M/H]$). The colours represent the probability scale, as labelled at right in the plot. Top and medium panels show the results when fitting the models to the HR-R and HR-I observed spectrum, respectively. The effective temperature obtained when using $\chi^{2}_{min}$ are similar $\rm T_{eff}$~=~6000~K (from the fit to HR-R spectrum) and 6500~K (HR-I), and a similar abundance ($\rm [M/H]$~=~$-$1.0 and $-$0.5) is set for each setup. However, the value of $\rm \log{g}$ obtained from the fit to the observed spectra is very different in HR-R ($\rm \log{g}$~=~1.5) and HR-I ($\rm \log{g}$~=~5.0). We have then run our fit taking both spectra and doing a single fit to the combined HR-R and HR-I spectrum. The $\chi^{2}_{min}$ fitting gives in this case values of $\rm T_{eff}$~=~6250~K, $\rm \log{g}$~=~4.5 and $\rm [M/H]$~=~$-$0.5, closer to the literature point ($\rm T_{eff}$~=~5728~K, $\rm \log{g}$~=~4.1 and $\rm [M/H]$~=~$-$0.36) plotted as a green dot. These results are shown in the bottom panel of the same Fig.~\ref{prob}.

We have also carried out a second analysis by choosing those models with $P\,>\,0.99$, within a small region of the parameter space. In the case of BD$+$08~3095 we have obtained 36, 78 and 42 compliant models when using HR-R spectrum only, HR-I spectrum only, or the combination of both set-ups in a single spectrum, respectively. These models are over-plotted with small grey dots over the blue region where best models are located, and a large black cross indicates the averaged stellar parameters with their dispersion given by an ellipse, obtained with those models in the top, middle and bottom panels, respectively. 

For the top panel (fitting to HR-R spectrum), these averaged values are $\rm T_{eff}$~=~5972~$\pm$~285~K and $\rm \log{g}$~=~1.8~$\pm$~1.1~dex. For the middle panel (fitting to HR-I spectrum), the averaged results are $\rm T_{eff}$~=~5356~$\pm$~663~K and $\rm \log{g}$~=~4.0~$\pm$~0.9~dex. The relative error for $\rm \log{g}$ is very high (61 and 23 per cent) compared with the one found for $\rm T_{eff}$ (5 and 12 per cent). This can be attributed to the larger step size in $\rm \log{g}$ in the MUN05 models, given a much smaller number of models with different $\rm \log{g}$ than the ones with eligible values of $\rm T_{eff}$. However, the result shown in the bottom panel for the fitting to the combined spectrum of both HR-R and HR-I spectral ranges gives $\rm T_{eff}$~=~5917~$\pm$~356~K, $\rm \log{g}$~=3.6~$\pm$~1.2 and $\rm [M/H]$~=~$-$0.88~$\pm$~0.33, closer, as before, to those found by other authors and reported in the literature (green point in the bottom panel). Anyhow, differences can still be attributed mostly to the model-error dominated by the large step size in all physical parameters in the MUN05 grid.

The results are presented as histograms for $\rm T_{eff}$, $\rm \log{g}$ and $\rm [M/H]$ in Figure~\ref{hist}, which shows the values resulted from the fit of the models to the observed spectrum of BD$+$08~3095 in HR-R (left column), HR-I (middle column) and the combined spectrum with the two spectral windows HR-R and HR-I (right column) for an easy comparison. We have plotted a Gaussian function, blue dashed line, showing the averaged and the dispersion values obtained for each stellar parameter. The results obtained with the $\chi^{2}_{min}$ model are shown with a green arrow. To complete the figure, we have over-plotted with a cyan short-dashed arrow the stellar parameters obtained from the {\sc SP\_ACE} model (see section~\ref{ace}) that best fits each observed spectrum. The values from the literature, usually obtained from a spectrum with a wider wavelength range and much lower spectral resolution, are represented as a magenta dot-dashed arrow.

Figure~\ref{hist} illustrates the discrepancies found in the derived stellar parameter when using different methods, and the sensitivity to spectral resolution, mathematical algorithms - fitting and parameter's selection - and model set (among other effects). The difficulty of the stellar parameter (mainly gravity) determination is higher in the HR-I setup, for which we cannot find {\sc SP\_ACE}'s solutions for many of the stars in our sample.  We will come back to this discussion in sections section~\ref{analisis} and \ref{ace}, after estimating the stellar parameters of the 97 stars from three different samples presented in this work. 

We emphasise the importance of the spectral range involved in the fitting. In the case of individual fits to HR-R and HR-I spectrum, the resulted average parameters are not very different in both setups, for $T_{\rm eff}$ and a [M/H], within the error bars. However, this is not the case of $\log{g}$, for which an important discrepancy is obtained. This can be explained due to the very different information in the spectral lines existing in the two (and short) spectral ranges. The effect is magnified due to the high spectral resolution. To try to get advantage from the whole information in our observed spectra, we repeated the model fitting to allow the code to use  the spectral information in the two spectral ranges HR-R and HR-I simultaneously. The average parameters are then found to be in better agreement with the ones from the literature. This conclusion reinforce the decision on making the spectral library by observing always all stars in both High-Resolution MEGARA setups.

\subsubsection{Stellar parameters for the commissioning stars}
\label{com-res}
We have applied the technique described in the previous section to our 21 COM stars, in order to assign the stellar parameters $\rm T_{eff}$, $\rm \log{g}$ and $\rm [M/H]$ with the $\chi^{2}_{min}$ model fitted to the observed spectrum. We have also obtained the mean values $<\rm T_{eff}>$, $<\rm \log{g}>$ and $<\rm [M/H]>$, as the average of the parameters of $N$ models with $\chi^{2}$ similar to $\chi^{2}_{min}$, within the allowed probability value. 

Table~\ref{parameters-com} summarises the results for the 21 commissioning stars, sorted by spectral type, from the hottest (top) to the coolest (bottom). Column 1 displays the star name; column 2 shows the $\chi^{2}_{min}$ obtained from the fitting process of the observed normalised spectrum to the MUN05 theoretical catalogue; column 3 displays the associated maximum probability, $\rm P_{max}$ associated to the $\chi^{2}_{min}$. Columns 4, 5 and 6 show the derived stellar parameters, $\rm T_{eff}$, $\rm \log{g}$, and $\rm [M/H]$, from the MUN05 model corresponding to $\chi^{2}_{min}$ model. Column 7 gives the number of models, $N$, in the likelihood region $R_{\nu,\alpha}$. Columns 8, 9 and 10 have the average stellar parameters: $\langle T_{\rm eff}\rangle$, $\langle\log{g}\rangle$ and $\langle \rm [M/H]\rangle$ respectively, obtained as the averaged values of the set of $N$ models, with their corresponding errors. Finally, columns 11, 12 and 13 give the stellar parameters, $\rm T_{eff,lit}$, $\rm \log{g_{lit}}$ and $\rm [M/H]_{lit}$, from the literature whenever available, given the reference code in column 14. The stellar parameters come from Holgado et al (2019, priv. comm) for HD~218915 and HD~192281; GOSSS catalogue \citep{sota14} for the spectral type of Schulte~9, whose $\rm T_{eff}$ has been obtained from \citet{blomme13} and $\rm log{g}$ from the calibration for O stars \citep{martins05}; MILES stellar parameters \citep{cen01b} for HD~220575, HD~224458 and HD~220954; XSL, XSHOOTER Spectral Library \citep{chen14} for HD~019445, HD~174350 and HD~216219; ELODIE \citep{pru04} for BD+17~4708 and HD~026630 and INDO-US, \citep{val04} for HD~020123, HD~025975 and HD~027971. Values for BD+40~4032 come from \citet{cam03}. Stars from MILES were re-calibrated by \cite{pru11} and the new values appear in an additional row. Table~\ref{parameters-com} shows that the $<\rm T_{eff}>$ values obtained are in agreement with those from the literature -- we will revisit this point in section~\ref{analisis}~-- except in three cases (HD~218915, HD~220575 and HD~174350). The metallicity when available is in general in agreement but there are important discrepancies in the $\langle\rm [M/H]\rangle$ obtained from both setups. 
The $\rm T_{eff}$ difference found in the hottest stars between our estimates and the values from the literature might come, on the one hand, due to the lack of {He}{i} and {He}{\textsc ii} lines in the HR-I spectral range and, on the other hand, due to the less dense grid in the MUN05 models for the larger values of $\rm T_{eff}$. In the particular case of Schulte~9, we also know that this star is a SB2-type binary \citep{naze12, lorenzo19}.

Figure~\ref{fits-hot} and Figure~\ref{fits-cool} show the observed spectra (in red) from the commissioning pilot program for hot and cool stars, respectively. The name of each star and its spectral type are given in each panel. The $\chi^{2}_{min}$ fitted model is displayed as green continuum-line. The averaged spectrum is obtained from the MUN05 set, by selecting those models with the closest values to the averaged stellar parameters according to the likelihood criterion of $ L~\ge~L_{max}-0.01$. For example, if we obtain $\langle \rm T_{eff}\rangle$~=~4300~K, $\rm \log{g}$~=~2.6~dex and $\langle\rm [M/H]\rangle$~=~-0.3~dex, we take the spectra  corresponding to $\rm T_{eff}$~=~4000~K and 4500~K, $\rm \log{g}$~=~2.5 and 3.0~dex and $\langle\rm [M/H]\rangle$~=~-0.5 and 0.0~dex. With these 8 models, and interpolating between each two among them, we obtain the spectrum corresponding to the averaged stellar parameters plotted as the blue dashed-line. This method is the same used in next sections for the M15 and OT stars. Both models are almost indistinguishable among them. The bottom panel shows the residuals (the difference between the observed and theoretical spectra). Each panel also displays the stellar parameters derived from the literature (black, whenever available), the $\langle \chi^{2} \rangle$ model (blue) and the model with $\chi^{2}_{min}$ (green). 

\begin{figure*}
\includegraphics[width=0.22\textwidth,angle=-90]{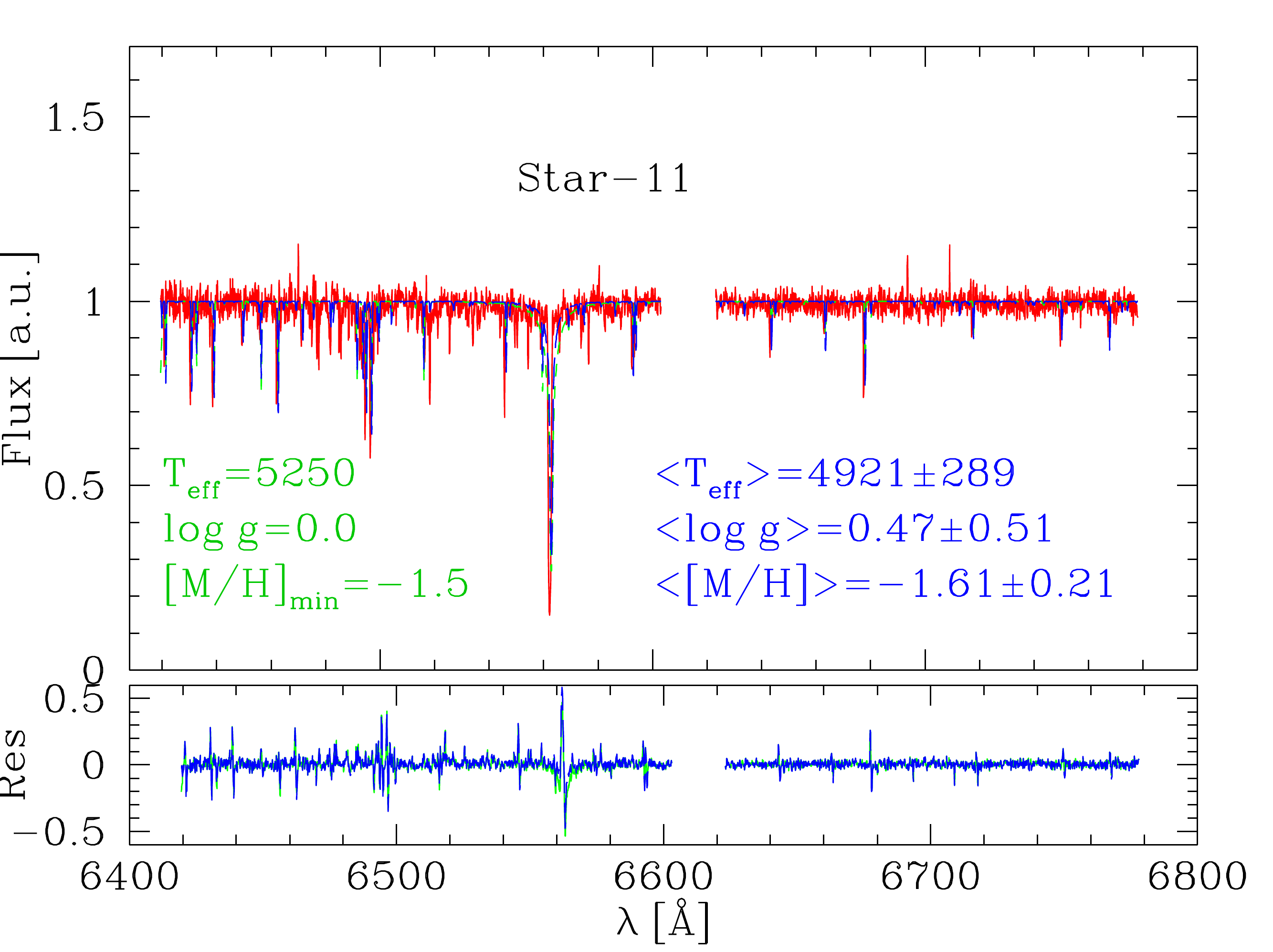}
\includegraphics[width=0.22\textwidth,angle=-90]{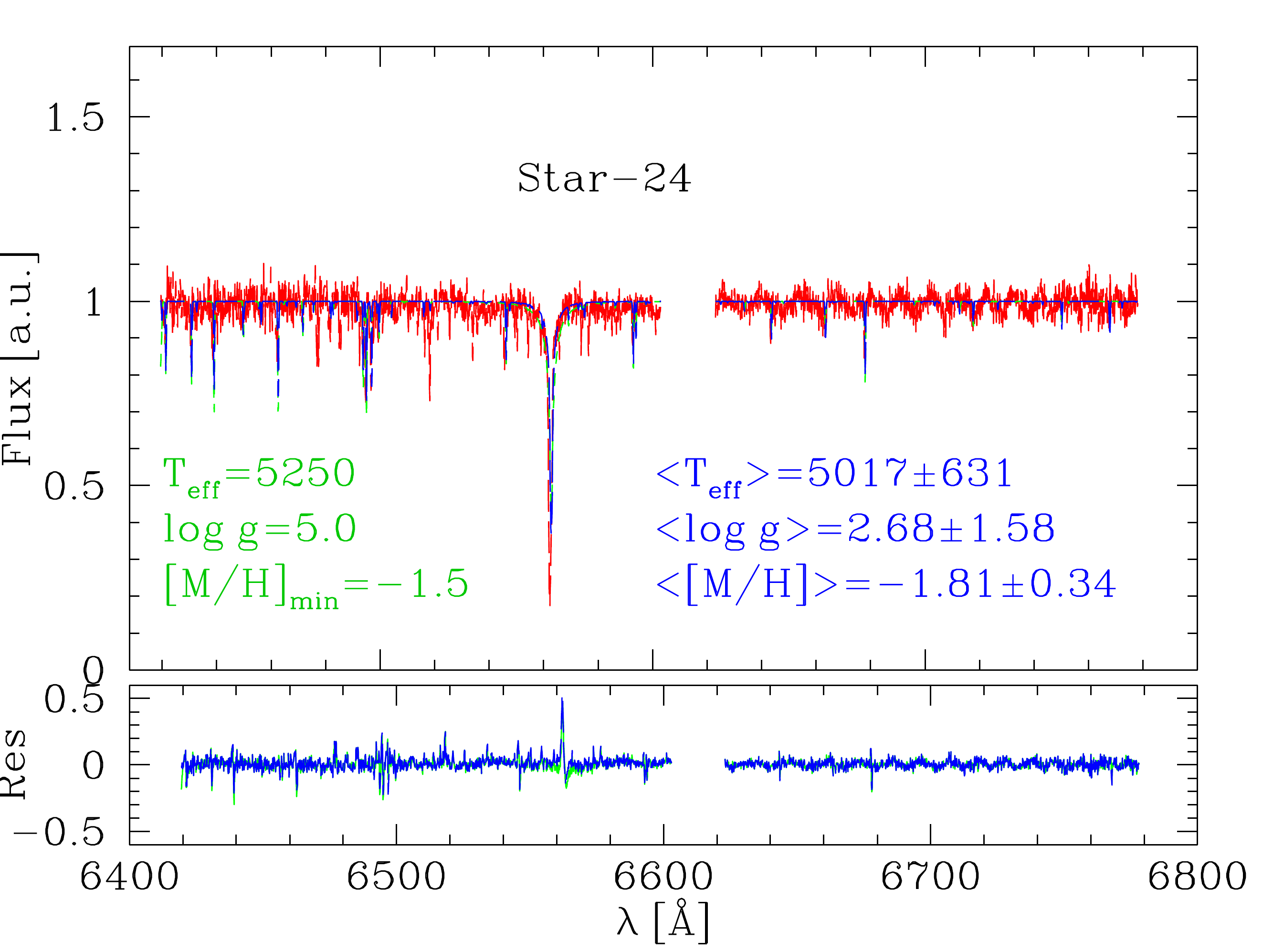}
\includegraphics[width=0.22\textwidth,angle=-90]{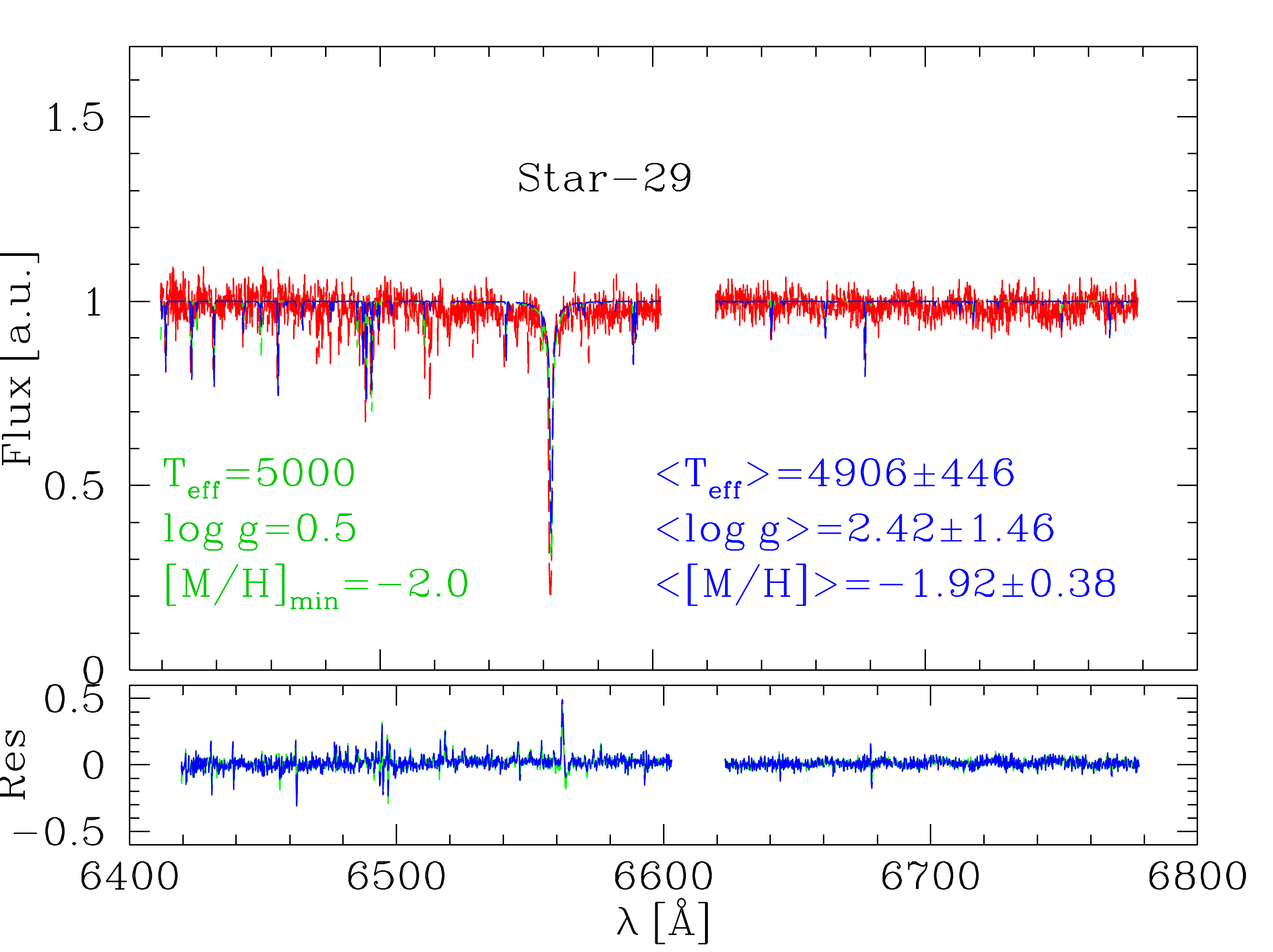}
\includegraphics[width=0.22\textwidth,angle=-90]{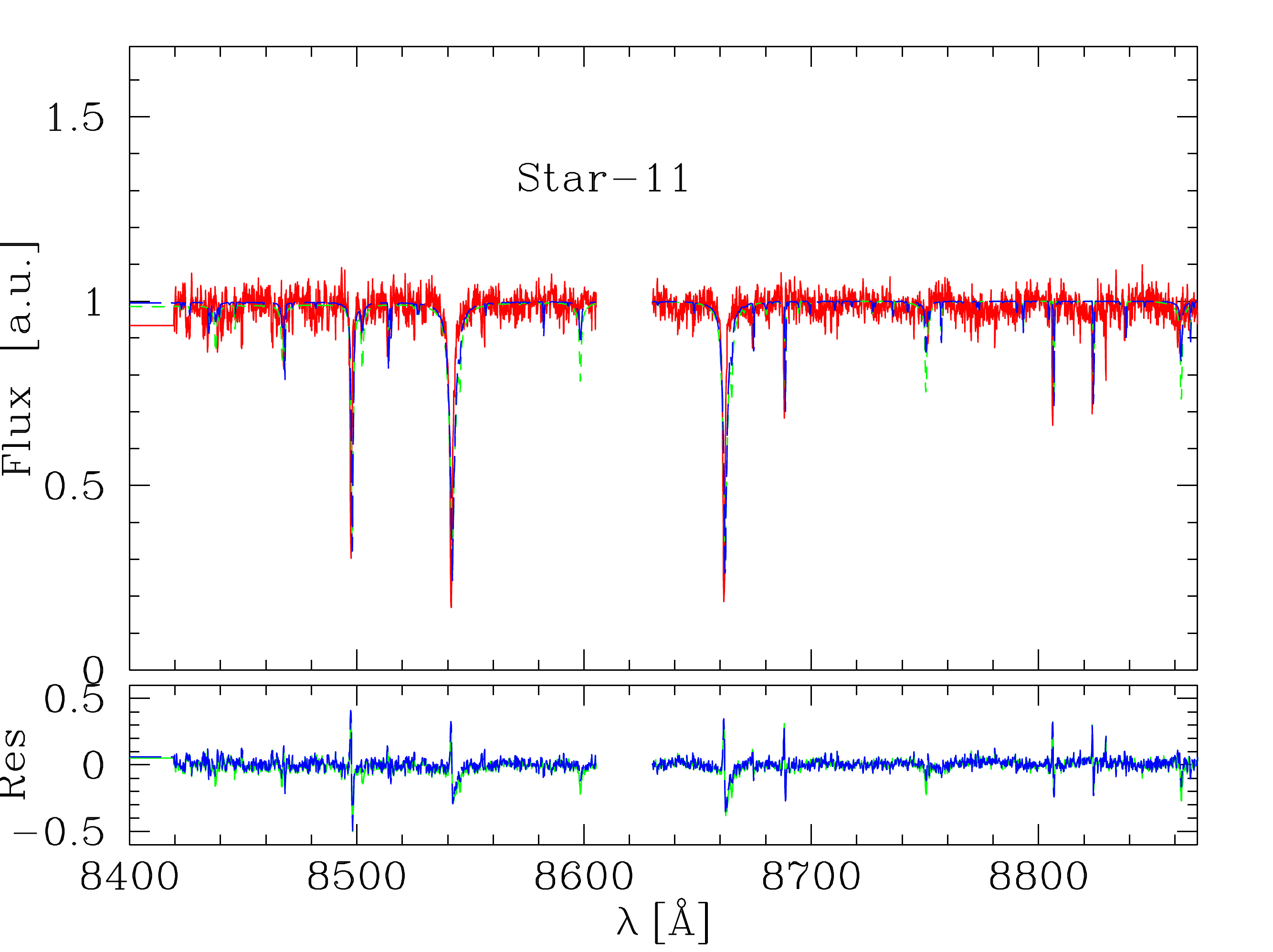}
\includegraphics[width=0.22\textwidth,angle=-90]{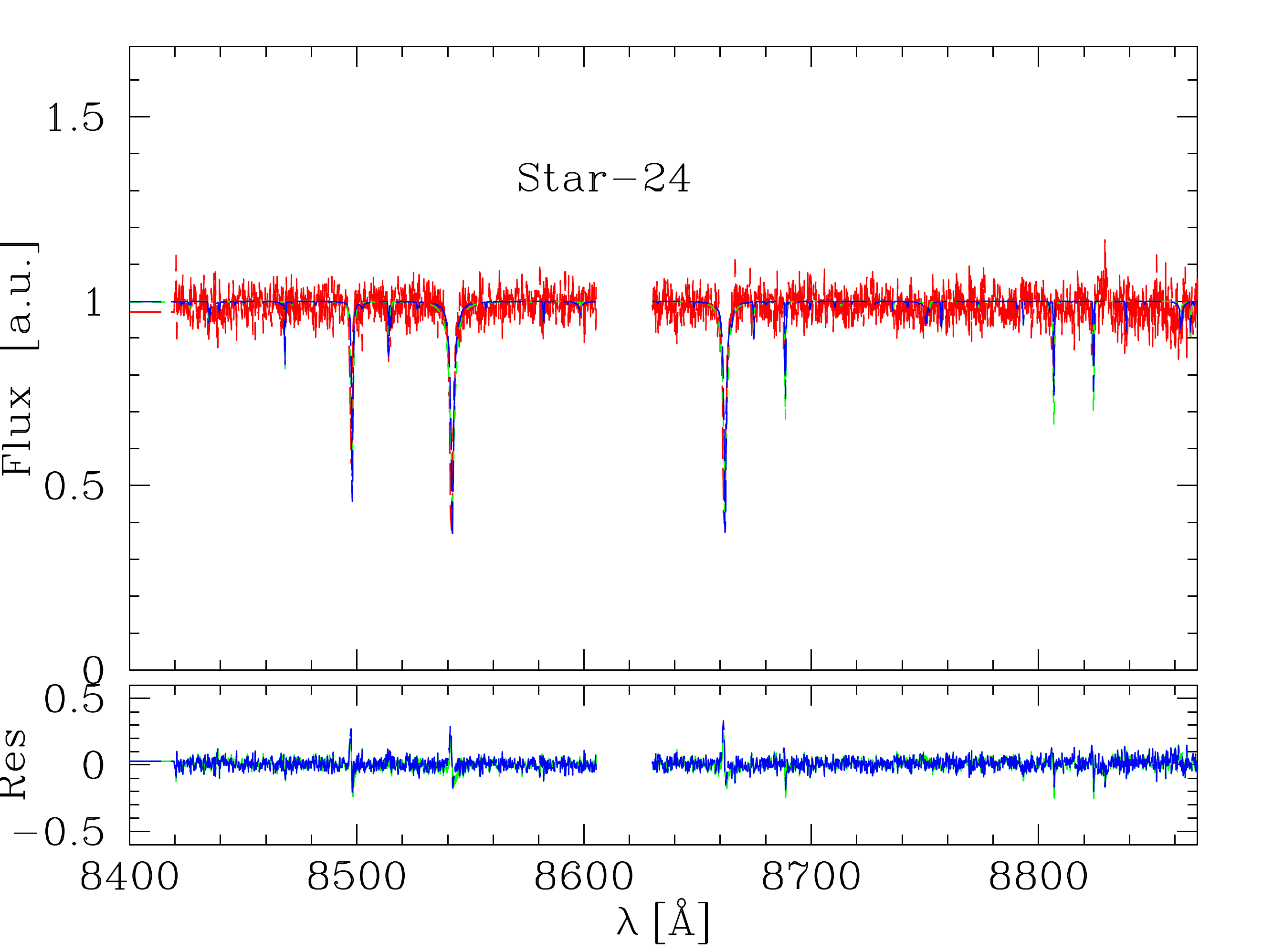}
\includegraphics[width=0.22\textwidth,angle=-90]{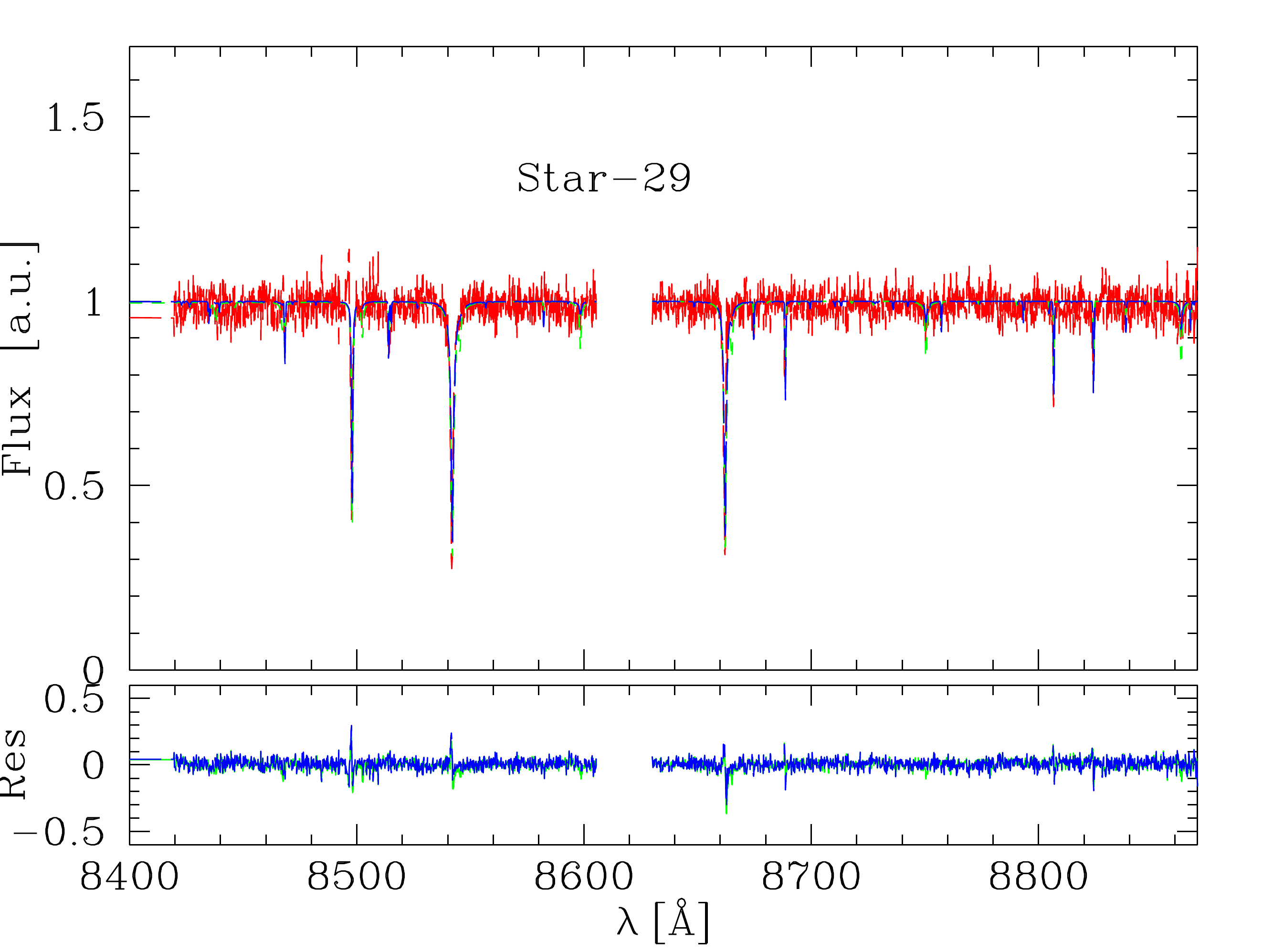}
\includegraphics[width=0.22\textwidth,angle=-90]{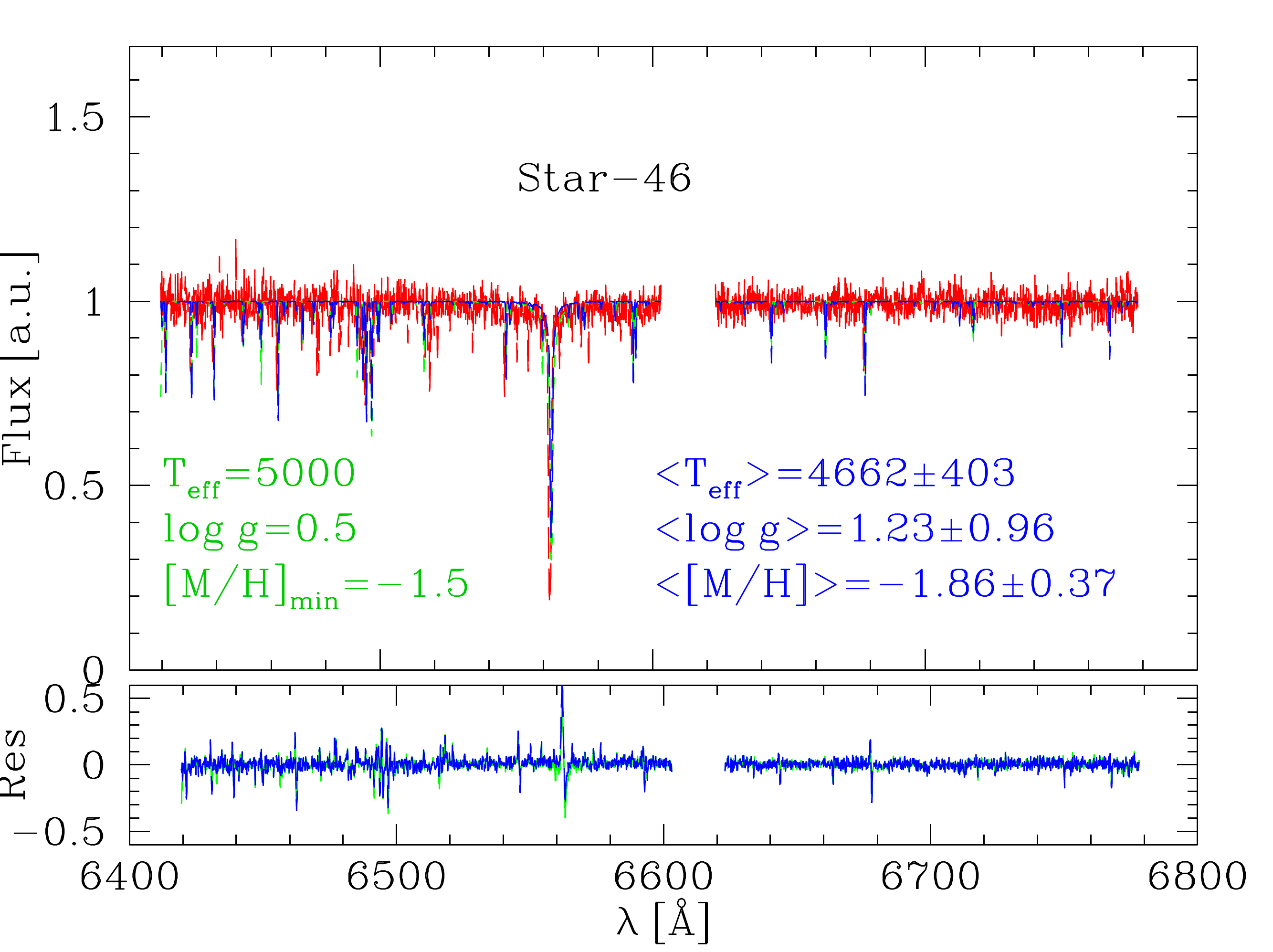}
\includegraphics[width=0.22\textwidth,angle=-90]{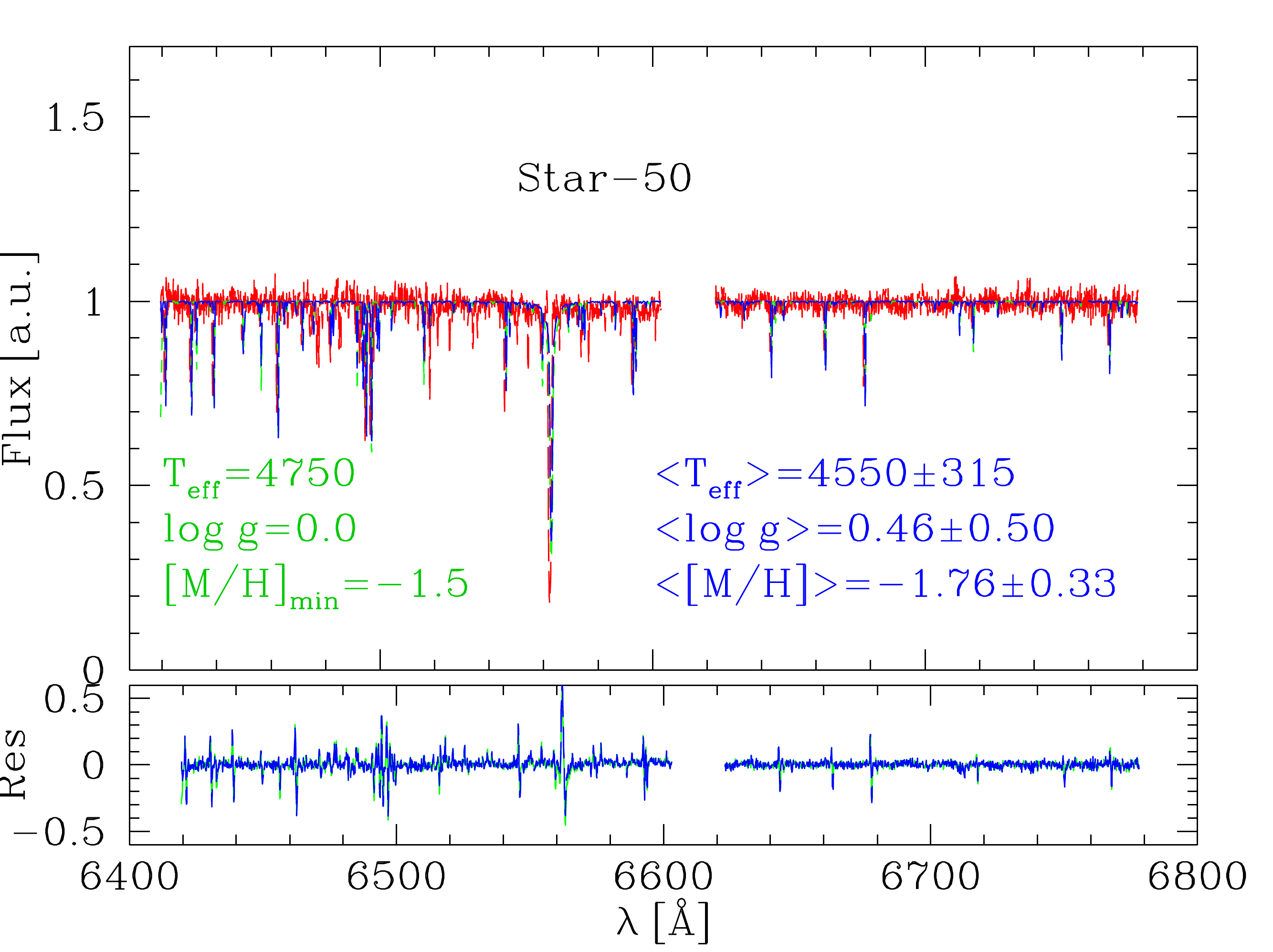}
\includegraphics[width=0.22\textwidth,angle=-90]{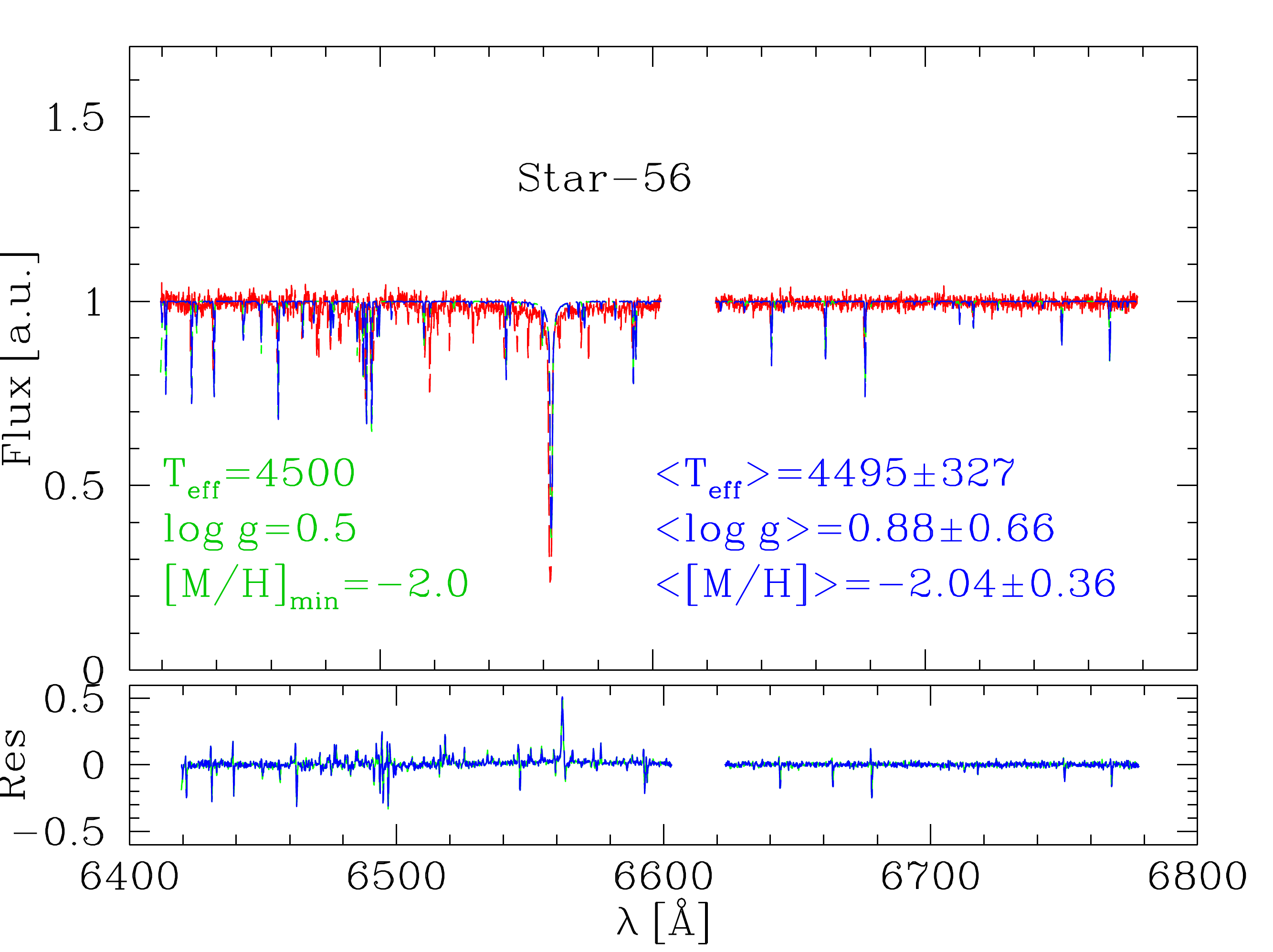}
\includegraphics[width=0.22\textwidth,angle=-90]{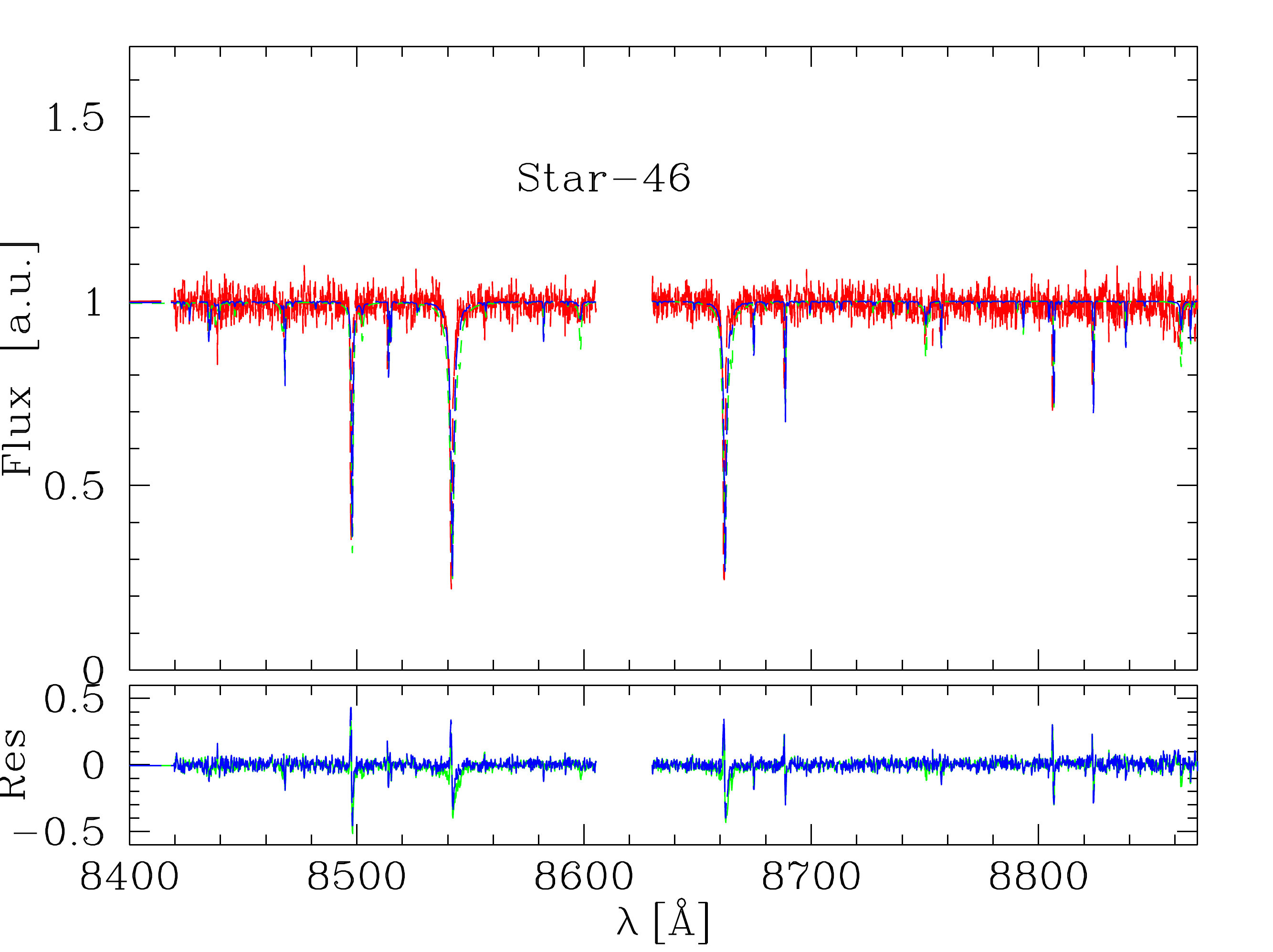}
\includegraphics[width=0.22\textwidth,angle=-90]{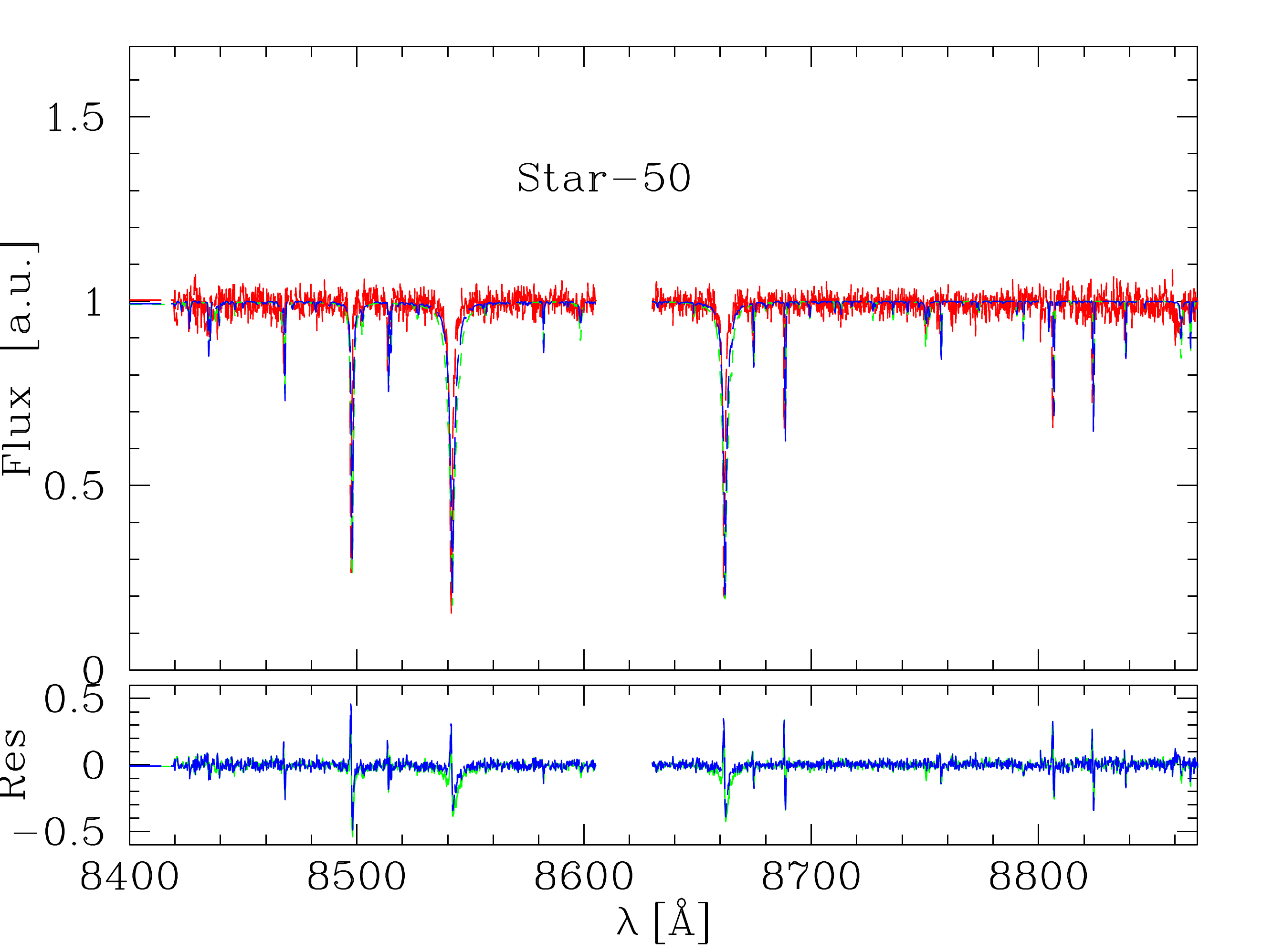}
\includegraphics[width=0.22\textwidth,angle=-90]{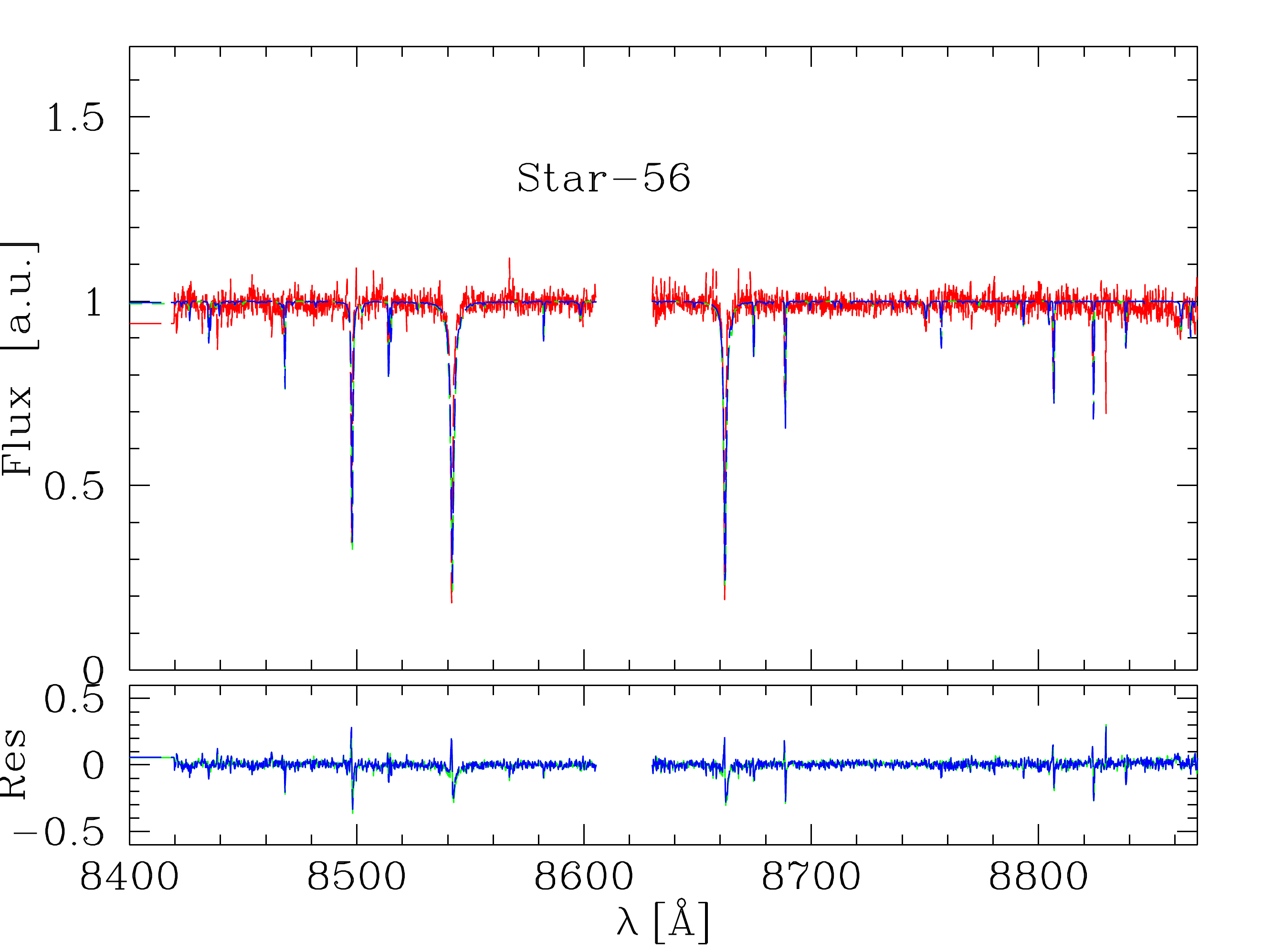}
\caption{Six selected M15 stars - from the sample of 56 - have been chosen to show the fitting results. The best-fitting MUN05 model is over-plotted to the observed spectrum for each star. The top panel of each star shows the observed spectrum (red solid line), the averaged fitted model (blue long-dashed line) and the $\chi^{2}_{min}$ model (green short-dashed line). The bottom sub-panel shows the residuals as Flux$_{obs}$ - Flux$_{mod}$ (blue and green lines as the corresponding fitted models). For each set of two rows we present the spectra of three stars: in the top panel the HR-R spectrum of the star is shown and just below the corresponding HR-I spectrum. The stellar parameters from the average model (blue) and the $\chi^{2}_{min}$ model (green) are shown in each panel. The spectra and their fitting results for the rest of the M15 stars are available online (Figure~A~1, Appendix A, online version).}
\label{fits-M15}
\end{figure*}

\begin{table*}
\scriptsize
\caption{Stellar parameters for the M15 stars derived from the fit of the theoretical models to the observed spectra in HR-R, HR-I and the simultaneous fitting to the combined spectrum with HR-R and HR-I. The table description is in subsection~\ref{M15-res}. The full table with the results for the 56 stars is available online (Table~A~1, Appendix A, online version).}
\label{parameters-M15}
\begin{tabular}{rccrccrccc}
\hline
Star & $\chi^{2}_{min}$ & P$_{\rm max}$ &  \multicolumn{3}{c}{Results with minimum $\chi^{2}$} &  \multicolumn{4}{c}{Average results with N models}
\\
  & & & T$_{\rm eff}$ & $\log{g}$ & $\rm [M/H]$ & $N_{p}$ & $\langle \rm T_{ eff} \rangle$ & $\langle\log{g}\rangle$ & $\langle\rm [M/H]\rangle$\\ 
\hline 
& \multicolumn{9}{c}{HR-R} \\ 
   1	&	5.251	&	0.15	&	7000	&	0.5	&	-1.5	&	10	&	7125	$\pm$	132	&	0.50	$\pm$	0.00	&	-1.50	$\pm$	0.00	\\
2	&	0.731	&	0.87	&	7250	&	1.5	&	-1.5	&	246	&	7494	$\pm$	489	&	1.59	$\pm$	0.74	&	-1.88	$\pm$	0.39	\\
3	&	0.910	&	0.82	&	5250	&	0.0	&	-2.0	&	102	&	5551	$\pm$	271	&	0.81	$\pm$	0.63	&	-1.93	$\pm$	0.39	\\
4	&	1.935	&	0.59	&	5750	&	0.0	&	-1.5	&	10	&	5800	$\pm$	197	&	0.20	$\pm$	0.26	&	-1.50	$\pm$	0.00	\\
5	&	1.960	&	0.58	&	6000	&	0.0	&	-2.0	&	60	&	6117	$\pm$	289	&	0.68	$\pm$	0.54	&	-1.90	$\pm$	0.38	\\    
  & \multicolumn{9}{c}{HR-I} \\ 
   1 & 0.884 & 0.83  & 6250  & 0.5 & 1.50  & 52  & 6457  $\pm$ 504 & 1.3 $\pm$ 1.0 & -1.73 $\pm$ 0.30  \\
2 & 0.558 & 0.91  & 6750  & 3.0 & 1.50  & 65  & 6673  $\pm$ 470 & 2.5 $\pm$ 1.0 & -1.98 $\pm$ 0.41  \\
3 & 0.710 & 0.87  & 4500  & 0.0 & 2.50  & 131 & 4616  $\pm$ 545 & 2.1 $\pm$ 1.3 & -2.00 $\pm$ 0.39  \\
4 & 0.614 & 0.89  & 4250  & 0.0 & 2.50  & 73  & 4497  $\pm$ 481 & 1.3 $\pm$ 0.9 & -1.95 $\pm$ 0.39  \\
5 & 1.144 & 0.77  & 5000  & 0.0 & 1.50  & 17  & 4750  $\pm$ 375 & 0.3 $\pm$ 0.4 & -1.59 $\pm$ 0.20  \\
  & \multicolumn{9}{c}{HR-R and HR-I in a combined single spectrum} \\ 
 1	&	375.217	&	0.00	&	7250	&	0.5	&	-1.5	&	2	&	7125	$\pm$	177	&	0.50	$\pm$	0.00	&	-1.50	$\pm$	0.00	\\
2	&	11.126	&	0.09	&	7500	&	1.0	&	-1.5	&	14	&	7304	$\pm$	223	&	1.43	$\pm$	0.65	&	-1.57	$\pm$	0.18	\\
3	&	6.303	&	0.36	&	5750	&	0.0	&	-1.5	&	11	&	5477	$\pm$	284	&	0.27	$\pm$	0.41	&	-1.64	$\pm$	0.23	\\
4	&	10.614	&	0.16	&	6750	&	0.5	&	-1.5	&	7	&	6357	$\pm$	690	&	0.36	$\pm$	0.24	&	-1.50	$\pm$	0.00	\\
5	&	5.322	&	0.31	&	6000	&	0.0	&	-1.5	&	3	&	5750	$\pm$	250	&	0.00	$\pm$	0.00	&	-1.50	$\pm$	0.00	\\  
\hline
\end{tabular}
\end{table*}
\normalsize

\begin{figure}
\includegraphics[width=0.48\textwidth,angle=0]{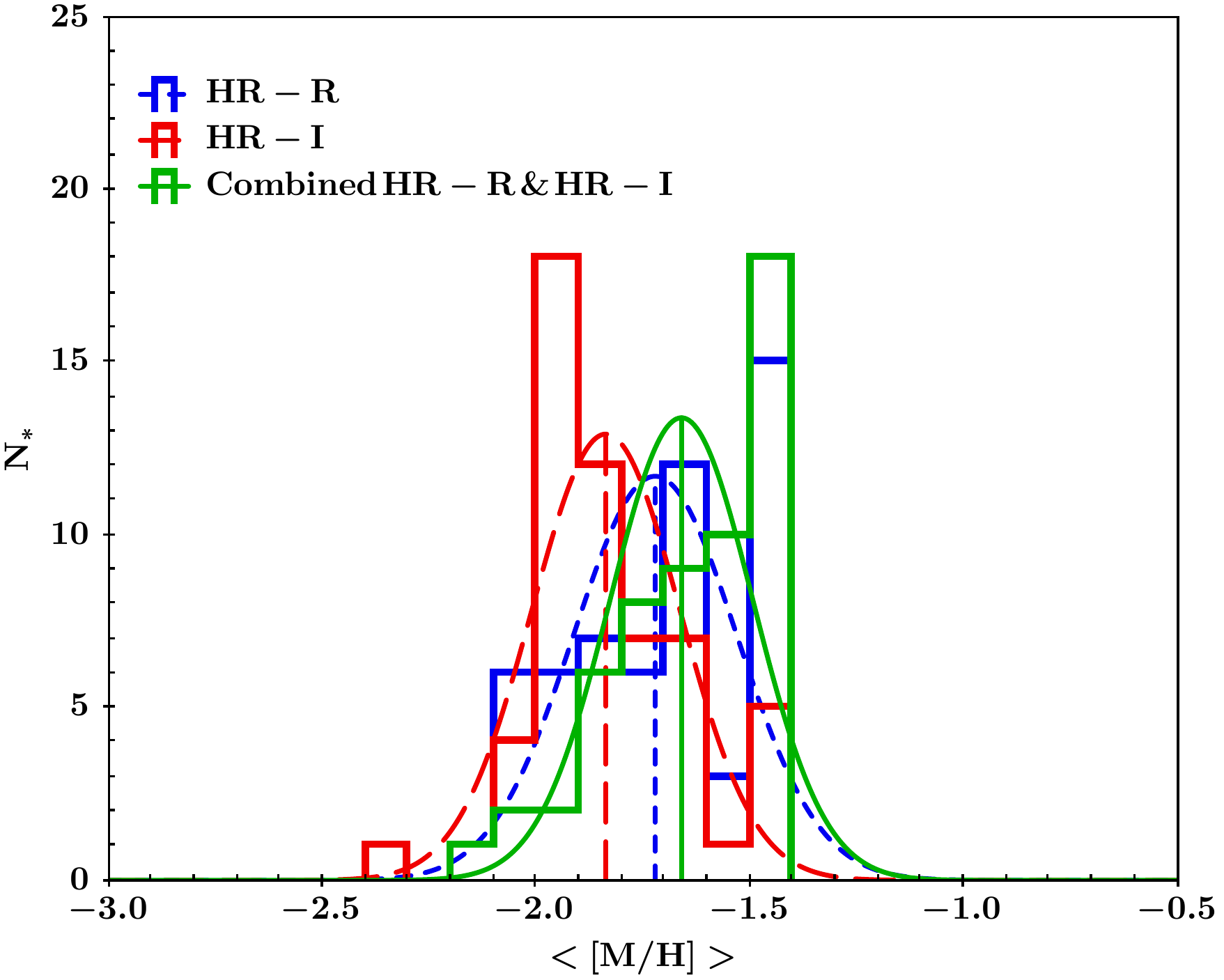}
\caption{Distribution of averaged [M/H] obtained for the M15 stars after using the $\chi^{2} $ technique for all of them. The blue lines correspond to the HR-R results while the red ones are  the fit to HR-I spectra. The green line represents the results for the fit to the combined spectrum with HR-R and HR-I. We have added a Gaussian fit for each set of data.}
\label{mh-m15}
\end{figure}

\subsubsection{Stellar parameters for the M15 stars}
\label{M15-res}
We have repeated the process described in the previous section for deriving the stellar parameters of the 56 stars obtained with MEGARA MOS within the central 12 arcsec region of the M15 cluster. We have not found in the literature any identification of these stars either their stellar parameters, except for the global metallicity of the cluster, which is estimated between  $\rm [Fe/H]\sim -2.15$ \citep{mcnamara04}, and  $-2.30$ \citep{sneden00,caretta09}. A detailed discussion on M15 abundance is presented in \citet{sobeck11}.

Taking into account this M15 metallicity commonly agreed in the literature, we have restricted the possible MUN05 models to the three sets with the lowest value of $\rm [M/H]$ (-2.5, -2.0 and -1.0). We have done the fit of the models to the observed spectra in each of the two different setups, HR-R and HR-I, obtaining two sets of physical stellar parameters. Also, as described in section~\ref{star-par} and also done for OT stars (subsection ~\ref{OT-res}), we have repeated the fit to a combined spectrum containing the two spectral ranges (HR-R and HR-I).

Table~\ref{parameters-M15} summarises the results (this table is given in electronic format but we show here some rows as an example). Column 1 displays the star number; column 2 the $\chi^{2}_{min}$ obtained from the fitting process of the observed normalised spectrum to the MUN05 theoretical catalogue; column 3 displays the associated maximum probability, $P_{\rm max}$. Columns 4, 5 and 6 show the derived stellar parameters, $\rm T_{eff}$, $\rm \log{g}$, and $\rm [M/H]$, from the MUN05 model corresponding to that $\chi^{2}_{min}$. Column 7 gives the number of models, $N$, in the likelihood region $R_{\nu,\alpha}$. Columns 8, 9 and 10 give the mean stellar parameters: $\langle \rm T_{eff}\rangle$, $\langle \rm \log{g}\rangle$ and $\langle \rm [M/H]\rangle$, with their corresponding errors, obtained as the averaged values from the set of $N$ models. The first 5 rows correspond to our results for HR-R; next 5 rows give the results of the fitting to the HR-I spectra. Finally, the last 5 rows show the parameters when fitting the models to the combination of HR-R and HR-I observations into a single spectrum. In this case, as said before, there are not stellar parameters from the literature to compare with, except the average cluster abundance, found to be around $\rm [M/H]$~$\sim$~-2~dex. We will analyse these results in section~\ref{analisis}. 

Figure~\ref{fits-M15} shows the observed spectra of six selected M15 stars (labelled as Star-11, 24, 29, 46, 50 and 56). The first row of this figure shows the panels with HR-R spectra and their fitted models of the stars 11, 24 and 29, while the second row displays the corresponding HR-I spectra of stars whose HR-R spectrum are shown just above. The sequence is repeated with rows 3 (HR-R) and 4 (HR-I) for stars 46, 50 and 56. We have over-plotted the $\chi^{2}_{min}$ model (displaying in green both the line fitting and the derived stellar parameters), and the average model (in blue). The fits represented in these figures are those to the combined spectra of HR-R and HR-I, with the two spectral windows in a single spectrum. The observed and fitted models for the 56 stars are given in Figure~A~1 of Appendix A, in the online version.

Figure~\ref{mh-m15} displays the histograms of the stellar metallicity distribution for all the 56 stars in our sample of M15 resulting from the fit of the models to HR-R (blue line) and HR-I (red line). We have over-plotted a Gaussian fit to the data of each setup, obtaining mean abundance values of $\rm [M/H]=-1.72~\pm$~0.19\,dex, and $\rm [M/H]=-1.84~\pm$~0.17\,dex, for HR-R and HR-I respectively. We have also fitted the model to the combined spectrum of HR-R and HR-I for each star obtaining $\rm [M/H]=-1.66~\pm$~0.16\,dex. The resulted abundance we derive for the 56 stars in the centre of the cluster is slightly higher than the average value for the whole cluster ($\rm [M/H]=-2.15$,dex), claimed in previous published papers \citep{sobeck11}. 

The purpose of including the M15 stars commissioning has been primarily to increase the number of stars in the sample analysed in this piece of work. However, a complete analysis of M15 with these observations, and all the setups throughout the complete optical spectra range, is being carried out by the MEGARA commissioning team.

\subsubsection{Stellar parameters for the 20 OT stars}
\label{OT-res}

We have also applied our fitting method to the HR-R and HR-I spectra of the 20 OT stars in our MEGARA-GTC library sample. In this case, each star may have different stellar parameters (not as for M15 stars, whose members are expected to share a common metallicity and age, implying close values of their physical parameters). As in the case of the COM stars, and as it will be the case in all the stars of the MEGARA-GTC library, the stars from the OT sub-sample have reported estimates of the stellar parameters usually obtained from spectra with a wider spectral range and lower spectral resolution than MEGARA spectra. The observations in HR-R and HR-I, with much higher spectral resolution can substantially change the results of the previous estimates. 

\begin{figure*}
\includegraphics[width=0.24\textwidth,angle=-90]{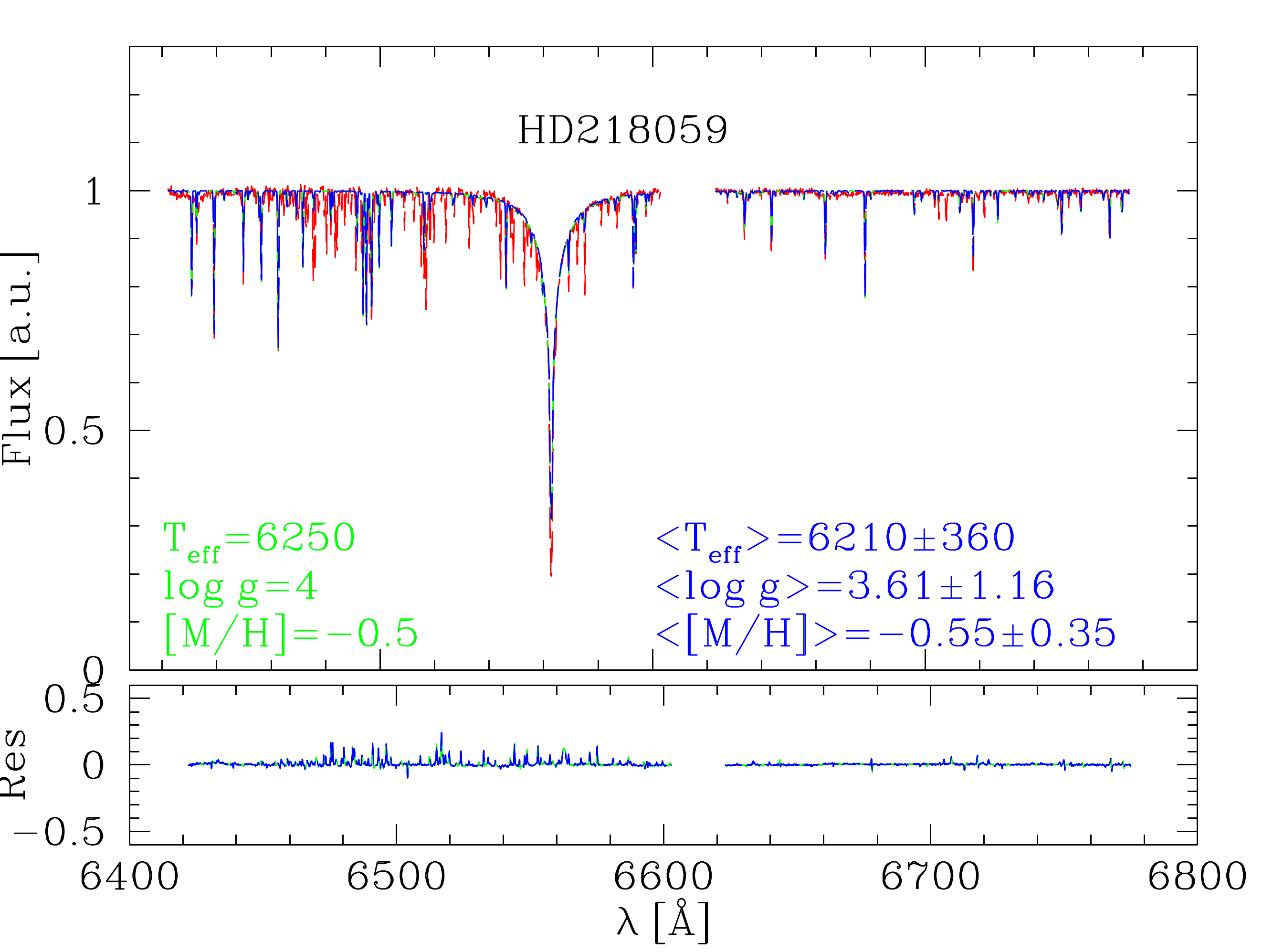}
\includegraphics[width=0.24\textwidth,angle=-90]{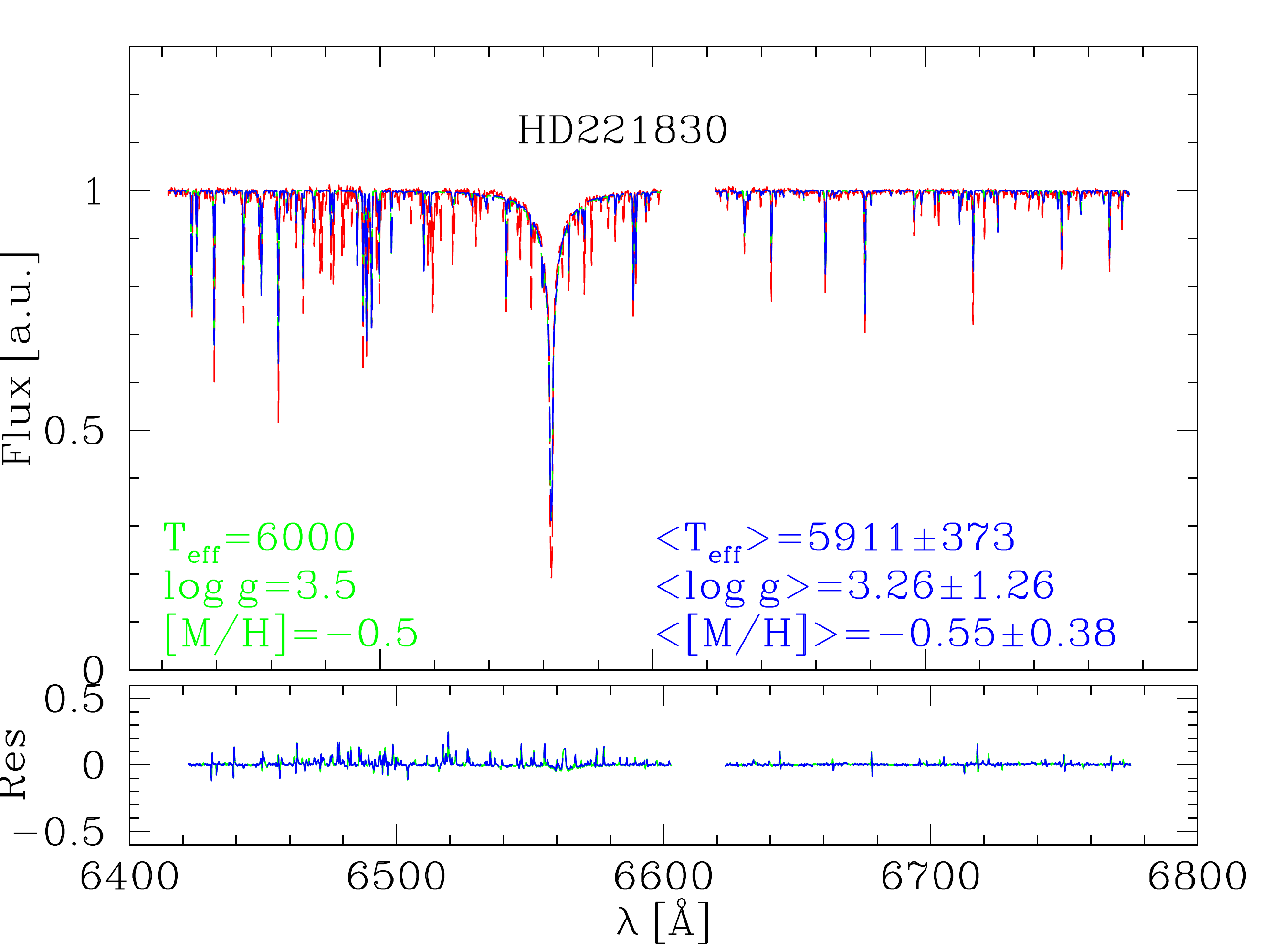}
\includegraphics[width=0.24\textwidth,angle=-90]{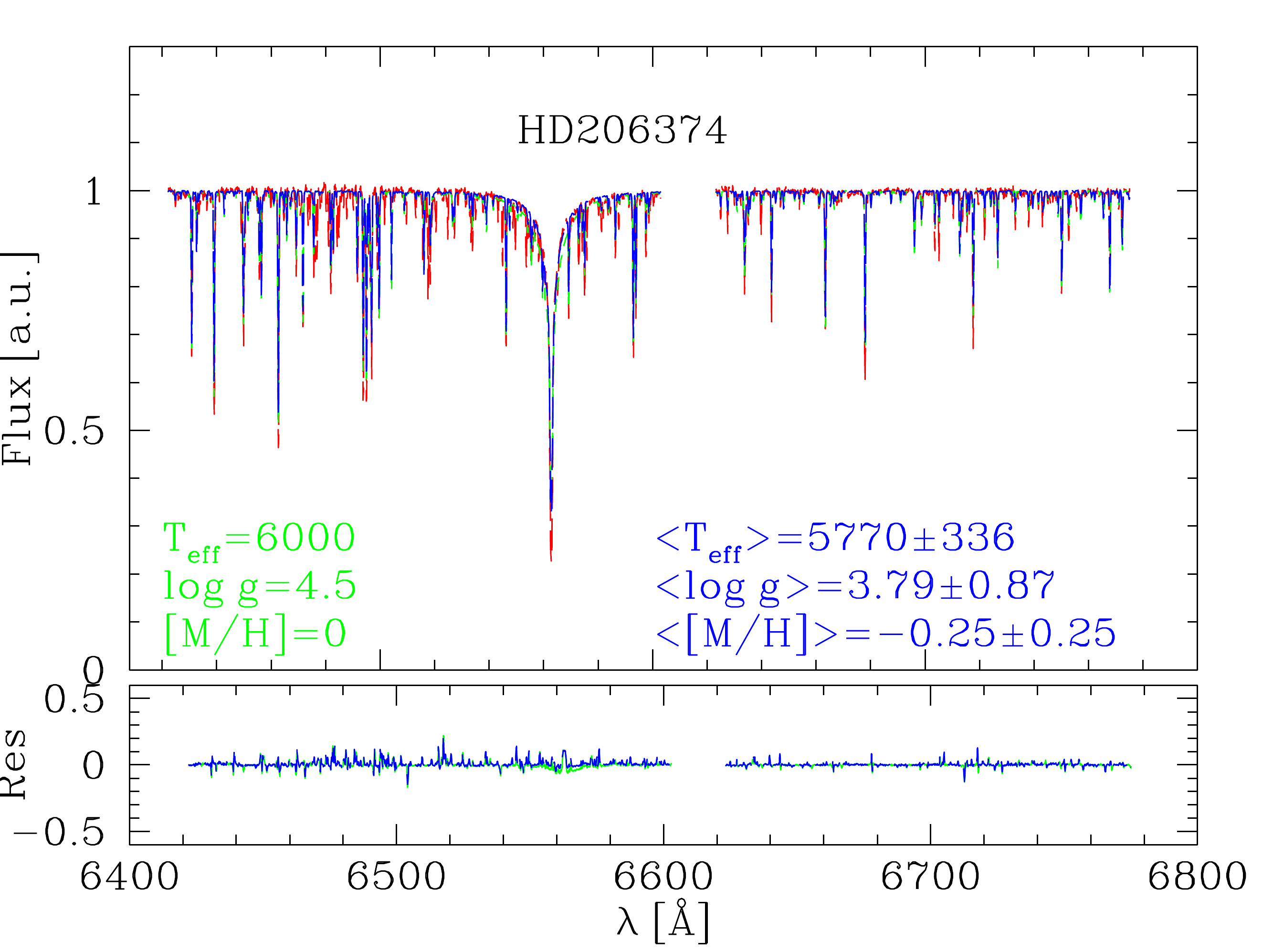}
\includegraphics[width=0.24\textwidth,angle=-90]{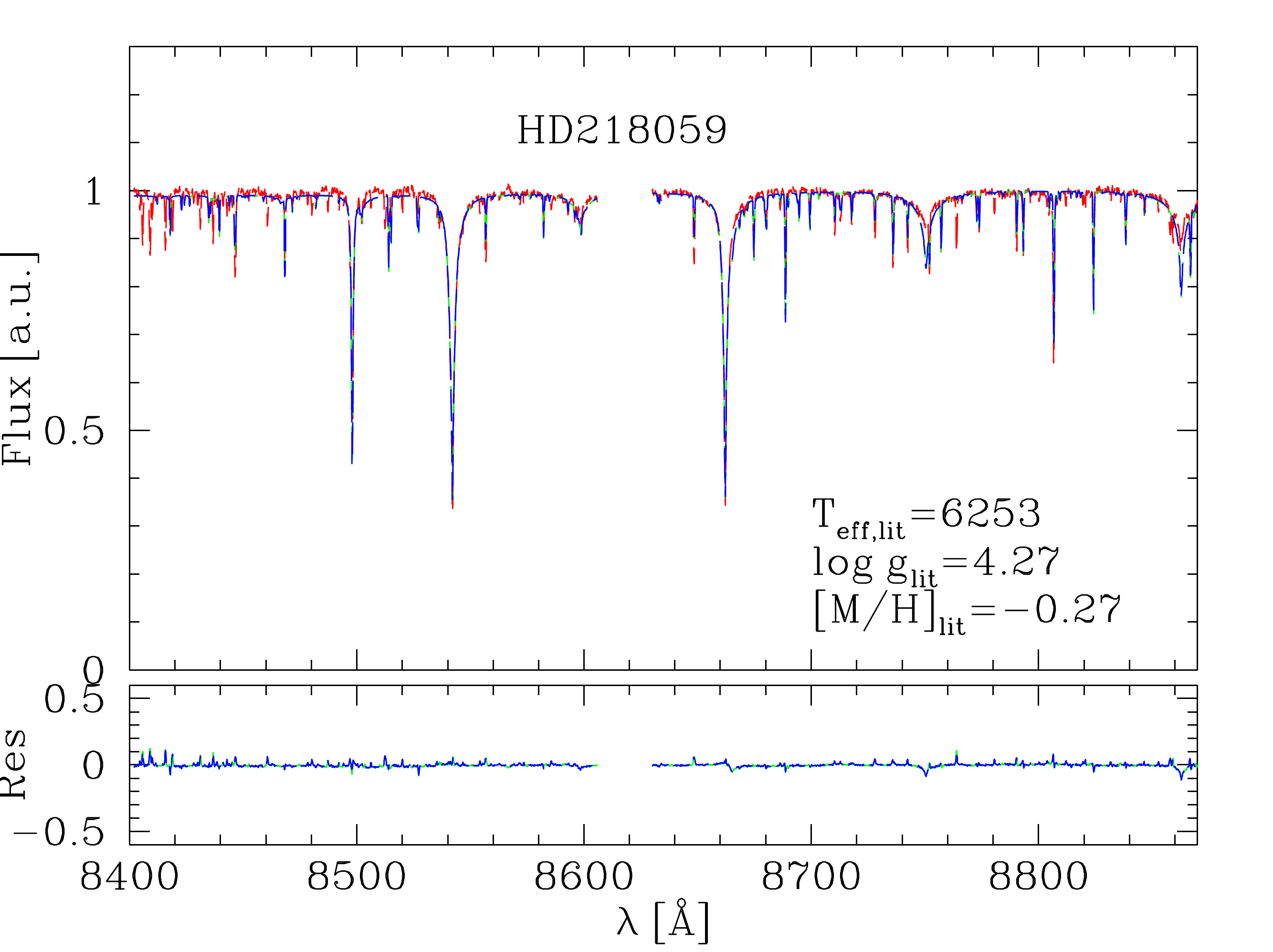}
\includegraphics[width=0.24\textwidth,angle=-90]{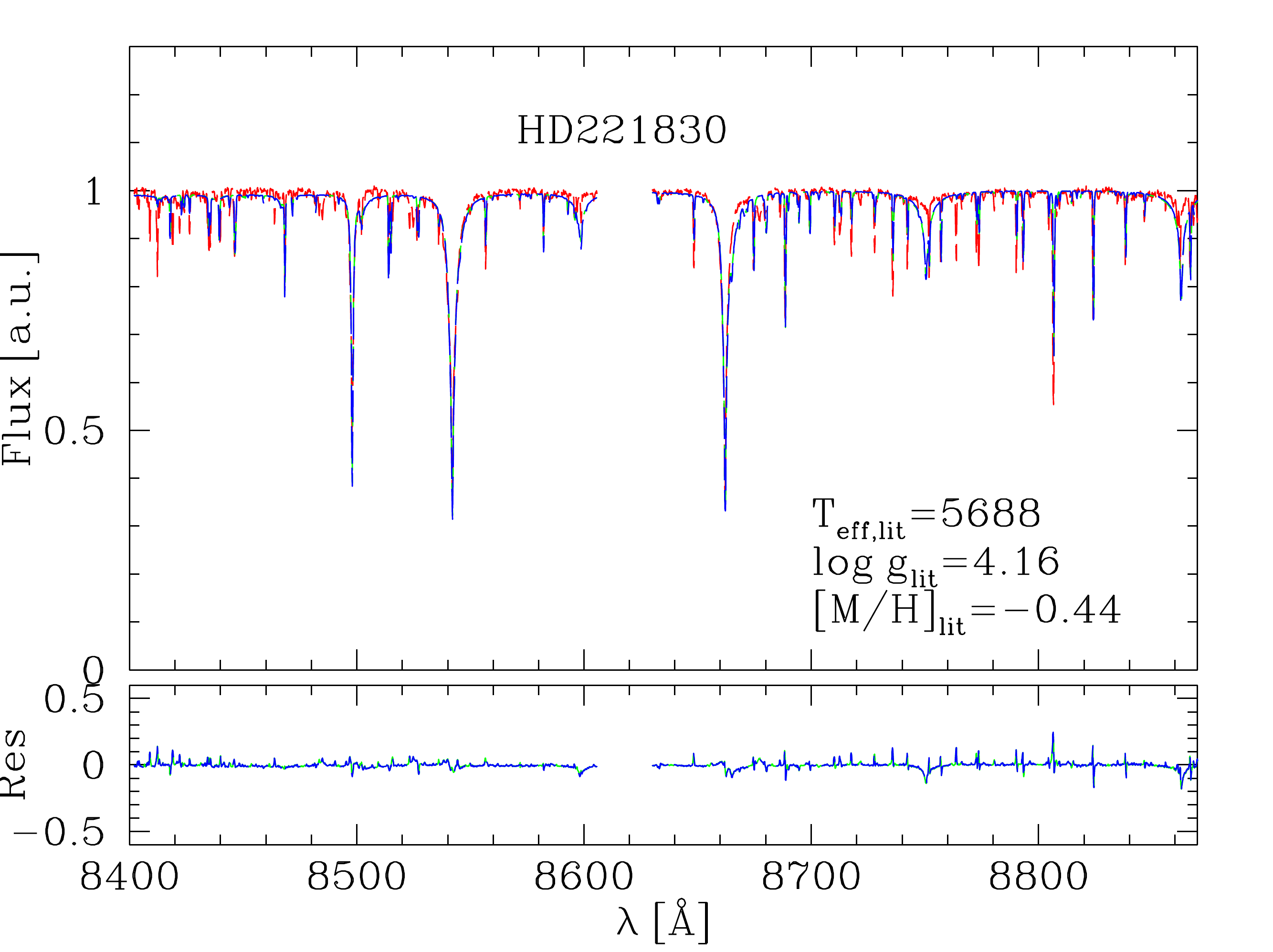}
\includegraphics[width=0.24\textwidth,angle=-90]{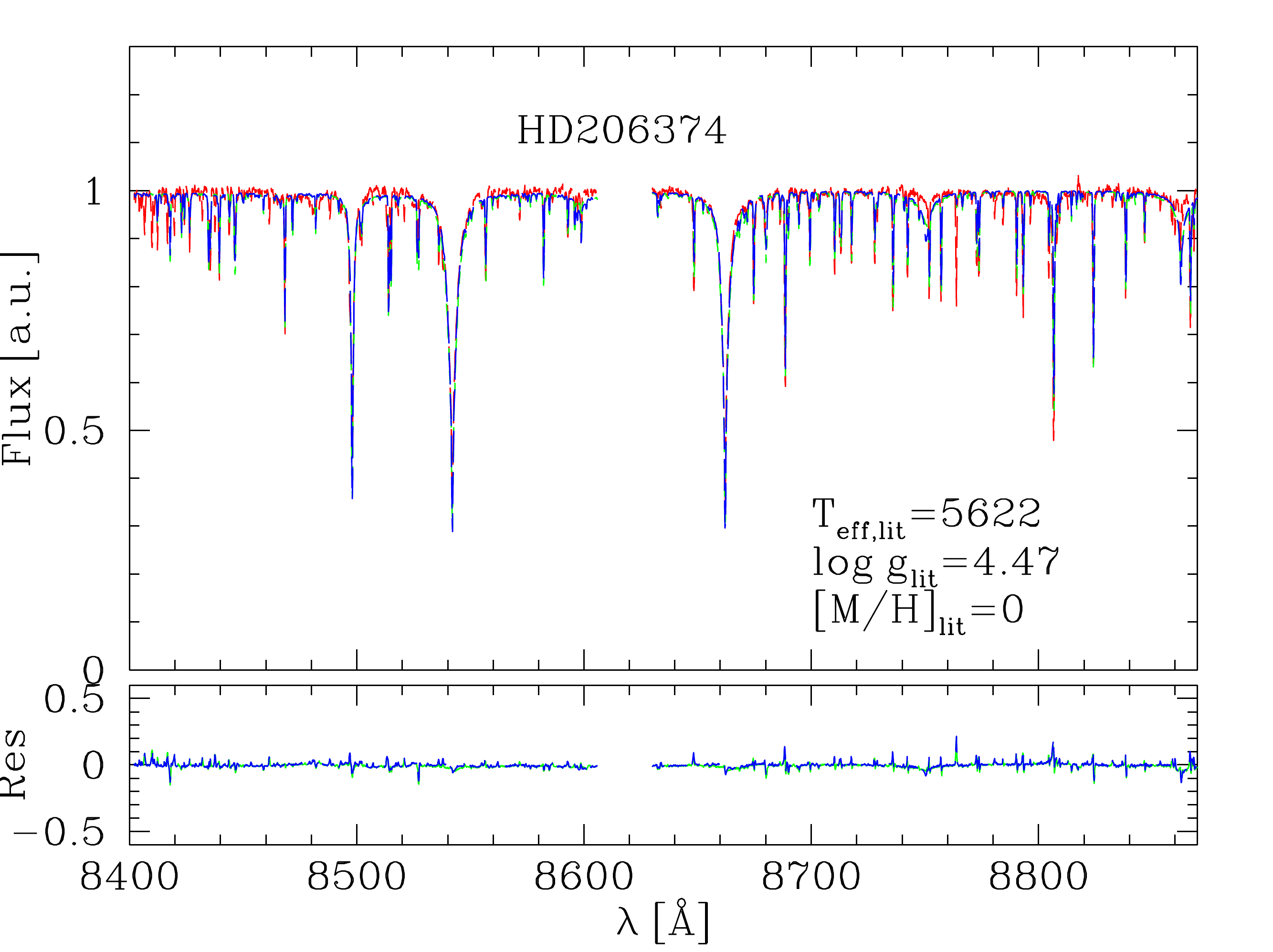}
\caption{The best MUN05 model fit is over plotted to each observed spectrum for three selected OT stars. The top panel of each star shows the observed spectrum (red solid line), the averaged fitted model (blue long-dashed line) and the minimum $\chi^{2}$ model (green short-dashed line). The corresponding bottom panel shows the residuals as Flux$_{obs}$ - Flux$_{mod}$ (blue and green lines as the corresponding fitted models). Three spectra in HR-R are shown in the top row  while the ones in HR-I are just below. The spectra and their fitting results for the rest of the OT stars are available online (Figure~B~1, Appendix B, online version).}
\label{fits-OT}
\end{figure*}
Table~\ref{parameters-OT} summarises our results for these stars. In the upper panel, star name is given in column 1. Columns 2 to 10 contain the results for the fitting to HR-R spectra while columns 11 to 19 have the corresponding values when using the HR-I observed spectra.  The value of $\chi^{2}_{min}$ in column 2 and 11, the corresponding maximum probability P in column 3 and 12, the stellar parameters which correspond to the $\chi^{2}_{min}$ model in columns 4, 5 and 6 (13, 14 and 15 for HR-I). Then, we have in column 7 the number of models $N$ with similar $\chi^{2}$ as $\chi^{2}_{min}$, and the averaged values of stellar parameters obtained with these $N$ models in columns 8, 9 and 10, with their corresponding errors; and the equivalent parameters for HR-I fittings in columns 16, 17, 18 and 19. The stellar parameters from the literature are given in Table~\ref{opentimestars}. The lower panel shows the values when fitting the models to the combined spectrum with the information of both, HR-R and HR-I set-ups, since these spectra are two windows of a single spectrum. Again, star name is given in column 1; columns 2 to 10 contain the results for the fitting to the combined spectra, corresponding to the parameters as labelled in the table; columns 11 to 13 repeat the physical parameters from the literature obtained with spectra of resolving Power, R, shown in column 14. 

Figure~\ref{fits-OT} shows the fitting to the HR-R (upper panel) and to the HR-I (lower panel) observed spectra of three stars from the OT sub-sample. The fits represented in these figures are those to the combined spectra of HR-R and HR-I, with the two spectral windows in a single spectrum. We have repeated the analysis for the whole OT star sub-sample and the fittings are shown in Figure~B~1 of Appendix B, in the online version.. 

\subsection{Analysis of the stellar parameter estimates}
\label{analisis}
We have found physical parameters in the literature for some of the stars in our COM and OT sub-samples. These parameters have been derived from spectra with lower resolution and wider wavelength range than our observations taken with MEGARA and HR-R/HR-I setups. We have, however, used these data to check the correlation between the stellar parameters derived from the literature and our estimates from the model fittings. The comparison presented in this section is between the previous published values and the ones we obtain when fitting the models to the HR-I spectrum only (in the case of the COM sub-sample) and to the combined spectrum with HR-R and HR-I, in the OT sub-sample. The wider the spectral range available for the fitting, the more reliable values of the physical parameters, as discussed in sub-section{star-par}.

\begin{figure}
\centering
\includegraphics[width=0.35\textwidth,angle=-90]{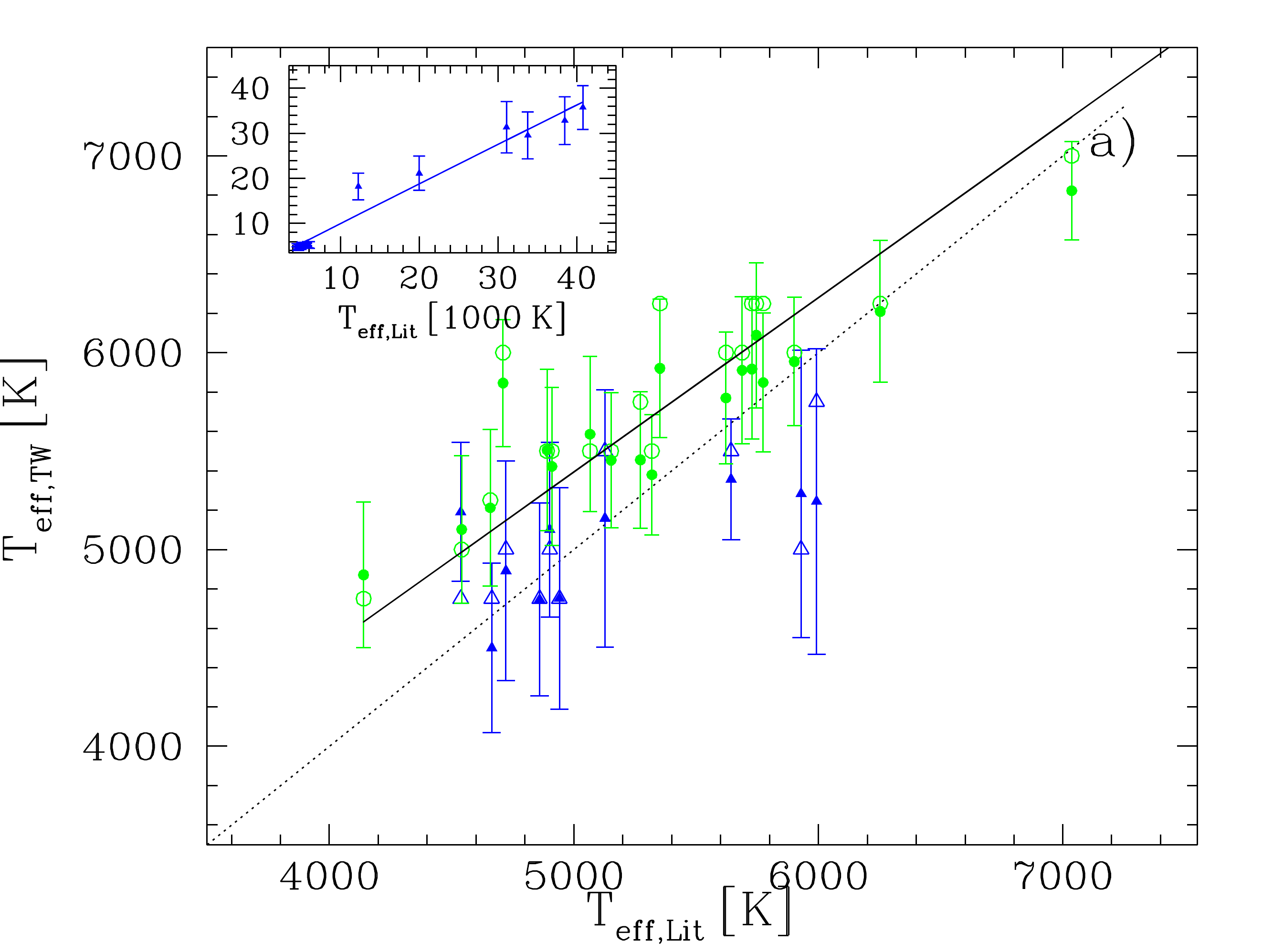}
\includegraphics[width=0.35\textwidth,angle=-90]{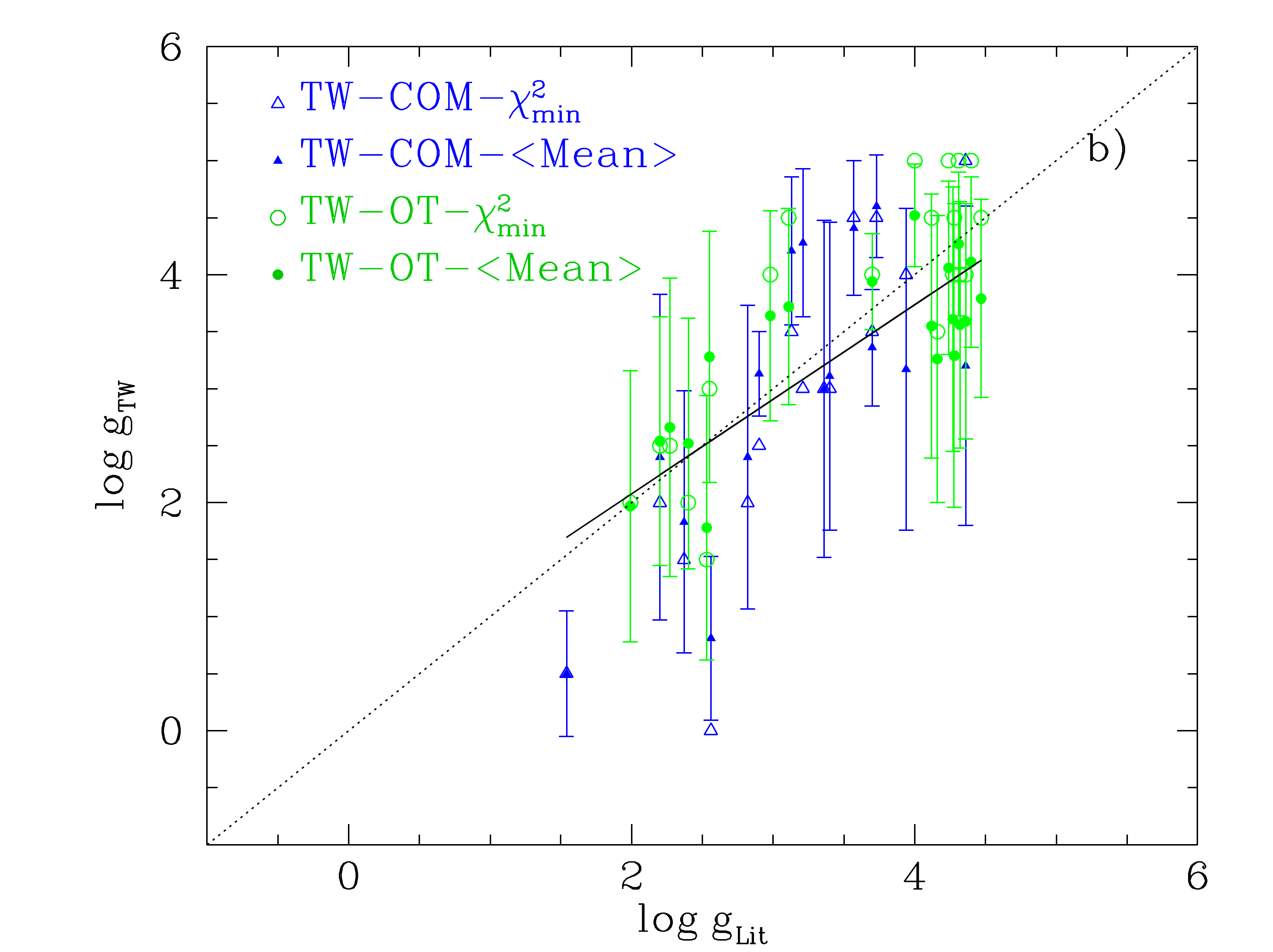}
\includegraphics[width=0.35\textwidth,angle=-90]{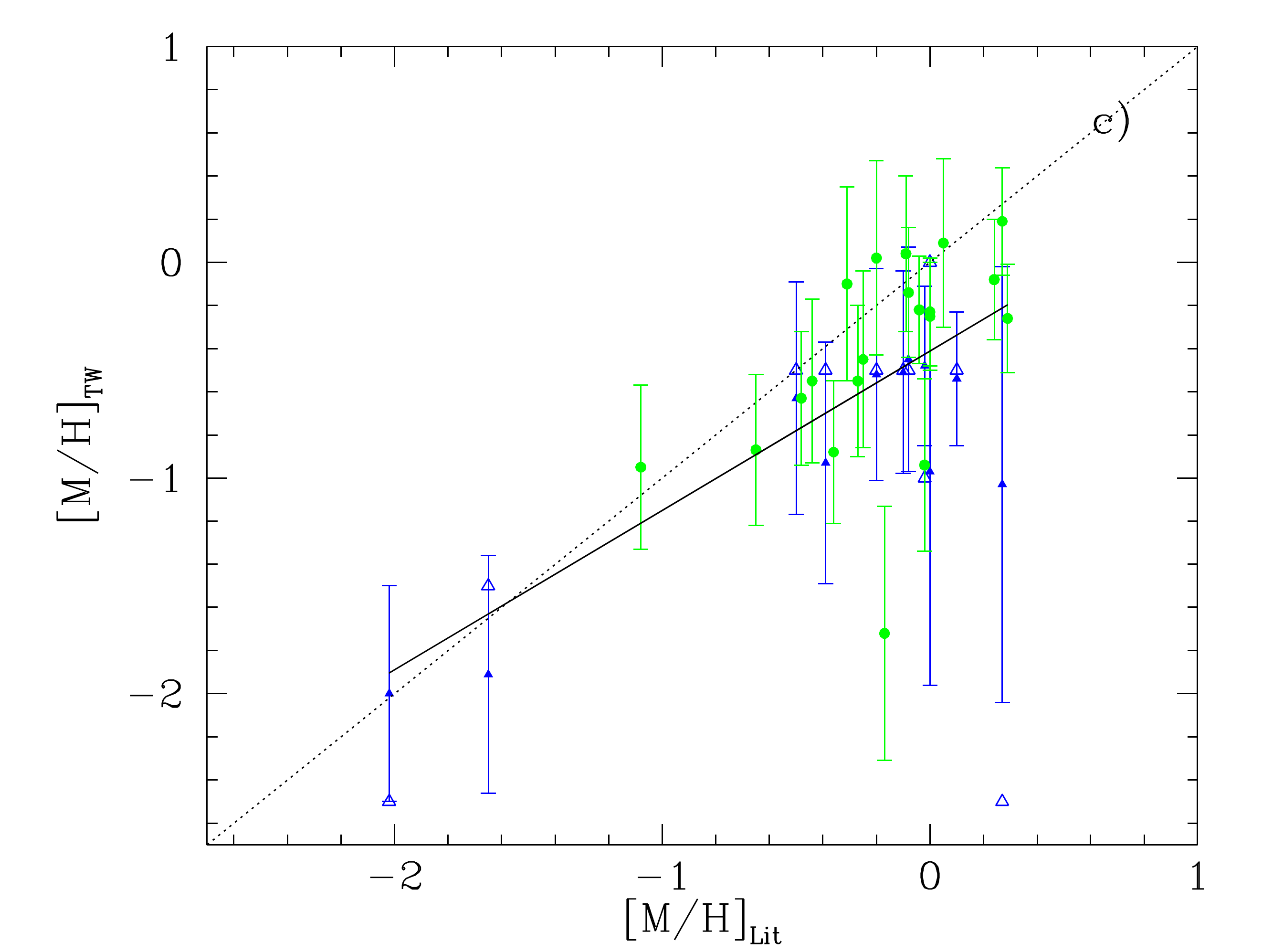}
\caption{Values of T$_{\rm eff}$, $\log{\rm g}$ and $\rm [M/H]$ resulting from our fitting compared with those from the literature for COM and OT stars. Open symbols correspond to the $\chi^{2}_{min}$ parameters, while the solid dots with error bars refer to the averaged values; blue triangles and green dots correspond to COM and OT stars. The minimum square straight lines in black is set for the whole set of points. The dotted identity line is in all panels.}
\label{compara}
\end{figure}

Figure~\ref{compara} shows the comparison between the stellar parameters found in this work after fitting the observed spectra (y-axis) with those obtained by previous works (x-axis). Open symbols represent the parameters of the $\chi^{2}_{min}$ model and the full ones the averaged parameters from the models whose $\chi^{2}$ fulfils Eq.~\ref{condition}. We plot the results for the COM stars as blue triangles, and the OT stars as green dots. For COM stars we have parameters from the fitting to HR-I spectra only, while in the case of OT stars sub-sample, we have plotted the fit to the combined  HR-R and HR-I spectra, that is using the complete spectral information (see values in
Table~\ref{parameters-com} and Table~\ref{parameters-OT}).

Panel a) of Figure~\ref{compara} shows our estimates of $\rm T_{eff}$ versus the ones from the literature.  When all points are used together, we obtained the minimum squares straight line shown by the black solid line. This fit follows closely the 1:1 line within the error bars. The embedded small figure shows the same plot including the hottest stars which reach 40000~K, where there are only a certain number of COM stars. The blue line is is the fit obtained only for these COM stars. 
The correlation between the averaged values of $\rm \log{g}$ and the ones from the literature (panel b) is in general quite good for the stellar parameters derived from the fitting to the observed spectra, with which we compute the black line fit. 
Finally, panel c) shows that the metallicity derived from the models also follows a clear trend, showing values similar to the ones from the literature ($\rm [M/H]_{Lit}$), within the errors. It is necessary to remind that the theoretical catalogue (MUN05 in our case) provides a discrete sampling of $\rm T_{eff}$, $\rm \log{g}$ and $\rm [M/H]$, which introduces an important source of uncertainty. 

\normalsize
\begin{table*}
\scriptsize
\caption{Stellar parameters for the 20 OT stars derived from the fitting with the theoretical models and from the literature for the HR-R setup. Table description in subsection~\ref{OT-res}.}
\label{parameters-OT}
\begin{tabular}{p{0.75cm}p{0.20cm}p{0.15cm}rp{0.10cm}rrrp{0.80cm}rp{0.20cm}p{0.15cm}rp{0.15cm}rrrp{0.80cm}p{1.05cm}}
\hline
& \multicolumn{9}{c}{HR-R} & \multicolumn{9}{c}{HR-I}\\ 
Star & $\chi^{2}_{min}$ & P$_{\rm max}$ &  \multicolumn{3}{c}{Results with ${\chi^{2}}_{min}$} & $N$ & \multicolumn{3}{c}{Average results with $N$ models} &
$\chi^{2}_{min}$ & P$_{\rm max}$ &  \multicolumn{3}{c}{Results with ${\chi^{2}}_{min}$} & $N$ & \multicolumn{3}{c}{Average results with $N$ models}
\\
Name & & & T$_{\rm eff}$ & $\log{g}$ & $[M/H]$ & & $\langle \rm T_{ eff} \rangle$ & $\langle\log{g}\rangle$ & $\langle\rm [M/H]\rangle$ & & & T$_{\rm eff}$ & $\log{g}$ & $[M/H]$ & & $\langle \rm T_{ eff} \rangle$ & $\langle\log{g}\rangle$ & $\langle\rm [M/H]\rangle$  \\ 
\hline
HD~147677	&	1.240	&	0.74	&	5250	&	1.0	&	0.0	&	25	&	5360	$\pm$	307	&	1.6	$\pm$	1.0	&	0.18	$\pm$	0.32	&	1.005	&	0.80	&	7250	&	5.0	&	0.0	&	8	&	7156	$\pm$	352	&	4.8	$\pm$	0.3	&	0.00	$\pm$	0.00	\\
HD~174912	&	0.963	&	0.81	&	6250	&	2.0	&	-0.5	&	40	&	6088	$\pm$	292	&	1.3	$\pm$	1.0	&	-0.86	$\pm$	0.36	&	0.428	&	0.93	&	6250	&	5.0	&	-0.5	&	42	&	5548	$\pm$	539	&	4.3	$\pm$	0.8	&	-0.87	$\pm$	0.38	\\
HD~200580	&	1.784	&	0.62	&	6000	&	0.0	&	-2.5	&	39	&	6910	$\pm$	1436	&	1.1	$\pm$	0.7	&	-1.51	$\pm$	0.97	&	1.150	&	0.77	&	6000	&	5.0	&	-0.5	&	26	&	5760	$\pm$	541	&	4.6	$\pm$	0.5	&	-0.71	$\pm$	0.40	\\
HD~206374	&	0.466	&	0.93	&	6000	&	3.0	&	0.0	&	27	&	5806	$\pm$	313	&	2.3	$\pm$	1.2	&	-0.20	$\pm$	0.25	&	0.112	&	0.99	&	5500	&	5.0	&	-0.5	&	79	&	5472	$\pm$	602	&	4.0	$\pm$	0.9	&	-0.55	$\pm$	0.37	\\
HD~211472	&	0.503	&	0.92	&	5500	&	4.5	&	0.0	&	18	&	5444	$\pm$	251	&	3.6	$\pm$	1.1	&	-0.03	$\pm$	0.21	&	0.227	&	0.97	&	5000	&	5.0	&	-0.5	&	58	&	5379	$\pm$	644	&	4.4	$\pm$	0.6	&	-0.44	$\pm$	0.36	\\
HD~218059	&	0.366	&	0.95	&	6250	&	2.0	&	-0.5	&	39	&	6192	$\pm$	295	&	1.6	$\pm$	1.1	&	-0.58	$\pm$	0.34	&	0.187	&	0.98	&	6250	&	4.0	&	-0.5	&	59	&	5661	$\pm$	551	&	3.8	$\pm$	1.1	&	-0.84	$\pm$	0.40	\\
HD~220182	&	3.347	&	0.34	&	6500	&	3.5	&	-1.5	&	218	&	7409	$\pm$	2355	&	3.0	$\pm$	1.4	&	-1.40	$\pm$	0.86	&	0.418	&	0.94	&	6000	&	5.0	&	0.0	&	62	&	5657	$\pm$	659	&	4.2	$\pm$	0.7	&	-0.35	$\pm$	0.33	\\
HD~221585	&	0.842	&	0.84	&	5500	&	1.0	&	0.0	&	31	&	5718	$\pm$	340	&	1.8	$\pm$	1.2	&	0.10	$\pm$	0.30	&	0.606	&	0.90	&	6500	&	4.0	&	0.0	&	52	&	6062	$\pm$	634	&	4.2	$\pm$	0.9	&	-0.03	$\pm$	0.30	\\
HD~221830	&	0.511	&	0.92	&	5750	&	1.5	&	-0.5	&	27	&	5815	$\pm$	299	&	1.7	$\pm$	1.1	&	-0.41	$\pm$	0.31	&	0.336	&	0.95	&	6000	&	5.0	&	-0.5	&	107	&	5493	$\pm$	697	&	3.9	$\pm$	1.1	&	-0.81	$\pm$	0.48	\\
BD~08$+$3095	&	0.397	&	0.94	&	6000	&	1.5	&	-1.0	&	36	&	5972	$\pm$	285	&	1.8	$\pm$	1.1	&	-1.06	$\pm$	0.39	&	0.117	&	0.99	&	6500	&	5.0	&	-0.5	&	78	&	5356	$\pm$	663	&	4.0	$\pm$	0.9	&	-1.21	$\pm$	0.48	\\
HD~100696	&	0.731	&	0.87	&	5500	&	2.0	&	-0.5	&	37	&	5615	$\pm$	304	&	2.3	$\pm$	1.2	&	-0.49	$\pm$	0.34	&	1.296	&	0.73	&	5250	&	0.5	&	-1.0	&	17	&	5382	$\pm$	332	&	1.2	$\pm$	0.9	&	-0.71	$\pm$	0.36	\\
HD~101107	&	0.233	&	0.97	&	9750	&	2.5	&	-2.5	&	106	&	9316	$\pm$	1843	&	3.2	$\pm$	0.9	&	-1.73	$\pm$	0.73	&	0.704	&	0.87	&	6750	&	5.0	&	-0.5	&	24	&	6469	$\pm$	332	&	4.3	$\pm$	0.6	&	-0.77	$\pm$	0.25	\\
HD~104985	&	1.042	&	0.79	&	5250	&	1.0	&	0.0	&	21	&	5405	$\pm$	349	&	1.5	$\pm$	1.2	&	0.07	$\pm$	0.33	&	0.873	&	0.83	&	4500	&	0.5	&	-1.0	&	29	&	4931	$\pm$	433	&	1.1	$\pm$	0.9	&	-0.67	$\pm$	0.47	\\
HD~113002	&	1.692	&	0.64	&	5750	&	0.5	&	-1.5	&	30	&	5867	$\pm$	243	&	0.8	$\pm$	0.7	&	-1.52	$\pm$	0.50	&	0.369	&	0.95	&	5000	&	0.5	&	-1.5	&	34	&	5066	$\pm$	400	&	1.5	$\pm$	1.1	&	-1.15	$\pm$	0.47	\\
HD~115136	&	0.390	&	0.94	&	5000	&	0.5	&	0.0	&	13	&	5212	$\pm$	247	&	1.5	$\pm$	0.9	&	0.27	$\pm$	0.33	&	0.159	&	0.98	&	4250	&	0.5	&	-1.0	&	33	&	4462	$\pm$	415	&	1.7	$\pm$	1.1	&	-0.44	$\pm$	0.51	\\
HD~117243	&	0.227	&	0.97	&	6000	&	2.0	&	0.0	&	26	&	5942	$\pm$	319	&	1.9	$\pm$	1.1	&	-0.15	$\pm$	0.24	&	0.336	&	0.95	&	5500	&	2.5	&	-0.5	&	73	&	5479	$\pm$	520	&	3.2	$\pm$	1.5	&	-0.47	$\pm$	0.46	\\
HD~131111	&	0.106	&	0.99	&	6250	&	4.0	&	0.0	&	30	&	6158	$\pm$	191	&	3.7	$\pm$	0.9	&	-0.23	$\pm$	0.25	&	0.063	&	1.00	&	4750	&	3.0	&	-0.5	&	62	&	4819	$\pm$	460	&	3.4	$\pm$	1.2	&	-0.52	$\pm$	0.44	\\
HD~131507	&	1.384	&	0.71	&	5250	&	1.5	&	0.5	&	19	&	5184	$\pm$	261	&	1.6	$\pm$	1.1	&	0.32	$\pm$	0.25	&	2.053	&	0.56	&	5500	&	0.0	&	-0.5	&	14	&	5357	$\pm$	389	&	0.5	$\pm$	0.5	&	-0.32	$\pm$	0.25	\\
HD~144206	&	0.301	&	0.96	&	7250	&	2.5	&	-1.5	&	201	&	8141	$\pm$	1221	&	2.2	$\pm$	0.9	&	-1.15	$\pm$	0.88	&	0.393	&	0.94	&	21000	&	3.0	&	0.5	&	50	&	17825	$\pm$	6607	&	3.2	$\pm$	0.4	&	-0.97	$\pm$	1.15	\\
HD~175535	&	0.552	&	0.91	&	5500	&	1.5	&	0.0	&	29	&	5638	$\pm$	331	&	2.2	$\pm$	1.2	&	0.10	$\pm$	0.31	&	0.207	&	0.98	&	4750	&	2.0	&	-0.5	&	49	&	5138	$\pm$	487	&	3.0	$\pm$	1.3	&	-0.18	$\pm$	0.49	\\
\hline
\end{tabular}

\begin{tabular}{lccrcrrccrrrrr}

\hline
& \multicolumn{9}{c}{Simultaneous fitting of HR-R \& HR-I} & \multicolumn{4}{c}{}\\ 
Star & $\chi^{2}_{min}$ & P$_{\rm max}$ &  \multicolumn{3}{c}{Results with ${\chi^{2}}_{min}$} & $N$ & \multicolumn{3}{c}{Average results with $N$ models} & \multicolumn{4}{c}{Data from Literature}\\
Name & & & T$_{\rm eff}$ & $\log{g}$ & $[M/H]$ & & $\langle \rm T_{ eff} \rangle$ & $\langle\log{g}\rangle$ & $\langle\rm [M/H]\rangle$ & $\langle \rm T_{ eff} \rangle$ & $\langle\log{g}\rangle$ & $\langle\rm [M/H]\rangle$ & R \\ 
\hline
HD~147677	&	0.627	&	0.89	&	5500	&	4.0	&	0.0	&	29	&	5422	$\pm$	402	&	3.6	$\pm$	0.9	&	-0.14	$\pm$	0.30	&	4910	&	3.0	&	-0.08	&	5000	\\
HD~174912	&	0.288	&	0.96	&	6250	&	4.0	&	-0.5	&	34	&	6088	$\pm$	369	&	3.6	$\pm$	1.1	&	-0.63	$\pm$	0.31	&	5746	&	4.3	&	-0.48	&	2000	\\
HD~200580	&	0.292	&	0.96	&	6250	&	4.5	&	-0.5	&	43	&	5849	$\pm$	354	&	3.3	$\pm$	1.3	&	-0.87	$\pm$	0.35	&	5774	&	4.3	&	-0.65	&	2000	\\
HD~206374	&	0.202	&	0.98	&	6000	&	4.5	&	0.0	&	38	&	5770	$\pm$	336	&	3.8	$\pm$	0.9	&	-0.25	$\pm$	0.25	&	5622	&	4.5	&	0.00	&	42000	\\
HD~211472	&	0.209	&	0.98	&	5500	&	5.0	&	0.0	&	27	&	5380	$\pm$	305	&	4.1	$\pm$	0.8	&	-0.22	$\pm$	0.25	&	5319	&	4.4	&	-0.04	&	42000	\\
HD~218059	&	0.237	&	0.97	&	6250	&	4.0	&	-0.5	&	31	&	6210	$\pm$	360	&	3.6	$\pm$	1.2	&	-0.55	$\pm$	0.35	&	6253	&	4.3	&	-0.27	&	42000	\\
HD~220182	&	0.479	&	0.92	&	5750	&	5.0	&	0.0	&	28	&	5455	$\pm$	347	&	4.3	$\pm$	0.6	&	-0.23	$\pm$	0.25	&	5272	&	4.3	&	0.00	&	42000	\\
HD~221585	&	0.684	&	0.88	&	6250	&	5.0	&	0.5	&	35	&	5921	$\pm$	352	&	4.1	$\pm$	0.8	&	0.19	$\pm$	0.25	&	5352	&	4.2	&	0.27	&	42000	\\
HD~221830	&	0.299	&	0.96	&	6000	&	3.5	&	-0.5	&	56	&	5911	$\pm$	373	&	3.3	$\pm$	1.3	&	-0.55	$\pm$	0.38	&	5688	&	4.2	&	-0.44	&	2000	\\
BD~08$+$3095	&	0.191	&	0.98	&	6250	&	4.5	&	-0.5	&	42	&	5917	$\pm$	356	&	3.6	$\pm$	1.2	&	-0.88	$\pm$	0.33	&	5728	&	4.1	&	-0.36	&	80000	\\
HD~100696	&	0.337	&	0.95	&	5500	&	2.5	&	-0.5	&	55	&	5505	$\pm$	410	&	2.7	$\pm$	1.3	&	-0.45	$\pm$	0.41	&	4890	&	2.3	&	-0.25	&	5000	\\
HD~101107	&	0.384	&	0.94	&	7000	&	5.0	&	-0.5	&	24	&	6823	$\pm$	250	&	4.5	$\pm$	0.5	&	-0.94	$\pm$	0.40	&	7036	&	4.0	&	-0.02	&	1000	\\
HD~104985	&	0.262	&	0.97	&	5250	&	2.5	&	0.0	&	26	&	5212	$\pm$	398	&	2.5	$\pm$	1.1	&	-0.10	$\pm$	0.45	&	4658	&	2.2	&	-0.31	&	5000	\\
HD~113002	&	0.180	&	0.98	&	5500	&	1.5	&	-1.0	&	38	&	5454	$\pm$	343	&	1.8	$\pm$	1.2	&	-0.95	$\pm$	0.38	&	5152	&	2.5	&	-1.08	&	1000	\\
HD~115136	&	0.243	&	0.97	&	5000	&	2.0	&	0.0	&	27	&	5102	$\pm$	375	&	2.5	$\pm$	1.1	&	0.09	$\pm$	0.39	&	4541	&	2.4	&	0.05	&	5000	\\
HD~117243	&	0.171	&	0.98	&	6000	&	4.0	&	0.0	&	33	&	5955	$\pm$	327	&	3.6	$\pm$	1.0	&	-0.08	$\pm$	0.28	&	5902	&	4.4	&	0.24	&	5000	\\
HD~131111	&	0.163	&	0.98	&	6000	&	4.5	&	0.0	&	39	&	5846	$\pm$	323	&	3.7	$\pm$	0.9	&	-0.26	$\pm$	0.25	&	4710	&	3.1	&	0.29	&	5000	\\
HD~131507	&	0.632	&	0.89	&	4750	&	2.0	&	0.0	&	29	&	4871	$\pm$	370	&	2.0	$\pm$	1.2	&	0.02	$\pm$	0.45	&	4140	&	2.0	&	-0.20	&	5000	\\
HD~144206	&	0.344	&	0.95	&	7500	&	4.0	&	-2.0	&	27	&	7417	$\pm$	240	&	3.9	$\pm$	0.4	&	-1.72	$\pm$	0.59	&	11957	&	3.7	&	-0.17	&	2000	\\
HD~175535	&	0.163	&	0.98	&	5500	&	3.0	&	0.0	&	38	&	5586	$\pm$	395	&	3.3	$\pm$	1.1	&	0.04	$\pm$	0.36	&	5066	&	2.6	&	-0.09	&	2000	\\
\hline
\end{tabular}

\end{table*}
\normalsize

\begin{figure}
\centering
\includegraphics[width=0.35\textwidth,angle=-90]{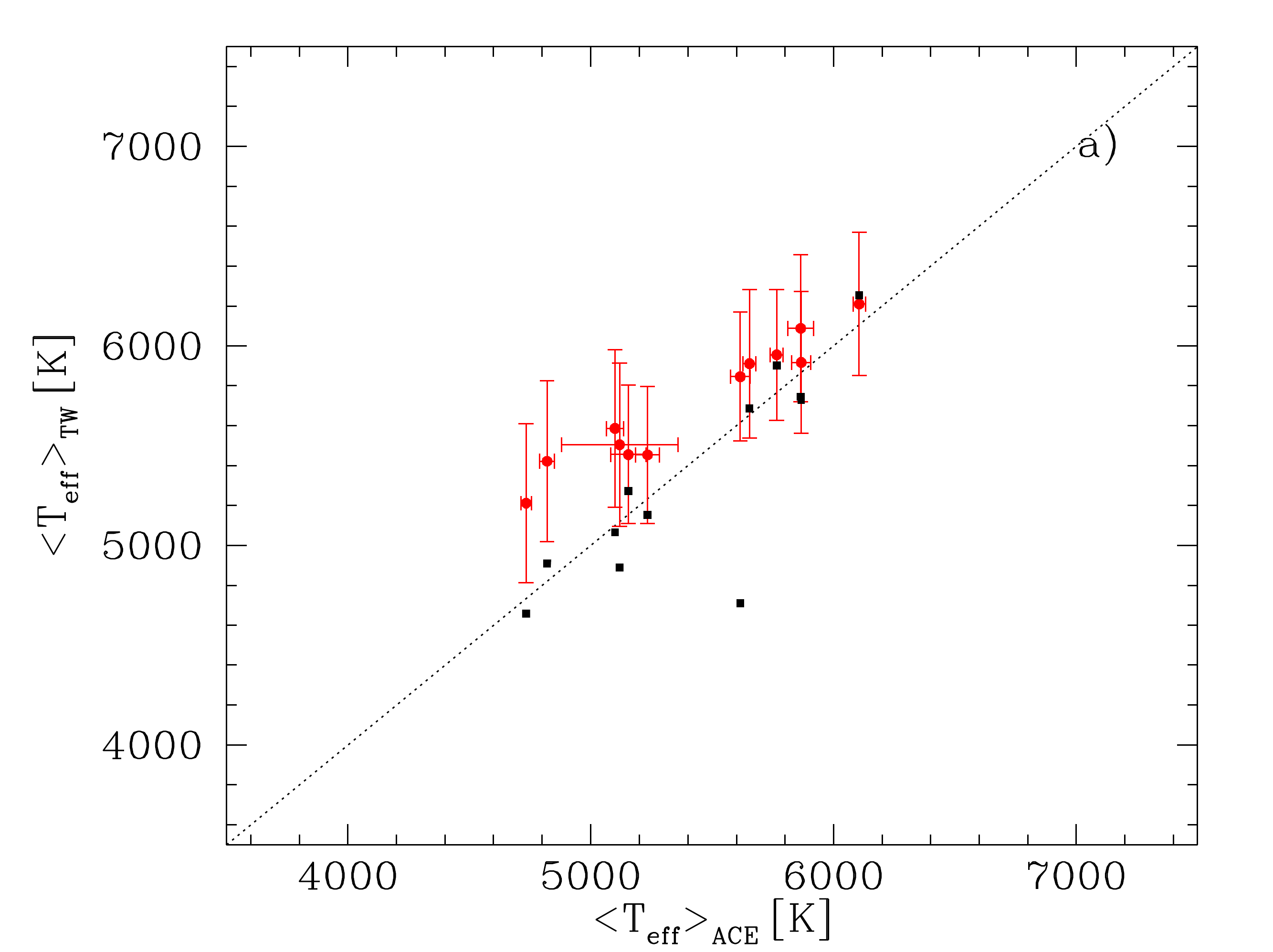}
\includegraphics[width=0.35\textwidth,angle=-90]{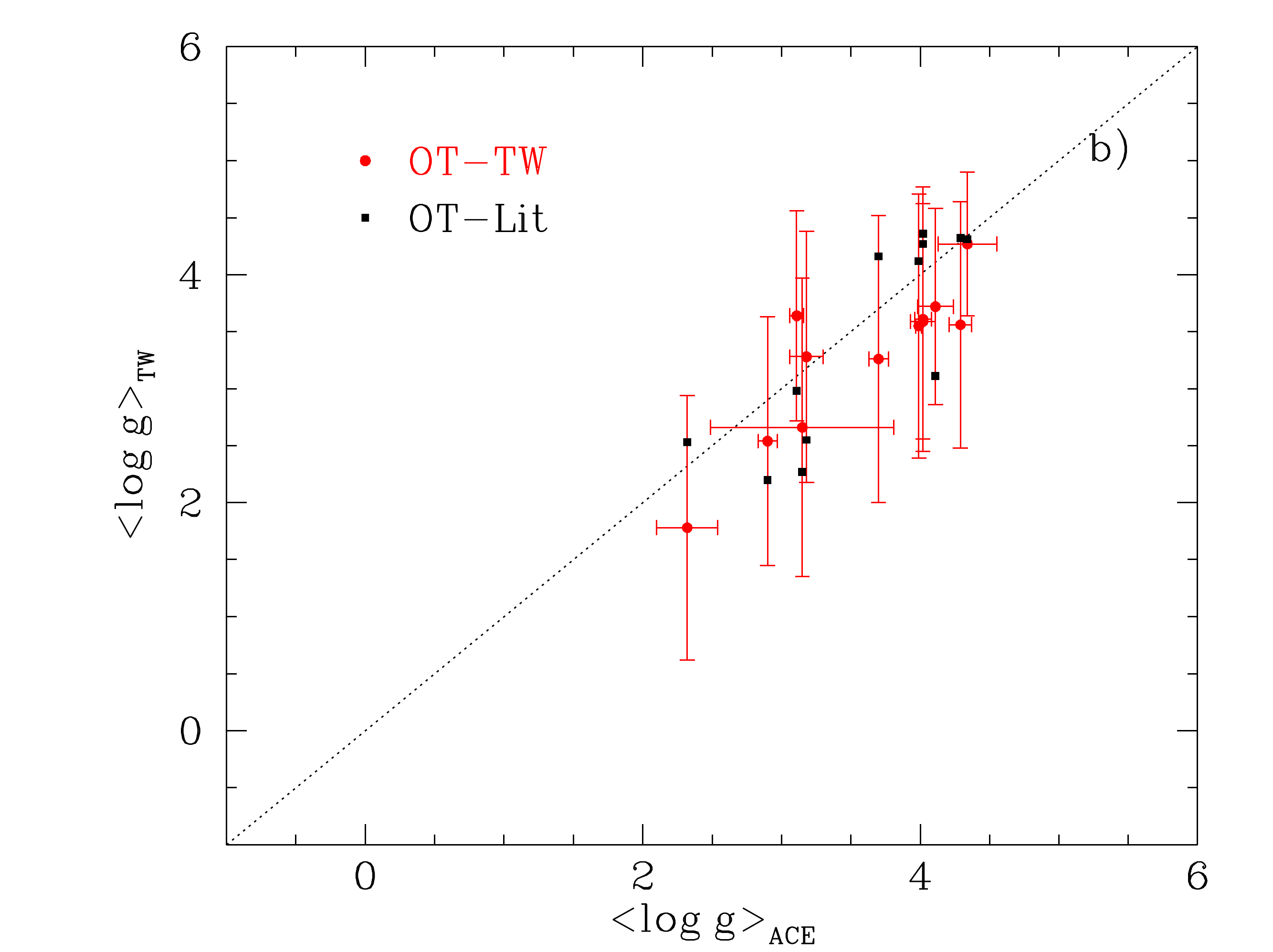}
\includegraphics[width=0.35\textwidth,angle=-90]{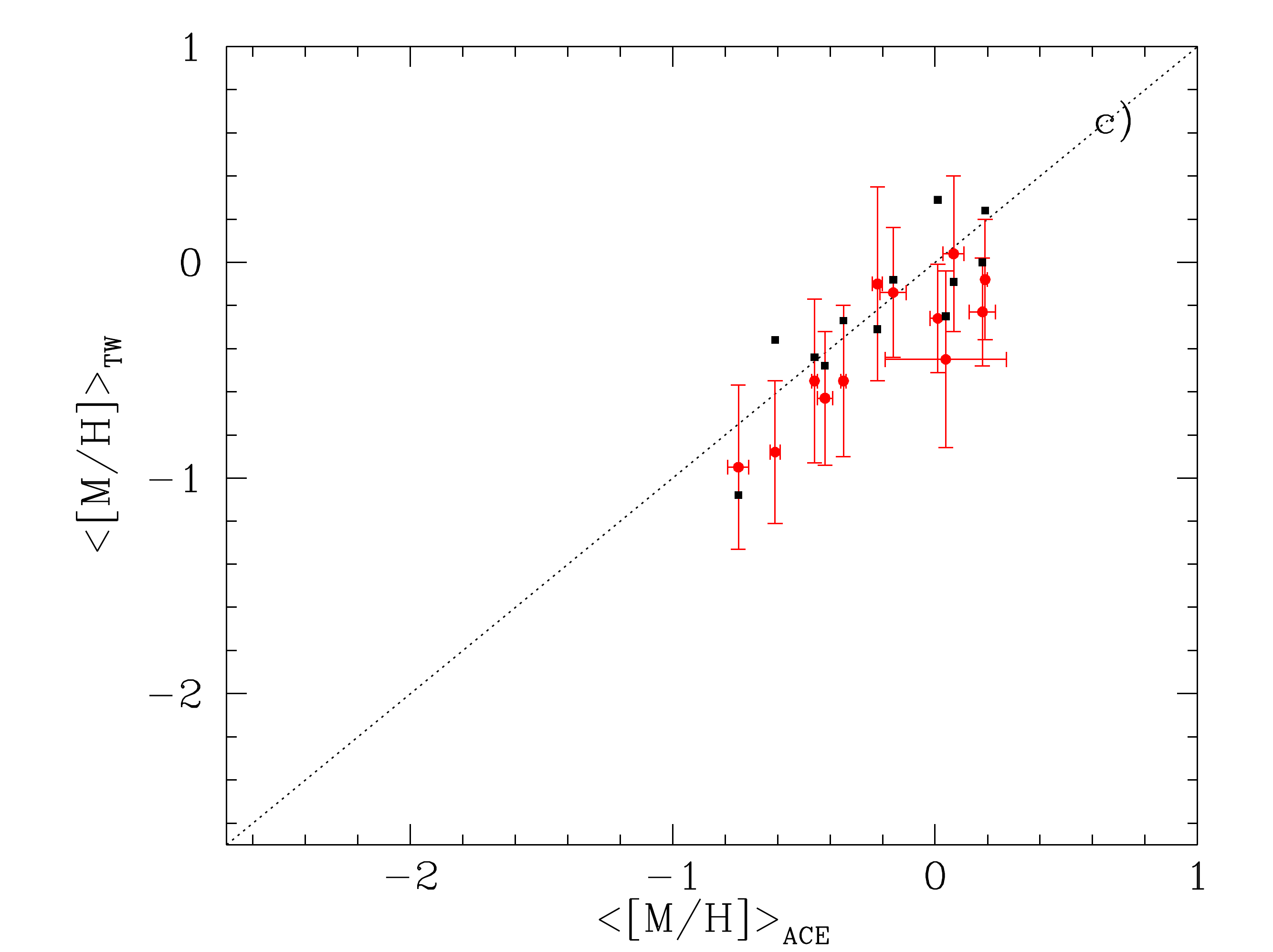}
\caption{Comparison of the stellar parameters estimated in this work for the OT stars sub-sample (red circles) with the results from the {\sc SP\_ACE} code: a) $\rm T_{eff}$; b) $\rm \log{g}$; and c) $\rm [M/H]$. Black filled squares are the estimates given in the literature.}
\label{ace-comp}
\end{figure}

For example, $\rm T_{eff}$ reaches a maximum value of 47500~K. In  summary, the stellar parameters obtained from our fitting are in good agreement with the ones from the literature, and near the 1:1 line slope, better for the OT stars, in which the models have been fitted to a  spectral range (combination of HR-R and HR-I) wider than the COM stars.  We will extend this analysis with a statistically significant sample of the MEGARA-GTC Library stars in Paper~II.

\subsection{Comparison with SP\_ACE  model}
\label{ace}

There are in the literature a certain number of methods \citep{heiter15, texe17, jofre18} to derive stellar physical parameters for different combinations of spectral range and resolution. Also there are public codes, e.g. {\sc FERRE } by \citet{allendeprieto15} and {\sc SP\_ACE}, Stellar Parameters And Chemical Abundances Estimator, by \citet{boecheetgrebel6}, with which it is possible to compute the best fit among a set of models (complete spectra or equivalent widths line catalogues) that reproduces the observed data and, simultaneously, yields stellar parameters with good precision. In this subsection  we compare our derived stellar parameters with those obtained with the {\sc SP\_ACE} whenever this code gives a solution.

We have compared our stellar parameters results with the estimates obtained with the {\sc SP\_ACE} code \citep{boecheetgrebel6}. This model is available on-line (\url{http://dc.g-vo.org/SP\_ACE}), offering a friendly Graphical User Interface, GUI, in which the input spectrum is introduced as a  two-columns table text file. The model is based on the generation of an Equivalent Widths (EW) library of 4363 absorption lines created as a function of the stellar parameters and abundances, and find the best-fitting applying a $\chi^{2}_{min}$ technique. This code computes the estimated spectral parameters for spectra in the ranges 5212~--~6860\,\AA\ and 8400~--~8924\,\AA\, resolving power 2000--20000 and is highly optimised for the fitting of the FGK-type stars, which implies stars cooler than $\rm T_{eff} \le 7000$\,K.

We have used this code to compute the stellar spectra for all stars from the OT sub-sample. The code {\sc SP\_ACE} fails when trying to fit some spectra. We have obtained the parameters obtained from {\sc SP\_ACE} using the combined spectrum of HR-R and HR-I for the OT stars sub-sample, finding a solution for 13 of them. 

Figure~\ref{ace-comp} shows the comparison of the averaged values obtained in this work (y-axis, and plotted as red circles) against the {\sc SP\_ACE} model results (x-axis) of: a) $\rm <T_{eff}>$, b) $\rm <\log{g}>$, and c) $\rm <[M/H]>$, respectively. The stellar parameters existing in the literature are plotted as black squares. We find in general a good correlation. However, as explained throughout this paper, this sample is not statistically significant to derive conclusions, and does not cover the complete stellar parameter space. 
\begin{figure*}
\centering
\includegraphics[width=0.35\textwidth,angle=-90]{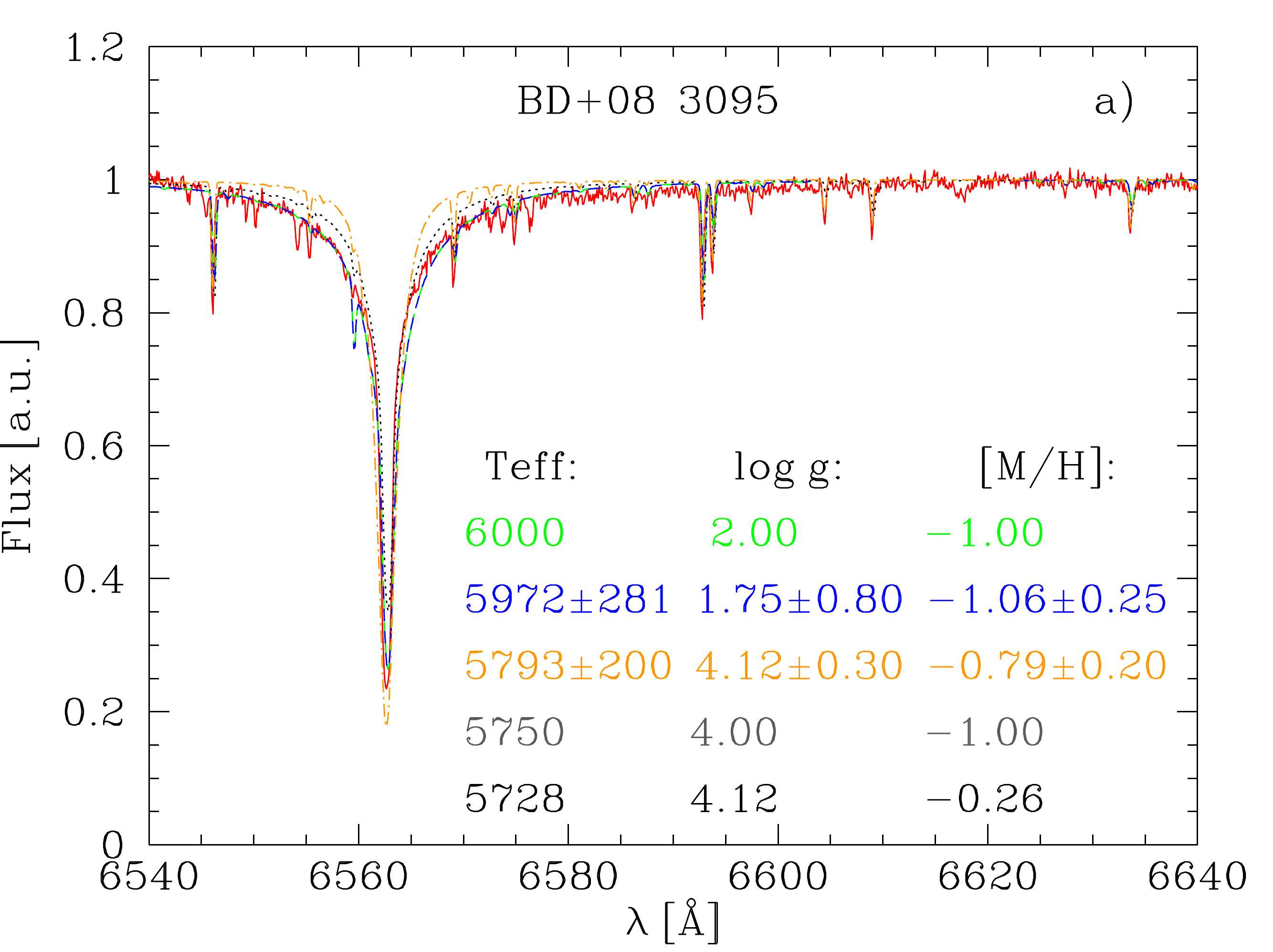}
\includegraphics[width=0.35\textwidth,angle=-90]{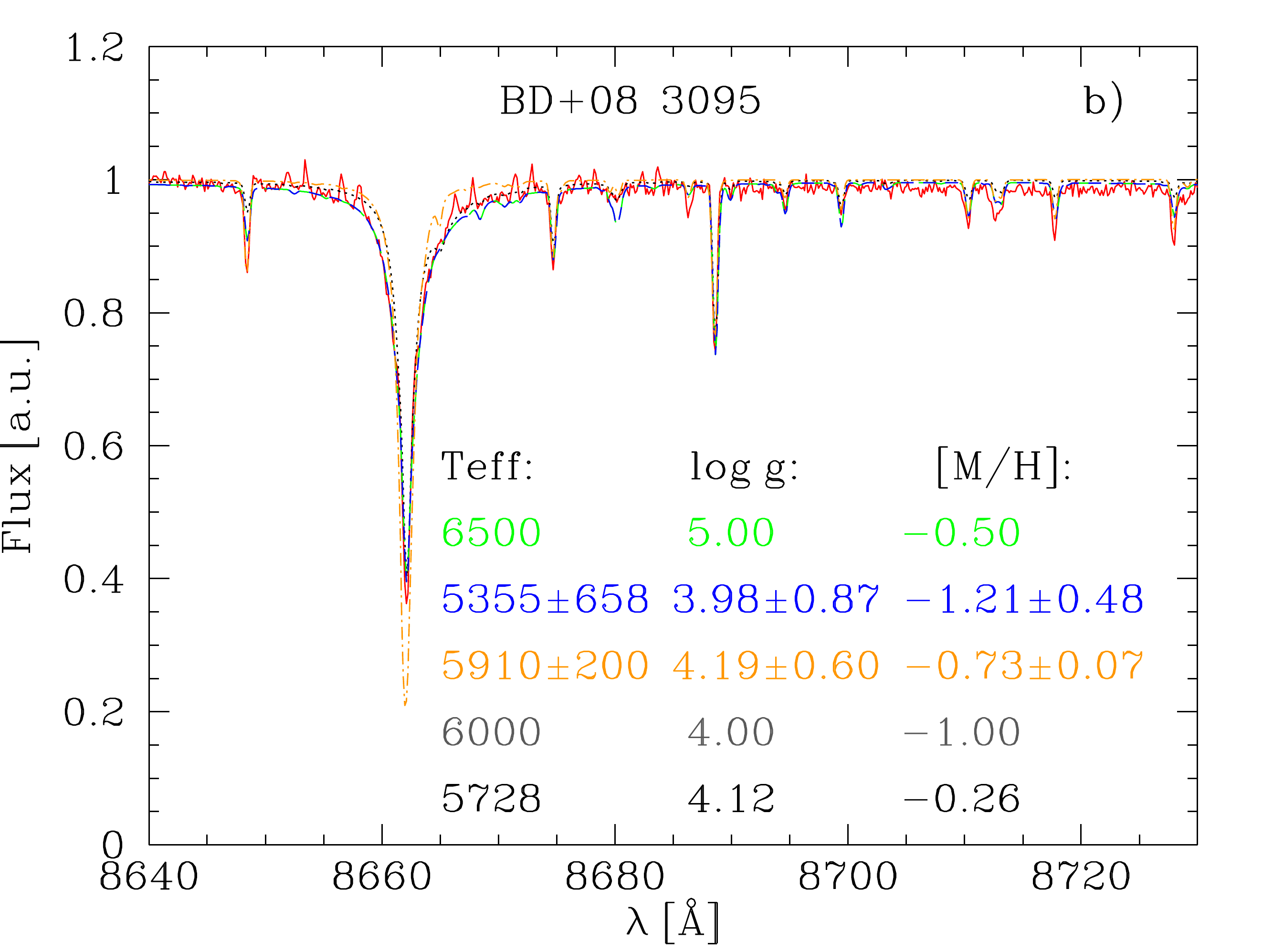} \includegraphics[width=0.35\textwidth,angle=-90]{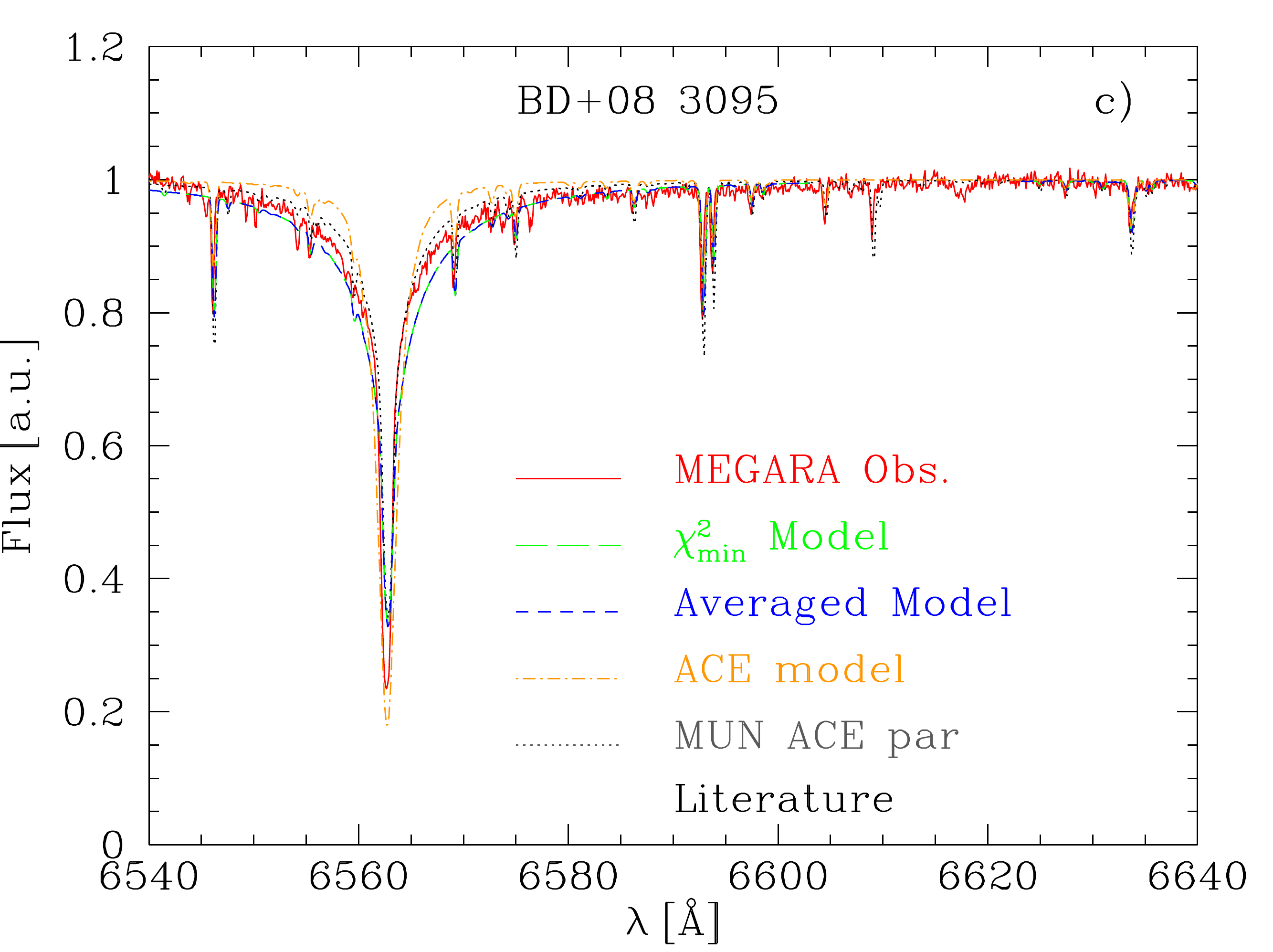}
\includegraphics[width=0.35\textwidth,angle=-90]{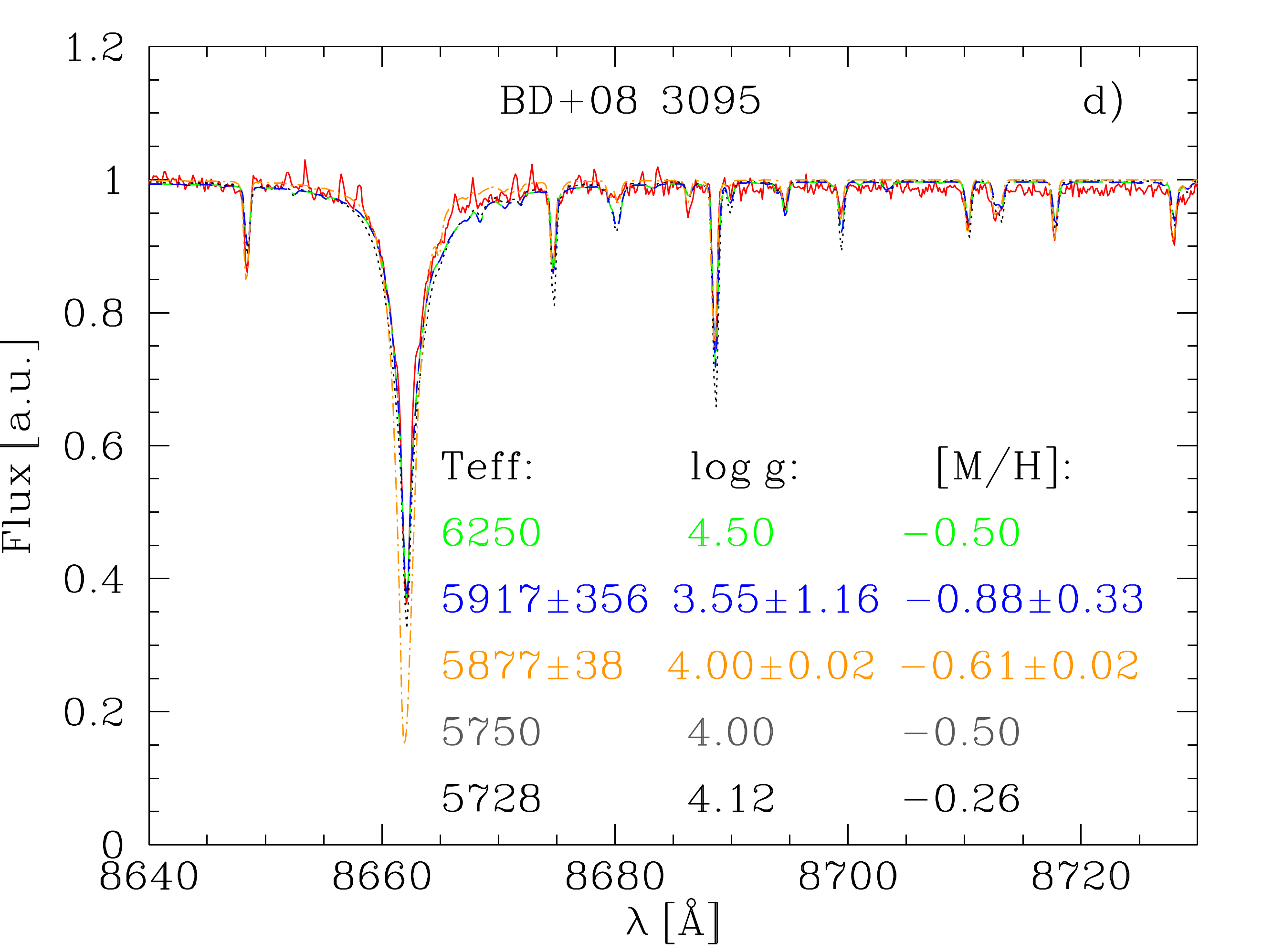}
\caption{The spectra for the OT star BD$+$08~3095 --red line --a) and c) in HR-R , b) and d) in HR-I, compared with our best models (the ones obtained with the $\chi^{2}_{min}$ --green long dashed line-- and the averaged --blue  short dashed line), and with the spectrum fitting using the {\sc SP\_ACE} model --orange dot-dashed line--. The dotted black line is the MUNARI model for the closest stellar parameters to the {\sc SP\_ACE} model estimates. Top panels represent the best fits done separately for each setup, while the bottom ones represent the best fit obtained simultaneously with both HR-R and HR-I spectra.}
\label{sp-ace}
\end{figure*}

As an example to study the fitting, stellar parameter determination and the results comparison, we have done a carefully study for the star BD$+$08~3095 (OT sub-sample), classified as GO~V in the literature, so within the range of optimisation of the {\sc SP\_ACE} code. We have not considered the correction from the star velocity profile. 

Figure~\ref{sp-ace} shows the detail of the observed spectrum in HR-R (panel a) and HR-I (panel b), respectively, around two strong absorption lines: H~$\alpha$ in HR-R and the strongest line of the {Ca}{\textsc ii} triplet in HR-I. These top panels represent the best fits done separately for each setup. In both panels the MEGARA observed spectra are plotted as a red line. The best MUN05 models obtained when fitting to the spectrum of each individual set-up, HR-R in (a) and HR-I in (b), are in green long-dashed line when applying our $\chi^{2}_{min}$ method - and in blue short-dashed line for our {\it average} model. The {\sc SP\_ACE} fits obtained for our spectra are shown as a dashed orange line. The MUN05 model corresponding to the stellar parameters resulting from the {\sc SP\_ACE} model is displayed as a dotted-black line. 

We see in panel a) that both MUN05 models fitted with our method are quite deep and the fitting to the peak of the observed spectrum is good enough. Both models have a lower level of continuum out of the H~$\alpha$ spectral window fitting well this level. The {\sc SP\_ACE} model model is, however, deeper and wider than the observations. The MUN05 model corresponding to the estimates of the {\sc SP\_ACE} is, in turn, less deep than the observed data. This outlines the fact that the MUN05 models corresponding to the stellar parameters derived from the {\sc SP\_ACE code} and ours are different, what comes from the different fitting methods ({\sc SP\_ACE} is based on the EW fitting while our method fits the entire continuum spectrum).

The problem to interpret this plot comes of the difficulty of fitting simultaneously the deep and the width of the H~$\alpha$ line: if we use the {\sc SP\_ACE} stellar parameters and compare the corresponding spectra of MUN model, the H~$\alpha$ profile is not well reproduced (at least in this example). Taking into account that the increase of the gravity reduces the deep of the line, it results difficult to understand how to create such deep profile with a large value as $\log{\rm g} \sim 4$. In this example, our resulted spectrum, either because our method or the models we have used, produces a better fit to the spectral lines, although the {\sc SP\_ACE} estimated stellar parameters have a good agreement with those previously obtained in the literature. 
 
Something similar occurs in panel b) with the {Ca}{\textsc ii} line. In this case the {\sc SP\_ACE} model is deeper as the observed profile, as in panel a), but it is also narrower. For this setup, however, all the stellar parameters are in good agreement among them. 

In the bottom ones, we represent the same fits best obtained using both HR-R and HR-I spectra in one only fit. In that case the stellar parameters are the same in both windows, panels c) and d). 
In panel c) we see that our fits (green or blue lines) are not deep enough compared with the observed spectrum, while the {\sc SP\_ACE} model is deeper than necessary. Moreover, the wings of the H$\alpha$ absorption feature are better fitted with our method, while the {\sc SP\_ACE} one has too wide line. For what refers to panel d) the whole spectrum is well fit for all cases, except the depth of the three CaT absorption lines which are so strong in the the {\sc SP\_ACE} model.
\begin{figure*}
\includegraphics[width=0.485\textwidth,angle=0]{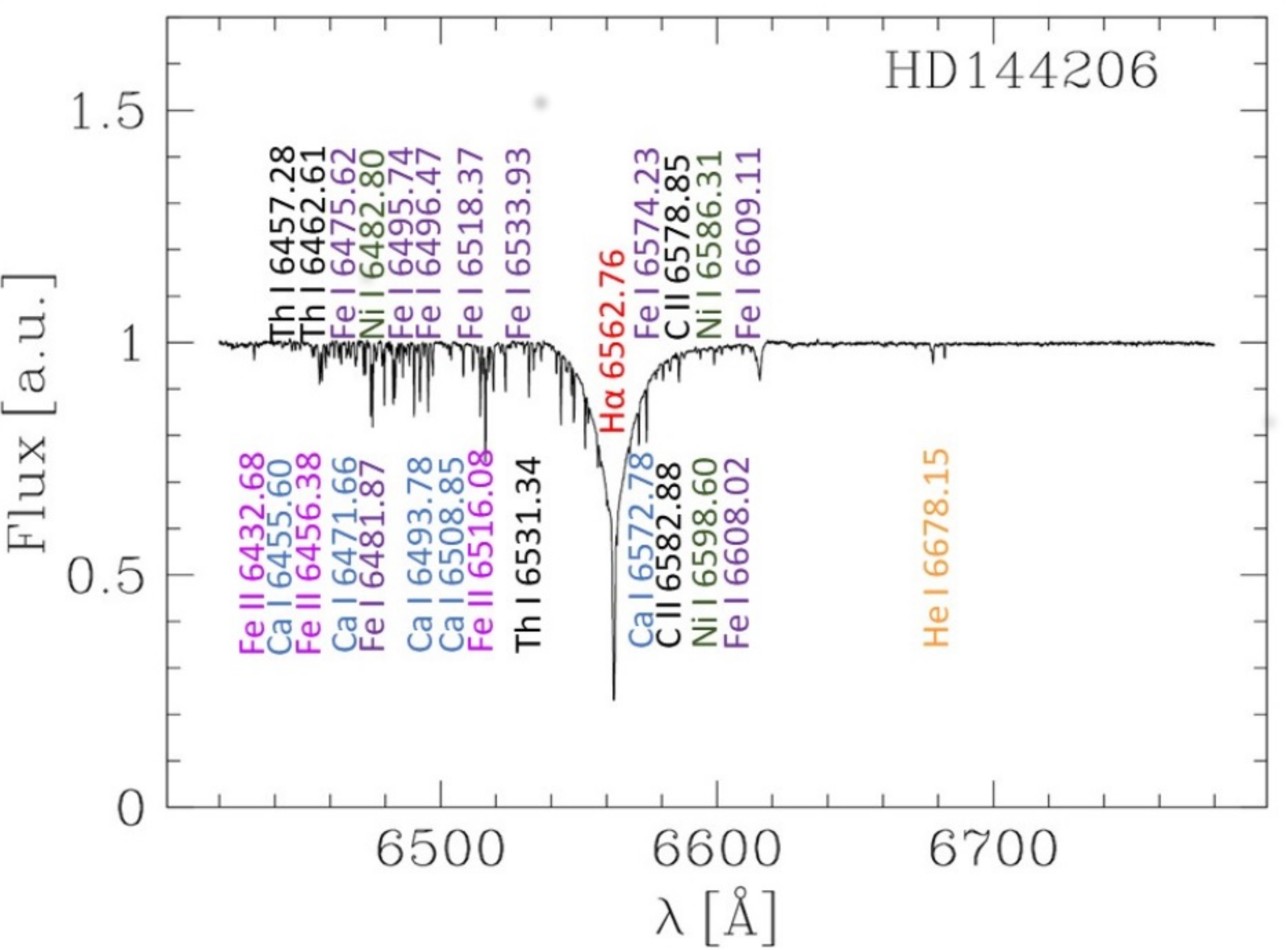}
\includegraphics[width=0.485\textwidth,angle=0]{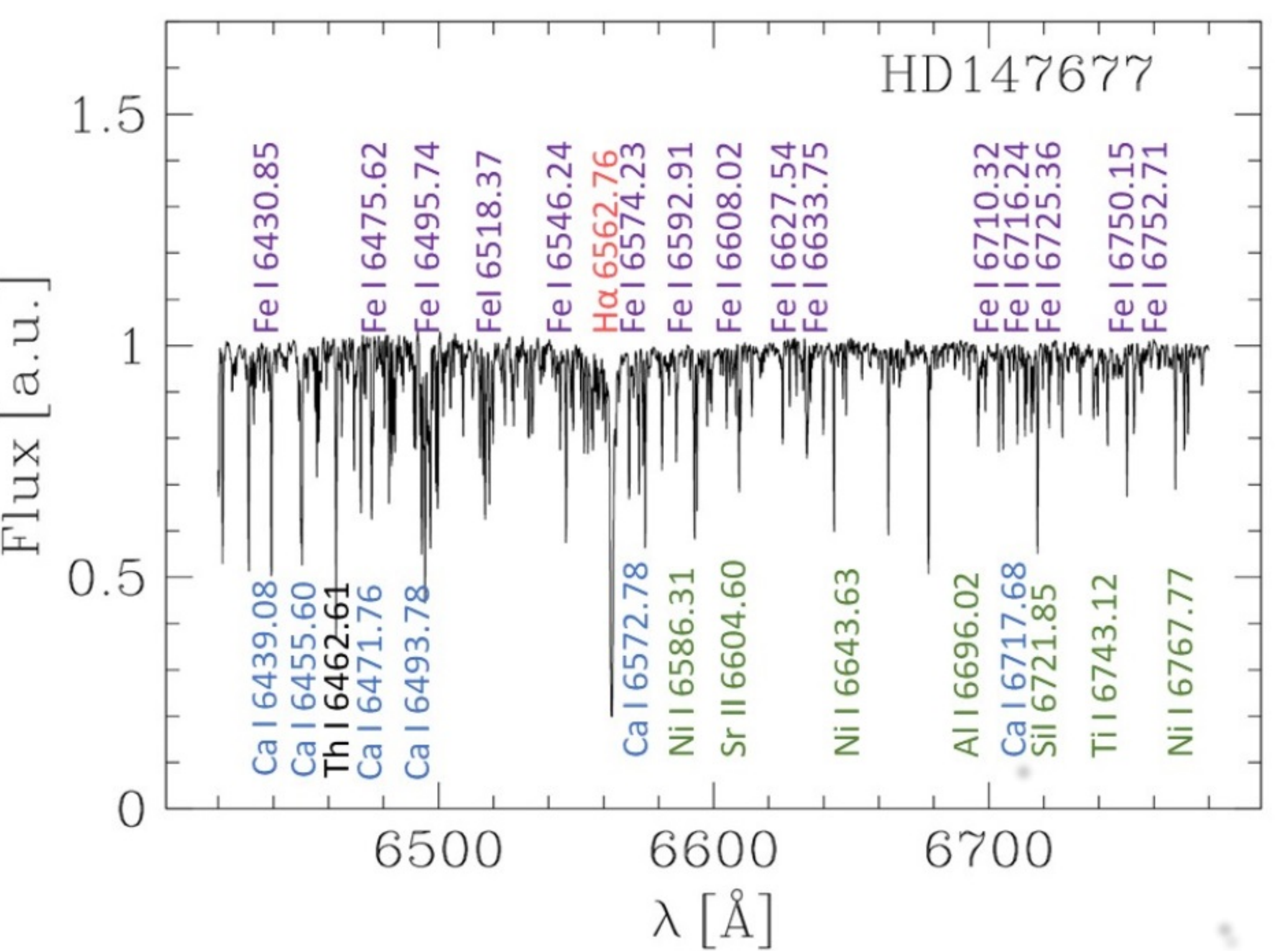}
\caption{Left) Spectrum of the HD~144206 star (B0~III) with the identification of the strongest spectral lines. Right) Spectrum of the HD~147677 (K0~III) with the identification of a representative number of lines.}
\label{linesHRR}
\end{figure*}

Probable causes for the discrepancies found can be the different theoretical model set used in each case and the limitations of the models themselves. 
{\sc SP\_ACE} EW library can model many weak lines while MUN05 models are not as good for this purpose. In return, {\sc SP\_ACE} models are mostly limited by the adoption of LTE 1D atmosphere models, but also by the microturbulence treatment and other minor effects discussed in detail \citep{boecheetgrebel6}. The high spectral resolution of MEGARA setups HR-R and HR-I, allows to obtain the information from the detailed absorption line structure, which might be very important to study non-LTE effects and its impact in the kinematics of low-velocity dispersion composed stellar populations, for whose understanding is crucial to have reliable models at this high resolution.

We will derive parameters for all the stars of the first release of the MEGARA-GTC library, expected in 2020, and that will contain between 200 and 300 stars \citep{cmgbg20}. 

The purpose of the stellar parameters determination in this piece of work was not a precise and absolute determination of these values, but the developing of a fitting code that allows the classification of the MEGARA-GTC library stars and cluster members with a uniform criterion and based on the same MEGARA high resolution data (HR-R and HR-I). The final goal is to find a reliable technique to be able to assign stellar parameters in an homogeneous way to all the observed stars of the MEGARA-GTC library, and then use these values to assign the right star to each point of the synthetic Hertzsprung-Russell diagrams when computing the isochrones in the next generation of high resolution  {\sc PopStar} models.

\subsection{HR-R spectra}\label{HRRspec}

We have identified the main absorption lines detected in the HR-R spectra of the observed stars. We have labelled the strongest lines in Figure~\ref{linesHRR}, left and right panels, for the stars HD~144206 (B0~III) and HD~147677 (K0~III) respectively, both belonging to the OT sub-sample. The HR-R hottest stars spectra are dominated by the helium ({He}{i} 6678.15\AA\ and 6867.48\AA; {He}{\textsc ii}~6560.10\AA) and hydrogen H~${\alpha}$~6562.76\AA\ (6562.71, 6562.72, 6562.85\AA). For spectral types later than A, the range starts to be populated with metal lines from {Fe}{i} (6430.85\, 6469.19, 6475.62, 6481.87, 6495.74, 6496.47, 6498.94, 6518.37, 6533.93, 6546.24, 6574.23, 6581.21, 6591.31, 6592.91, 6593.87, 6597.54, 6608.02, 6609.11, 6627.54, 6633.41, 6633.75, 6703.57, 6710.32, 6713.74, 6716.24, 6725.36, 6750.15 and 6752.71\AA ); {Ca}{i} (6439.08, 6449.81, 6455.60, 6471.66, 6493.78, 6499.65, 6508.85, 6572.78, 6717.68 and 6798.48\AA); {Al}{i} (6696.02 and 6698.67\AA); {Si}{i} (6721.85 and 6741.63\AA); {Ti}{i} (6497.68, 6554.22, 6599.10 and 6743.12\AA); {Co}{i} (6454.99 and 6771.03\AA); {Ni}{i} (6482.80, 6586.31, 6598.60, 6635.12, 6643.63, 6767.77 and 6772.31\AA); {V}{i} (6504.16\AA); {Cr}{i} (6537.92, 6630.01, 6537.92 and 6630.01\AA); {Th}{i} (6457.28, 6462.61, 6531.34 and 6989.65\AA), and single ionised lines like {Fe}{\textsc ii} (6516.08, 6432.68 and 6456.38\AA); {Ti}{\textsc ii} (6491.57\AA),  {Sc}{\textsc ii} (6604.60\AA), {Mg}{\textsc ii} (6545.97\AA). 

\subsection{HR-I spectra}\label{HRIspec}

\begin{figure*}
\includegraphics[width=0.48\textwidth,angle=0]{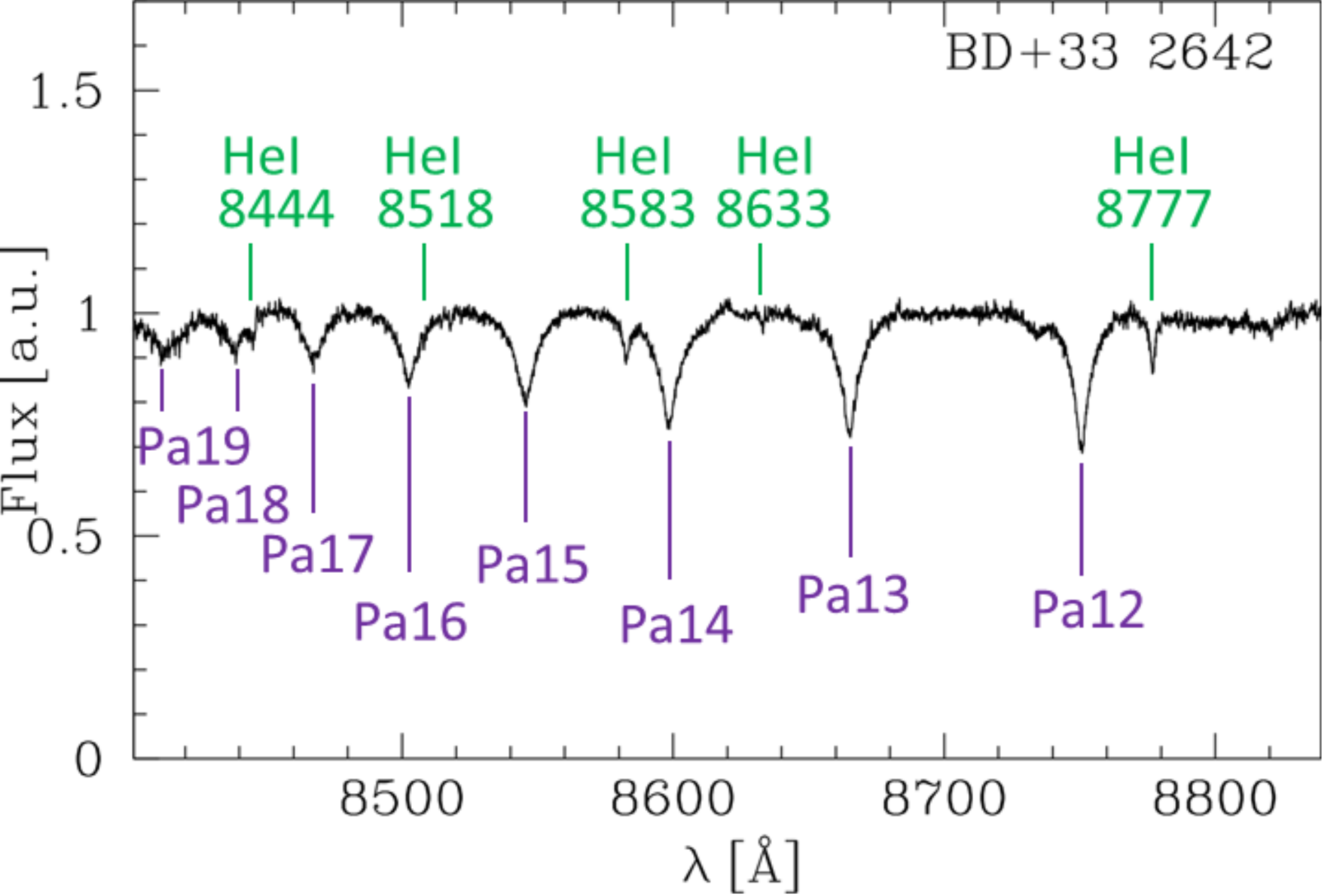}
\includegraphics[width=0.48\textwidth,angle=0]{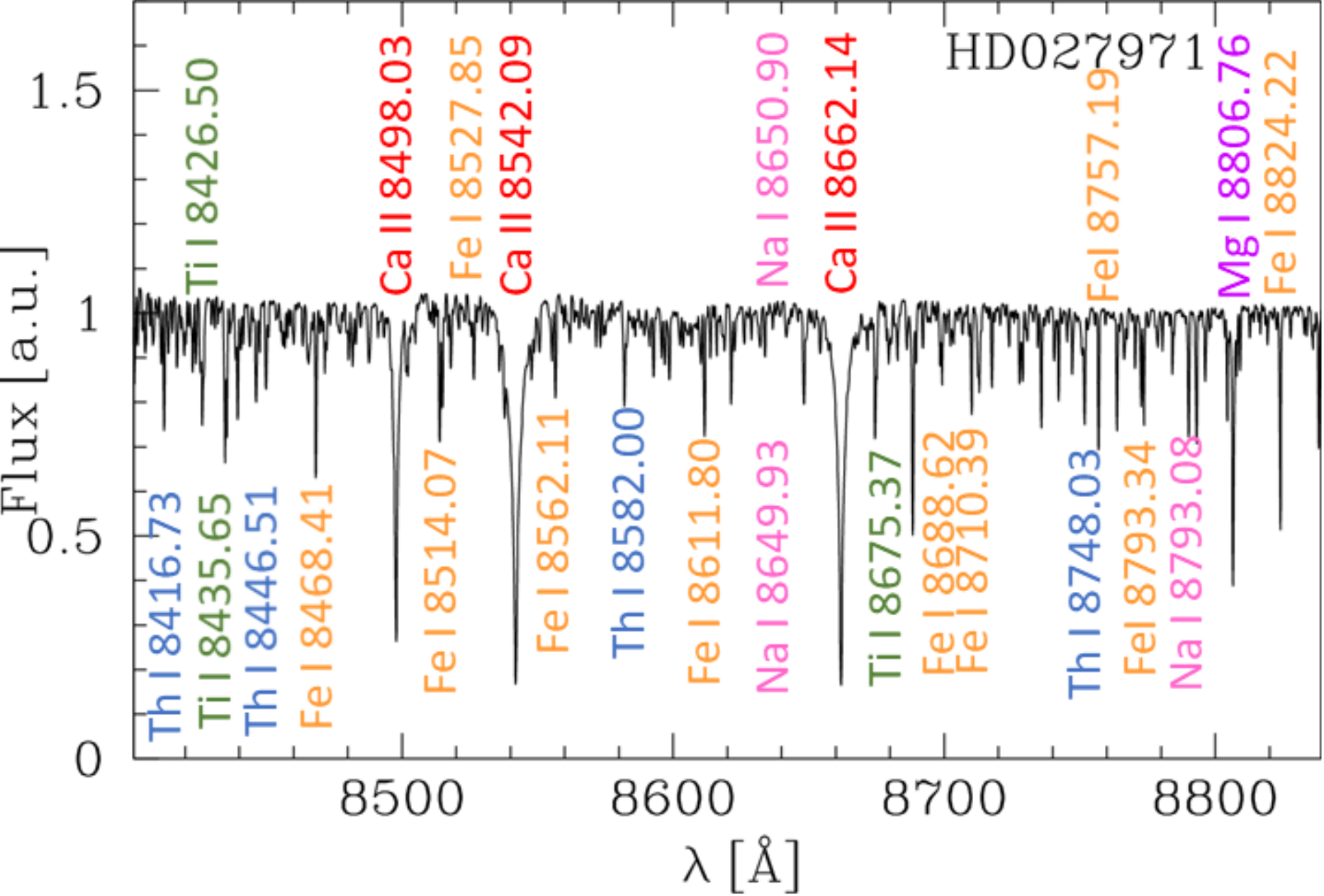}
\caption{Left) Spectrum of the hot star BD+33~2642 (O7pD) and spectral lines identified. Pa series and {He}{i} lines are clearly identified. Right) Spectrum of the cool star HD~027971 (K1~III). The strongest lines of {Ca}{\textsc ii}, {Fe}{i}, {Th}{i} and {Ti}{i} are identified.}
\label{linesHRI}
\end{figure*}

We have also identified the main spectral lines in the HR-I setup that appear in the observed stars. Figure~\ref{linesHRI}, left and right panels, show the spectrum of a hot and a cool star respectively from the COM sub-sample.

The observed COM hot stars shown in  Figure~\ref{extractedspectra} (a) shows hydrogen (Paschen) and helium lines in the spectral-type sequence from O4 to A0. The Paschen series H~I lines (Pa19 8413.33, Pa18 8437.96, Pa17 8467.27, Pa16 8502.50, Pa15 8545.39, Pa14 8598.40, Pa13 8665.03 and Pa 12 8750.47\AA) are clear and strong along the whole sequence, with maximum strength and width for giant B stars. The {He}{i} lines are identified at 8444\AA\ (3{\rm p}o-3{\rm d} at 8444.44, 8444.46, 8444.65\AA); 8481\AA\ (3{\rm p}0-3{\rm s} at 8480.67, 8480.68, 8480.88\AA), 8518\AA\ (1{\rm S}-1{\rm p}0 at 8518.04\AA), 8531\AA\ (1{\rm d}-1{\rm p}0 at 8530.93\AA), 8532\AA\ (3{\rm d}-3{\rm p}0 at 8532.10, 8532.11, 8532.13\AA), 8583\AA\ (3{\rm p}0-3{\rm d} at 8582.51, 8582.52\AA), 8633\AA\ (3{\rm p}0-3{\rm s} at 8632.71, 8632.73, 8632.93\AA), 8777\AA\ (3{\rm p}0-3{\rm d} at 8776.65, 8776.67, 8776.88\AA) and 8849\AA\ (3{\rm p}0-3{\rm s} at 8849.16, 8849.37\AA). 

The observed cool stars from the commissioning single-star sub-sample shown in  Figure~\ref{extractedspectra} (b) shows the {Ca}{\textsc ii} and {Mg}{i} strong lines and a large number of {Fe}{i}, {Th}{i}, {Ti}{i} and {Na}{i} lines. The spectral type sequence ranges from F8 to K4. The Calcium triplet lines ({Ca}{\textsc ii} 2{\rm d}-2{\rm p}0 at 8498.03, 8542.09, 8662.14\AA) are the most prominent ones of the spectra in all cool stars; followed in intensity by {Mg}{i} 1{\rm p}0-1{\rm d} 8806.76. {Fe}{i} lines are easily identified in the spectra at 8468.41, 8514.07, 8611.80, 8661.90, 8674.75, 8688.62, 8757.19, 8763.97, 8793.34, 8824.22 and 8838.43\AA. The strongest {Th}{i} lines can be detected at 8416.73, 8421.22, 8446.51, 8478.36, 8748.03 and 8758.24\AA. {Ti}{i} detected lines are at 8412.36, 8426.50, 8434.96, 8435.65 and 8675.37\AA. Finally, {Na}{i} lines are at 8649.93, 8650.90 and 8793.08\AA.  

We have measured the indices of {Ca}{\textsc ii} (Ca1, Ca2, Ca3), centred on the lines 8498.03, 8542.09 and 8662.14\AA\ respectively, and Pa (Pa1, Pa2, Pa3), centred on the series lines P17, P14 and P12, as defined by \citet{cen01a}, using the same line and continuum windows in the MEGARA spectra. From these measurements, we have derived the composed index $\rm CaT = \rm Ca1 + \rm Ca2 + \rm Ca3$. We have measured this index on the 97 stars using their spectra normalised to the continuum. 

For the CaT lines we have re-defined the windows of both continuum and lines to decontaminate the indices from the contribution of other metallic lines that become only apparent at the high spectral resolution of MEGARA HR-I. The new proposed continuum windows for {\textsc ii}{Ca}{\textsc ii} lines are (wavelength in \AA), for Ca1 and Ca2 lines, 8450.0 - 8460.0 (blue) and 8565.5 - 8575.0 (red), while for Ca3 we use 8619.5 - 8642.5 (blue) and 8700.5 - 8710.0 (red). The feature windows are 8482.0 - 8512.0 (Ca1), 8531.0 - 8554.0 (Ca2) and 8650.0 - 8673.0 (Ca3).

Table~\ref{linemeasurements} summarises the measurement of {Ca}{\textsc ii} lines equivalent widths (in \AA) using \citet{cen01a} windows and the new windows defined in this paper, for the 21 COM stars and the 20 OT stars. This table also includes the {Mg}{\textsc i} and sTiO indices as defined in \citet{cen09}, for which we have kept the same spectral windows. Table~\ref{linemeasurementsM15} has the same information as Table~\ref{linemeasurements} but for the M15 stars sub-sample.

The columns in Table~\ref{linemeasurements} and Table~\ref{linemeasurementsM15} are as follows: first column identifies the star; column 2 gives the continuum SNR averaged over the whole spectrum as defined by \citet{stoehr08}, columns 3 to 5 are the three {Ca}{\textsc ii} lines equivalent widths as defined by \citet{cen01a} in \AA; column 6 is the {Mg}{\textsc i} equivalent width (\AA); columns 7 to 9 are the Pa lines equivalent width, columns 10 to 12 are the three {Ca}{\textsc ii} lines equivalent widths, in \AA\, with the windows re-defined in this work. The last column is the sTiO color, as defined by \citet{cen09}. All the values include the random errors due to continuum and are related to the spectra SNR. These errors are below 3 per cent in all cases (S/N~$\geq$~100). 

The results of all measurements and the conclusions will be included in Paper~II with a statistically significant sample and stars within a wide range of physical stellar parameters.

\begin{figure}
\includegraphics[width=0.35\textwidth,angle=-90]{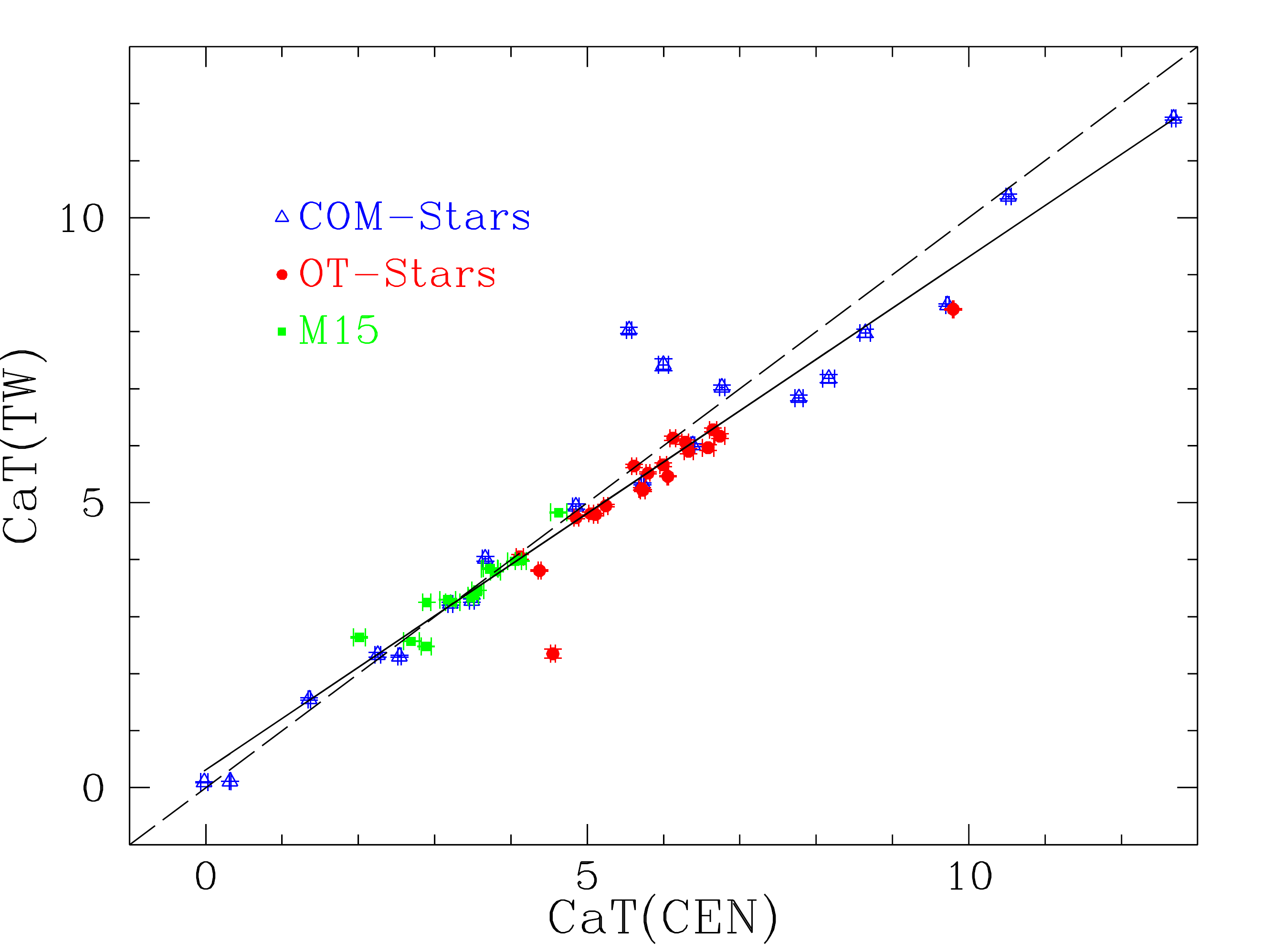}
\caption{CaT index measured in all the HR-I spectra presented in This Work (TW) vs. the same index CaT measured following the \citet{cen01a} definition. Different sample stars are drawn with different symbol as labelled in the plot. The dashed line represents the 1:1 relation while the solid line is the actual fit to the observed data.}
\label{CaT}
\end{figure}

Figure~\ref{CaT} plots in the y-axis the values of the re-defined CaT index in this work, CaT(TW), against the results (x-axis) when measuring the index with \citet{cen01a} windows, CaT(CEN), for all the 97 stars. The dashed line represents the 1:1 relation while the solid line is the actual fitting to the observations, showing values of the new CaT index slightly lower than the ones measured with CaT(CEN). This is particularly noticeable for the largest index values obtained in giant and supergiant stars. The reason is the contamination of the CaT(CEN) index with other metallic lines that however do not fall in the line window defined in the new (high resolution) index. 

\begin{table*}
\caption{{Ca}{\textsc ii}, {Mg}{\textsc i}, Pa lines equivalent widths, in \AA, as defined by \citet{cen01a} and sTiO colour defined by \citet{cen09} for the sub-sample of 21 COM stars observed in HR-I and the 20 stars of the OT sub-sample. Table description in subsection~\ref{HRIspec}.}
\label{linemeasurements}
\begin{turn}{90}
 \resizebox{22cm}{!} {
\begin{tabular}{lrccccccccccc}
\hline
Star Name & SNR & Ca1CEN & Ca2CEN & Ca3CEN & Mg~I & Pa1 & Pa2 & Pa3 & Ca1MEG & Ca2MEG & Ca3MEG & sTiO \\
\hline
Schulte~9 &   548 & $-$0.181 $\pm$ 0.006 & $-$0.030 $\pm$ 0.002 & $-$0.109 $\pm$ 0.003 & +0.089 $\pm$ 0.002 & $-$0.177 $\pm$ 0.003 & $-$0.214 $\pm$ 0.003 & $-$0.720 $\pm$ 0.005 & $-$0.024 $\pm$ 0.002 & $-$0.038 $\pm$ 0.005 & $-$0.123 $\pm$ 0.002 & +0.635 \\
HD192281 &   280 & $-$0.104 $\pm$ 0.011 & $-$0.037 $\pm$ 0.008 & +0.161 $\pm$ 0.045 & $-$0.245 $\pm$ 0.000 & $-$0.070 $\pm$ 0.009 & $-$0.848 $\pm$ 0.010 & +0.289 $\pm$ 0.038 & $-$0.015 $\pm$ 0.005 & $-$0.020 $\pm$ 0.007 & $-$0.060 $\pm$ 0.006 & +0.636 \\ 
BD+25~4655 &    71 & $-$0.159 $\pm$ 0.047 & +0.444 $\pm$ 0.045 & +0.201 $\pm$ 0.116 & +0.052 $\pm$ 0.017 & +0.014 $\pm$ 0.023 & +0.244 $\pm$ 0.042 & +0.025 $\pm$ 0.001 & $-$0.034 $\pm$ 0.000 & +0.465 $\pm$ 0.040 & +0.285 $\pm$ 0.035 & +0.626 \\ 
BD+33~2642 &   105 & $-$1.360 $\pm$ 0.026 & $-$1.911 $\pm$ 0.032 & $-$1.580 $\pm$ 0.023 & +0.112 $\pm$ 0.745 & $-$0.871 $\pm$ 0.015 & $-$2.150 $\pm$ 0.030 & $-$1.893 $\pm$ 0.029 & $-$1.284 $\pm$ 0.028 & $-$1.843 $\pm$ 0.024 & $-$1.810 $\pm$ 0.020 & +0.634 \\ 
HD218915 &   256 & $-$0.735 $\pm$ 0.012 & $-$0.897 $\pm$ 0.014 & $-$0.908 $\pm$ 0.010 & $-$0.011 $\pm$ 0.006 & $-$0.235 $\pm$ 0.007 & $-$1.884 $\pm$ 0.012 & $-$1.150 $\pm$ 0.013 & $-$0.645 $\pm$ 0.012 & $-$0.729 $\pm$ 0.011 & $-$0.930 $\pm$ 0.009 & +0.633 \\ 
BD+40~4032 &   146 & $-$0.507 $\pm$ 0.020 & $-$1.039 $\pm$ 0.023 & $-$1.942 $\pm$ 0.019 & +0.046 $\pm$ 0.009 & $-$0.192 $\pm$ 0.012 & $-$1.841 $\pm$ 0.021 & $-$1.959 $\pm$ 0.024 & $-$0.578 $\pm$ 0.021 & $-$1.015 $\pm$ 0.019 & $-$1.686 $\pm$ 0.015 & +0.630 \\ 
HD220575 &   232 & $-$1.725 $\pm$ 0.012 & $-$2.544 $\pm$ 0.013 & $-$5.451 $\pm$ 0.011 & $-$0.123 $\pm$ 0.001 & $-$0.659 $\pm$ 0.007 & $-$4.529 $\pm$ 0.012 & $-$5.208 $\pm$ 0.013 & $-$1.711 $\pm$ 0.013 & $-$2.377 $\pm$ 0.011 & $-$4.377 $\pm$ 0.008 & +0.624 \\ 
BD+42~3227 &   111 & $-$1.170 $\pm$ 0.025 & $-$1.264 $\pm$ 0.028 & $-$1.230 $\pm$ 0.020 & $-$0.106 $\pm$ 0.011 & $-$0.238 $\pm$ 0.020 & $-$2.001 $\pm$ 0.025 & $-$1.209 $\pm$ 0.023 & $-$1.221 $\pm$ 0.026 & $-$1.419 $\pm$ 0.023 & $-$1.378 $\pm$ 0.018 & +0.630 \\ 
BD+12~237  &   171 & $-$0.693 $\pm$ 0.016 & $-$1.144 $\pm$ 0.019 & $-$1.370 $\pm$ 0.017 & +0.007 $\pm$ 0.005 & $-$0.360 $\pm$ 0.011 & $-$2.039 $\pm$ 0.017 & $-$2.362 $\pm$ 0.018 & $-$0.737 $\pm$ 0.018 & $-$1.210 $\pm$ 0.015 & $-$1.281 $\pm$ 0.012 & +0.623 \\ 
BD+17~4708 &   172 & $-$0.492 $\pm$ 0.022 & $-$0.909 $\pm$ 0.019 & $-$0.852 $\pm$ 0.019 & $-$0.131 $\pm$ 0.007 & $-$0.061 $\pm$ 0.011 & $-$0.092 $\pm$ 0.018 & $-$0.190 $\pm$ 0.018 & $-$0.548 $\pm$ 0.026 & $-$0.991 $\pm$ 0.017 & $-$0.789 $\pm$ 0.014 & +0.625 \\ 
HD026630 &   235 & $-$2.371 $\pm$ 0.016 & $-$6.029 $\pm$ 0.016 & $-$4.285 $\pm$ 0.014 & $-$0.512 $\pm$ 0.006 & $-$0.357 $\pm$ 0.010 & $-$1.198 $\pm$ 0.013 & $-$0.545 $\pm$ 0.017 & $-$2.526 $\pm$ 0.019 & $-$5.417 $\pm$ 0.010 & $-$3.795 $\pm$ 0.010 & +0.638 \\ 
HD216219 &   346 & $-$1.154 $\pm$ 0.011 & $-$2.625 $\pm$ 0.011 & $-$1.945 $\pm$ 0.009 & $-$0.282 $\pm$ 0.004 & $-$0.150 $\pm$ 0.006 & $-$0.342 $\pm$ 0.008 & $-$0.254 $\pm$ 0.011 & $-$1.125 $\pm$ 0.011 & $-$2.457 $\pm$ 0.008 & $-$1.749 $\pm$ 0.007 & +0.632 \\ 
HD011544 &   147 & $-$2.070 $\pm$ 0.021 & $-$4.754 $\pm$ 0.022 & $-$3.697 $\pm$ 0.022 & $-$0.698 $\pm$ 0.012 & $-$0.385 $\pm$ 0.015 & +0.051 $\pm$ 0.015 & $-$0.683 $\pm$ 0.026 & $-$2.409 $\pm$ 0.027 & $-$4.747 $\pm$ 0.016 & $-$3.218 $\pm$ 0.015 & +0.616 \\ 
HD019445 &   227 & $-$0.246 $\pm$ 0.010 & $-$0.819 $\pm$ 0.017 & $-$0.292 $\pm$ 0.007 & $-$0.131 $\pm$ 0.006 & $-$0.007 $\pm$ 0.001 & +0.150 $\pm$ 0.160 & +0.025 $\pm$ 0.006 & $-$0.339 $\pm$ 0.016 & $-$0.861 $\pm$ 0.014 & $-$0.353 $\pm$ 0.007 & +0.624 \\ 
HD020123 &   134 & $-$1.345 $\pm$ 0.020 & $-$3.418 $\pm$ 0.023 & $-$2.002 $\pm$ 0.017 & $-$0.457 $\pm$ 0.010 & $-$0.122 $\pm$ 0.008 & +0.071 $\pm$ 0.010 & $-$0.017 $\pm$ 0.002 & $-$1.751 $\pm$ 0.028 & $-$3.506 $\pm$ 0.018 & $-$1.765 $\pm$ 0.013 & +0.615 \\ 
HD224458 &   225 & $-$1.286 $\pm$ 0.015 & $-$2.760 $\pm$ 0.016 & $-$2.335 $\pm$ 0.015 & $-$0.613 $\pm$ 0.007 & $-$0.286 $\pm$ 0.010 & $-$0.407 $\pm$ 0.015 & $-$0.177 $\pm$ 0.022 & $-$1.215 $\pm$ 0.015 & $-$2.669 $\pm$ 0.012 & $-$2.126 $\pm$ 0.011 & +0.627 \\ 
HD220954 &    66 & $-$1.276 $\pm$ 0.044 & $-$4.290 $\pm$ 0.062 & $-$2.600 $\pm$ 0.049 & $-$0.734 $\pm$ 0.026 & $-$0.116 $\pm$ 0.014 & $-$1.539 $\pm$ 0.046 & $-$0.259 $\pm$ 0.049 & $-$1.123 $\pm$ 0.042 & $-$3.494 $\pm$ 0.040 & $-$2.559 $\pm$ 0.038 & +0.619 \\ 
HD025975 &   110 & $-$1.205 $\pm$ 0.029 & $-$3.980 $\pm$ 0.038 & $-$2.590 $\pm$ 0.033 & $-$0.664 $\pm$ 0.014 & $-$0.149 $\pm$ 0.013 & $-$0.448 $\pm$ 0.029 & +0.042 $\pm$ 0.019 & $-$1.168 $\pm$ 0.031 & $-$3.384 $\pm$ 0.027 & $-$2.284 $\pm$ 0.023 & +0.634 \\ 
HD027971 &    88 & $-$1.519 $\pm$ 0.041 & $-$4.050 $\pm$ 0.044 & $-$3.074 $\pm$ 0.042 & $-$0.657 $\pm$ 0.019 & $-$0.277 $\pm$ 0.025 & $-$0.457 $\pm$ 0.038 & $-$0.722 $\pm$ 0.040 & $-$1.533 $\pm$ 0.044 & $-$3.674 $\pm$ 0.031 & $-$2.768 $\pm$ 0.031 & +0.618 \\ 
HD174350 &    75 & $-$1.543 $\pm$ 0.042 & $-$2.959 $\pm$ 0.046 & $-$1.497 $\pm$ 0.027 & $-$0.549 $\pm$ 0.019 & +0.244 $\pm$ 0.029 & $-$0.298 $\pm$ 0.151 & +1.065 $\pm$ 0.335 & $-$2.533 $\pm$ 0.071 & $-$2.991 $\pm$ 0.036 & $-$1.895 $\pm$ 0.027 & +0.632 \\ 
HD185622 &   136 & $-$1.656 $\pm$ 0.022 & $-$2.497 $\pm$ 0.018 & $-$1.394 $\pm$ 0.013 & $-$0.204 $\pm$ 0.005 & +0.161 $\pm$ 0.011 & +0.400 $\pm$ 0.054 & +1.480 $\pm$ 0.027 & $-$2.857 $\pm$ 0.037 & $-$3.536 $\pm$ 0.019 & $-$1.631 $\pm$ 0.012 & +0.623 \\ 
HD100696 &   212 & $-$0.878 $\pm$ 0.012 & $-$2.598 $\pm$ 0.016 & $-$2.321 $\pm$ 0.016 & $-$0.477 $\pm$ 0.007 & $-$0.444 $\pm$ 0.009 & $-$0.625 $\pm$ 0.017 & $-$0.670 $\pm$ 0.016 & $-$0.477 $\pm$ 0.008 & $-$2.691 $\pm$ 0.013 & $-$2.351 $\pm$ 0.012 & +0.645 \\ 
HD101107 &   690 & $-$1.090 $\pm$ 0.005 & $-$2.332 $\pm$ 0.005 & $-$2.633 $\pm$ 0.005 & $-$0.395 $\pm$ 0.002 & $-$0.378 $\pm$ 0.003 & $-$1.048 $\pm$ 0.005 & $-$1.780 $\pm$ 0.005 & $-$0.760 $\pm$ 0.004 & $-$2.260 $\pm$ 0.004 & $-$2.439 $\pm$ 0.003 & +0.636 \\ 
HD104985 &   104 & $-$0.899 $\pm$ 0.024 & $-$2.692 $\pm$ 0.032 & $-$2.402 $\pm$ 0.030 & $-$0.618 $\pm$ 0.015 & $-$0.431 $\pm$ 0.035 & $-$0.602 $\pm$ 0.036 & $-$0.706 $\pm$ 0.035 & $-$0.467 $\pm$ 0.014 & $-$2.798 $\pm$ 0.027 & $-$2.399 $\pm$ 0.025 & +0.645 \\ 
HD113002 &   174 & $-$0.853 $\pm$ 0.018 & $-$2.147 $\pm$ 0.020 & $-$1.854 $\pm$ 0.018 & $-$0.280 $\pm$ 0.008 & $-$0.374 $\pm$ 0.027 & $-$0.489 $\pm$ 0.112 & $-$0.641 $\pm$ 0.020 & $-$0.537 $\pm$ 0.013 & $-$2.285 $\pm$ 0.017 & $-$1.916 $\pm$ 0.015 & +0.642 \\ 
HD115136 &    59 & $-$0.910 $\pm$ 0.038 & $-$3.004 $\pm$ 0.058 & $-$2.670 $\pm$ 0.055 & $-$0.725 $\pm$ 0.026 & $-$0.488 $\pm$ 0.026 & $-$0.760 $\pm$ 0.082 & $-$0.810 $\pm$ 0.067 & $-$0.364 $\pm$ 0.018 & $-$2.979 $\pm$ 0.045 & $-$2.621 $\pm$ 0.045 & +0.647 \\ 
HD117243 &   138 & $-$1.052 $\pm$ 0.023 & $-$2.732 $\pm$ 0.025 & $-$2.502 $\pm$ 0.026 & $-$0.587 $\pm$ 0.012 & $-$0.382 $\pm$ 0.014 & $-$0.610 $\pm$ 0.029 & $-$0.828 $\pm$ 0.023 & $-$0.770 $\pm$ 0.019 & $-$2.824 $\pm$ 0.020 & $-$2.467 $\pm$ 0.018 & +0.642 \\ 
HD131111 &   130 & $-$0.837 $\pm$ 0.020 & $-$2.573 $\pm$ 0.028 & $-$2.307 $\pm$ 0.028 & $-$0.528 $\pm$ 0.012 & $-$0.417 $\pm$ 0.031 & $-$0.634 $\pm$ 0.028 & $-$0.649 $\pm$ 0.029 & $-$0.390 $\pm$ 0.011 & $-$2.561 $\pm$ 0.022 & $-$2.291 $\pm$ 0.021 & +0.644 \\ 
HD131507 &    65 & $-$0.928 $\pm$ 0.033 & $-$3.133 $\pm$ 0.051 & $-$2.677 $\pm$ 0.049 & $-$0.705 $\pm$ 0.024 & $-$0.565 $\pm$ 0.025 & $-$0.726 $\pm$ 0.034 & $-$0.829 $\pm$ 0.185 & $-$0.249 $\pm$ 0.010 & $-$3.218 $\pm$ 0.041 & $-$2.702 $\pm$ 0.040 & +0.651 \\ 
HD144206 &   378 & $-$1.929 $\pm$ 0.007 & $-$2.877 $\pm$ 0.009 & $-$4.991 $\pm$ 0.006 & $-$0.597 $\pm$ 0.002 & $-$0.686 $\pm$ 0.005 & $-$3.042 $\pm$ 0.008 & $-$4.855 $\pm$ 0.008 & $-$1.643 $\pm$ 0.008 & $-$2.509 $\pm$ 0.006 & $-$4.241 $\pm$ 0.005 & +0.619 \\ 
HD147677 &   109 & $-$0.241 $\pm$ 0.010 & $-$2.376 $\pm$ 0.044 & $-$1.932 $\pm$ 0.032 & $-$0.297 $\pm$ 0.014 & $-$0.143 $\pm$ 0.014 & $-$0.294 $\pm$ 0.085 & $-$0.462 $\pm$ 0.034 & +0.926 $\pm$ 0.054 & $-$1.376 $\pm$ 0.022 & $-$1.900 $\pm$ 0.024 & +0.643 \\ 
HD174912 &   210 & $-$0.856 $\pm$ 0.016 & $-$2.159 $\pm$ 0.017 & $-$2.092 $\pm$ 0.016 & $-$0.253 $\pm$ 0.007 & $-$0.260 $\pm$ 0.021 & $-$0.583 $\pm$ 0.027 & $-$1.161 $\pm$ 0.016 & $-$0.551 $\pm$ 0.012 & $-$2.226 $\pm$ 0.014 & $-$2.013 $\pm$ 0.011 & +0.639 \\ 
HD175535 &   110 & $-$1.046 $\pm$ 0.024 & $-$2.964 $\pm$ 0.031 & $-$2.638 $\pm$ 0.031 & $-$0.520 $\pm$ 0.015 & $-$0.476 $\pm$ 0.015 & $-$0.719 $\pm$ 0.045 & $-$0.811 $\pm$ 0.030 & $-$0.625 $\pm$ 0.017 & $-$3.034 $\pm$ 0.025 & $-$2.617 $\pm$ 0.024 & +0.647 \\ 
HD200580 &   232 & $-$0.611 $\pm$ 0.012 & $-$1.960 $\pm$ 0.016 & $-$1.804 $\pm$ 0.014 & $-$0.328 $\pm$ 0.007 & $-$0.265 $\pm$ 0.019 & $-$0.384 $\pm$ 0.013 & $-$0.693 $\pm$ 0.014 & $-$0.141 $\pm$ 0.003 & $-$1.881 $\pm$ 0.013 & $-$1.786 $\pm$ 0.012 & +0.637 \\ 
HD206374 &   164 & $-$0.838 $\pm$ 0.016 & $-$2.503 $\pm$ 0.021 & $-$2.272 $\pm$ 0.020 & $-$0.589 $\pm$ 0.010 & $-$0.315 $\pm$ 0.022 & $-$0.182 $\pm$ 0.035 & $-$0.493 $\pm$ 0.019 & $-$0.760 $\pm$ 0.015 & $-$2.667 $\pm$ 0.016 & $-$2.214 $\pm$ 0.016 & +0.643 \\ 
HD220182 &   145 & $-$0.782 $\pm$ 0.017 & $-$2.704 $\pm$ 0.022 & $-$2.241 $\pm$ 0.022 & $-$0.617 $\pm$ 0.010 & $-$0.438 $\pm$ 0.025 & $-$0.716 $\pm$ 0.031 & $-$0.777 $\pm$ 0.026 & $-$0.340 $\pm$ 0.009 & $-$2.662 $\pm$ 0.018 & $-$2.212 $\pm$ 0.016 & +0.641 \\ 
HD211472 &   144 & $-$0.612 $\pm$ 0.014 & $-$2.508 $\pm$ 0.024 & $-$2.123 $\pm$ 0.020 & $-$0.706 $\pm$ 0.010 & $-$0.381 $\pm$ 0.021 & $-$0.324 $\pm$ 0.052 & $-$0.692 $\pm$ 0.027 & $-$0.278 $\pm$ 0.007 & $-$2.521 $\pm$ 0.017 & $-$2.146 $\pm$ 0.016 & +0.640 \\ 
HD218059 &   201 & $-$0.851 $\pm$ 0.018 & $-$2.143 $\pm$ 0.020 & $-$2.055 $\pm$ 0.018 & $-$0.342 $\pm$ 0.008 & $-$0.286 $\pm$ 0.009 & $-$0.565 $\pm$ 0.000 & $-$0.848 $\pm$ 0.017 & $-$0.560 $\pm$ 0.013 & $-$2.190 $\pm$ 0.015 & $-$2.057 $\pm$ 0.015 & +0.639 \\ 
HD221585 &   100 & $-$1.237 $\pm$ 0.036 & $-$3.152 $\pm$ 0.033 & $-$1.940 $\pm$ 0.033 & $-$0.388 $\pm$ 0.013 & $-$0.417 $\pm$ 0.018 & $-$0.307 $\pm$ 0.034 & +1.130 $\pm$ 0.052 & $-$0.749 $\pm$ 0.025 & $-$3.132 $\pm$ 0.026 & $-$2.023 $\pm$ 0.025 & +0.675 \\ 
HD221830 &   220 & $-$1.367 $\pm$ 0.021 & $-$2.415 $\pm$ 0.017 & $-$2.341 $\pm$ 0.018 & $-$0.377 $\pm$ 0.006 & $-$0.243 $\pm$ 0.018 & $-$0.179 $\pm$ 0.012 & $-$0.545 $\pm$ 0.013 & $-$1.542 $\pm$ 0.026 & $-$2.553 $\pm$ 0.013 & $-$2.035 $\pm$ 0.012 & +0.648 \\ 
BD+08~3095 &   124 & $-$0.697 $\pm$ 0.026 & $-$1.740 $\pm$ 0.028 & $-$1.679 $\pm$ 0.028 & $-$0.327 $\pm$ 0.013 & $-$0.243 $\pm$ 0.015 & $-$0.429 $\pm$ 0.025 & $-$0.582 $\pm$ 0.027 & $-$0.459 $\pm$ 0.021 & $-$1.865 $\pm$ 0.025 & $-$1.725 $\pm$ 0.023 & +0.637 \\ 
\hline
\end{tabular}
}
\end{turn}{90}
\end{table*}

\begin{table*}
\caption{{Ca}{\textsc ii}, {Mg}{\textsc i}, Pa lines equivalent widths, in \AA, as defined by \citet{cen01a} and sTiO colour defined by \citet{cen09} for the 56 M15 stars sub-sample observed in HR-I during MEGARA commissioning. Table description in subsection~\ref{HRIspec}.}
\label{linemeasurementsM15}
\begin{turn}{90}
\resizebox{22cm}{!} {
\begin{tabular}{lrccccccccccc}
\hline
Star Name & SNR & Ca1CEN & Ca2CEN & Ca3CEN & Mg~I & Pa1 & Pa2 & Pa3 & Ca1MEG & Ca2MEG & Ca3MEG & sTiO \\
\hline
M15\_01 &   7 & $-$0.152 $\pm$ 0.160 & $-$0.534 $\pm$ 0.356 & +0.007 $\pm$ 0.005 & $-$0.580 $\pm$ 0.419 & $-$0.058 $\pm$ 0.130 & +0.877 $\pm$ 0.000 & $-$1.495 $\pm$ 1.207 & $-$0.972 $\pm$ 0.447 & $-$2.231 $\pm$ 0.383 & +0.138 $\pm$ 0.076 & +0.606 \\
M15\_02 &  12 & $-$2.318 $\pm$ 0.230 & $-$2.397 $\pm$ 0.272 & $-$4.151 $\pm$ 0.246 & $-$0.539 $\pm$ 0.096 & $-$0.897 $\pm$ 0.140 & $-$2.240 $\pm$ 0.267 & $-$4.315 $\pm$ 0.280 & $-$2.624 $\pm$ 0.251 & $-$3.110 $\pm$ 0.210 & $-$3.509 $\pm$ 0.169 & +0.625 \\ 
M15\_03 &  15 & $-$0.657 $\pm$ 0.268 & $-$1.381 $\pm$ 0.349 & $-$0.532 $\pm$ 0.137 & $-$0.035 $\pm$ 0.023 & $-$0.226 $\pm$ 0.227 & +0.063 $\pm$ 0.050 & +0.108 $\pm$ 0.173 & $-$0.644 $\pm$ 0.282 & $-$1.550 $\pm$ 0.295 & $-$0.613 $\pm$ 0.130 & +0.646 \\ 
M15\_04 &  14 & $-$0.580 $\pm$ 0.249 & $-$0.499 $\pm$ 0.115 & $-$0.184 $\pm$ 0.049 & +0.112 $\pm$ 0.066 & $-$0.256 $\pm$ 0.176 & +0.284 $\pm$ 0.893 & +0.124 $\pm$ 0.254 & $-$1.258 $\pm$ 0.534 & $-$1.379 $\pm$ 0.239 & $-$0.433 $\pm$ 0.091 & +0.633 \\ 
M15\_05 &  11 & +0.028 $\pm$ 0.014 & +0.025 $\pm$ 0.007 & +0.543 $\pm$ 0.209 & $-$0.187 $\pm$ 0.099 & +0.086 $\pm$ 0.156 & +0.419 $\pm$ 0.538 & $-$0.281 $\pm$ 0.205 & $-$0.575 $\pm$ 0.273 & $-$1.299 $\pm$ 0.279 & +0.206 $\pm$ 0.071 & +0.619 \\
M15\_06 &  11 & $-$0.068 $\pm$ 0.049 & $-$1.713 $\pm$ 0.599 & $-$3.565 $\pm$ 0.251 & $-$0.775 $\pm$ 0.100 & +0.278 $\pm$ 0.114 & $-$3.000 $\pm$ 0.287 & $-$1.444 $\pm$ 0.267 & +0.282 $\pm$ 0.183 & $-$0.568 $\pm$ 0.173 & $-$3.422 $\pm$ 0.178 & +0.605 \\
M15\_07 &  12 & +0.087 $\pm$ 0.045 & $-$0.926 $\pm$ 0.214 & $-$0.714 $\pm$ 0.156 & +0.048 $\pm$ 0.042 & +0.109 $\pm$ 0.196 & +2.326 $\pm$ 0.208 & +0.549 $\pm$ 0.096 & +0.277 $\pm$ 0.149 & $-$1.254 $\pm$ 0.229 & $-$0.405 $\pm$ 0.078 & +0.643 \\ 
M15\_08 &   9 & $-$0.320 $\pm$ 0.270 & $-$1.995 $\pm$ 0.427 & $-$1.969 $\pm$ 0.661 & $-$0.290 $\pm$ 0.161 & $-$0.483 $\pm$ 0.191 & $-$1.245 $\pm$ 0.270 & $-$0.330 $\pm$ 0.169 & +0.813 $\pm$ 0.001 & $-$0.620 $\pm$ 0.316 & $-$1.533 $\pm$ 0.443 & +0.643 \\
M15\_09 &  16 & $-$0.551 $\pm$ 0.211 & $-$1.225 $\pm$ 0.228 & $-$0.500 $\pm$ 0.099 & $-$0.116 $\pm$ 0.057 & $-$0.194 $\pm$ 0.098 & $-$0.160 $\pm$ 0.094 & $-$0.405 $\pm$ 0.191 & $-$0.335 $\pm$ 0.143 & $-$1.492 $\pm$ 0.220 & $-$0.606 $\pm$ 0.095 & +0.632 \\ 
M15\_10 &   7 & +0.213 $\pm$ 0.227 & +0.323 $\pm$ 0.129 & +1.198 $\pm$ 0.506 & $-$0.199 $\pm$ 0.116 & +0.064 $\pm$ 0.041 & +2.241 $\pm$ 0.522 & +0.063 $\pm$ 0.087 & $-$0.912 $\pm$ 0.931 & $-$1.186 $\pm$ 0.346 & +0.953 $\pm$ 0.353 & +0.622 \\ 
M15\_11 &  35 & $-$0.531 $\pm$ 0.074 & $-$1.495 $\pm$ 0.096 & $-$0.923 $\pm$ 0.065 & +0.042 $\pm$ 0.010 & $-$0.231 $\pm$ 0.042 & +0.367 $\pm$ 1.908 & +0.133 $\pm$ 0.065 & $-$0.567 $\pm$ 0.081 & $-$1.616 $\pm$ 0.083 & $-$0.891 $\pm$ 0.053 & +0.646 \\ 
M15\_12 &   8 & $-$0.051 $\pm$ 0.045 & $-$0.116 $\pm$ 0.052 & +1.086 $\pm$ 0.138 & +0.005 $\pm$ 0.004 & $-$0.430 $\pm$ 0.315 & +1.883 $\pm$ 3.746 & +1.766 $\pm$ 0.321 & $-$0.902 $\pm$ 0.798 & $-$1.351 $\pm$ 0.413 & +1.347 $\pm$ 1.631 & +0.645 \\ 
M15\_13 &  17 & $-$0.445 $\pm$ 0.176 & $-$0.843 $\pm$ 0.170 & $-$0.700 $\pm$ 0.143 & $-$0.169 $\pm$ 0.095 & $-$0.182 $\pm$ 0.088 & +0.283 $\pm$ 0.407 & +0.088 $\pm$ 0.246 & $-$0.568 $\pm$ 0.239 & $-$0.975 $\pm$ 0.154 & $-$0.400 $\pm$ 0.069 & +0.640 \\ 
M15\_14 &  17 & $-$0.637 $\pm$ 0.234 & $-$1.156 $\pm$ 0.208 & $-$0.896 $\pm$ 0.172 & $-$0.463 $\pm$ 0.122 & $-$0.109 $\pm$ 0.124 & +0.288 $\pm$ 0.166 & $-$0.586 $\pm$ 0.163 & $-$0.318 $\pm$ 0.139 & $-$1.244 $\pm$ 0.177 & $-$0.379 $\pm$ 0.062 & +0.632 \\ 
M15\_15 &  30 & $-$0.445 $\pm$ 0.082 & $-$1.371 $\pm$ 0.121 & $-$0.927 $\pm$ 0.088 & $-$0.006 $\pm$ 0.002 & $-$0.235 $\pm$ 0.050 & $-$0.252 $\pm$ 0.095 & $-$0.347 $\pm$ 0.123 & $-$0.370 $\pm$ 0.075 & $-$1.560 $\pm$ 0.105 & $-$1.182 $\pm$ 0.090 & +0.636 \\ 
M15\_16 &  22 & $-$0.305 $\pm$ 0.074 & $-$1.054 $\pm$ 0.131 & $-$0.758 $\pm$ 0.093 & +0.099 $\pm$ 0.042 & $-$0.247 $\pm$ 0.077 & +0.297 $\pm$ 0.117 & $-$0.141 $\pm$ 1.023 & $-$0.494 $\pm$ 0.133 & $-$1.495 $\pm$ 0.140 & $-$1.163 $\pm$ 0.121 & +0.640 \\ 
M15\_17 &  22 & $-$0.308 $\pm$ 0.075 & $-$1.048 $\pm$ 0.132 & $-$0.763 $\pm$ 0.093 & +0.096 $\pm$ 0.039 & $-$0.248 $\pm$ 0.075 & +0.306 $\pm$ 0.124 & $-$0.125 $\pm$ 66.06 & $-$0.499$\pm$ +0.13 & $-$1.496 $\pm$ 0.135 & $-$1.157 $\pm$ 0.111 & +0.640 \\ 
M15\_18 &  20 & $-$0.126 $\pm$ 0.036 & $-$0.637 $\pm$ 0.089 & $-$0.652 $\pm$ 0.101 & $-$0.143 $\pm$ 0.065 & +0.186 $\pm$ 0.000 & +1.359 $\pm$ 0.625 & +0.094 $\pm$ 0.189 & $-$0.056 $\pm$ 0.017 & $-$1.099 $\pm$ 0.125 & $-$0.613 $\pm$ 0.078 & +0.633 \\ 
M15\_19 &  20 & $-$0.231 $\pm$ 0.070 & $-$0.949 $\pm$ 0.153 & $-$0.590 $\pm$ 0.111 & $-$0.141 $\pm$ 0.058 & $-$0.064 $\pm$ 0.144 & +0.234 $\pm$ 0.000 & $-$0.149 $\pm$ 0.086 & $-$0.246 $\pm$ 0.082 & $-$1.264 $\pm$ 0.163 & $-$0.578 $\pm$ 0.089 & +0.629 \\ 
M15\_20 &  15 & +0.433 $\pm$ 0.190 & $-$0.763 $\pm$ 0.140 & $-$0.710 $\pm$ 0.138 & $-$0.205 $\pm$ 0.132 & +0.073 $\pm$ 0.130 & +1.682 $\pm$ 0.177 & +0.345 $\pm$ 0.559 & +0.666 $\pm$ 0.330 & $-$1.006 $\pm$ 0.161 & $-$0.778 $\pm$ 0.129 & +0.632 \\ 
M15\_21 &  33 & $-$1.069 $\pm$ 0.122 & $-$2.065 $\pm$ 0.117 & $-$0.991 $\pm$ 0.062 & +0.030 $\pm$ 0.005 & $-$0.588 $\pm$ 0.053 & $-$0.782 $\pm$ 0.179 & +0.358 $\pm$ 0.077 & $-$0.380 $\pm$ 0.049 & $-$2.013 $\pm$ 0.087 & $-$1.560 $\pm$ 0.075 & +0.669 \\ 
M15\_22 &  42 & $-$0.471 $\pm$ 0.049 & $-$1.658 $\pm$ 0.084 & $-$1.104 $\pm$ 0.063 & $-$0.131 $\pm$ 0.028 & $-$0.234 $\pm$ 0.040 & $-$0.252 $\pm$ 0.053 & $-$0.244 $\pm$ 0.074 & $-$0.310 $\pm$ 0.038 & $-$1.721 $\pm$ 0.069 & $-$1.221 $\pm$ 0.054 & +0.638 \\ 
M15\_23 &  51 & $-$0.459 $\pm$ 0.038 & $-$1.802 $\pm$ 0.073 & $-$1.305 $\pm$ 0.059 & $-$0.172 $\pm$ 0.019 & $-$0.101 $\pm$ 0.024 & $-$0.003 $\pm$ 0.002 & $-$0.578 $\pm$ 0.127 & $-$0.316 $\pm$ 0.028 & $-$1.884 $\pm$ 0.060 & $-$1.264 $\pm$ 0.045 & +0.633 \\ 
M15\_24 &  27 & $-$0.318 $\pm$ 0.060 & $-$1.617 $\pm$ 0.149 & $-$0.809 $\pm$ 0.074 & $-$0.047 $\pm$ 0.016 & $-$0.125 $\pm$ 0.003 & $-$0.116 $\pm$ 1.147 & $-$0.537 $\pm$ 0.118 & $-$0.040 $\pm$ 0.009 & $-$1.493 $\pm$ 0.106 & $-$0.948 $\pm$ 0.072 & +0.637 \\ 
M15\_25 &  41 & $-$0.458 $\pm$ 0.056 & $-$1.257 $\pm$ 0.073 & $-$0.978 $\pm$ 0.064 & $-$0.099 $\pm$ 0.018 & $-$0.189 $\pm$ 0.041 & +0.185 $\pm$ 0.256 & $-$0.190 $\pm$ 0.082 & $-$0.159 $\pm$ 0.022 & $-$1.339 $\pm$ 0.067 & $-$1.071 $\pm$ 0.061 & +0.639 \\ 
M15\_26 &  39 & $-$0.747 $\pm$ 0.096 & $-$1.656 $\pm$ 0.116 & $-$1.381 $\pm$ 0.098 & $-$0.144 $\pm$ 0.026 & $-$0.256 $\pm$ 0.044 & $-$0.798 $\pm$ 11.78 & $-$0.294$\pm$ +0.08 & $-$0.018$\pm$ +0.00 & $-$1.276 $\pm$ 0.068 & $-$1.525 $\pm$ 0.083 & +0.645 \\ 
M15\_27 &  57 & $-$0.524 $\pm$ 0.036 & $-$2.153 $\pm$ 0.075 & $-$1.452 $\pm$ 0.050 & $-$0.178 $\pm$ 0.020 & $-$0.122 $\pm$ 0.033 & $-$0.038 $\pm$ 0.026 & $-$0.619 $\pm$ 0.062 & $-$0.426 $\pm$ 0.034 & $-$2.146 $\pm$ 0.062 & $-$1.408 $\pm$ 0.041 & +0.633 \\ 
M15\_28 &  63 & $-$0.639 $\pm$ 0.040 & $-$1.968 $\pm$ 0.062 & $-$1.190 $\pm$ 0.041 & $-$0.225 $\pm$ 0.020 & $-$0.266 $\pm$ 0.062 & +0.046 $\pm$ 0.033 & $-$0.333 $\pm$ 0.059 & $-$0.530 $\pm$ 0.037 & $-$2.025 $\pm$ 0.050 & $-$1.235 $\pm$ 0.033 & +0.644 \\ 
M15\_29 &  30 & $-$0.410 $\pm$ 0.077 & $-$1.346 $\pm$ 0.126 & $-$0.614 $\pm$ 0.074 & $-$0.071 $\pm$ 0.022 & $-$0.130 $\pm$ 0.131 & +0.360 $\pm$ 8.544 & +0.335 $\pm$ 0.583 & $-$0.351 $\pm$ 0.073 & $-$1.382 $\pm$ 0.102 & $-$0.508 $\pm$ 0.053 & +0.641 \\ 
M15\_30 &  34 & $-$0.186 $\pm$ 0.029 & $-$1.229 $\pm$ 0.089 & $-$0.803 $\pm$ 0.063 & $-$0.151 $\pm$ 0.029 & $-$0.115 $\pm$ 0.063 & +0.251 $\pm$ 88.55 & $-$0.314$\pm$ +0.11 & $-$0.381$\pm$ +0.06 & $-$1.675 $\pm$ 0.090 & $-$0.922 $\pm$ 0.059 & +0.631 \\ 
M15\_31 &  21 & $-$0.029 $\pm$ 0.008 & $-$0.748 $\pm$ 0.122 & $-$0.275 $\pm$ 0.045 & $-$0.133 $\pm$ 0.052 & +0.164 $\pm$ 0.082 & +0.503 $\pm$ 0.000 & $-$0.450 $\pm$ 0.143 & $-$0.020 $\pm$ 0.006 & $-$1.129 $\pm$ 0.144 & $-$0.283 $\pm$ 0.039 & +0.622 \\ 
M15\_32 &  51 & $-$0.667 $\pm$ 0.057 & $-$1.874 $\pm$ 0.079 & $-$1.189 $\pm$ 0.050 & $-$0.278 $\pm$ 0.032 & $-$0.272 $\pm$ 0.031 & $-$0.057 $\pm$ 0.056 & $-$0.242 $\pm$ 0.051 & $-$0.625 $\pm$ 0.060 & $-$1.865 $\pm$ 0.059 & $-$1.354 $\pm$ 0.050 & +0.642 \\ 
M15\_33 &  28 & +1.305 $\pm$ 0.111 & +0.810 $\pm$ 0.070 & +2.430 $\pm$ 0.108 & +0.253 $\pm$ 0.046 & +1.175 $\pm$ 0.072 & +2.384 $\pm$ 0.141 & +2.101 $\pm$ 1.147 & +0.994 $\pm$ 0.094 & +0.688 $\pm$ 0.049 & +2.499 $\pm$ 0.096 & +0.619 \\ 
M15\_34 &  31 & $-$0.665 $\pm$ 0.101 & $-$1.712 $\pm$ 0.129 & $-$0.800 $\pm$ 0.065 & $-$0.064 $\pm$ 0.014 & $-$0.040 $\pm$ 0.029 & +0.065 $\pm$ 0.286 & +0.009 $\pm$ 0.000 & $-$0.648 $\pm$ 0.104 & $-$1.749 $\pm$ 0.107 & $-$1.083 $\pm$ 0.075 & +0.640 \\ 
M15\_35 &  56 & $-$0.770 $\pm$ 0.054 & $-$2.119 $\pm$ 0.073 & $-$1.159 $\pm$ 0.044 & $-$0.163 $\pm$ 0.023 & $-$0.411 $\pm$ 0.030 & +0.009 $\pm$ 0.007 & +0.022 $\pm$ 0.019 & $-$0.558 $\pm$ 0.045 & $-$2.113 $\pm$ 0.055 & $-$1.301 $\pm$ 0.041 & +0.652 \\ 
M15\_36 &  62 & $-$0.368 $\pm$ 0.026 & $-$1.481 $\pm$ 0.049 & $-$1.047 $\pm$ 0.039 & $-$0.216 $\pm$ 0.020 & $-$0.216 $\pm$ 0.042 & +0.079 $\pm$ 0.185 & $-$0.279 $\pm$ 0.119 & $-$0.270 $\pm$ 0.020 & $-$1.794 $\pm$ 0.049 & $-$1.186 $\pm$ 0.035 & +0.637 \\ 
M15\_37 & 125 & $-$0.524 $\pm$ 0.018 & $-$1.725 $\pm$ 0.029 & $-$1.227 $\pm$ 0.022 & $-$0.196 $\pm$ 0.010 & $-$0.318 $\pm$ 0.014 & $-$0.205 $\pm$ 0.054 & $-$0.405 $\pm$ 0.027 & $-$0.232 $\pm$ 0.009 & $-$1.758 $\pm$ 0.023 & $-$1.345 $\pm$ 0.020 & +0.641 \\ 
M15\_38 &  27 & $-$0.371 $\pm$ 0.091 & $-$1.291 $\pm$ 0.136 & $-$0.941 $\pm$ 0.109 & $-$0.163 $\pm$ 0.048 & $-$0.175 $\pm$ 0.049 & +0.149 $\pm$ 0.516 & $-$0.239 $\pm$ 0.117 & $-$0.384 $\pm$ 0.102 & $-$1.460 $\pm$ 0.123 & $-$1.083 $\pm$ 0.101 & +0.635 \\ 
M15\_39 &  74 & $-$0.565 $\pm$ 0.038 & $-$1.765 $\pm$ 0.056 & $-$1.175 $\pm$ 0.040 & $-$0.192 $\pm$ 0.017 & $-$0.176 $\pm$ 0.022 & $-$0.050 $\pm$ 0.092 & $-$0.164 $\pm$ 0.040 & $-$0.481 $\pm$ 0.036 & $-$1.717 $\pm$ 0.042 & $-$1.173 $\pm$ 0.034 & +0.640 \\ 
M15\_40 &  44 & $-$0.717 $\pm$ 0.068 & $-$1.874 $\pm$ 0.087 & $-$1.134 $\pm$ 0.055 & $-$0.216 $\pm$ 0.029 & $-$0.303 $\pm$ 0.085 & $-$0.109 $\pm$ 0.069 & $-$0.093 $\pm$ 0.023 & $-$0.644 $\pm$ 0.069 & $-$1.907 $\pm$ 0.071 & $-$1.286 $\pm$ 0.051 & +0.646 \\ 
M15\_41 &  19 & $-$0.171 $\pm$ 0.063 & $-$0.554 $\pm$ 0.113 & +0.060 $\pm$ 0.014 & +0.055 $\pm$ 0.085 & $-$0.079 $\pm$ 0.044 & +1.024 $\pm$ 0.034 & $-$0.136 $\pm$ 0.111 & $-$0.485 $\pm$ 0.185 & $-$1.178 $\pm$ 0.186 & +0.002 $\pm$ 0.000 & +0.632 \\ 
M15\_42 &  45 & $-$0.219 $\pm$ 0.024 & $-$1.633 $\pm$ 0.087 & $-$1.037 $\pm$ 0.058 & $-$0.079 $\pm$ 0.013 & $-$0.229 $\pm$ 0.072 & $-$0.271 $\pm$ 0.109 & $-$0.389 $\pm$ 0.076 & +0.166 $\pm$ 0.021 & $-$1.496 $\pm$ 0.061 & $-$1.150 $\pm$ 0.050 & +0.635 \\ 
M15\_43 &  13 & $-$0.227 $\pm$ 0.117 & $-$0.586 $\pm$ 0.158 & $-$0.062 $\pm$ 0.020 & $-$0.209 $\pm$ 0.083 & +0.067 $\pm$ 0.054 & +0.036 $\pm$ 0.052 & $-$0.406 $\pm$ 0.224 & $-$0.710 $\pm$ 0.402 & $-$1.556 $\pm$ 0.309 & $-$0.266 $\pm$ 0.070 & +0.622 \\ 
M15\_44 &  41 & $-$0.656 $\pm$ 0.056 & $-$2.144 $\pm$ 0.076 & $-$1.823 $\pm$ 0.073 & $-$0.372 $\pm$ 0.028 & $-$0.194 $\pm$ 0.071 & $-$0.154 $\pm$ 0.054 & $-$0.452 $\pm$ 0.084 & $-$0.375 $\pm$ 0.036 & $-$2.341 $\pm$ 0.065 & $-$2.111 $\pm$ 0.065 & +0.641 \\ 
M15\_45 &  18 & $-$0.361 $\pm$ 0.127 & $-$0.959 $\pm$ 0.184 & $-$0.511 $\pm$ 0.093 & $-$0.118 $\pm$ 0.066 & $-$0.108 $\pm$ 0.086 & +0.129 $\pm$ 0.099 & $-$0.521 $\pm$ 0.201 & $-$0.304 $\pm$ 0.117 & $-$1.182 $\pm$ 0.173 & $-$0.666 $\pm$ 0.101 & +0.630 \\ 
M15\_46 &  33 & $-$0.407 $\pm$ 0.068 & $-$1.232 $\pm$ 0.105 & $-$0.842 $\pm$ 0.074 & $-$0.081 $\pm$ 0.030 & $-$0.238 $\pm$ 0.060 & $-$0.325 $\pm$ 0.093 & $-$0.425 $\pm$ 0.136 & $-$0.250 $\pm$ 0.046 & $-$1.383 $\pm$ 0.099 & $-$0.937 $\pm$ 0.071 & +0.633 \\ 
M15\_47 &  10 & +0.249 $\pm$ 0.216 & +0.840 $\pm$ 0.341 & $-$0.045 $\pm$ 0.018 & $-$0.141 $\pm$ 0.128 & +0.217 $\pm$ 0.317 & +1.046 $\pm$ 25.21 & $-$0.815$\pm$ +0.38 & $-$0.839$\pm$ +0.72 & $-$1.293 $\pm$ 0.370 & $-$0.228 $\pm$ 0.073 & +0.599 \\ 
M15\_48 &  18 & $-$0.620 $\pm$ 0.214 & $-$1.134 $\pm$ 0.182 & $-$0.662 $\pm$ 0.128 & $-$0.156 $\pm$ 0.041 & $-$0.242 $\pm$ 0.205 & $-$0.211 $\pm$ 0.212 & $-$0.333 $\pm$ 0.160 & $-$0.567 $\pm$ 0.205 & $-$1.286 $\pm$ 0.169 & $-$0.533 $\pm$ 0.080 & +0.635 \\ 
M15\_49 &  15 & $-$0.834 $\pm$ 0.339 & $-$1.041 $\pm$ 0.205 & $-$0.243 $\pm$ 0.049 & $-$0.098 $\pm$ 0.111 & $-$0.198 $\pm$ 0.121 & +0.202 $\pm$ 0.153 & $-$0.197 $\pm$ 0.152 & $-$1.010 $\pm$ 0.414 & $-$1.577 $\pm$ 0.243 & $-$0.322 $\pm$ 0.053 & +0.642 \\ 
M15\_50 &  43 & $-$0.498 $\pm$ 0.056 & $-$1.570 $\pm$ 0.084 & $-$1.102 $\pm$ 0.067 & $-$0.139 $\pm$ 0.023 & $-$0.266 $\pm$ 0.039 & $-$0.239 $\pm$ 0.075 & $-$0.296 $\pm$ 0.172 & $-$0.383 $\pm$ 0.047 & $-$1.648 $\pm$ 0.069 & $-$1.265 $\pm$ 0.059 & +0.639 \\ 
M15\_51 &   5 & $-$1.905 $\pm$ 0.496 & $-$1.433 $\pm$ 0.441 & $-$3.244 $\pm$ 0.483 & $-$0.294 $\pm$ 0.188 & $-$1.279 $\pm$ 0.286 & $-$0.881 $\pm$ 0.295 & $-$5.944 $\pm$ 0.567 & $-$3.978 $\pm$ 0.541 & $-$4.143 $\pm$ 0.465 & $-$2.416 $\pm$ 0.344 & +0.599 \\ 
M15\_52 &  20 & $-$0.001 $\pm$ 0.000 & $-$0.961 $\pm$ 0.149 & +0.167 $\pm$ 0.027 & +0.078 $\pm$ 0.064 & +0.011 $\pm$ 0.012 & +1.915 $\pm$ 0.201 & +0.184 $\pm$ 0.422 & $-$0.285 $\pm$ 0.096 & $-$1.476 $\pm$ 0.174 & +0.371 $\pm$ 0.054 & +0.635 \\ 
M15\_53 &  31 & $-$0.299 $\pm$ 0.056 & $-$1.027 $\pm$ 0.105 & $-$0.200 $\pm$ 0.024 & $-$0.056 $\pm$ 0.013 & $-$0.043 $\pm$ 0.028 & +0.249 $\pm$ 0.161 & $-$0.220 $\pm$ 1.329 & $-$0.356 $\pm$ 0.072 & $-$1.349 $\pm$ 0.110 & $-$0.450 $\pm$ 0.044 & +0.632 \\ 
M15\_54 &  24 & +0.213 $\pm$ 0.059 & $-$0.498 $\pm$ 0.070 & $-$0.093 $\pm$ 0.014 & $-$0.019 $\pm$ 0.009 & 0.079 $\pm$ 0.063 & 1.801 $\pm$ 2.050 & +0.096 $\pm$ 0.181 & $-$0.186 $\pm$ 0.056 & $-$1.263 $\pm$ 0.134 & +0.022 $\pm$ 0.003 & +0.631 \\ 
M15\_55 &  15 & +0.475 $\pm$ 0.209 & $-$0.496 $\pm$ 0.103 & +0.898 $\pm$ 0.227 & $-$0.037 $\pm$ 0.037 & +0.246 $\pm$ 0.000 & +1.808 $\pm$ 0.272 & +0.129 $\pm$ 0.261 & $-$0.307 $\pm$ 0.133 & $-$1.676 $\pm$ 0.249 & +0.748 $\pm$ 0.169 & +0.619 \\ 
M15\_56 &  45 & $-$0.356 $\pm$ 0.048 & $-$1.114 $\pm$ 0.076 & $-$0.544 $\pm$ 0.040 & $-$0.100 $\pm$ 0.021 & $-$0.139 $\pm$ 0.034 & +0.022 $\pm$ 0.007 & $-$0.142 $\pm$ 0.123 & $-$0.544 $\pm$ 0.080 & $-$1.429 $\pm$ 0.079 & $-$0.663 $\pm$ 0.040 & 0.630 \\
\hline
\end{tabular}
}
\end{turn}{90}
\end{table*}

\section{MEGARA-GTC Library Database}\label{database}

\begin{figure*}
\includegraphics[width=0.9\textwidth,angle=0]{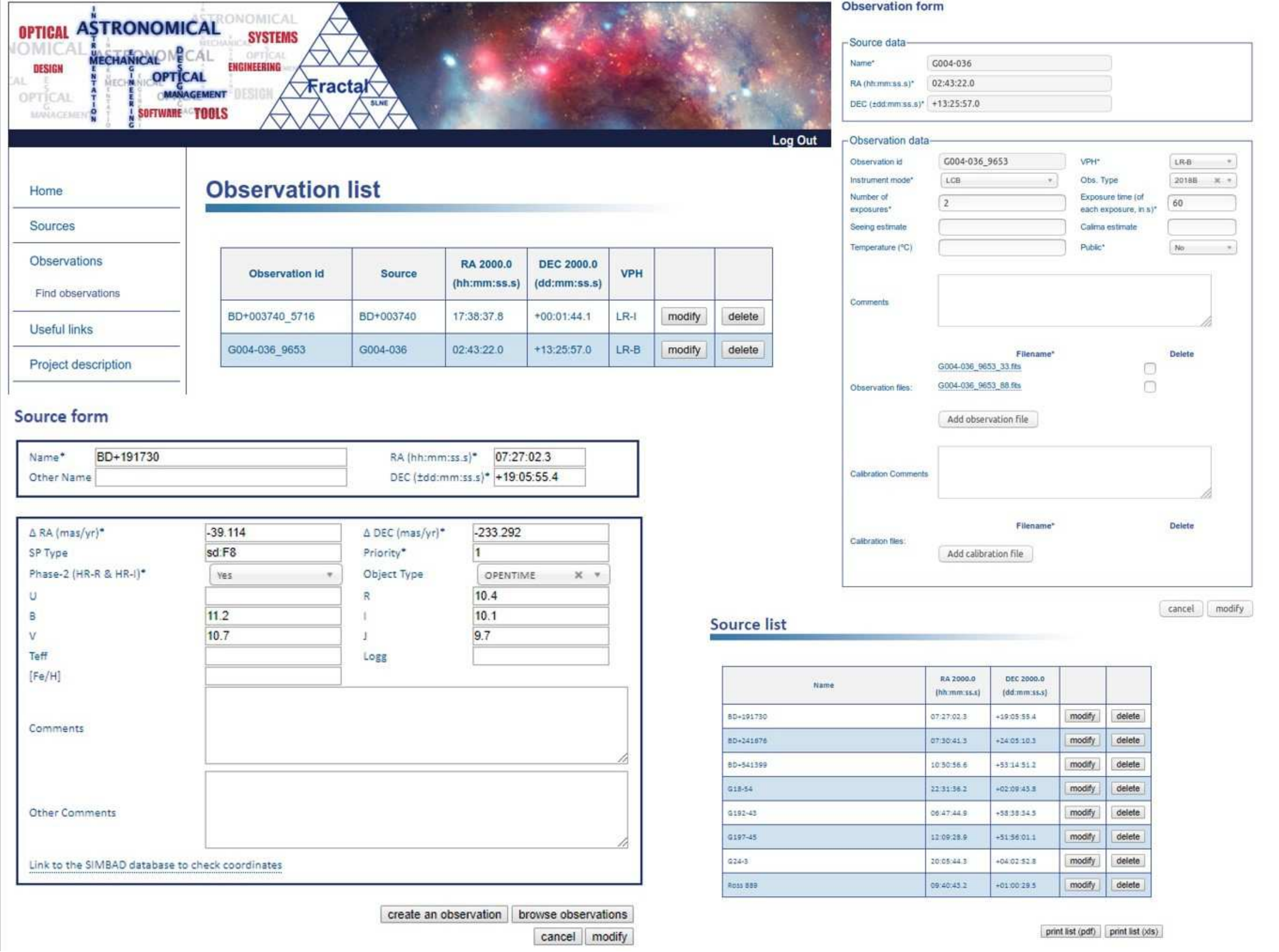}
\caption{MEGARA GTC-Library Database. The figure shows some views of the web-based tool. Top left: Home page and main menus; Top right: Observation form where the proper observed data can be handled; Bottom left: Source form to handle library stars; Bottom right: Source list after a searching process with the options of examining the content of each record in more detail and printing filtered list in different formats for further analysis.}
\label{image}
\end{figure*}

The goals of the MEGARA-GTC stellar library are to compile input spectra for {\sc PopStar} models and to produce a useful, public and accessible database of fully reduced and calibrated star spectra for other MEGARA users. Handling several thousands of stellar data and their GTC observations in different setups, which have to be prepared for execution, and be reduced and analysed, requires a specific software tool. We have developed a database in MySQL with all library data and a web-based tool that allows the management of stellar parameters and observed spectra. There are several permission levels allowing different actions on the database, from manipulating and updating both the library and the observations, to just making queries and retrieving reduced spectra. 

The database also supports the MEGARA-GTC library team for preparing and uploading the OBs to the GTC-Phase 2 tool. To prepare a new OB set, we search unobserved stars in the MEGARA-GTC library filtered by a certain magnitude range in both R and I bands and/or by spectral type, considering that all stars within that group and for a given setup will have a similar SNR when choosing the appropriate exposure time in each setup. The tool has the capability of exporting a file suited to the GTC-Phase 2 format. Due to the GTC Phase-2 flexibility for changing, adding or removing OBs, we use the database functionality to make decisions on the queued observations, tuning the priorities to guarantee a good coverage of the stellar parameters.

Figure~\ref{image} shows some views of the GUI. The {\it Source} form functionality allows the authorised users to upload, change and delete library stars and/or their properties; while the {\it Observation} menu includes the functions of listing, searching, modifying, deleting, uploading and retrieving fully reduced library spectra obtained with MEGARA at the GTC. The tool also includes a {\it Statistics} menu with information on the observations distribution as a function of the physical stellar parameters and MEGARA setups, and a {\it Download} Menu for getting the released observations. A complete description of the tool will come in Paper~II. The MEGARA-GTC stellar library database and the web-based tools reside on \url{https://www.fractal-es.com/megaragtc-stellarlibrary/}.

\section{Conclusions}\label{conclusions}

MEGARA, a GTC 10m-telescope facility instrument, is a fibre-fed spectrograph with medium-high spectral resolution (R = 6000, 12000, 20000), covering the range 3650~--~9750~\AA. The instrument was successfully commissioned at the Observatorio del Roque de los Muchachos (La Palma, Spain) in the summer of 2017 and is in operation since July 2018. In this paper, we have introduced the MEGARA-GTC spectral library, an empirical star catalogue whose spectra shall be used as the seed of the new generation of {\sc PopStar} models for interpretation of the contribution of the stellar populations in a wide range of MEGARA observations with the same instrument configuration. The maximum priority has been given to the HR-R and HR-I setups, centred, in rest-frame, at H~${\alpha}$ and the brightest line of {Ca}{\textsc ii} triplet, respectively.

We have presented a first sample of 97 stars, 21 of them observed in the HR-I setup during the MEGARA commissioning as a pilot program to demonstrate the feasibility of the MEGARA-GTC Library project, 56 stars at both HR-R and HR-I of the centre of M15 cluster and 20 stars, also in HR-R and HR-I, obtained from our ongoing GTC filler Open Time program. 

For all these 97 stars we have derived the stellar parameters: effective temperature, $\rm T_{eff}$, surface gravity, $\rm \log{g}$ and metallicity $\rm [M/H]$. To obtained them, we have developed a code that uses a $\chi^{2}$ technique to fit the theoretical modelled spectra by \citet{mun05} to the high resolution MEGARA observations, both normalised to their respective continuum. The best results are found when using the complete information in both HR-R and HR-I spectral windows, combining them in a single spectrum to which the model is compared. This allows us to use all the spectral information. We have used this combined spectrum to derive the stellar parameters. We have compared our results with the parameters published in the literature, when available, and with the ones obtained when using the ´{\sc SP\_ACE} code (when offering a solution), finding in general a very good agreement in the values derived for $\rm T_{eff}$, $\rm \log{g}$ and $\rm [M/H]$ from all these different methods and sources, within the error bars and the intrinsic model limitations, whose grid's density puts a limit on the errors of the derived stellar parameters. The goodness of our method is probed in this sample of observations, and will be applied it to the MEGARA-GTC Library, sampling a wider range of stellar parameters.

We have measured known stellar indices (based on measurements of equivalent widths) and we have proposed updated indices based on new spectral windows more suitable to the MEGARA spectral resolution. We have also tested the reliability of measuring these indices with non-flux calibrated observations. The complete indices study and the relationship with stellar parameters will come in future papers.

We are running a filler-type program with MEGARA at the GTC that has been granted so far with 175 hr of observing time in the semesters 2018B, 2019A and 2019B, which should provide around 300 stars observed in both HR-R and HR-I setups. All spectra are being reduced with the MEGARA DRP and pass through proper quality control, analysis and classification processes. The MEGARA IFU mode has an enormous advantage for these filler observations with respect to a long-slit instrument, especially in bad observing conditions: MEGARA data have always identical spectral resolution since the pupil stop is at the fibre at the spectrograph pseudo-slit, keeping it independent from sky conditions, and flux is always recovered by adding spaxels according to the actual seeing. 

To manage the MEGARA-GTC library, We have put together a MySQL database residing on  \url{https://www.fractal-es.com/megaragtc-stellarlibrary/}, to handle both data and observations. We will upload all quality checked spectra to deliver a first release of the MEGARA spectral library in 2020, which will allow users to retrieve the spectra along with the evolutionary synthesis models to support the observations interpretation.

We will use these observations to produce the first synthetic spectra with {\sc PopStar} for the MEGARA highest resolution modes, which will be refined as far as the library observations progress.

\section{Acknowledgements}
This work has been supported by MINECO-FEDER grants AYA2016-75808-R and AYA2016-79724-C4-3-P, and has been partially funded by FRACTAL, INAOE and CIEMAT. This work is based on observations made with the Gran Telescopio Canarias (GTC), installed in the Spanish Observatorio del Roque de los Muchachos, in the island of La Palma. This work is based on data obtained with MEGARA instrument, funded by European Regional Development Funds (ERDF), through {\it Programa Operativo Canarias FEDER 2014-2020}. The authors thank the support given by Dr. Antonio Cabrera and Dr. Daniel Reverte, GTC Operations Group staff, during the preparation and execution of the observations at the GTC. The authors acknowledge the anonymous referee for her/his comments which have substantially improved the manuscript.

\end{document}